\documentclass[a4paper,11pt,fleqn]{book}

\usepackage[T1]{fontenc}
\usepackage[utf8]{inputenc}
\usepackage[french,german,english]{babel}
\usepackage{textcomp}

\usepackage{setspace}

\usepackage{graphicx,color}

\usepackage{booktabs}
\usepackage{lipsum}
\usepackage{microtype}
\usepackage{url}
\usepackage[final]{pdfpages}

\usepackage{units}
\usepackage{float}
\usepackage{caption}
\usepackage{subcaption}

\usepackage{amsmath,amssymb,exscale}
\usepackage{mathrsfs}
\usepackage{hyperref}
\hypersetup{pdfborder={0 0 0},
	colorlinks=true,
	linkcolor=black,
	citecolor=cyan,
	urlcolor=magenta}
\usepackage{blkarray}
\usepackage{cleveref}
\usepackage{enumerate}
\usepackage{cite}
\usepackage{tikz}
\usetikzlibrary{arrows,matrix,positioning}
\usepackage{enumitem}

\newcommand{\scp}[1]{\scriptstyle#1}
\DeclareMathOperator{\rank}{rank}
\input epsf
\allowdisplaybreaks
\newcommand\numberthis{\addtocounter{equation}{1}\tag{\theequation}}

\newcommand{\cmuf}{%
  \fontshape{sl}\fontsize{12}{12}\selectfont}

\usepackage{fancyhdr}

\fancypagestyle{plain}{
	\fancyhf{}

	\fancyfoot[EL,OR]{\thepage}}
\fancypagestyle{addpagenumbersforpdfimports}{
	\fancyhead{}
	
	\fancyfoot{}
	\fancyfoot[RO,LE]{\thepage}
}

\makeatletter
\def\cleardoublepage{\clearpage\if@twoside \ifodd\c@page\else
    \hbox{}
    \thispagestyle{empty}
    \newpage
    \if@twocolumn\hbox{}\newpage\fi\fi\fi}
\makeatother \clearpage{\pagestyle{plain}\cleardoublepage}

\usepackage{color}
\usepackage{titlesec}

\titleformat{\chapter}[display]
  {\normalfont\Huge\bfseries}{Chapter \thechapter}{1em}{\Huge}
\titlespacing*{\chapter}{0pt}{50pt}{30pt}
\titlespacing*{\section}{0pt}{13.2pt}{*0}  
\titlespacing*{\subsection}{0pt}{13.2pt}{*0}
\titlespacing*{\subsubsection}{0pt}{13.2pt}{*0}

\makeatletter
\def\resetMathstrut@{%
  \setbox\z@\hbox{%
    \mathchardef\@tempa\mathcode`\(\relax
      \def\@tempb##1"##2##3{\the\textfont"##3\char"}%
      \expandafter\@tempb\meaning\@tempa \relax
  }%
  \ht\Mathstrutbox@1.2\ht\z@ \dp\Mathstrutbox@1.2\dp\z@
}
\makeatother

\def\a{\alpha}
\def\b{\beta}
\def\B{\Box}

\def\d{\delta}
\def\D{\Delta}
\def\e{\varepsilon}
\def\ep{\epsilon}
\def\eps{\varepsilon}
\def\f{\frac}
\def\g{\gamma}

\def\G{\Gamma}
\def\k{\kappa}
\def\s{\sigma}

\def\l{\left}

\def\mc{\mathcal}

\def\m{\mu}
\def\n{\nu}

\def\nn{\nonumber}

\def\p{\partial}

\def\r{\right}
\def\s{\sigma}
\def\t{\theta}

\def\x{\chi}
\def\j{\xi}

\def\z{\zeta}

\def\be{\begin{equation}}
\def\ee{\end{equation}}

\def\bea{\begin{eqnarray}}
\def\eea{\end{eqnarray}}

\def\ba{\begin{array}}
\def\ea{\end{array}}

\def\bc{\begin{center}}
\def\ec{\end{center}}

\def\bl{\begin{flushleft}}
\def\el{\end{flushleft}}

\def\br{\begin{flushright}}
\def\er{\end{flushright}}

\def\bi{\begin{itemize}}
\def\ei{\end{itemize}}

\def\bt{\begin{tabular}}
\def\et{\end{tabular}}

\newtheorem{question}{Question}
\def\bq{\begin{question}}
\def\eq{\end{question}}

\newtheorem{definition}{Def}
\def\bd{\begin{definition}}
\def\ed{\end{definition}}

\newtheorem{answer}{Answer}
\def\ban{\begin{answer}}
\def\ean{\end{answer}}

\newtheorem{possibleanswer}{Possible answer}
\def\bpa{\begin{possibleanswer}\normalfont}
\def\epa{\end{possibleanswer}}

\newtheorem{theorem}{Theorem}
\def\bth{\begin{theorem}}
\def\eth{\end{theorem}}

\begin{document}   

\frontmatter
\begin{titlepage}
\begin{center}

\null\vspace{2cm}
{\large \textbf{POINCAR\'E, SCALE AND CONFORMAL SYMMETRIES: GAUGE PERSPECTIVE AND COSMOLOGICAL RAMIFICATIONS}} \\

\vspace{2cm}

THÈSE 
\\

\leavevmode\\

SOUMISE \`A  LA FACULT\'E DES SCIENCES DE BASE\\
LABORATOIRE DE PHYSIQUE DES PARTICULES ET DE COSMOLOGIE\\
PROGRAMME DOCTORAL EN PHYSIQUE\\

\vspace{1cm}

{\textbf{\large{\'ECOLE POLYTECHNIQUE F\'ED\'ERALE DE LAUSANNE}}}\\

\vspace{1cm}

POUR L'OBTENTION DU GRADE DE DOCTEUR \`ES SCIENCES

\vspace{1cm}

PAR

\vspace{.5cm}

{\large{\textbf{\textsc{Georgios Konstantinou Karananas}}}}

\vfill

{\large{{\textsc{Adviser: Prof. Mikhail Shaposhnikov}}}}

\vfill

Suisse, juillet 2016

\vfill
\end{center}
\end{titlepage}

\thispagestyle{empty}

\vspace*{\fill}

    \textcopyright~Georgios Karananas, 2016

\clearpage

\setcounter{page}{0}    
\cleardoublepage
\chapter*{Abstract}
\addcontentsline{toc}{chapter}{Abstract} 

Symmetries are omnipresent and play a fundamental role in the description of Nature. Thanks to them, we have at our disposal nontrivial selection rules that dictate how a theory should be constructed. This thesis, which is naturally divided into two parts, is devoted to the broad physical implications that spacetime symmetries can have on the systems that posses them. 

In the first part, we focus on local symmetries. We review in detail the techniques of a self-consistent framework -- the coset construction -- that we employed in order to discuss the dynamics of the theories of interest. The merit of  this approach lies in that we can make the (spacetime) symmetry group act internally and thus, be effectively separated from coordinate transformations. We investigate under which conditions it is not needed to introduce extra compensating fields to make relativistic as well as nonrelativistic theories invariant under local spacetime symmetries and more precisely under scale (Weyl) transformations. In addition, we clarify the role that the field strength associated with shifts (torsion) plays in this context. We also highlight the difference between the frequently mixed concepts of Weyl and conformal invariance and we demonstrate that not all conformal theories (in flat or curved spacetime), can be coupled to gravity in a Weyl invariant way. Once this ``minimalistic'' treatment for gauging symmetries is left aside, new possibilities appear. Namely, if we consider the Poincar\'e group, the presence of the compensating modes leads to nontrivial particle dynamics. We investigate in detail their behavior and we derive constraints such that the theory is free from pathologies. 

In the second part of the thesis, we make clear that even when not gauged, the presence of spontaneously broken (global) scale invariance can be quite appealing. First of all, it makes possible for the various dimensionful parameters that appear in a theory to be generated dynamically and be sourced by the vacuum expectation value of the Goldstone boson of the nonlinearly realized symmetry -- the dilaton. If the Standard Model of particle physics is embedded into a scale-invariant framework, a number of   
interesting implications for cosmology arise. As it turns out, the early inflationary stage of our Universe and its present-day acceleration become linked, a connection that might give us some insight into the dark energy dynamics. Moreover, we show that in the context of gravitational theories which are invariant under restricted coordinate transformations, the dilaton instead of being introduced ad hoc, can emerge from the gravitational part of a theory. Finally, we discuss the consequences of the nontrivial way this field emerges in the action.

{\textbf{Keywords:}} Poincar\'e invariance, scale invariance, conformal invariance, gauge theory, gravity, cosmology.

\cleardoublepage
\chapter*{R\'esum\'e}
\addcontentsline{toc}{chapter}{R\'esum\'e} 

Les sym\'etries sont omnipr\'esentes et jouent un r\^ole fondamental dans la description de la nature. Gr\^ace \`a elles, nous avons \`a notre disposition des r\`egles de s\'election non triviales qui dictent la construction des th\'eories physiques. 

Cette th\`ese, qui est naturellement divis\'e en deux parties, est consacr\'ee aux vastes implications physiques que les sym\'etries d'espace-temps peuvent avoir sur les syst\`emes qui les poss\`edent.  Dans la premi\`ere partie, nous nous concentrons sur les sym\'etries locales. Nous examinons en d\'etail les techniques d'un cadre auto-coh\'erent -- coset construction -- que nous avons employ\'e pour examiner les th\'eories qui nous int\'eressent. Le m\'erite de cette approche r\'eside dans le fait que nous pouvons faire agir le  groupe de sym\'etrie en interne et donc, \^etre efficacement s\'epar\'e des transformations de coordonn\'ees. Nous \'etudions les conditions dans lesquelles il n'est pas n\'ecessaire d'introduire des champs de compensation suppl\'ementaires pour rendre une th\'eorie relativiste ou non-relativiste invariante par rappord aux sym\'etries d'espace-temps locales et plus pr\'ecis\'ement par les transformations de changement d'\'echelle (transformation de Weyl). En outre, nous clarifions le r\^ole que le tenseur du champ associ\'e aux d\'eplacements (torsion) joue dans ce contexte. Nous soulignons \'egalement la diff\'erence entre les concepts souvent mixtes de transformation de Weyl et d'invariance conforme et nous d\'emontrons que toutes les th\'eories conformes (en espace-temps plat ou courb\'e), ne peuvent pas \^etre  couples \`a la gravit\'e d'une mani\`ere invariante sous transformation de Weyl. Si ce traitement `` minimaliste '' pour jauger les sym\'etries est laiss\'e de c\^ote, de nouvelles possibilit\'es apparaissent. Si nous consid\'erons le groupe de Poincar\'e, la pr\'esence des modes de compensation conduit \`a une dynamique des particules non triviale. Nous \'etudions en d\'etail leur comportement et nous en d\'eduisons des contraintes pour que la th\'eorie soit exempte de pathologies.

Dans la deuxi\`eme partie de la th\`ese, nous montrons clairement que m\^eme lorsque l'invariance globale en changements d'\'echelle n'est pas jaug\'e, une brisure  spontan\'ee de cette sym\'etrie peut \^etre tr\`es attrayante. Tout d'abord, il est possible de g\'en\'erer dynamiquement les diff\'erents param\`etres dimensionels qui apparaissent dans la th\'eorie. Ceux-ci proviennent ensuite de la valeur moyenne dans le vide du boson de Goldstone de la sym\'etrie bris\'ee - le dilaton. Si le mod\`ele standard de la physique des particules est int\'egr\'e dans un cadre invariant par les changements d'\'echelle, un certain nombre d' implications int\'eressantes pour la cosmologie se posent. Il se trouve, que la phase d'inflation de notre Univers et son acc\'el\'eration actuelle deviennent li\'ee, une connexion qui pourrait nous donner un aper\c{c}u de la dynamique de l'\'energie sombre. Enfin, nous montrons que dans le cadre des th\'eories de la gravitation invariantes sous  les transformations de coordonn\'ees restreintes, le dilaton peut sortir du secteur gravitationnel d'une th\'eorie au lieu d'\^etre introduit ad hoc. Nous discutons les cons\'equences int\'eressantes de la fa\c{c}on dont ce champ \'emerge dans l'action.

{\textbf{Mots cl\'es:}} invariance de Poincar\'e , invariance en changements d'\'echelle, invariance conforme, th\'eorie de jauge, gravitation, cosmologie.

\chapter*{Acknowledgements}

\addcontentsline{toc}{chapter}{Acknowledgements} 

I am very grateful to my adviser Misha Shaposhnikov for his constant support and guidance throughout these years. 

I am also very grateful to Fedor Bezrukov, Sasha Monin and Javier Rubio for the very fruitful collaboration. 

I would like to express my gratitude to Yuri Obukhov, David Pirtskhalava and Ivo Sachs for accepting to be members of the jury.

Many thanks go to Alexey Boyarsky, Kostas Farakos, Alex Kehagias, Riccardo Rattazzi, Oleg Ruchayskiy and Sergey Sibiryakov. 

Last but not least, I would like to thank  for all their love and support my parents Maria and Kostas, my brothers Vasilis and Themos, my girlfriend Anna, and my friends Aggeliki, Agis, Andreas, Gerasimos, Giorgos, Lorenzo, Raphael, Stefanos and Velisaris.

\bigskip
 
\hfill Giorgos

\tableofcontents 
\cleardoublepage  
\phantomsection

\setlength{\parskip}{1em}

\mainmatter

\chapter{Introduction}

The Standard Model (SM) of particle physics that describes the electroweak and strong interactions has enabled us to explain in a self-consistent manner a plethora of phenomena. Especially after the discovery of the Higgs boson, SM could be thought of as a predictive effective field theory valid up to energies comparable to the gravitational scale $M_P=2.435 \times 10^{18}$ GeV.

Despite its unprecedented success, we now understand that the SM is not a complete theory for a number of reasons. From the experimental point of view, it is not possible to address in its context several well established observational facts, like for example the neutrino masses and oscillations, the baryon asymmetry of the Universe and the origin of dark matter. From the theoretical point of view, the SM suffers -- among others -- from two severe fine-tuning problems, namely the hierarchy and cosmological constant problems. For the former, according to the rules of the effective field theory, the Higgs mass receives large radiative corrections, making it very sensitive to whatever physics lie beyond the SM.  Therefore, its smallness requires an extreme fine-tuning in order to compensate for these contributions that are related to the ultraviolet dynamics. For the latter, its predicted value is approximately $M_P^4$, which is by many orders of magnitude larger than the one observed. 

Although these problems do not pose a threat to the consistency of theory, there is no (satisfactory) explanation on what could be the underlying principle making the electroweak scale and the cosmological constant so small as compared to $M_P$. 

One of the directions towards a possible resolution of the aforementioned theoretical puzzles is to allow certain parameters to be small provided that the symmetry of the theory is enhanced when these are set to zero. If this line of reasoning is applied to the SM, one observes that responsible for the smallness of the Higgs mass and the cosmological constant term could be the presence of exact scale and/or conformal invariance.

In general, theories exhibiting scale and conformal invariance (see, for example,~\cite{DiFrancesco:1997nk, Rychkov:2016iqz}) constitute a very interesting and rich subject for investigations. They appear ubiquitously for describing physical systems, whenever a separation of scales exists. The presence of these symmetries restricts sufficiently the dynamics, so that many properties of the system can be inferred and in some cases, the theory can even be solved completely. Thus, they give an important handle on quantum field theory (for recent progress see~\cite{Komargodski:2011vj,Komargodski:2011xv,Luty:2012ww,Dymarsky:2013pqa, ElShowk:2012ht}).

In particular, if the SM is considered as part of a larger scale or conformally invariant framework, the resulting theory should ultimately be confronted with observations. For it to be phenomenologically viable -- apart from incorporating gravity in a consistent with all the symmetries manner -- the additional symmetry that it enjoys should be spontaneously broken; this leads to the appearance of a Goldstone boson, the dilaton. As a result, all the scales (at the classical level) can have a common origin: the vacuum expectation value of the dilaton.

It is well known that all classical considerations concerning scale and conformal theories might not survive at the quantum level. This is almost a trivial statement, since a mass scale that explicitly breaks the classical symmetry is introduced when a theory is regularized. However, if this mass scale is related to the vacuum expectation value of the dilaton, then the symmetries of the system remain anomaly-free when quantum corrections are taken into account~\cite{Englert:1976ep,Shaposhnikov:2008xi,Armillis:2013wya,Gretsch:2013ooa}. Notice though that if such a regularization prescription is used, the loop expansion will generate an infinite number of divergences, therefore an infinite number of counter-terms (not necessarily with the same functional form as the terms in the tree-level theory) will be needed in order to account for them.\footnote{For scale-invariant theories, the scale invariance of the (regularized) quantum effective action follows trivially from dimensional analysis. For theories with conformal symmetry on the other hand, investigating what is the fate of the symmetry at the quantum level is more subtle, see~\cite{Armillis:2013wya,Gretsch:2013ooa} for details.} As we already mentioned, gravity should be part of any realistic model and so, the requirement of renormalizability has to be abandoned in any case. Notice though, that this should not be considered as a drawback, as long as we end up with a predictive effective field theory.

Our purpose in this thesis is to investigate several formal (part~\ref{part:theor_cons}) as well as phenomenological (part~\ref{part:pheno}) aspects associated with theories that possess spacetime symmetries. More specifically, in part~\ref{part:theor_cons}, we address Poincar\'e, scale and conformal invariance from a gauge perspective and we discuss in great detail the role of the compensating fields that have to be introduced. Part~\ref{part:pheno}, deals with the cosmological ramifications of scale invariance and with various properties of the dilaton field. For the convenience of the reader, each part contains its own outline.

\part{Theoretical considerations}
\label{part:theor_cons}

\chapter{Outline of Part I}

This part of the thesis is exclusively devoted to formal aspects of gauged spacetime symmetries. In chapter~\ref{sec:coset_constr}, we show that a natural way to get a handle on the dynamics and implications of the systems under consideration is provided by the coset construction. This technique is extremely powerful and very useful, for it allows to systematically build invariant (effective) actions using only symmetry arguments. When dealing with spacetime symmetries, this method makes it possible to completely disentangle the gauge (internal) transformations from the coordinate ones, in complete analogy with the situation in Yang-Mills theories~\cite{Yang:1954ek}.

One illustrative example for realizing the potential of the coset construction, is to consider the gauging of the Poincar\'e group and consequently the emergence of the gravitational interaction. Conventionally, gravity is treated in the context of Einstein's theory of General Relativity (GR). However, one can follow the paradigm of the SM and take the gauge approach as a guiding principle.

Even though this approach to gauging is certainly not unique, it is more practical than the conventional one, see for example~\cite{Utiyama:1956sy,Brodsky:1961,Sciama:1962,Kibble:1961ba}; the gauge field associated with translations (vielbein) is automatically guaranteed to have an inverse, and more importantly, both field strengths -- curvature $\omega$ and torsion $T$ -- transform covariantly under the group operations. Therefore, the Lagrangian describing the dynamics of the theory can be straightforwardly written down by considering all possible invariants constructed from curvature and torsion at a given order in derivatives
\begin{equation}
\label{introact}
\mathscr L = \mathscr L_0+\mathscr L_1(\omega,T)+\mathscr L_2(\omega,T)+\ldots \
\end{equation}
where $\mathscr L_0, \mathscr L_1,\ldots,$ contain terms with zero derivatives (cosmological constant), one derivative (scalar curvature, Holst term) etc. 

The theory in which both curvature and torsion are present, is known in the literature as Poincar\'e Gravitational Theory (PGT). One might wonder if the presence of the degrees of freedom associated with the connection is a desirable feature. We will be back to this point in a while. It should be noted that if the goal is to eliminate the extra modes and recover the Einstein-Hilbert action from the PGT, then the connection should be expressed in terms of derivatives of the vielbein by imposing the \emph{covariant} constraint of vanishing torsion.\footnote{This is equivalent to integrating out the connection by using its equation of motion (at the lowest order in derivatives)~\cite{Delacretaz:2014oxa}.} This fact should not come as a surprise, since it can well be the case that the number of fields needed to gauge a spacetime symmetry is smaller than what would be expected. The investigation of when this is actually possible has to be carried out in a systematic way, especially when conformal theories are considered. Let us explain why this is the case. A very powerful tool for studying these systems is coupling them to a nondynamical metric~\cite{Komargodski:2011vj,Luty:2012ww,Gretsch:2013ooa}. In an even more general setup, all the couplings are considered as background sources~\cite{Baume:2014rla}. It is usually assumed that a conformally invariant theory can be embedded in a curved background in a Weyl invariant manner. It is necessary that a theory be conformal in flat spacetime, in order to couple it to gravity in a Weyl-invariant way. It has been shown that the condition becomes sufficient, only if actions with at most one derivative of conformally variant fields are considered~\cite{Iorio:1996ad} (see also \cite{O'Raifeartaigh:1996hf}). However, to the best of our knowledge, there is no proof for the condition to be sufficient in general.

The authors of the interesting work~\cite{Iorio:1996ad} proceed as follows. Given a scale-invariant theory in flat spacetime, it can be made Weyl invariant by gauging dilatations with the help of an additional field $W _ \m$ (Weyl gauging).\footnote{Throughout this thesis, we use greek letters $(\mu,\nu,\ldots)$ for spacetime indices.} It so happens that the Weyl variation of a certain combination of the gauge field\,\footnote{The metric-compatible covariant derivative $\nabla_\m$, as well as the Christoffel symbols $\G_{\m\n}^\lambda$ are defined in Appendix~\ref{Christoffel}.}
\be
\Theta _ {\m \n} = \nabla _ \m W _ \n - W _ \m W_ \n + \f {1} {2} g _ {\m \n} W ^ \s W _ \s \ ,
\label{tensor_A}
\ee
where $\nabla$ denotes the standard covariant derivative and $g_{\m\n}$ the metric,
does not depend on $W _ \m$. It is proportional to the variation of the Schouten tensor
\be
S _ {\m \n} = R _ {\m \n} - \f {R} {2 (n-1)} g _ {\m \n} \ ,
\label{tensor_R}
\ee
with the following convention for the curvatures
\be
\label{ricci-conv}
R = R ^ \m _ \m,~~R _ {\m \n} = R ^ {~~~\s}_ { \s \m ~ \n }~~\text{and}~~R ^ {~~~\s}_ { \lambda \m ~ \n } = \p _ \lambda \G _ {\m \n }^ \s - \p _ \m \G _ {\lambda \n }^ \s + 
\G _ {\lambda \rho} ^ \s \G _ {\m \n} ^ \rho - \G _ {\m \rho} ^ \s \G _ {\lambda \n} ^ \rho \ .
\ee
Therefore, if the gauge field enters the Lagrangian only in the combination~\eqref{tensor_A}, it is possible to trade it for the expression in~\eqref{tensor_R}, leaving all the symmetries intact. As a result, the theory becomes Weyl invariant and no additional degrees of freedom are introduced. The authors call this procedure \emph{Ricci gauging}. Lastly, they prove that for a theory without higher derivatives of conformally variant fields, the described Weyl gauging leads necessarily to the appearance of the tensor~\eqref{tensor_A}, provided the theory is conformal. Consequently, these theories can be made Weyl invariant when coupled to gravity.

The tensor composed of the Weyl gauge field and possessing the transformation properties of~\eqref{tensor_A} can be found by trial and error, but a systematic recipe can be easily provided by the coset construction, as we show in chapter~\ref{ch:Weyl_Ricci}. When this formalism is applied to the Poincar\'e group plus dilatations, the aforementioned relation between $\Theta_{\m\n}$ and $S_{\m\n}$ follows immediately from the requirement (or better say the covariant constraint) of vanishing torsion. Meanwhile, if one does not insist on having a torsionless theory, then $W_\m$ can be shown to be related to one of the irreducible pieces of the torsion tensor, something that was realized many years ago in~\cite{Obukhov:1982zn}.

There is a number of questions that arise at this point. To start with, it is natural to wonder whether Ricci gauging can be applied to higher-derivative conformal theories as well. It turns out that its range of applicability is quite vast, even though there are certain subtleties that arise due to the presence of more than one derivatives. Actually, if we consider for example a quartic in derivatives theory of a scalar field in an arbitrary number of spacetime dimensions $n>2$, it is a straightforward (although a bit algebraically involved) exercise to couple it to gravity in a Weyl invariant manner using this procedure.

What is interesting is that certain terms in the Lagrangian of the resulting theory, blow up at the limit $n\to 2$. In chapter~\ref{ch:Weyl_vs_Conf}, we demonstrate that is an indication that the Weyl invariant generalization of a conformal higher-derivative theory does not exist in two spacetime dimensions. Actually, this ``obstruction'' does not appear only in $n=2$, but is present in all (even) dimensions, if the number of derivatives acting on a field exceeds $n$. This implies that it is not always possible for a conformally invariant theory to be made Weyl invariant. Even though the two notions of Weyl and conformal symmetry are used interchangeably, it should be stressed that the former is not just the curved-space generalization of the latter, but rather a different concept. To put in other words, for a theory invariant under diffeomorphisms $\times$ Weyl, its flat limit automatically produces a theory which is conformal; the opposite is not always true.

Yet another point worth investigating is whether nonrelativistic theories can also be coupled to a curved background in a Weyl invariant way using only the geometrical data. Notice that there has been renewed interest in these theories in the context of many body systems/condensed matter physics, which has been partially sparked by~\cite{Son:2005rv,Son:2008ye}. As we illustrate in chapter~\ref{ch:NR_Weyl}, the role of torsion here is indispensable, since for the concept of Weyl invariance to even exist, these theories must necessarily be torsionful. Moreover, it is always possible to express the spatial part of the Weyl gauge field in terms of degrees of freedom already present in the theory. As for the temporal part, whether or not it can be eliminated depends on the (nonrelativistic) symmetry group under consideration. For the Lifshitz algebra plus dilatations, there is no obstacle to its elimination, therefore the situation is similar to what occurs with Lorentz-invariant theories. On the other hand, for a theory invariant under the centrally extended Galilei algebra plus dilatations this is not the case, because the presence of  boosts complicates considerably the situation. However, even in this case, as long as the temporal part of the gauge field is absent, such a theory is going to be automatically Weyl-symmetric.

Up until this point, we have been exclusively interested on how to achieve invariance of a theory under a symmetry group by keeping the minimal number of compensating gauge fields. However, new and quite interesting possibilities appear if the extra degrees of freedom are not eliminated. Coming back to the PGT, certain torsionful theories~\cite{Ho:2011qn,Ho:2011xf,Ho:2015ulu}, have attracted considerable attention, since they are free from pathologies and have very interesting cosmological phenomenology. In general, not all theories in which torsion is propagating are ghost and tachyon free. In chapter~\ref{chapt:Poincaregrav}, we have carried out a detailed analysis of the spectrum of the most general theory that results from the gauging of the Poincar\'e group and contains terms at most quadratic in the field strengths. We have allowed for parity-odd terms in the action and we have derived the conditions for absence of ghosts and tachyons.

\chapter{Gauging spacetime symmetries}
\label{sec:coset_constr}

The necessary ingredients for building an effective field theory are the field/particle content and symmetries. The latter impose constraints on a Lagrangian, for it (or better to say the action) should be a singlet under the symmetry transformations. Once all the symmetries of a system are known, the number of free parameters in the Lagrangian is reduced.

The reason why it may be needed to go from rigid symmetries to gauged ones is twofold. On the one hand, the background gauge fields act like sources for the corresponding conserved currents. Gauge invariance in this case puts severe constraints (selection rules) on the partition function:  integrating out dynamical fields leads -- in the absence of anomalies -- to a gauge-invariant partition function. On the other hand, the gauge field theories are an appropriate language to talk about massless vector and tensor degrees of freedom, e.g. photons and gravitons.

Any global symmetry group can be made local by introducing a sufficient number of corresponding compensators (gauge fields) with appropriate transformation properties.\footnote{Strictly speaking, this is true only when the symmetry is not anomalous.} A question that naturally arises is whether this number can be smaller than the number of generators of the symmetry group considered. For internal symmetries (the ones that commute with the generators of spacetime translations), this does not seem to be the case. However, for spacetime symmetries the gauging may not require as many fields as there are generators. For example, as we will demonstrate later, the Poincar\'e group can be made local without introducing the spin connection as an independent field, but rather as a function of the vielbein (at least for torsionless theories). We will also show that some Weyl invariant theories do not require the introduction of a gauge field to account for the local scale transformations, since its role can be played by a certain combination of curvature tensors or torsion. 

The action of spacetime symmetries on the fields, obtained as an induced representation, is related to the nonlinear realization of symmetries. Therefore, when talking about certain physical systems we find that the coset construction provides the appropriate language. It allows one to circumvent certain difficulties related to the transformation properties of the fields under the corresponding symmetry group, automatically providing the necessary building blocks. In this introductory chapter, based mainly on~\cite{Karananas:2016hrm,Karananas:2015eha,Delacretaz:2014oxa}, we review in detail the basic ingredients of this approach and discuss its relevance for gauging spacetime symmetries.

\section{Internal symmetries}

 The nonlinear realization of internal symmetries (the ones that commute with the generators of spacetime translations) in flat spacetime was introduced in~\cite{Coleman:1969sm,Callan:1969sn} and it is used to obtain the building blocks for a theory that exhibits a specific symmetry breaking pattern $S \to S _ 0$. In other words, it allows one to construct the most general action of a group $S$ such that when restricted to its subgroup $S _ 0$, it becomes a linear representation. 

The procedure can be described as follows. For the symmetry breaking pattern, one realizes the action of the group $S$ on the coset space $S/S _ 0$ by left multiplication. Choosing the coset representative as 
\be
\Omega = e ^ {i \pi T} \in S \ ,
\label{coset_rep}
\ee
where $T$ is the set of all broken generators and $\pi$ (Goldstone fields) constitutes a parametrization of the coset,\footnote{For brevity we suppress all the indices corresponding to the Lie algebra.} one gets the transformation
\be
s \Omega = \Omega ' \bar s _ 0\ , ~~~\text{with}~~~ \bar s_0\equiv \bar s _ 0 (\pi,s) \in S _ 0 \ .
\label{g_action_coset}
\ee 
Central role to this approach plays the Maurer-Cartan form
\be
\Omega ^ {-1} \p _ \m \Omega \ ,
\label{MC_form}
\ee
that is calculated using the commutation relations of the group under consideration. If we denote by $t$ all the unbroken generators, this expression can be written as 
\be
\Omega ^ {-1} \p _ \m \Omega = i \nabla _ \m \pi \, T + i \omega _ \m t \ ,
\ee
and it is easy to check that under~\eqref{g_action_coset}, it transforms as
\be
\l ( \Omega ^ {-1} \p _ \m \Omega\r ) ' = \bar s _ 0 \l( \Omega ^ {-1} \p _ \m \Omega \r ) \bar s _ 0 ^ {-1}  + \bar s _ 0 \p _ \m \bar s _ 0 ^ {-1} \ .
\ee
For compact groups, the above translates into the corresponding transformations of  $\nabla_\m \pi$ and $\omega_\m$
\be
\begin{aligned}
\nabla_\m \pi ' T & =  \bar s _ 0  \, \nabla_\m \pi ' T \, \bar s _ 0 ^ {-1}  \ ,  \\
i\omega _ \m ' t & =  \bar s _ 0  \, i\omega _ \m  t \, \bar s _ 0 ^ {-1} + \bar s _ 0 \p _ \m \bar s _ 0 ^ {-1} \ ,
\label{h_trans_coset}
\end{aligned}
\ee
which can be used to write automatically $S$-invariant Lagrangians by constructing singlets of the subgroup $S _ 0$.\footnote{It should be mentioned that with this procedure, the resulting Lagrangians will only contain terms that are exactly invariant under the symmetry transformations.} 

The gauging of the group 
$S$ within this framework goes along the standard lines; it is achieved by promoting the partial derivative $\p_\m$ in~\eqref{MC_form} to a covariant derivative $\tilde D_\m$ including gauge fields that correspond to each generator of the symmetry group and under the action of 
$S$ transform as 
\be
\tilde A ' _ \m = s \tilde A _ \m s ^ {-1} + s \p _ \m s ^ {-1} \ .
\label{gauge_fields_trans}
\ee

\section{Spacetime symmetries}

The difference between internal and spacetime symmetries is that the latter are usually (if not necessarily) realized on the infinite dimensional spaces of fields. These infinite dimensional representations are induced representations that are defined in the following way. For a group $K$, its subgroup $K _ 0 \subset K$ that is realized on a linear space $V$, there is a natural action of the group $K$ on the coset $K  / K _ 0$ by left multiplications.\footnote{Usually the coset $K  / K _ 0$ is isomorphic to the spacetime manifold.} 
For example, let us take $K$ to be the $n$-dimensional Poincar\'e group and $K _ 0$ to be the Lorentz group. It is clear that in this case $K / K _ 0 = \mathbb{R} ^ n$. The action of $K$ on the coset is as follows\,\footnote{Lorentz indices are denoted with capital latin letters $(A,B,\ldots)$. We use the Landau-Lifshitz signature for the Minkowski metric, $\eta_{AB}=\text{diag}(+,-,-,\ldots)$.}
\be
k e ^ {i P y} = e ^ {i P (\Lambda y + a)} \bar k _ 0 (k) \ ,
\ee
where $k \in K$, $\bar k _ 0 (k) \in K _ 0$, $P_A$ are momenta, $\Lambda^A_{\ B}$ correspond to Lorentz rotations, $y_A$ are Cartesian coordinates on the coset~$\mathbb{R} ^ n$ and $a_A$ are parameters of the translations. Considering a representation of the Lorentz group
\be
\begin{aligned} 
\rho: K _ 0 & \to & {GL} ( V ) \ ,  \\
T _ {\bar k _ 0} \Psi & = & \rho (\bar k_0) \Psi \ ,
\end{aligned}
\ee
we define the induced representation of the full Poincar\'e group according to
\be
\l ( T _ k \Psi \r ) (y') = \rho (\bar k _ 0 (k)) \Psi (y) \ ,
\ee
which corresponds to the standard transformation of a field 
\be
\Psi _ {(\Lambda,a)} (y) = D (\Lambda) \Psi (\Lambda ^ {-1} y - a) \ .
\ee

Even though the generators $P _ A$, which correspond to the coset $K / K _ 0$, are not  broken and are realized linearly on the space of fields, the very construction of this representation makes it natural to include the momenta in the coset~\eqref{coset_rep} when discussing the breaking and/or gauging of spacetime symmetries. Consequently, for the symmetry group $G$ (with algebra $\mathfrak g$) that includes both internal and spacetime symmetries and that is broken down to a subgroup $H$ (with algebra $\mathfrak h$), one gets the coset in the form
\be
\Omega = e ^ {i P x} e ^ {i \pi (x) T} \ ,
\ee
where by $T$ we denote all the broken generators (not only the internal ones). The way to introduce a different set of coordinates on the spacetime manifold  is to have them appearing in the coset representative through the auxiliary functions $y^A (x)$, which means that in general we may write
\be
\Omega = e ^ {i P y (x)} e ^ {i \pi (x) T} \ .
\ee
Under the action of the spacetime symmetry group $K$, the coordinates transform according to
\be
k e ^ {i P y (x)} = e ^ {i{P y (x')}} \bar k _ 0 \ ,~~~\text{with}~~~ \bar k _ 0\equiv \bar k _ 0(x,k) \ .
\ee
These transformations may be viewed in a different way, namely, keeping the coordinates $x$ unchanged while transforming the functions $y^A (x) \to y^{'A} (x)$ \footnote{For example, in a two-dimensional Euclidean space, one may choose polar coordinates corresponding to $(y^1 (r,\varphi), y^2(r,\varphi)) = (r \sin \varphi, r \cos \varphi)$. Then the transformation under rotations 
\be
(y^1, y^2) \to (r \cos (\varphi + \a), r \sin (\varphi + \a)) \ , \nn
\ee
can be equivalently viewed either like
$\varphi \to \varphi' = \varphi + \a$, or as a change of the functional form $y ^ {'1} (r,\varphi)= r \cos (\varphi + \a)$, and similar for $y ^ {'2}$.}
\be
k e ^ {i P y (x)} = e ^ {iP y' (x)} \bar k _ 0 \ .
\ee
The reason for this choice becomes clear when the gauging of a spacetime symmetry group is considered, for in this case one does not have to take into account the transformation of the fields due to the change of coordinates $x$ and the gauging goes along the lines of that for internal symmetries. However, by doing so, the additional functions $y^A(x)$, with very specific transformation properties, had to be introduced. Of course, they are not physical and should be dispensed with. This is easily achieved by simply demanding that the resulting theory is invariant under diffeomorphisms as well. 

In a sense, introducing these additional spurious fields allows us to decouple the diffeomorphisms from the (local) transformations under the spacetime symmetry group. The gauge fields $ \tilde A _ \m$ transform in the standard way~\eqref{gauge_fields_trans} under the local spacetime transformations and separately under the diffeomorphisms $x \to x'$,
\be
\tilde A '_ \m (x') = \tilde A _ \n (x) \f {\p x ^ \n} {\p x ^{' \m}} \ . 
\label{gauge_diffeomorphisms}
\ee
The Maurer-Cartan form can now be written as
\be
\label{coset_rep_y}
\Omega ^ {-1} \tilde D _ \m \Omega 
 =  i  e_\m^A P_A+ i \nabla _ \m \pi \, T + i \omega _ \m t \ ,
\ee
where as before $P _ A$ are momenta, whereas $t$ and $T$ are the rest of the unbroken and broken generators respectively. For symmetry groups with the following schematic structure of commutation relations
\be
\begin{aligned}
\l [ t,t \r ] & = t \ , \\
\l [ t, P \r ] & =  P \ ,  \\
\l [ t, T \r ] & =  T \ , 
\label{representation_commut}
\end{aligned}
\ee
and upon using the definition of the transformation of the coset representative
\be
g \Omega = \Omega' \bar h (y,g) \ ,
\label{coset_rep_trans}
\ee
we find that the transformations of $\nabla _ \m \pi$, $\omega _ \m$ and $e_ \m ^ A$, are given by
\be
\begin{aligned}
\nabla_\m \pi ' T & =  \bar h \, \nabla_\m \pi ' T \, \bar h^{-1} (\pi,s) \ , \\
i\omega _ \m ' t & =  \bar h \, i \omega _ \m  t \, \bar h ^ {-1}+ \bar h \p _ \m \bar h ^ {-1} \ ,  \\
e ^ {'A} _ \m P _ A & =  e ^ {A} _ \m \, \bar h (\pi,g) P _ A \bar h ^ {-1} (\pi,g) \ ,
\label{space_time_h_trans_coset}
\end{aligned}
\ee
The coefficients $e_ \m ^ A$ due to their specific transformation properties under the diffeomorphisms~\eqref{gauge_diffeomorphisms} can be thought of as the vielbein.

As a result, we have the necessary tools to analyze a system with spontaneously broken symmetries. For example, any $H$-invariant function of $\nabla _ \m \pi$ would produce a Lagrangian which is ``secretly'' $G$ invariant, if one also uses $e_ \m ^ \n$ to build an invariant measure. Similarly, the connection can be used to construct higher derivative terms and/or coupling to matter fields.

It should be noted that the main feature of the nonlinear realization of spacetime symmetries -- as compared to internal ones --  is the counting of degrees of freedom. For the case of internal symmetries, the number of Goldstone modes is always equal to the number of broken generators. For spacetime symmetries, this is not always true, since it is not rare that a smaller number of Goldstone bosons is enough to realize a symmetry breaking pattern. This happens because the fluctuations produced by the action of all broken generators on the vacuum are not independent. From the physical point of view, this phenomenon manifests itself through the equations of motion, when at low energies certain modes may become gapped and, therefore, can be explicitly integrated out. From a more formal perspective, it can be understood with the help of the inverse Higgs mechanism, which consists of imposing covariant (consistent with all symmetries) constraints on the system and solving them algebraically, thus, reducing the number of necessary fields~\cite{Ivanov:1975zq,Low:2001bw,Endlich:2013vfa,Brauner:2014aha}. Notice, however, that the constraints that can be solved are those for which the commutator of a  broken generator $T$ with the momentum contains another broken generator $T'$ 
\be
\l [ P , T \r ] \supset T ' \ . 
\label{stand_inv_hig}
\ee
In this case, the Goldstone corresponding to $T$ is expressed in terms of the derivatives of other fields, by solving $\nabla _ \m \pi _ {T'} = 0$.

\subsection{Coset construction and the Poincar\'e group}
\label{sec:cosPoincareg}

The \emph{Poincar\'e group} is the semi-direct product of translations $P_A$ and Lorentz transformations $J_{AB}$ and its algebra is defined by the following commutation relations between the generators
\be
\begin{aligned}
\label{poin_cr}
\l [P _ A, P_B \r ] & =  0 \ ,  \\
\l [ J _ {AB}, P _ C \r ] & =  i \l ( \eta _ {BC} P _ A - \eta _ {AC} P _ B \r ) \ ,  \\
\l [ J _ {AB}, J _ {CD} \r ] & =  i \l ( J _ {AD} \eta _ {BC} + J _ {BC} \eta _ {AD} -
J _ {BD} \eta _ {AC} - J _ {AC} \eta _ {BD} \r ) \ .  \\
\end{aligned}
\ee

The role of this group in particle physics is fundamental and twofold. On one hand, it dictates the symmetries of the underlying Minkowski spacetime of Special Relativity. On the other hand, particle states in quantum field theories are classified according to the unitary irreducible representations of this particular group~\cite{Wigner:1939cj,Weinberg:1995mt}.

The pursuit of a gravitational theory with better microscopic behaviour that GR, as well as the fact that Yang-Mills theories enjoyed big success, initiated investigations~\cite{Utiyama:1956sy,Brodsky:1961} that eventually lead to the formulation of a gravitational theory that results from the gauging of the Poincar\'e group~\cite{Sciama:1962,Kibble:1961ba}. As we have already mentioned, within the framework of the coset construction, gravity is obtained by promoting the 10 (constant) parameters of the group to depend arbitrarily on position, and at the same time to demand that the theory be invariant under general coordinate transformations (diffeomorphisms)~\cite{Ivanov:1981wn,Ivanov:1981wm,Delacretaz:2014oxa}. Since we want Lorentz rotations to be unbroken, then according to the discussion in the previous chapter, the coset representative contains only the momenta~\cite{Delacretaz:2014oxa}
\be
\label{coset_rep_poinc}
\Omega = e ^ {i P _ A y ^ A} \ .
\ee
Using the standard formulas for two operators $X$ and $Y$
\bea
e ^ {-Y} X e ^ Y & = & X + \l [ X, Y \r] + \f {1}{2}\l [ \l [ X, Y \r], Y \r ] + \dots \ , \\
e ^ {-Y} \p_\m e ^ Y & = & \p_\m Y + \f {1}{2} \l [ \p_\m Y, Y \r ] + \f {1}{6} \l [ \l [ \p_\m Y, Y \r], Y \r ] + \dots  \ , 
\eea
we find that the Maurer-Cartan form~\eqref{coset_rep_y} becomes
\be
\Omega ^ {-1} \l ( \p _ \m + i \tilde e ^ A _ \m P _ A + \f {i} {2} \tilde \omega ^ {AB} _ \m J _ {AB} \r ) \Omega = 
i e _ \m ^ A P _ A + \f {i} {2} \omega ^ {AB} _ \m J _ {AB}\ .
\ee
Here $\tilde e_\m^A$ , $\tilde \omega_\m^{AB}$ are the 40 a priori independent gauge fields corresponding to translations and Lorentz rotations respectively, while their counterparts without the tilde are defined as
\be
\begin{aligned}
&e_\m ^A=\tilde e_\m^A+\p_\m y^A -\tilde \omega_\m^{AB} y_B \ ,~~~~~\omega_\m^{AB}=\tilde \omega_\m^{AB} \ .
\end{aligned}
\ee
It is straightforward to check that the transformation of $e _ \m ^ A$ under the action of the local Lorentz group is
\be
e _ \m ^ A \rightarrow \Lambda^A_{\ B} e_\m ^B \ ,
\ee
whereas under a diffeomorphism $x\rightarrow x'$,
\be
e _ \m ^ A(x) \rightarrow e _ \m ^ A(x)  \f {\p x ^ \n} {\p x ^{' \m}} \ .
\ee
Consequently, $e _ \m ^ A$ can be interpreted as a vielbein that is used to mix spacetime and Lorentz indices, to define the metric $g_{\m\n}=e_\m^A e_\n^B \eta_{AB}$, and to construct the diffeomorphism-invariant measure
\be
d ^ n x \, \det e_{\m}^A\equiv d ^ n x \, \det e \ .
\label{detemeasure}
\ee
Notice that if we do not require that the theory be invariant under the full group of diffeomorphisms, then the construction of the invariant measure is not necessary. For example one may be interested in theories invariant only with respect to transverse diffeomorphisms (TDiff), see for example~\cite{Buchmuller:1988wx, Alvarez:2006uu, Blas:2011ac} and references therein. In this case, the theory is invariant only under the subgroup of coordinate transformations with Jacobian equal to unity, thus we can allow for the presence of arbitrary powers of the vielbein (or equivalently the metric) determinant. We will discuss in more detail these theories in the last chapter of the thesis.

The field $\omega ^ {A B} _ \m$, in turn, under a Lorentz transformation behaves as 
\be
\omega ^ {A B} _ \m \rightarrow \omega ^ {CD} _ {\m} \Lambda _ {C} ^ {~A} \Lambda _ {D} ^ {~B}+\l(\Lambda\p_\m \Lambda^{-1}\r)^{AB} \ ,
\ee
thus it can be interpreted as the spin connection. 

The covariant derivative, as in any gauge theory,  
is defined by   
\be
D _ A  = E ^ \m _ A \l ( \p _ \m + \f {i} {2} \omega ^ {A B} _ \m J _ {AB} \r ) \ ,
\label{coset_covd}
\ee
where we denoted with $E^\m_A$ the inverse vielbein,\footnote{The inverse vielbein is defined as
$$E^\m _ A e_{\m B} =\eta_{AB} \ , $$
and its existence is guaranteed as long as $\det \l ( \p _ \m y ^ A \r ) \neq 0$.} and at this stage, $\omega _ \m ^ {AB}$ are considered as independent degrees of freedom. To express them in terms of the vielbein, as it is usually done for torsionless gravity, we should impose some constraints. We will be back to this point shortly.

As customary, the field strength tensors, torsion $T$ and curvature $\omega$, are readily obtained by considering the commutator of two covariant derivatives acting for example on a vector field. They are respectively given by
\begin{align}
\label{tor_tens_text}
&T_{\m\n}^A= \p _ \m e _ \n ^ A - \p _ \n e _ \m ^ A - \omega _ {\m B} ^ {A}  e _ \n ^B + \omega _ {\n B} ^ {A}  e _ \m ^B \ , \\
&\omega_{\mu\nu}^{AB}=\partial_\mu \omega_\nu^{AB}-\partial_\nu \omega_\mu^{AB}-\omega_\mu^{AC}\omega_{\nu C}^{B}+\omega_\nu^{AC}\omega_{\mu C}^{B}  \ ,
\end{align}
and it can be checked that they transform covariantly, as they should.

Even though vielbein and connection are independent degrees of freedom, it should be made clear that this need not necessarily be the case. As we argued before, localizing a spacetime symmetry may not require the introduction of as many (gauge) fields as there are generators. For the case at hand, the commutators of $P_A$ and $J_{AB}$ (see~\eqref{poin_cr}), suggest that the covariant constraints 
\be
T _ {\m \n} ^ A = 0 \ ,
\label{shift_IH}
\ee
could be solved.\footnote{This should be compared with the ``standard'' inverse Higgs mechanism~\eqref{stand_inv_hig}.} We end up with the well-known 
\be
\begin{aligned}
\label{spin-con4}
\omega ^ {AB} _ \m =\bar \omega^{AB}_\m&=-\frac{1}{2}\l[E ^ {\n A} \l ( \p _ \m e _ \n ^ B - \p _ \n e _ \m ^ B \r )  - E ^ {\n B} \l ( \p _ \m e _ \n ^ A - \p _ \n e _ \m ^ A \r ) \r.
\\
&\l.- e _ {\m C} E ^ {\n A} E ^ {\lambda B} \l ( \p _ \n e _ \lambda ^ C - \p _ \lambda e _ \n ^ C \r ) \r] \ ,
\end{aligned}
\ee
clearly exhibiting that for torsionless theories, the Poincar\'e group can be made local without the connection being an independent field.

\chapter{Weyl and Ricci gauging from the coset construction}
\label{ch:Weyl_Ricci}

In this chapter, following closely~\cite{Karananas:2015eha}, we show that Weyl and Ricci gauging can be carried out in a more systematic way, by employing the coset construction. We illustrate that the relation between the structure $\Theta_{\m\n}$ composed of the Weyl vector and the Schouten tensor $S_{\m\n}$ given in~\eqref{tensor_A} and~\eqref{tensor_R} respectively, can be obtained by the analog of the inverse Higgs constraints~\eqref{stand_inv_hig}. We use the word analog, because as we have already mentioned what is usually called inverse Higgs mechanism is a constraint that can be solved algebraically with respect to a certain field (or fields). In our case (for a theory without torsion) we find a constraint that leads to the relation 
\be
(n-2)\Theta_{\m\n} \simeq S_{\m\n} \ .
\label{Ricci_Weyl}
\ee
The reason we use the symbol ``$\simeq$'', is because we want to stress that the above expression is not an equality in the sense that the field $W _ \m$ can be expressed in terms of the metric; it is clear that this equation cannot be solved algebraically. Rather, what we imply is that the combination on the left-hand side of~\eqref{Ricci_Weyl} transforms identically to the one on the right-hand side. Therefore, it can be substituted by the latter in a consistent with all the symmetries way. We also show that once the requirement of having a torsionless theory is relaxed, $W_\m$ is found to be equal to one of the irreducible components of the torsion tensor.

It should also be noted that contrary to the standard gauging of internal symmetries, the Weyl gauge field $W_\m$ appears in the covariant derivative not only with the operator of dilatations, but with the generators of Lorentz transformations as well. This happens because scale invariance is a spacetime symmetry (which does not commute with spacetime translations). In our treatment, the form of the covariant derivative follows automatically.

This chapter is organized as follows. In Sec.~\ref{sec:local_scale_trans}, we gauge scale transformations and obtain the relation between $\Theta_{\m\n}$ and $S_{\m\n}$ . In Sec.~\ref{sec:examples}, we demonstrate how Ricci gauging works by considering two examples. The first one is the purely gravitational Weyl square theory in four dimensions, and the second one is the $n$-dimensional generalization of the Riegert theory. In Sec.~\ref{sec:altern}, we discuss how Weyl gauging can take place if torsion is present in the theory. Sec.~\ref{sec:conslusions2}, contains the conclusions.

\section{Local scale transformations}
\label{sec:local_scale_trans}

In the previous chapter we showed how the gravitational interaction emerges in the context of the coset construction by gauging the Poincar\'e group. Our goal here is to obtain a Weyl-invariant theory, consequently, we will gauge scale transformations as well. In this case, the coset representative is identical to the one in~\eqref{coset_rep_poinc} and does not contain generators other than the momenta. 

For the Maurer-Cartan form we obtain
\be
\Omega ^ {-1} \l ( \p _ \m + i \tilde e ^ A _ \m P _ A + \f {i} {2} \tilde \omega ^ {AB} _ \m J _ {AB} + i \tilde W _ \m D \r ) \Omega = 
i e _ \m ^ A P _ A + \f {i} {2} \omega ^ {AB} _ \m J _ {AB} + i W _ \m D \ ,
\ee
where, as before, we denoted with $\tilde e_\m^A$ and $\tilde \omega_\m^{AB}$ the gauge fields for translations and Lorentz transformations, and in addition we introduced the fields $\tilde W_\m$ which are associated with dilatations. Notice that in the presence of the new symmetry, the relation between the quantities without and with tilde are 
\be
\begin{aligned}
&e_\m ^A=\tilde e_\m^A+\p_\m y^A -\tilde \omega_\m^{AB} y_B+\tilde W_\m y^A \ ,~~~~~\omega_\m^{AB}=\tilde \omega_\m^{AB} \ , ~~~~~W_\m=\tilde W_\m \ .
\end{aligned}
\ee

Using the analog of~\eqref{coset_rep_trans}
\be
\Omega '  =  g \Omega \bar h ^ {-1} (y,g), ~~ \text{with} ~~ \bar h = e ^ {- i t \a (y,g)} \in H = SO (n-1,1) \times \mathbb R \ ,
\ee
and the commutation relations presented in Appendix~\ref{conf_group}, one finds the transformation properties of the gauge fields

\begin{table}[H]
\centering
\bt{c | ccc}
$ $ &$e ^ {'A} _ {\m}$ & $\omega ^ {' AB} _ {\m} $ & $W' _ {\m}$ \\
\hline
$J$ & $ e _ {\m} ^ B \Lambda _ B ^ {~A}$ & $\omega ^ {CD} _ {\m} \Lambda _ {C} ^ {~A} \Lambda _ {D} ^ {~B}+\l(\Lambda\p_\m \Lambda^{-1}\r)^{AB}$ & $W _ {\m}$ \\
$D$ & $ e ^ {-\a} e _ {\m} ^ A $ & $\omega ^ {AB} _ {\m}$ & $W _ {\m} + \p _ \m \a $
\et
\end{table}

The transformations of $e _ \m ^ A$, $\omega _ \m ^ {AB}$ and $W_ \m$ are precisely the ones for the vielbein, spin connection and the Weyl gauge field. According to the rules of the coset construction, the covariant derivative of a matter field $\psi$ is now given by
\be
D _ A \psi  = E^ \m _ A \l ( \p _ \m + \f {i} {2} \omega ^ {A B} _ \m J _ {AB} + i W _ \m D \r ) \psi \ .
\label{coset_covd}
\ee

It is clear that by analogy with the previous chapter, one can construct generalized field strength tensors corresponding to shifts, Lorentz and also scale transformations
\begin{align}
\centering
\label{tensor-dil-1}
e _{ \m \n } ^ A & =  \p _ \m e _ \n ^ A - \p _ \n e _ \m ^ A - \omega _ {\m B} ^ {A}  e _ \n ^B + \omega _ {\n B} ^ {A}  e _ \m ^B +
W _ \m e _ \n ^ A - W _ \n e _ \m ^ A\ ,  \\
\label{tensor-dil-2}
\omega ^ {AB} _ {\m \n} & =  \p _ \m \omega _ \n ^ {AB} - \p _ \n \omega _ \m ^ {AB} 
- \omega _ {\m C} ^ {A} \omega _ \n ^{CB} + \omega _ {\n C} ^ {A} \omega _ \m ^{CB}\ ,  \\
\label{tensor-dil-3}
W _ {\m \n} & =  \p _ \m W_ \n - \p _ \n W _ \m \ , 
\end{align}
that transform covariantly

\begin{table}[H]
\centering
\bt{c | cccc}
$ $ &$e ^ {'A} _ {\m \n}$ & $\omega ^ {' AB} _ {\m \n} $ & $W' _ {\m \n}$ \\
\hline
$J$ & $ e _ {\m \n} ^ B \Lambda _ B ^ {~A}$ & $\omega ^ {CD} _ {\m \n} \Lambda _ {C} ^ {~A} \Lambda _ {D} ^ {~B}$ & $W _ {\m \n}$ \\
$D$ & $ e ^ {-\a} e _ {\m \n} ^ A $ & $\omega ^ {AB} _ {\m \n}$ & $W _ {\m \n}$
\et
\end{table}

Inspection of the commutation relations given in Appendix~\ref{conf_group}, reveals that once we set
\be
e _ {\m \n} ^ A = 0 \ ,
\label{shift_IH}
\ee
we obtain
\be
\label{spin-con3}
\omega ^ {AB} _ \m = \bar \omega^{AB}_\m + \d\omega^{AB}_\m \ ,
\ee
where $ \bar \omega^{AB}_\m$ is the standard spin connection for a torsionless theory presented in~\eqref{spin-con4}, and
\be
\label{spin-con5}
\d\omega_\m^{AB}= I_{\m \n}^{A B} W^ \n \ , \ \ \  I_{\m \n}^{A B}= e_\n^A e_\m^B-e_\n^B e_\m^A \ .
\ee
Plugging the expression for $\omega$ to the definition of the covariant derivative~\eqref{coset_covd}, we find that it can rewritten as follows
\be
D _ A \psi  = E ^ \m _ A \l ( \p _ \m + \f {i} {2} \bar \omega ^ {A B} _ \m J _ {AB} - i e ^ A _ \m e ^ B _ \n W ^ \n J _ {A B} + i W _ \m D \r ) \psi \ .
\label{coset_covd_e}
\ee
In particular, for a vector field $V ^ A$ with scaling dimension $\D _ V$, we get
\be
e ^ B _ \m D _ B V ^ A = \p _ \m V ^ A - \bar \omega ^ A _ {\m B} V ^ B + ( e _ \m ^ A E _ B ^ {\n} - E ^ {\n A} e _ {\m B} )V ^ B W _ \n - \D _ V W _ \m V ^ A \ .
\label{covd_vector}
\ee
We can clearly see now the reason why the Weyl gauge field ``couples'' to  spin as well. Using the Christoffel symbols defined in Appendix~\ref{Christoffel}, one can show that the expression for the covariant derivative~\eqref{coset_covd_e} coincides with the one used in~\cite{Iorio:1996ad}.

Notice that the field strength tensor corresponding to shifts is not the only covariant structure. Even though imposing another constraint is not in the spirit of the standard inverse Higgs mechanism, it can be done consistently.\footnote{For pure Poincar\'e invariance that we studied in chapter~\ref{chapt:Poincaregrav}, no additional constraint could be imposed, since there are no candidates for elimination, provided one wants to obtain dynamical gravity.} The gauge field $\omega _ \m ^ {AB}$ depends on $W_ \m$; therefore, we may hope to relate certain structure depending on this vector to a tensor that depends only on the vielbein.

Plugging the expression (\ref{spin-con3}) to the formula (\ref{tensor-dil-2}), we get
\be
\omega ^ {AB} _ {\m \n} = \bar \omega ^ {AB} _ {\m \n} + \d \omega ^ {AB} _ {\m \n},
\label{omega_full}
\ee
with $\bar \omega _{\m\n}^{AB}$ given by~\eqref{barRiem} and repeated here for the convenience of the reader
\be
\label{barR}
\bar \omega _{\m\n}^{AB} = \p _ \m \bar\omega _ \n ^ {A B} - \p _ \n \bar\omega _ \m ^ {A B} 
-\bar \omega _ {\m C} ^ {A} \bar\omega _ \n ^{CB} +\bar \omega _ {\n C} ^ {A} \bar\omega _ \m ^{CB} \ ,
\ee
and
\be
\begin{aligned}
\label{del_omega}
\d \omega ^ {AB} _ {\m \n} & = I^{AB}_{\n\lambda}\nabla_\m W^\lambda -I^{AB}_{\m\lambda}\nabla_\n W^\lambda 
+\l(e^A_\m e^B_\n-e^B_\m e^A_\n\r)W^2 \\
&+
\l(e^A_\n W^B-e^B_\n W^A\r)W_\m - \l(e^A_\m W^B-e^B_\m W^A\r)W_\n \ , 
\end{aligned}
\ee
where we used the vielbein to manipulate the indices of $W _ \m$, so that $W^2=W_B W^B=W_\m W^\m$.

None of the constraints imposed on $\omega ^ {A B} _ {\m \n}$, although consistent with its transformation properties, can be solved algebraically with respect to $W _ \m$. Nevertheless, imposing
\be
\omega _ {\m \n} + \omega _ {\n \m} \simeq 0 \ , \ \ \  \text{with} \ \ \  \omega _ {\m \n} \equiv \omega ^ {A B} _ {\m \s} E ^ {\s} _ B e _ {\n A} \ ,
\label{omega_constr}
\ee
and using~\eqref{omega_full} leads to~\eqref{Ricci_Weyl}, which coincides with the expression obtained in~\cite{Iorio:1996ad}, except that we use a different convention for the Riemann curvature tensor, see~\eqref{ricci-conv}.

The substitution $S _ {\m \n}$ for $\Theta _ {\m \n}$ is similar in  spirit to the standard inverse Higgs phenomenon, according to which, certain degrees of freedom are not needed to realize a symmetry breaking pattern and as a result, they can be eliminated. Note, however, that the opposite substitution is not legitimate (at least not for arbitrary field configurations), since the Schouten tensor is subject to the Bianchi identity
\be
\nabla ^ \m S _ {\m \n} - \nabla _ \n S = 0 \ ,
\label{Bianchi_Schouten}
\ee
which is not satisfied by $\Theta _ {\m \n}$.

In~\cite{Iorio:1996ad}, it was shown that the substitution~\eqref{Ricci_Weyl} can always be made for conformal (in flat spacetime) theories with at most one derivative of conformally variant fields. In this case, the invariance under Weyl rescalings does not require the introduction of extra degrees of freedom, since the inhomogeneous pieces of the transformation that appear in the derivatives can be compensated for by curvature terms. 

It should also be noted that the constraint~\eqref{omega_constr} taken as an equality, only implies the equivalence between the Schouten tensor $S _ {\m \n}$ and the symmetric part of $\Theta _ {\m \n}$. However, in a weaker sense (that is, equivalence of the transformation properties), it is possible to relate 
$S _ {\m \n}$  to the full $\Theta_{\m\n}$. In fact, the antisymmetric part is given by
\be
2 \Theta _ {\m \n} ^ {\text{anti}} = \p _ \m W _ \n - \p _ \n W _ \m  = W _  {\m \n} \ ,
\ee
which is invariant under Weyl transformations and can be safely added to $\Theta _ {\m \n} ^ {\text{sym}}$, resulting in~\eqref{Ricci_Weyl}.

\section{Examples}
\label{sec:examples}

\subsection{Weyl tensor}

As a first example, we build the Weyl-invariant action for pure gravity in a four dimensional spacetime, without coupling to matter.  After imposing the constraint $e ^ A _ {\m \n} = 0$, we are left with three objects: two Weyl-invariant curvatures $\omega ^ {A B} _ {\m \n}$ and $W _ {\m \n}$, and the Weyl covariant vielbein $e ^ A _ \m$. In order to account for the noninvariance of the measure $d ^ 4 x \, \det e$,  it should be multiplied four times by the inverted vielbein
\be
\int d ^ 4 x \, \det e \, E ^ \m _ A \, E ^ \n _ B \, E ^ \lambda _ C \, E ^ \s _ D \ .
\label{measure_W}
\ee
The lowest-order (in derivatives) diffeomorphism-invariant action, which also respects the gauged scale and Poincar\'e symmetries 
can be obtained by all possible contractions of (\ref{measure_W}) with
\be
W _ {\m \n} W _ {\lambda \s} ~~~\text{and} ~~~\omega ^ {AB} _ {\m \n} \omega ^ {AB} _ {\lambda \s} \ .
\ee
The first term leads to the following obviously Weyl-invariant action (we do not assume parity invariance)
\be
S _ 1 = \int d ^ 4 x \, \det e \, \l ( c _ 1 W _ {\m \n} W ^ {\m \n} + c _ 2 \ep ^ {\m \n \lambda \s} W _ {\m \n} W_ {\lambda\s} \r ),
\label{A_W_inv}
\ee
 with $c_1$ and $c_2$ being constants and $\ep ^ {\m \n \lambda \s} = E ^ \m _ A \, E ^ \n _ B \, E ^ \lambda _ C \, E ^ \s_D \ep ^ {ABCD}$. The contractions with $\omega ^ {AB} _ {\m \n} \omega ^ {AB} _ {\lambda \s}$ can be simplified once the constraint~\eqref{omega_constr} is imposed. The antisymmetric part of $\omega _ {\m \n}$ from~\eqref{omega_constr} is proportional to $\p _ \m W _ \n - \p _ \n W _ \m$, which already has been taken into account in~\eqref{A_W_inv}. We may thus consider only configurations with $\omega ^ {A B} _ {\m \n} E ^ \n _ B = 0$. As a result, the only possible contractions are the following
\be
\label{conf-inv-act2}
\ep^{IJKL}E^\m_I E^\n_J E^\rho_K E^\s_L\ep_{ABCD}~\omega_{\m\n}^{AB}\omega_{\rho\s}^{CD} \ , 
\ee
and
\be
\label{conf-inv-act3}
\ep^{IJKL}E^\m_I E^\n_J  E^\rho_K E^\s_L~\eta_{AC}\eta_{BD} \omega_{\m\n}^{AB}\omega_{\rho\s}^{CD}\ , \ \ \ \ep^{IJKL}E^\m_I E^\n_J   E^\rho_A E^\s_B~\eta_{KC}\eta_{LD} \omega_{\m\n}^{AB}\omega_{\rho\s}^{CD} \ .
\ee 
Simplifying these expressions, leads to
\be
S _ 2 = \int d ^ 4 x \, \det e \, \l ( c _ 3 C _ {\m \n \lambda \s} C ^ {\m \n \lambda \s} + 
c _ 4 \ep ^ {\kappa \rho \lambda \s} C _ {\m \n \kappa \rho}  C ^ {\m \n} _ {~~~\lambda \s} \r ),
\ee
where $c_3$, $c_4$ are constants, and $C _ {\m \n \lambda \s}$ is the Weyl tensor 
\be
\begin{aligned}
\label{weyl_tens}
C _ {\m \n \lambda \s}&=R _ {\m \n \lambda \s}+\f{1}{n-2}\l( g_{\n\lambda}R_{\m\s}+g_{\m\s}R_{\n\lambda}-g_{\n\s}R_{\m\lambda}-g_{\m\lambda}R_{\n\s}\r)\\
&+\f{1}{(n-1)(n-2)}\l(g_{\m\lambda}g_{\n\s}-g_{\m\s}g_{\n\lambda} \r)R \ .
\end{aligned}
\ee

\subsection{Higher derivative action}
\label{sec:higher_der_coset}

Here, we wish to get a better grasp on the range of applicability of Ricci gauging. To be more precise, we want to understand whether or not the presence of more than one derivative of a conformally variant field constitutes an obstruction in the Ricci gauging. To achieve that, we consider a theory with a higher number of derivatives of a scalar field, namely, a conformally invariant theory in an $n$-dimensional flat spacetime given by the following action 
\be
S_{\Box^2}=\int d^n x ( \Box \phi ) ^ 2 \ .
\ee
According to the coset construction described previously, we introduce the covariant derivative~\eqref{coset_covd_e} for the field $\phi$ in the following way
\be
D _ A \phi = E ^ \m _ A \l ( \nabla _ \m \phi - \D_\phi W _ \m \phi \r ) \ ,
\ee
where $\D_\phi = \f {n} {2} - 2$ is the scaling (mass) dimension of $\phi$. Therefore,
\be
\begin{aligned}
e ^ B _ \m D _ B D _ A \phi &= \p _ \m D _ A \phi - \bar \omega _ {\m A } ^ {~~B} D _ B \phi + ( e _ {\m  A}E ^ {\n B} - E ^ {\n} _ A e _ {\m} ^ B ) D _ B \phi W _ \n\\
& - ( \D_\phi + 1 ) W _ \m D _ A \phi \ ,
\end{aligned}
\ee
where we used the fact that the scaling dimension of $D _ A \phi$ is equal to $\D_\phi+1$. As a result, the following substitution
\be
\Box \phi \to D _ A D ^ A \phi = \nabla ^ 2 \phi + 2 W _ \m \nabla ^ \m \phi - \l ( \f {n} {2} - 2 \r ) \l ( \nabla _ \m W ^ \m + \f {n} {2} W ^ \m W _ \m \r ) \ ,
\ee
where $\nabla ^ 2 = g ^ {\m \n} \nabla _ \m \nabla _ \n$, leads to the Weyl-invariant action
\be
S _ {\Box ^ 2} = \int d ^ n x \, \det e ( D _ A D ^ A \phi ) ^ 2 \ .
\label{Lagr_box_2}
\ee
The question we would like to address now is whether it is possible to use Ricci gauging [or, equivalently, the weak form of the constraint~\eqref{Ricci_Weyl}] to completely get rid of the field $W _ \m$. Lengthy but straightforward calculations lead to 
\be
\begin{aligned}
S_ {\Box ^ 2} & = \int d^nx \det e~\left\{\vphantom{\frac{A}{B}}( \nabla ^ 2 \phi ) ^ 2 - \l [ 4 \Theta _ {\m \n} - (n-2) \Theta g _ {\m \n} \r ] \nabla ^ \m \phi \nabla ^ \n \phi\right. \\
&\left. -  \phi ^ 2 \l [ \f {n-4} {2} \nabla ^ 2 \Theta + (n-4) \Theta _ {\m \n} \Theta ^ {\m \n} - \f {n (n-4)} {4} \Theta ^ 2 \r ]\right.  \\
&\left.- \phi ^ 2 (n - 4) W ^ \n \nabla ^ \m \l (  \Theta _ {\m \n} - g _{\m\n} \Theta \r ) \vphantom{\frac{A}{B}}\right\}\ ,
\end{aligned}
\ee
where $\Theta = \Theta ^ \m _ \m$. Notice that the dependence of the action on $W _ \m$ for $n= 4$ is only through the tensor $\Theta _ {\m \n}$ and Ricci gauging can be used without any trouble. Although, for general $n$, there is an explicit $W _ \m$ dependence in the last term, it is clear that after the substitution (we assume $n \neq 2$)
\be
\Theta _ {\m \n} \to \f {1} {n-2} S _ {\m \n} \ ,
\ee
this term drops out by virtue of the Bianchi identity~\eqref{Bianchi_Schouten}. Therefore, it is shown that the theory given by the Lagrangian (\ref{Lagr_box_2}) can be Ricci gauged in an arbitrary (not equal to two) number of dimensions. The resulting action can be written in the following form 
\be
S _ {\Box ^ 2} = \int d ^ n x \sqrt {g} \phi \mc Q_4 (g)  \phi,
\ee
with
\be
\begin{aligned}
\mc Q_4 (g) &= \nabla ^ 2 + \nabla ^ \m  \l[\l(\frac{4} {n-2} S _ {\m \n} -  g _ {\m \n}S \r ) \nabla ^ \n\r] - \frac{n-4} {2 (n-2)} \nabla ^ 2 S \\
&- \frac{n-4}{(n-2)^2} S _ {\m \n} S ^ {\m \n} + \frac{n (n-4)}{4 (n-2)^2} S ^ 2 \ , 
\label{pan-rieg}
\end{aligned}
\ee
being the Paneitz operator~\cite{Paneitz:1983_2008}, which is the Weyl covariant generalization of $\Box ^ 2$, see also Appendix~\ref{riegert}.\footnote{In a four dimensional  space-time, the Paneitz operator is also known as Paneitz-Riegert operator and it was constructed by different authors~\cite{Fradkin:1981jc,Fradkin:1981iu,Fradkin:1982xc,Riegert:1984kt}.}

At this point, it is natural to wonder what happens when $n\to 2$. In this limit, the coefficients in front of the Schouten tensor~\eqref{tensor_R} diverge. At the same time, the Schouten tensor itself vanishes due to the following relation between Ricci curvatures in two dimensions
\be
\label{ricc_2d}
R_{\m\n}=\f R 2 g_{\m\n} \ .
\ee
Therefore, this is a case that has to be examined separately. Actually, as we will prove in the next chapter, it is not possible to construct a fourth-order Weyl covariant operator in $n=2$ spacetime dimensions. Based on this observation, we will be able to show that even though the notions of Weyl and conformal invariance are used interchangeably, the former is not just the curved-space generalization of the latter, but rather a different concept.

\section{Torsionful theory}
\label{sec:altern}

The field strength corresponding to shifts $e^ A _ {\m \n}$ and the generalized spin connection $\omega ^ {A B} _ \m$ have the same symmetry properties; therefore, they have equal number of independent components. This is the reason why we were able to solve the inverse Higgs constraint~\eqref{shift_IH} with respect to the  $\omega ^ {A B} _ \m$ and express it in terms of the vielbein and the Weyl vector field $W_\m$. This way, we built a Weyl-invariant torsionless theory. Here we look for an alternative solution to this constraint. 

In order to understand what the possible solutions might be, we should analyze the structure of irreducible representations of $e ^ A _ {\m \n}$, since they can be set to zero independently. Any tensor that possesses the symmetries of the quantity $e ^ A _ {\m \n}$, admits the following decomposition in an $n$-dimensional spacetime (see also Appendix~\ref{irred_decomp}). A vector,
\be
\varepsilon _ \m =E ^ \n _ A e ^ A _ {\m \n} \ ,
\label{vect}
\ee
a completely antisymmetric tensor
\be
\mathscr A ^ { \s _ 1 \dots \s _ {n-3}} = \frac{1}{n} \,\ep^{\s_1\s_2\cdots \s_{n-3}\m\n\lambda}e_{\lambda A}\, e_{\m\n}^A \ ,
\label{antisym}
\ee
and a traceless tensor with mixed symmetries
\be
\mathscr E_{\m\n}^A=e_{\m\n}^A-\frac{3}{2(n-1)}\l(\varepsilon _\m e_\n^A- \varepsilon_\n e^A_\m \r)-\frac{1}{2}E^{\lambda A}\l(e^B_{\lambda\m} e_{\n B}-e^B_{\lambda\n} e_{\m B}\r) \ .
\label{traceless}
\ee
Written in this form, the constraints~\eqref{shift_IH}, make it clear that~\eqref{antisym} and~\eqref{traceless} can only be solved with respect to their counterparts contained in $\omega ^ {A B} _ \m$. However, for the vector part~\eqref{vect} there are two options. The first one, which has been chosen in the previous section, is to eliminate the vectorial part of the spin connection. The second one is to solve the constraint with respect to $W _ \m$, keeping $\omega ^ {AB} _ \m E^ \m _ B$ undetermined, which yields a torsionful theory.

We see from~\eqref{tensor-dil-1} that 
\be
\label{tors-A-1}
W_\m e_\n^A-W_\n e_\m^A= -T_{\m\n}^A \ ,
\ee
where the torsion tensor $T_{\m\n}^A$ was defined in~\eqref{tor_tens_text} and reads
\be
\label{tors-tens-1}
T_{\m\n}^A\equiv \p _ \m e _ \n ^ A - \p _ \n e _ \m ^ A - \omega _ {\m B} ^ {A}  e _ \n ^B + \omega _ {\n B} ^ {A}  e _ \m ^B \ , 
\ee
Tracing~\eqref{tors-A-1}, we obtain
\be
\label{tors-A-2}
W_\m=-\frac{1}{n-1} \upsilon_\m \ ,
\ee
where we denoted with  $\upsilon_\m$ the torsion vector
\be
\label{tors-A-4}
\upsilon_\m= E^\n_AT_{\m\n}^A= E^\n_A\l (\p _ \m e _ \n ^ A - \p _ \n e _ \m ^ A + \omega _ {\n B} ^ {A}  e _ \m ^B \r) \ .
\ee

It is straightforward to check that under Weyl rescalings the vector $\upsilon _ \m$ transforms exactly as the Weyl field, i.e. 
\be
\upsilon'_\m=\upsilon_\m-(n-1)\p_\m \alpha \ .
\ee
As a result, once we consider nonvanishing torsion, the degrees of freedom carried by $W_\m$ can be traded for the vector $\upsilon_\m$.

\section{Summary and Outlook}
\label{sec:conslusions2}

In this chapter we touched upon the question of whether the conformal invariance of a system in flat spacetime implies that the system can be coupled to gravity in a Weyl-invariant way. We used the prescription of the standard coset construction in order to gauge scale transformations (along with the Poincar\'e group), leading to a Weyl-invariant (in curved spacetime) theory. It was demonstrated that the main ingredient needed for Ricci gauging, namely the relation between the additional gauge field corresponding to the local scale transformations and the Ricci curvature -- first obtained in~\cite{Iorio:1996ad} -- can be extracted from the analog of the inverse Higgs constraint. 

This revealed that the two structures~\eqref{Ricci_Weyl} transform in the same way, and therefore, whenever the tensor $\Theta _ {\m \n}$ appears in the action, it can be substituted by its counterpart without any contradiction with the underlying symmetries. The answer to the question of whether such a prescription for conformally invariant theories leads to a complete elimination of the gauge field $W _ \m$ does not have a definite answer at the moment and can only be divined. 

We presented a couple of examples of how Ricci gauging works. First, we obtained the Weyl-invariant action for pure gravity in four spacetime dimensions, which is given, as is well known, by the square of the Weyl tensor. Next, we considered a theory with more than one derivative of a scalar field~\eqref{Lagr_box_2}. In a four dimensional spacetime, the Ricci gauging can be straightforwardly employed. However, it so happens that the scaling dimension of the field is zero in this case; thus, the field is actually conformally invariant. Notice that there is no contradiction with~\cite{Iorio:1996ad}, since the condition of having at most one derivative was only imposed on conformally variant fields.

Considering the system in $n \neq 2$, we showed that Ricci gauging can be applied even for theories with more than one derivative of conformally variant fields. In the example we considered, the procedure turned out to be a little bit subtle. Namely, the Weyl gauged Lagrangian cannot be written as a function depending only on $\Theta _ {\m \n}$, but rather, it also depends explicitly on $W _ \m$. However, this dependence drops out, once Ricci gauging is performed.

Finally, we also presented an alternative way of introducing the Weyl symmetry. We showed, by solving the inverse Higgs constraint, that the role of the gauge field associated with local scale transformations can be played by the vector part of the torsion tensor.

\chapter{Weyl vs. Conformal}
\label{ch:Weyl_vs_Conf}

\section{Introduction}

The purpose of this chapter, based to a large extent on~\cite{Karananas:2015ioa}, is to clarify the difference between the occasionally mixed notions of conformal and Weyl invariance. The conformal symmetry in a $n$-dimensional (not necessarily flat) space-time is defined as the group of coordinate transformations
\be
x' = F (x) \ ,
\ee
which leave the metric $g_{\m\n}$ invariant up to a conformal factor
\be
g _ {\m \n} (x) = \Omega (x') \, g _{\lambda \s}' (x') \, \f {\p F ^ \lambda} {\p x ^ \m} \f {\p F ^ \s} {\p x ^ \n} \ .
\label{metr_trans}
\ee
For the infinitesimal form of the transformations
\be
x'^{\m} = x ^ \m + f ^ \m \ ,
\ee
the relation~\eqref{metr_trans} leads to the conformal Killing equations
\be
\nabla _ \m f _ \s + \nabla _ \s f _ \m = \f {2} {n} g _ {\m \s} \nabla f \ ,
\label{conf_K}
\ee
where we used the shorthand notation $\nabla f = \nabla _ \m f ^ \m$, and we denoted with $\nabla$ the metric-compatible covariant derivative
\be
\nabla _ \m f _ \n = \p _ \m f _ \n - \G ^ \lambda _ {\m \n} f _ \lambda \ ,
\ee
with $\Gamma^\lambda_{\m\n}$ being the Christoffel symbols.

Here, we focus only on theories with scalars, leaving the investigation of fields with non-zero spin for elsewhere. The infinitesimal transformation of a scalar field with scaling (mass) dimension $\D_\phi$ under the full conformal group can be written in the following compact form
\be
\d _ c \phi = - \l ( f ^ \m \nabla _ \m \phi + \f {\D_\phi} {n} \nabla f \phi \r ).
\label{delta_c_phi}
\ee
A system is called conformally invariant if the variation of its action functional $S[g_{\m\n},\phi]$ under the full group of conformal transformations~\eqref{delta_c_phi} is zero, i.e.
\be
\d _ c S[g_ {\m \n}, \phi] = \int d ^ n x \, \f {\d S} {\d \phi} \, \d _ c \phi = 0 \ .
\label{conf_S}
\ee

Meanwhile, Weyl rescalings constitute another type of transformations, which are given by the simultaneous pointwise transformations of the metric and fields
\be
\begin{aligned}
\hat g _ {\m \n} (x) & = e ^ {2 \s (x)} g _ {\m \n} (x)~~~\text{and}~~~\hat \phi (x) & = e ^ {- \D_\phi \s} \phi (x) \ ,
\end{aligned}
\ee
with $\s$ being an arbitrary function. Writing the above expressions in their infinitesimal form as
\be
\d _ \s g _ {\m \n} = 2 \s g _ {\m \n} ~~~ \text {and} ~~~ \d _ \s \phi = - \D_\phi \s \phi \ ,
\label{Weyl_inf}
\ee
leads to the following condition for a theory to be Weyl invariant
\be
\d _\s S[g _ {\m \n}, \phi] = \int d ^ n x \, \s \l ( 2 \, \f {\d S} {\d g _ {\m \n}} g _ {\m \n} - \D_\phi \, \f {\d S} {\d \phi} \phi \r ) = 0 \ .
\label{conf_W}
\ee

Note that~\eqref{delta_c_phi} can be written as
\be
\d _ c \phi = \d _ d \phi + \d _ {\bar \s} \phi \ ,
\ee
where we denoted by $ \d _ {\bar \s}\phi$ the Weyl transformation corresponding to the  specific value of $\s = \bar \s \equiv \nabla f / n$, and $\d _ d \phi$ is the standard transformation of the scalar field under the general coordinate transformations
\be
\d _ d \phi = - f ^ \m \p _ \m \phi \ .
\ee
As a result, equation~\eqref{conf_S} can be rewritten as
\be
0 = \int d ^ n x \, \f {\nabla f} {n} \l ( 2 \, \f {\d S} {\d g _ {\m \n}} g _ {\m \n} - \D_\phi \, \f {\d S} {\d \phi} \phi \r ) \ ,
\ee
where we used the fact that the $\d _ d \phi$ transformations can be compensated for by the corresponding transformations of the metric (provided the theory is diffeomorphism invariant). It is clear that Weyl invariance implies conformal invariance, but not the other way around, since $\nabla f $ is not an arbitrary function of coordinates.

\section{Examples}

Let us present another way to understand why Weyl invariance  necessarily implies conformal invariance in flat space-time. The corresponding conformal Killing equations now read
\be
\p _ \m \eps _ \n + \p _ \n \eps _ \m = \f {2} {n} \eta _ {\m \n} \p_\lambda \eps^\lambda \ ,
\ee
with $\eta_{\m\n}=\text{diag}\l(1,-1,\ldots\r)$, the Minkowski metric, and $\e ^ \m$ being the flat space-time analog of~$f ^ \m$. This set of equations has the following $(n+1)(n+2)/2$ parametric solution for $n \neq 2$
\be
\eps ^  \m = a ^ \m + \omega ^ \m _ {~\n} x ^ \n + c x ^ \m + 2 (b\cdot x) x ^ \m - x ^ 2 b ^ \m \ .
\label{Killings_flat}
\ee
Here $a ^ \m$, $\omega _ {\m\n} = -\omega _ {\n \m}$, $c$ and $b ^ \m$ are constants corresponding to translations, Lorentz transformations, dilatations and special conformal transformations (SCT) respectively. In two dimensions, $\e ^ \m$ is given by an arbitrary generalized harmonic function.\footnote{An example of the integrated version of the equation~\eqref{delta_c_phi} is the transformation of a scalar field under the SCT which is given by
\be
\phi ' (x') = (1 - 2\, b\cdot x + b ^ 2 x^2)^ {\D_\phi} \phi (x) \ , ~~~\text{with}~~~ x'^{\m} = \f {x ^ \m - b ^ \m x^2} {1 - 2\, b\cdot x + b ^ 2 x^2} \ .
\ee}

The standard procedure allows one to build the energy-momentum tensor
\be
T^{\text{impr}} _ {\m \n} = 2 \f {\d S} {\d g ^ {\m \n}} \Big | _ {g _ {\m \n} = \eta _ {\m \n}} \ ,
\ee
which is automatically traceless on the equations of motion, see~\eqref{conf_W}. As a result, all currents of the form $j _ \m = T^{\text{impr}} _ {\m \n} \eps ^ \n$, with $\e ^ \m$ given in~\eqref{Killings_flat}, are conserved.  

Conversely, if a theory is conformally invariant, then according to~\cite{Wess:conf,Polchinski:1987dy}, it is possible to write all currents corresponding to the conformal group in the following way
\be
j _ \m = T _ {\m \n} \eps ^ \n - \p \eps K _ \m + \p ^ \n \p \eps L _ {\m \n} \ ,
\label{conserved_curr}
\ee
where $T _ {\m \n}$ is the energy-momentum tensor (not necessarily traceless), $K _ \m$ is a vector and $L _ {\m \n}$ is a rank-two tensor such that\be
\p _ \m T _ {\m \n} = 0\ , \ \ \  T _ {\m \n } = T _ {\n \m}\ , \ \ \ T _ {\m} ^ \m = n \, \p _ \m K ^ \m \ \ \ \text{and} \ \ \ K _ \m = \p ^ \n L _ {\n \m} \ .
\label{T_K_L_n}
\ee
Notice that for $n=2$, there is an additional restriction 
\be
L _ {\m \n} = \eta _ {\m \n} L \ ,
\label{T_K_L_2}
\ee
with $L$ being a scalar function.

The conditions presented above allow to construct the improved (traceless) energy momentum tensor $T^{\text{impr}} _ {\m \n}$. 

However, it is not guaranteed that the theory can be made Weyl invariant. In what follows, we will consider several examples of conformally invariant theories which cannot be made Weyl invariant when coupled to gravity. We should mention though,  that we will not consider theories with non-linearly realized space-time symmetries, like in the case of galileons~\cite{Nicolis:2008in}. There, the reason that the conformal invariance of a certain action for the galileon does not imply Weyl invariance, is associated with the fact that this action is actually a Wess-Zumino term, see also~\cite{Goon:2012dy}.

\subsection{$\Box$}
\label{subsec:box}

For the purposes of illustration, it is instructive to begin by considering the Lagrangian of a free massless field in a one-dimensional spacetime
\be
\label{box}
\mc L =\f{\dot\phi^2}{2} \ . 
\ee
If the scaling dimension of $\phi$ is $\D_\phi=-1/2$, then the theory is invariant under the one-dimensional conformal group
\be
\d\phi=-\l(\eps \dot\phi-\f{1}{2}\phi \dot\eps\r) \ ,
\ee
where
\be
\eps= a+b\, t+\f{c}{2} \, t^2 \ ,
\ee
with $a,b$ and $c$, constants. The conserved currents associated with translations, dilatations and special conformal transformations  can be written according to~\eqref{conserved_curr} as
\be
J = \frac{\dot \phi ^ 2}{2} \eps - \f{\phi\dot\phi}{2} \dot \eps + \f{\phi^2}{4} \ddot \eps.
\ee
Clearly, this theory cannot be made Weyl invariant, for there are no geometric structures in $n=1$ one could use to account for the non-invariance of $\dot \phi ^ 2$.

\subsection{$\Box ^ 2$}
\label{subsec:box2}

Let us now consider the theory given by the following Lagrangian
\be
\mc L _ {\Box^2} = \f {1} {2} ( \Box \phi ) ^2 \ ,
\label{box2}
\ee
with $\Box=\eta^{\m\n}\p_\m \p_\n$ the D'Alembertian. Using the flat space-time analog of formula~\eqref{delta_c_phi} with 
$\D_\phi = n/2 - 2$, it is straightforward to check that in $n \neq 2$, the variation of this Lagrangian is given by a total derivative
\be
\d \mc L _ {\B ^ 2} = - \p^ \m \l [ \eps _ \m \mc L _ {\Box^2} - \f {2} {n} \p ^ \n \p \e \l ( \p _ \m \phi \p _ \n \phi - \f {1} {2} \eta _ {\m \n} ( \p \phi ) ^ 2 \r ) \r ] \ .
\ee
In this case, using the following definitions
\be
\begin{aligned}
T _ {\m \n} & = \eta _ {\m \n} \l( \p _ \lambda \phi \p ^ \lambda \B \phi  + \f {1} {2}  \l ( \B \phi \r ) ^ 2\r ) - \p _ \m \B  \phi \p _ \n \phi - \p _ \n \B \phi \p _ \m \phi \ , \\
K _ \m & = \f {1} {2} \B \phi \p _ \m \phi + \f {\D_\phi} {n}  \phi \p _ \m \B  \phi \ , \\
L _ {\m \n} & = \f {1} {n} \l ( 2 \p _ \n \phi \p _ \m \phi - \eta _ {\m \n} \l ( \p \phi \r ) ^ 2 + \D_\phi \eta _ {\m \n} \phi \B  \phi \r ) \ ,
\label{K_L_b2}
\end{aligned}
\ee
it is straightforward to check that the relations presented in~\eqref{T_K_L_n} are satisfied. Therefore, the system is indeed conformally invariant for $n \neq 2$.

As we showed in detail in Sec.~\ref{sec:higher_der_coset}, the Weyl-invariant generalization of the theory~\eqref{box2}, reads\,\footnote{Obviously, we are not forced to resort to the coset construction in order to couple this theory to gravity in a Weyl invariant manner. It suffices to write down the most general action with four derivatives and demand that it be invariant under Weyl rescalings. Notice that by doing so, there will also be a contribution proportional to Weyl tensor squared. Since this term is invariant by itself, it need not be included.}
\be
S _ {\Box ^ 2} = \int d ^ n x \sqrt {g} \phi \, \mc Q_4 (g) \, \phi \ ,
\ee
with $Q_4 (g)$ the Paneitz operator that was defined in~\eqref{pan-rieg}. 
As we mentioned in the previous chapter, the above expression is not well defined for $n\to 2$. To understand what is going on in this limit, let us consider the most general ansatz for the operator $Q_4 (g)$ in two dimensions 
\be
\mc Q_4(g) =\nabla^4 +c_1\nabla^\m\l( R\nabla_\m\r)+c_2 \nabla^2 R +c_3 R^2 \ ,
\ee
with $c_1, c_2$ and $c_3$ constants. A straightforward calculation shows the Weyl variation of $\nabla^4$ will produce terms that cannot be cancelled by the variation of $R$-dependent terms, for example $\l ( \nabla^\m \nabla _ \n \s \r ) \nabla _ \m \nabla _ \n$. Therefore, for $n=2$ there is no Weyl covariant generalization of the fourth-order differential operator. Hence, in this case, the system~\eqref{box2}  cannot be coupled to gravity in a Weyl invariant way, although this does not come as a surprise, for as it is clear from~\eqref{K_L_b2}, the condition~\eqref{T_K_L_2} is not satisfied. One can say that the system at hand in a two dimensional space-time, is only invariant under global conformal transformations, which correspond to the six dimensional sub-algebra of the Virasoro algebra. Let us note that global conformal transformations are defined on the two dimensional sphere. The non-zero commutation relations are 
\begin{align*}
&\l [ l_{-1},l_0 \r ] = -l_{-1} \ , ~~~ \l [ l_{-1},l_1 \r ] = -2l_{0} \ , ~~~\l [ l_0,l_1 \r ] = -l_{1} \ ,\\
&\l [ \bar l_{-1},\bar l_0 \r ] = -\bar l_{-1} \ , ~~~ \l [ \bar l_{-1},\bar l_1 \r ] = -2\bar l_{0} \ , ~~~\l [ \bar l_0,\bar l_1 \r ] = -\bar l_{1} \ , 
\end{align*}
where the generators in terms of the complex coordinates $z$ and $\bar z$, read
\begin{align*}
\displaystyle& l_{-1}=-\p_z \ , ~~~\bar l_{-1}=-\p_{\bar z}\,~~~~\text{(translations)} \ ,\\
\displaystyle& l_{0}=-z\p_z \ , ~~~\bar l_{0}=-\bar z\p_{\bar z}\,~~~~~\text{(rotations and dilatations)} \ ,\\
\displaystyle& l_{1}=-z^2\p_z \ , ~~~\bar l_{1}=-\bar z^2\p_{\bar z}~~~\text{(SCT)} \ . \\
\end{align*}

\subsection{$\Box ^ 3$}

The fact that it is impossible to construct a Weyl invariant action for the system~\eqref{box2} in two dimensions, is a particular case of a more general result~\cite{GJMS,GJMS_2,GJMS_3}, see also~\cite{Nakayama:2013is}. This states that for even number of dimensions, there exist Weyl invariant generalizations of $\Box ^ k$ only for 
$k \leq \f {n} {2}$. Therefore, considering a theory with six derivatives in a four dimensional space-time
\be
\mc L _ {\B ^ 3}= \f {1} {2} \l ( \p _ \m{\B} \phi \r ) ^ 2 \ ,
\label{box3}
\ee
one is sure that it cannot be made Weyl invariant. This can be immediately seen by inspecting the Weyl covariant analog of the operator~\eqref{box3}.\footnote{Explicit expressions for the operator have been obtained in~\cite{Branson:1985,Osborn:2015rna}.} It contains terms proportional to
\be
\f{1}{(n-2)(n-4)}B_{\m\n}S^{\m\n} \ ,~~~\f{1}{n-4}\nabla^\m\l(B_{\m\n}\nabla^\n\r) \ ,
\label{singular_terms}
\ee
thus it does not exist in $n=2$ and $n=4$ dimensions for a non-zero Bach tensor $B_{\m\n}$ 
\be
B_{\m\n}=C_{\m\rho\n\s}S^{\rho\s} +\nabla^\rho \nabla_\m S_{\n\rho}-\nabla^2 S_{\m\n}\ ,
\label{Bach_tensor}
\ee
with $C_{\m\rho\n\s}$ being the Weyl tensor.

However, straightforward computations reveal -- taking into account that the scaling dimension of the field in this case is equal to $\D_\phi = n/2 - 3$ -- that the conformal variation of the Lagrangian~\eqref{box3}  is also a total derivative
\be
\d \mc L _ {\B ^ 3} = - \p ^ \m \l [ \eps _ \m \mc L _ {\Box^3} - 
\f {1} {n} \p ^ \n \p \e \l ( 4 \p _ \m \p _ \n \phi \B \phi  - \f {1} {2} ( \B \phi ) ^ 2 \l  ( \f {n} {2} + 3 \r ) \eta _ {\m \n} \r ) \r ] \ .
\ee
Moreover, one can build the energy-momentum tensor
\be
\begin{aligned}
T _ {\m \n} &= \Box^2 \phi \, \p _ \m \p _ \n \phi - \l ( \p _ \m \phi \, \p _ \n \B^2 \phi + \p _ \n \phi \, \p _ \m \B ^ 2 \phi \r ) \\
&+ \p ^\lambda \phi \, \p _ \m \p _ \n \p _ \lambda \B \phi + \B \phi \, \p _ \m \p _ \n \B \phi + \p ^ \lambda \B \phi \, \p _ \m \p _ \n \p _ \lambda \phi \\
&- \p _ \m \B \phi \, \p _ \n \B \phi - \eta _ {\m \n} \l [ \f {1} {2} \l( \p _ \lambda \B \phi \r ) ^ 2 + \p ^\lambda \p ^ \s \phi \, \p _ \lambda \p _ \s \B \phi \r ] \ ,
\end{aligned}
\ee
as well as the operators
\be
\begin{aligned}
K _ \m &= \a \, \p _ \m \p _ \n \phi \, \p _ \n \B  \phi - (n+\a) \p ^ \n \phi \,  \p _ \m \p _ \n \B  \phi - 
\l ( \f {n} {2} + \a \r ) \p _ \m \B \phi \, \B  \phi  \\
&+ \l ( \a + \f {n} {2} + 2 \r ) \p _ \m \phi \, \B ^ 2 \phi + \l ( \f {n} {2} - 3 \r ) \phi \, \p _ \m \B ^ 2 \phi \ , 
\end{aligned}
\ee
and
\be
\begin{aligned}
L _ {\m \n} &=& \l ( \a - \f {n-10} {4} \r ) \p _ \m \phi \, \p _ \n \B  \phi - \l ( \a + \f {3 n - 10 } {4} \r ) \p _ \n \phi \, \p _ \m \B  \phi  \nn \\
&+& \f {n-10} {4} \p _ \m \p _ \n \phi \, \B  \phi - \f {n+10} {4} \phi \, \p _ \m \p _ \n \B  \phi 
+ \f {3n - 2} {4} \eta _ {\m \n} \phi \, \B ^ 2 \phi \ .
\end{aligned}
\ee
The above satisfy~\eqref{T_K_L_n} for arbitrary values of the constant $\a$,  therefore, the theory is conformal in flat space-time. Notice, though, that for $L_{\m\n}$ to be symmetric, we have to set $\a=-n/4$.

\subsection{Curved space-time}

In order to further expose the difference between the concepts of Weyl and conformal symmetries we consider the curved space-time counterpart of $\B^3$. It is obvious that the sixth-order Weyl covariant operator for $n\neq 2$ and $n \neq 4$ is also conformally invariant for an arbitrary metric. It may happen though that there are no conformal Killings for a specific background to start with. To guarantee that the conformal group is not empty, we stick to Einstein manifolds only, for which
\be
\label{ein-man-1}
R_{\m\n}=\f {R}{n} g_{\m\n} \ . 
\ee
It is easy to check that the Bach tensor~\eqref{Bach_tensor} in this case vanishes identically.\footnote{To make this point clear, we proceed as follows. For Einstein manifolds, the Schouten tensor~\eqref{tensor_R} becomes
$$
S_{\m\n}=\f{n-2}{2n(n-1)} R g_{\m\n} \ .
$$
Upon plugging the above into the definition of the Bach tensor~\eqref{Bach_tensor} and recalling that the Weyl tensor is traceless in all of its indices, we find that 
$$
B_{\m\n}=\f{n-2}{2n(n-1)}\l(\nabla_\m\nabla_\n R-g_{\m\n}\nabla^2 R \r) \ ,
$$
which is zero for all $n$. This follows trivially from the (contracted) Bianchi identities, which yield that the scalar curvature $R$ is constant (for $n\neq 2$).} Therefore, the dangerous terms~\eqref{singular_terms} disappear, thus the limit 
$n \to 4$ of the conformally invariant curved space analog of $\Box ^3$, can be safely considered. In doing so, one obtains a conformally invariant operator with leading term $\nabla ^ 6$.

To illustrate the procedure in more detail, we consider the Paneitz operator, which for Einstein manifolds becomes regular at $n=2$. It is straightforward to check using the relation
\be
\label{ein-man-conf_K}
\nabla_\m \nabla_\n \nabla f + \f{1}{2n(n-1)} g_{\m\n}\l(n\, f^\s \nabla_\s R+2\,R\nabla f\r) = 0 \ ,
\ee
following from the conformal Killing equations for $n\neq 2$, that the corresponding action
\be
\begin{aligned}
\label{ein-man-3}
S=\int d^n x\sqrt{g}\Bigg[\nabla^2\phi\nabla^2\phi&-\f{4-n(n-2)}{2n(n-1)}R\l(\nabla\phi\r)^2  -\f{n-4}{4(n-1)}\phi^2\nabla^2 R \\
&+\f{(n-2)(n+2)(n-4)}{16n(n-1)^2}\phi^2R^2\Bigg]\ ,
\end{aligned}
\ee
is invariant under the ($n \neq 2$) conformal transformations, as it should. The limit $n \to 2$ in turn is regular
\be
\begin{aligned}
S=\int d^n x\sqrt{g}\Bigg[\nabla^2\phi\nabla^2\phi&-R\l(\nabla\phi\r)^2  +\f{1}{2}\phi^2\nabla^2 R \Bigg],
\end{aligned}
\ee
and is invariant under global conformal transformations.
The reason it is not invariant under the full conformal group is that the relation~\eqref{ein-man-conf_K} does not follow automatically for two dimensional theories. Rather, it has to be imposed by hand, reducing the conformal group to its subgroup of global transformations. Clearly this is a peculiarity of two dimensions.

\section{Generalization}

The examples we considered clearly show that not any conformally invariant (both in flat and curved space-time) theory can be made Weyl invariant. In fact, there is a whole class of theories not allowing Weyl invariant generalizations. Indeed, as it was mentioned before, according
to~\cite{GJMS,GJMS_2,GJMS_3}, the Weyl covariant analogs of $\Box^k$ exist unless the number of space-time dimensions $n$ is even and less than $k/2$. The impossibility to construct the corresponding operators in even number of dimensions manifests itself through the presence of terms singular at $n = 2,4,6,\ldots$ However, it seems plausible that similar to the situation described in the previous section those terms vanish (or at least become regular) once the geometry is restricted to that of Einstein spaces. As a result, the corresponding limit $n \to 4,6,\ldots$ exists and is invariant under conformal transformations (or only global conformal transformations for $n \to 2$).

Since flat spaces are a particular case of Einstein ones, according to the above argument, the theories whose dynamics is described by the Lagrangian in flat space-time
\be
\mc L _ {\Box ^ {k}} = \f {1} {2} \phi \Box ^ k \phi \ 
\label{conf_k}
\ee
are conformal (for $n \neq 2$). We can convince ourselves that this is the case by considering the variation of the Lagrangian with respect to conformal transformations. For $k = 2 m$ and $k = 2 m + 1$, the Lagrangian can be rewritten as
\be
\mc L _ {\Box ^ {2m}} = \f {1} {2} ( \Box ^ m \phi ) ^ 2 ~~ \text {and} ~~ \mc L _ {\Box ^ {2m+1}} = \f {1} {2} ( \p_\m \Box ^ m \phi ) ^ 2,
\ee
while the variations are respectively given by
\be
\begin{aligned}
\d _ c \mc L _ {\Box ^ {2 m}} =& - \p _ \m \Bigg \{ \e ^ \m \mc L _ {\Box ^ {2 m}} - \f {2 m^2} {n} \, \p _ \n \p \e \, \bigg [ \p ^ \m \Box ^ {m-1} \phi \p ^ \n \Box ^ {m-1} \phi \\
&- \f {1} {2} \eta ^ {\m \n} \l ( \p \Box ^ {m-1} \phi \r ) ^ 2 \bigg ] \Bigg \} \ ,
\label{conf_2m}
\end{aligned}
\ee
and
\be
\label{conf_2m1}
\begin{aligned}
\d _ c \mc L _ {\Box ^ {2 m + 1}} = &- \p _ \m \Bigg \{ \e ^ \m \mc L _ {\Box ^ {2 m + 1}} - \f {1} {n} \, \p _ \n \p \e  \bigg [ {2 m (m+1)} \p ^ \m \p ^ \n \Box ^ {m-1} \phi \Box ^ {m} \phi  \\
& - \f {1} {2} \eta ^ {\m \n} \l ( \f {n} {2} - 1 + {2 m (m+1)} \r ) \l ( \Box ^ {m} \phi \r ) ^ 2 \bigg ] \Bigg \} \ . 
\end{aligned}
\ee
At the same time, according to~\cite{GJMS_3}, the Lagrangian~\eqref{conf_k} cannot be made Weyl invariant in an even number of dimensions if $n < 2 k$.

Similarly, it can be proven that for manifolds with vanishing Ricci tensor, the theories given by the Lagrangian
\be
\mc L _ {\nabla ^ {2k}} = \f {1} {2} \phi \nabla ^ {2k} \phi \ ,
\ee
are also conformally invariant.

\chapter{Gauging nonrelativistic field theories using the coset construction}
\label{ch:NR_Weyl}

\section{Introduction}

After the excursion into the details of Weyl and conformal symmetries we took previously, we now turn to the gauging of nonrelativistic spacetime symmetries, namely the centrally extended Galilei algebra (also known as Bargmann algebra) and the Lifshitz algebra. This chapter follows closely~\cite{Karananas:2016hrm}.

Nonrelativistic theories coupled to curved backgrounds appear naturally in Lorentz violating modifications of gravity, like Ho\v{r}ava-Lifshitz gravity~\cite{Horava:2009uw}, as well as in holographic duals of nonrelativistic systems~\cite{Balasubramanian:2008dm}. 
 Even though there is a large number of papers dedicated to studying these systems~\cite{Son:2013rqa,Geracie:2014nka,Iorio:2014pwa,Andringa:2010it,Christensen:2013lma,Bergshoeff:2014uea,Hartong:2015zia,Hartong:2015wxa,Hartong:2014pma,Hartong:2014oma,Brauner:2014jaa,Jensen:2014aia,Banerjee:2014pya,Banerjee:2014nja,Geracie:2015xfa,Geracie:2015dea}, we nevertheless believe that our approach allows one to clarify some subtleties. For example, it will become clear that for theories with local Galilei invariance the condition for vanishing spatial torsion is not consistent unless the temporal part of the torsion is set to zero as well.

One of our goals is to try to generalize the results of chapter~\ref{ch:Weyl_Ricci} for the case of theories exhibiting local nonrelativistic invariance. Namely, we wish to understand the conditions under which a theory can be rendered Weyl invariant without introducing 
an additional gauge field $W_\m$ corresponding to local scale transformations. With the coset construction, we were able to show that if for Lorentz invariant theories the field $W _ \m$ appears only in the very specific combination~\eqref{tensor_A}, it can be traded for the Schouten tensor~\eqref{tensor_R}.

Using the same approach, we address this question for the case of nonrelativistic theories coupled to a curved background. Considering first the centrally extended Galilei algebra, we show that the mere notion of Weyl invariance can be introduced only for torsionful theories. We show that for twistless torsionful theories, it is always possible to express the spatial components of the Weyl vector in terms of torsion, which in turn is a function of the vielbein.

Next, we turn to the Lifshitz algebra. In this case, there is no obstacle to the complete elimination of the Weyl gauge field; thus, any scale invariant theory in flat space can be coupled to a curved background in a Weyl invariant way, provided one allows for nonvanishing torsion. This is  similar to the situation occurring with Lorentz invariant theories,  where torsion may play the role of an additional degree of freedom making a theory Weyl invariant.

This chapter is organized as follows. In Sec.~\ref{sec:Galilei}, we gauge the Galilei algebra and we demonstrate how matter fields can be coupled systematically to curved backgrounds. Moreover, we show what the constraints leading to torsionless and torsionful geometries are. In Sec.~\ref{sec:gal}, we study the scale invariant generalizations of the Galilei as well as the Lifshitz algebras. For the former, by solving the inverse Higgs constraint, we express the spatial part of the vector field associated with scale transformations in terms of the vielbein. In addition, we demonstrate that locally Lifshitz-invariant theories can always be made Weyl invariant without introducing the corresponding independent gauge field. We present our conclusions in Sec.~\ref{sec:conclusions3}.

\section{Galilei algebra}
\label{sec:Galilei}

The centrally extended Galilei algebra (sometimes called Bargmann algebra) in a $n$-dimensional spacetime can be obtained from the Poincar\'e one using the standard \.In\"on\"u-Wigner contraction~\cite{Inonu:1953sp}. Let us briefly outline the procedure. The first step is to express the algebra of the Poincar\'e group~\eqref{poin_cr} in terms of components\,\footnote{In what follows, we will use lowercase latin letters to denote spatial Lorentz indices.}
\be
\begin{aligned}
&\l[ P_0,P_0\r]=\l[P_0, P_i\r]=\l[ P_i, P_j\r]=0 \ ,\\
&\l[J_{0i},  P_0\r]=-i \bar P_i \ , \l[J_{0i},  P_j\r]=-i \d_{ij} P_0 \ ,\\
&\l[J_{ij}, P_0\r]=0 \ ,  \l[ J _ {i j},  P _ {k} \r ] = i \l ( \d _ {i k} P _ j - \d _ {j k} P _ i  \r ) \ , \\
&\l[J_{0i},J_{0j}\r]=-i J_{ij} \ ,\l [ J _ {i j}, J_ {0k} \r ] = i \l ( \d _ {i k} J _{ 0j} - \d _ {j k} J _ {0i}  \r ) \ ,\\
&\l [ J _ {i j}, J _ {k l} \r ] = i \l ( J _ {j l} \d _ {i k} + J _ {i k} \d _ {j l} - J _ {i l} \d _ {j k} - J _ {j k} \d _ {i l}  \r ) \ . 
\end{aligned}
\ee
Next, considering the redefinitions
\be
\begin{aligned}
 P _ 0 & = M c ^ 2 + H \ ,  \\
P _ i & =  c P _ i \ , \\
J _ {0 i} & =  c K _ i \ ,
\end{aligned}
\ee
and taking the limit $c \to \infty$, we get -- provided $M$ commutes with $K _ i$ and $P _ i$ and therefore plays the role of a central charge -- the following non-zero commutation relations
\be
\begin{aligned}
&\l [ J _ {i j}, J _ {k l} \r ]  =  i \l ( J _ {j l} \d _ {i k} + J _ {i k} \d _ {j l} - J _ {i l} \d _ {j k} - J _ {j k} \d _ {i l}  \r ) \ ,  \\
&\l [ J _ {i j}, P _ {k} \r ] = i \l ( \d _ {i k} P _ j - \d _ {j k} P _ i  \r ) \ ,  \\
&\l [ J _ {i j}, K _ {k} \r ]  =  i \l ( \d _ {i k} K _ j - \d _ {j k} K _ i  \r ) \ ,  \\
&\l [ K _ {i}, P _ {j} \r ] = - i \d _ {i j} M \ ,  \\
&\l [ K _ {i}, H \r ]  =  - i P _ i \ .
\label{Galilei_cr}
\end{aligned}
\ee
In the above -- although there is little room for confusion -- $J$ correspond to (spatial) rotations, $K$ correspond to boosts, $H$ and $P$ correspond to temporal and spatial translations respectively, and $M$ is the central extension corresponding to the particle number operator or the mass.

The coset construction techniques have been used to gauge the Galilei group $\mathrm{Gal} (n)$ in~\cite{Brauner:2014jaa}, where Goldstone bosons for boosts were introduced. To build a theory with local Galilei invariance but without spontaneously breaking any symmetry, we consider the coset space of the full $\mathrm{Gal} (n)$ group over its subgroup generated by $J$, $K$ and $M$.\footnote{This possibility was mentioned in~\cite{Brauner:2014jaa} and partly worked out in \cite{Jensen:2014aia}.} Following the logic described in the previous chapters, we take the coset representative in the form
\be
\Omega = e ^ {i H z + i P _ i y ^ i}\ .
\label{coset_momenta}
\ee
Introducing the gauge fields $\tilde n_\m$ and $\tilde e_\m^i$ for temporal and spatial translations, respectively, $\tilde \omega^i_\m$ for boosts, $\tilde \t ^ {i j} _ \m$ for $SO(n-1)$ rotations, and $\tilde A_\m$ for the particle number $U (1)$, we find that the Maurer-Cartan form is given by the following expression,
\be
\Omega ^ {-1} \tilde D _ \m \Omega 
 =  i n _ \m H +i e_\m^i P_i+ i \omega ^ i _ \m K _ i+ \f {i} {2} \t ^ {i j} _ \m J _ {i j} + i A _ \m M \ ,
\ee
where the quantities without the tilde could be thought of as the fields in the unitary gauge. According to the procedure described in the previous section, the fields $n _ \m$ and $e _ \m ^ i$ are identified with the temporal and spatial components of the vielbein. For later convenience, we also define the inverse vielbein, $ V^\m\equiv E ^ \m _ 0$ and $E^\m_i$, such that
\be
V^\m n_\m=1 \ ,~~~V^\m e_\m^i =0 \ ,~~~n_\m E^\m_i=0 \ ,~~~e_{\m i} E^\m_j=\d_{ij} \ ,~~~e_\m^i E^\n_i=\d^\n_\m-n_\m V^\n \ .
\ee
The transformation properties of the fields can be obtained from the transformation of the coset representative~\eqref{coset_rep_trans}. However, unlike what we have encountered so far, the structure of the commutation relations of the Galilei group~\eqref{Galilei_cr} is not the one presented in~\eqref{representation_commut}. This fact results in the mixing of the $U (1)$ gauge field with the vielbein under boosts. In the following table, we present the transformation properties of the fields under rotations $J$, boosts $K$ and $U(1)$ with parameters $ R _ {i j}$, $\eta_i$, and $\a$ correspondingly.
\begin{table}[H]
\centering
\bt{c | cccc}
 &$J$  & $K$ & $M$\\
\hline
$n ' _ {\m}$ & $n _ {\m}$   &  $n _ \m$ & $n _ {\m}$ \\
$V ^ {' \m}$ & $V^ {\m}$   &  $V^\m+ \eta^ i E ^ \m _ i$ & $V _ {\m}$ \\
$e ^ {'i} _ {\m}$ & $R_{ij } e^ {j} _ {\m}$ & $e _ \m ^ i - \eta ^ i n _ \m$ & $e _ \m ^ i$ \\
$E^{'\m}_i$ & $R _ {i j} E^ {\m} _j$ & $E^\m _i$ & $E ^ \m _ i$ \\
$\t ^ {'i j} _ {\m}$ & $R _ {i k} R_ {j l} \t ^ {kl} _ {\m} + \l ( R \p _ \m R ^ {-1}\r ) _ {i j}$ & $\t ^ {i j} _ {\m}$ & $\t ^ {i j} _ {\m}$ \\
$\omega ^ {'i} _ {\m}$ & $R _ {i j} \omega ^ {j} _ {\m}$ & $\omega ^ {i} _ {\m} + \t ^ {i j} _ {\m} \eta _ j + \p _ \m \eta _ i$ & $\omega ^ {i} _ {\m}$ \\
$A _ \m '$ & $A _ \m $ & $A _ \m - \eta _ i e ^ i _ \m + \f {1} {2} \eta^2 n _ \m$ & $A _ \m + \p _ \m \a $
\et
\label{gfield_trans_1}
\end{table}
It should be noted that the actual transformation properties of $A_\m$ are different from the ones presented in  the above table.
Indeed, using the commutation relations of the Galilei group, it is straightforward to show that
\be
e ^ {-i K \eta} e ^ {i P y} = e ^ {i P y'} e ^ {-i K \eta} e ^ {-i M f} \ ,~~~\text{ with} ~~~f = \eta _ i y ^ i  - \f {1} {2} \eta ^ 2 z  \ .
\ee
Hence, the ``honest'' transformation of the $U (1)$ gauge field under $K$ is given by
\be
A ' _ \m = A _ \m - \eta _ i e ^ i _ \m + \f {1} {2} \eta^2 n _ \m + \p _ \m f \ .
\ee
The last term in the expression above was dropped in the previous table, since  it has precisely the form of the gauge transformation of $A_\m$.  

The standard definition of the field strengths leads to
\be
\begin{aligned}
\label{field_str}
 n _ {\m \n} & = \p _ {\m}n _ {\n}-\p _ {\n}n _ {\m} \ , \\
 e ^ i _ {\m \n} & =  \p _ {\m}e ^ i _ {\n} -\p _ {\n}e ^ i _ {\m} + \t _ {\m} ^ {i j} e _ {\n  j} - \t _ {\n} ^ {i j} e _ {\m  j}
+\omega _ {\m} ^ {i} n _{ \n}-\omega _ {\n} ^ {i} n _{ \m} \ , \\
\t ^ {i j} _ {\m \n} & =  \p _ {\m} \t ^ {i j} _ {\n} - \p _ {\n} \t ^ {i j} _ {\m}+ \t ^i_ {\m k} \t^{kj} _ \n-  \t ^i_ {\n k} \t^{kj} _ \m \ , \\
\omega ^ {i} _ {\m \n} & =  \p _ {\m} \omega ^ {i} _ {\n}- \p _ {\n} \omega ^ {i} _ {\m}
+ \t _ {\m} ^ {i j} \omega _ {\n j} -\t _ {\n} ^ {i j} \omega _ {\m j} \ , \\
A _ {\m \n} & =  \p _ {\m} A _{ \n}-\p _ {\n} A _{ \m} + \omega _ {\m} ^ {i} e   _ {\n i}- \omega _ {\n} ^ {i} e   _ {\m i}\ .
\end{aligned}
\ee
A straightforward calculation reveals that
\begin{table}[H]
\centering
\bt{c | cccc}
 &$J$ & $K$ & $M$\\
\hline
$n' _ {\m \n}$ & $n _ {\m \n}$   &  $n _ {\m  \n}$ &  $n _ {\m  \n}$ \\
$e ^ {'i} _ {\m  \n}$ & $R _ {i j} e^ {j} _ {\m  \n}$ & $e _ {\m  \n} ^ i - \eta ^ i n _ {\m  \n}$ &  $e _ {\m  \n} ^ i$ \\
$\t ^ {'i j} _ {\m  \n}$ & $R _ {i k} R _ {j l} \t ^ {kl} _ {\m  \n} $ & $\t ^ {i j} _ {\m  \n}$ & $\t ^ {i j} _ {\m  \n}$ \\
$\omega ^ {'i} _ {\m  \n}$ & $R _ {i j} \omega ^ {j} _ {\m  \n}$ & $\omega ^ {i} _ {\m  \n} + \t ^ {i j} _ {\m  \n} \eta _ j $ &  $\omega ^ {i} _ {\m  \n}$ \\
$A _ {\m  \n} '$ & $A _ {\m  \n} $ & $A _ {\m  \n} - \eta _ i e ^ i _ {\m  \n} + \f {1} {2} \eta _ i \eta _ i n _ {\m \n}$ & $A _ {\m  \n}$
\et
\label{gfield_trans_2}
\end{table}

\subsection{Coupling to matter}

With the gauge fields at our disposal, we can build the temporal and spatial \emph{covariant derivatives} of a matter field $\Psi$ belonging to an irreducible\,\footnote{These are induced by representations of the $SO(n-1)$ rotation group. In our case the action of boosts on matter fields is trivial.} representation of the Galilei group as
\be
\begin{aligned}
\nabla _ t \Psi &= V^ \m \l ( \p _ \m \Psi + \f {i} {2} \t ^ {i j} _ \m \rho (J _ {i j}) \Psi + i m A _ \m \Psi \r ) \ ,\\
\nabla _ i \Psi &= E ^ \m _ i \l ( \p _ \m \Psi + \f {i} {2} \t ^ {j k} _ \m \rho (J _ {j k}) \Psi + i m A _ \m \Psi \r ) \ ,
\label{covariant_D}
\end{aligned}
\ee
where  $\rho$ is the representation of the $\mathfrak{so}(n-1)$ the field belongs to, and $m$ is the charge of  $\Psi$ under $U(1)$. It can be easily shown that at the leading order in $\eta$
\be
\begin{aligned}
( \nabla _ t \Psi ) ' & =  \nabla _ t \Psi + \eta _ i \nabla _ i \Psi \ , \\
( \nabla _ i \Psi ) ' & =  \nabla _ i \Psi - i m \, \eta _ i \nabla _ t \Psi\ .
\label{matter_boosts}
\end{aligned}
\ee
Even though the derivatives defined above do not actually transform covariantly under local boosts it is still true that any Lagrangian that is invariant under the Galilei group in flat space -- which corresponds to the limit where all gauge fields vanish -- can be made locally Galilei invariant by substituting all partial derivatives by covariant ones, i.e. $\p _ t \to \nabla _ t$ and $\p_i\to\nabla_i$. The local invariance of the Lagrangian under rotations and $U (1)$ is clear, for under their action, the covariant derivative transforms covariantly. The only nontrivial point is the transformation with respect to boosts, which in the flat background has the form
\be
\Psi ' (t,x)= e ^ {-i m \, x ^ i v ^ i} \Psi (t,x _ i + v _ i t) \ , ~~~\text{with}~~~ v _ i = const \ . \\
\ee
Let us consider a Lagrangian that is invariant under boosts, i.e.
\be
\begin{aligned}
&\mc  L \l [ \p _ t \Psi, \p _ i \Psi, \Psi \r] =\\
&\mc  L \l [e ^ {-i m \, x ^ i v ^ i} \l ( \p _ t \Psi + v _ i \p _ i \Psi \r ),e ^ {-i m \, x ^ i v ^ i} \l ( \p _ i \Psi - i m \, v _ i \p _ t \Psi \r ), \Psi (t,x) \r] \ ,
\label{Lagr_flat_boost}
\end{aligned}
\ee

It follows automatically that the Lagrangian with all partial derivatives substituted by covariant ones is invariant under local boosts. Indeed, the fact that the transformations~\eqref{matter_boosts} coincide in the flat limit [up to the $U (1)$ factor that we dropped] with the ones presented implicitly in~\eqref{Lagr_flat_boost} guarantees the cancellation of all factors containing $\eta$ (there are no terms that contain derivatives of $\eta$).

For example, consider the theory of a field $\psi$ with spin $s$ in a $2+1$-dimensional flat spacetime whose dynamics is described by the following Lagrangian 
\be
\mc L = \f {i} {2} \bar \psi \overset{\leftrightarrow}{\p} _ t \psi - \f {1} {2 m} \partial _ i \bar \psi \partial _ i  \psi \ ,
\label{flat_free_Lagr}
\ee
with $\bar \psi \overset{\leftrightarrow}{\p} _ t \psi = \bar \psi \p _ t  \psi - \p _ t \bar \psi \psi$. Promoting partial derivatives to covariant ones and multiplying by the determinant of the temporal and spatial vielbeins (denoted collectively by $\det e$), we obtain the action that is locally Galilei and diffeomorphism invariant,
\be
S = \int d t d^2 x \,  \det e \, \l ( \f {i} {2} \bar \psi \overset{\leftrightarrow}{\nabla} _ t \psi - \f {1} {2 m} \nabla _ i \bar \psi \nabla _ i  \psi \r ) \ .
\label{free_curved}
\ee
It should be stressed that  had we chosen a Lagrangian with the time derivative appearing in the nonsymmetric form, i.e.
\be
\mc L _ {\text{nonsym}}= i \bar \psi \p _ t  \psi - \f {1} {2 m} \partial _ i \bar \psi \partial _ i  \psi \ ,
\ee
which differs from~\eqref{flat_free_Lagr} by $-\f {i} {2} \p _ t \l ( \bar \psi \psi \r )$, the procedure would not have worked. The reason is that the Lagrangian in this case is not invariant under boosts, but rather it shifts by a total derivative. From~\eqref{Lagr_flat_boost}, it follows that $\D \mc L _ {\text{nonsym}} = -\f {i} {2} v _ i \p _ i \l ( \bar \psi \psi \r )$, which  cannot be written as a total derivative upon promoting $v _ i$ to $\eta _ i (x)$, since $\p \eta$ terms do appear in this case.

\subsection{Torsionless geometry}

At the moment, we have all the building blocks for constructing a theory with local Galilei symmetry. However, it appears that there are many more degrees of freedom than are actually needed in order to accomplish our goal. As we have seen, the standard way to eliminate redundancies within the coset construction is to impose covariant constraints that can be solved algebraically. 

Using the transformation properties of the fields, we see that the only covariant quantity is the temporal component of the torsion $n _ {\m \n}$. Meanwhile, both the spatial torsion $e ^ i _ {\m \n}$ and the $U (1)$  field strength $A _ {\m \n}$ transform covariantly under all group operations, apart from boosts. However, the mixing of $e ^ i _ {\m \n}$ and $A _ {\m \n}$ with $n _ {\m \n}$ can be eliminated by imposing
\be
n _ {\m \n} = 0 \ .
\label{temporal_torsion}
\ee
It is clear that in this case $n_\m$ corresponds to a closed form, i.e. $n _ \m = \p _ \m \tau$ where $\tau$ is some function that can be identified with global time. With this condition, the other two constraints,
\be
e ^ i _ {\m \n} = 0 
\label{spatial_torsion}
\ee
and
\be
A _ {\m \n} = 0 \ ,
\ee
become covariant and they can be used to specify completely the  $\mathfrak{so}(n-1)$ part of the connection
\be
\begin{aligned}
\label{thet_2}
\t^{ij}_\m = \bar \t^{ij}_\m &\equiv -\f{1}{2}\l[\vphantom{\frac{a}{b}} E ^ {\n} _ i(\p _ {\m } e _ { \n } ^ j-\p _ {\n } e _ { \m } ^ j) - E ^ {\n } _ j(\p _ {\m } e _ {\n } ^ i-\p _ {\n } e _ {\m } ^ i  )\r. \\
&\l.- e_{\m k} E ^{\rho } _ iE ^{ \s } _ j( \p _ {\rho } e _ {\s} ^ k-\p _ {\s } e _ {\rho} ^ k) +n_\m E ^ {\rho} _ i E ^{ \s} _ j (\p _ {\rho} A _ {\s}-\p _ {\s} A _ {\rho})\r.\\
&\l.-e_{\m i}E^\rho_{j}V^\s(\p_{\rho}n_{\s}-\p_{\s}n_{\rho})+e_{\m j}E^\rho_{i}V^\s(\p_{\rho}n_{\s}-\p_{\s}n_{\rho}) \vphantom{\frac{a}{b}}\r]\ ,
\end{aligned}
\ee
as well as the connection that corresponds to boosts
\be
\begin{aligned}
\label{omeg_2}
\omega ^ i _ \m = \bar \omega ^ i _ \m& \equiv E ^ {\s} _ i V ^ \n (\p _ \s A _\n-\p _ \n A _\s) n _ \m + \f{1}{2}e ^ j _ \m E ^ {\s} _ i E ^ {\n} _ j (\p _ \s A _ \n-\p _ \n A _ \s )\\
&+  \f{1}{2}\l(\p _\s  e _ \n ^ { i}E ^ {j\s}+\p _\s  e _ \n ^ { j}E ^ {i\s}-\p _\n  e _ \s  ^ { i}E ^ { j\s}-\p _\n  e _ \s  ^ { j}E ^ { i\s} \r) V ^ \n e ^ j _ \m \ .
\end{aligned}
\ee

Continuing with the example that we started previously, we see that the term corresponding to the interaction of the spin and the magnetic field  appears naturally in the action~\eqref{free_curved}. Indeed, using the expression~\eqref{thet_2}, we see from the first term in~\eqref{free_curved} that the derivative of the gauge field $A _ \m$ couples to $\bar \psi \psi $
as
\be
 \f {i} {2} \bar \psi \overset{\leftrightarrow}{\nabla} _ t \psi \supset - \f {s} {4} \e _ {i j}  E ^ {\m} _ i E ^ {\n} _ j (\p _ \m  A _  \n -\p _ \n  A _ \m )\bar\psi \psi \ .
\ee
Upon an appropriate rescaling of the fields, the coupling constant $g _s$ appears in front of this term. There is no need for a redefinition of the transformation properties of the gauge field $A _ \m$ in order to make the theory invariant under the general coordinate transformations, as was done for example in~\cite{Geracie:2014nka,Jensen:2014aia}. A somewhat similar approach was suggested in~\cite{Andreev:2013qsa}.

\subsection{Torsionful theory}

It should be stressed that it is not consistent to impose the spatial torsionlessness condition~\eqref{spatial_torsion} without having the temporal torsion be zero as well, for the condition $e ^ i _ {\m \n} = 0$ alone is not invariant under boosts. However, there is still an alternative to what was done in the previous section. According to the coset construction, any covariant constraint can be imposed without contradicting the symmetry breaking pattern. The tensor $n _ {\m \n}$ can be naturally decomposed into representations of the $\mathfrak{so}(n-1)$, namely, 
$E ^ \m _ i E ^ \n _ j n _ {\m \n}$ and $E ^ \m _ i V ^ \n n _ {\m \n}$. However, only the first one is a singlet with respect to the boosts and thus can be safely set to zero,
\be
E ^ \m _ i E ^ \n _ j n _ {\m \n} = 0 \ .
\label{ttnc_n}
\ee
The constraints consistent with the above condition are the following:
\be
e ^ i _ {\m \n} E^{\m}_i E^\n_j = 0 ~~~ \text {and} ~~~ E^{\m}_i E^\n_j A_{\m\n} = 0 \ .
\ee
Consequently, the spin connection $\t ^ {ij} _ \m$ and $\omega ^ i _ \m$ can be fixed only partly, since we can express in terms of the vielbein and the U(1) gauge field only $(n-1)^2(n-2)/2 + (n-2)(n-1)/2$ components. These  correspond to $\t^{ij}_\m E ^ \m _ k$ and $\omega^ {i } _ \m E ^ { j \m}-\omega^ {j } _ \m E ^ {i \m}$, respectively.

We should also note that the condition~\eqref{ttnc_n} coincides with the one imposed on the temporal torsion in the case of the twistless torsional Newton-Cartan (TTNC) geometry discussed in a number of papers~\cite{Christensen:2013lma,Hartong:2014oma,Hartong:2014pma,Bergshoeff:2014uea,Hartong:2015wxa,Hartong:2015zia}. Contrary to our case, the authors of~\cite{Hartong:2015wxa,Hartong:2014oma} were able to fully determine the connections associated with spatial rotations and boosts. This was made possible by introducing a ``St\"uckelberg field," thus requiring that the $U(1)$ symmetry be realized nonlinearly.

\section{Adding dilatations}
\label{sec:gal}

\subsection{Galilei algebra}

It is interesting to investigate under what conditions a theory that is scale invariant in flat space can be promoted to a Weyl invariant one without   introducing a gauge field corresponding to the local scale transformations. Notice that the nonzero commutators of the dilatation generator $D$ and the Galilei ones are 
\be
\begin{aligned}
&\l [ D, H \r ]  = - 2 i H \ , \l [ D, P _ i \r ]  =  - i P _ i \ , \l [ D, K _ i \r ]  =  i K _ i \ .
\end{aligned}
\ee
As one can see, the scaling of space and time for theories that are not Lorentz invariant does not have to be homogeneous, which is manifest due to the factor $2$.

At this point we have to decide what geometry to consider. It is rather obvious that the standard transformation
\be
n_ \m \to e ^ {- 2 \s} n _ \m \ ,
\ee
is not consistent with the torsionlessness condition $\p _ {\m}n _ {\n} -\p _ {\n}n _ {\m}= 0$.  The other option is~\eqref{ttnc_n}, which as we saw leads to additional -- as compared to the Newton-Cartan data~\cite{Son:2013rqa} -- independent degrees of freedom.

As we argued previously, the coset construction provides the natural language to speak about  local scale transformations as well. The only modification one has to make to the procedure used for gauging the Galilei algebra is to introduce yet another gauge field $W _ \m$ that corresponds to the dilatations, in complete analogy with the relativistic case of chapter~\ref{ch:Weyl_Ricci}. The transformation properties of the fields under the Galilei group are not changed and are given in the tables of the previous section. The scaling properties may be found using the commutation relations presented previously. The ones that are not singlets are as follows:
\be
\begin{aligned}
&\hat n _ {\m} =  e ^ {2 \s} n_ \m \, , ~\, \hat V ^ {\m} = e ^ {- 2 \s} V^ \m \, , ~\, \hat e ^ {i} _ {\m} = e ^ {\s} e ^ {i} _ {\m} \, , \\
&\hat E^{\m}_i = e ^ {\s} E ^ \m _i \, , ~\, \hat \omega ^ {i} _ {\m} = e ^ {-\s} \omega ^ {i} _ {\m} \, , ~\, \hat W _ \m = W _ \m - \p _ \m \s \, .
\label{scaling_gauge}
\end{aligned}
\ee
Similarly, for the (modified) field strengths
\be
\begin{aligned}
\label{mod_field_str}
n _ {\m \n} & =  \p _ {\m}n _ {\n}-\p _ {\n}n _ {\n}+ 2(W _ {\m} n _ {\n}-W _ {\n} n _ {\m}) \ , \\
e ^ i _ {\m \n} & =  \p _ {\m}e ^ i _ {\n} -\p _ {\n}e ^ i _ {\m} + \t _ {\m} ^ {i j} e _ {\n  j} - \t _ {\n} ^ {i j} e _ {\m  j}
+\omega _ {\m} ^ {i} n _{ \n}-\omega _ {\n} ^ {i} n _{ \m}  + W _{ \m} e ^ i _ {\n} -W _{ \n} e ^ i _ {\m} \ , \\
\t ^ {i j} _ {\m \n} & =  \p _ {\m} \t ^ {i j} _ {\n} - \p _ {\n} \t ^ {i j} _ {\m} + \t ^i_ {\m k} \t^{kj} _ \n-  \t ^i_ {\n k} \t^{kj} _ \m \ , \\
\omega ^ {i} _ {\m \n} & =  \p _ {\m} \omega ^ {i} _ {\n}- \p _ {\n} \omega ^ {i} _ {\m}
+ \t _ {\m} ^ {i j} \omega _ {\n j}- \t _ {\n} ^ {i j} \omega _ {\m j} - W _ {\m} \omega ^ i _ {\n}+W _ {\n} \omega ^ i _ {\m} \ , \\
W _ {\m \n} & =  \p _ {\m }W _ {\n}-\p _ {\n }W _ {\m} \  ,  \\
A _ {\m \n} & =  \p _ {\m} A _{ \n}-\p _ {\n} A _{ \m} + \omega _ {\m} ^ {i} e   _ {\n i}-\omega _ {\n} ^ {i} e   _ {\m i}\ ,
\end{aligned}
\ee
we find that  
\be
\hat n  _ {\m \n} = e ^ {2 \s} n _ {\m \n}\ , ~~~ \hat e ^ {i} _ {\m \n} = e ^ {\s} e ^ {i} _ {\m \n}\ , ~~~ \hat \omega ^ {i} _ {\m \n} = \hat e ^ {-\s} \omega ^ {i} _ {\m \n} \ .
\label{scaling_strength}
\ee
Imposing the constraint~\eqref{temporal_torsion} does not lead to the torsionless geometry, i.e. the 1-form $n_\m$ is not forced to be closed, but rather it satisfies the TTNC condition~\eqref{ttnc_n}, which is compatible with the scaling transformations~\eqref{scaling_gauge}. On top of that, the constraint on the temporal torsion allows us to express the spatial part of the Weyl gauge field $W _ \m$ in terms of the vielbein. We readily obtain
\be
W _ i \equiv W _ \m E ^ \m _ i =-\f{1}{2}E ^ \m _ i V ^ \n (\p _ {\m} n _ {\n }-\p _ {\n} n _ {\m}) \ .
\label{spatial_W}
\ee

Solving the other two constraints~\eqref{spatial_torsion} in this case produces
\begin{align}
\label{spatial_theta_W}
\t^{ij}_\m & =  \bar \t^{ij}_\m +  e _ \m ^ { i } W ^ { j } -e _ \m ^ { j } W ^ { i } \ ,  \\
\omega ^ i _ \m & =  \bar \omega ^ i _ \m + W _ t e ^ i _ \m \ ,
\end{align}
where $\bar \t^{ij}_\m$ and $\bar \omega ^ i _ \m$ are given respectively by~\eqref{thet_2} and~\eqref{omeg_2}, and we defined the temporal component of the Weyl gauge field as $W _ t = V ^ \m W _ \m$. Having no other covariant quantities that we can use in order to eliminate $W _ t$, we can conclude that for generic curvature $\t ^ {i j} _ {\m \n}$, it is impossible to express the temporal part of the Weyl field in terms of the vielbein and $A _ \m$, so it stays an independent degree of freedom. However, this does not necessarily mean that a theory cannot be made Weyl invariant without introducing this additional degree of freedom.

Indeed, as before [see~\eqref{covariant_D}], the covariant derivative can be defined as
\be
D _ t \Psi = \nabla _ t \Psi - \D_\Psi W _ t \Psi, ~~ D _ i \Psi = \nabla _ i \Psi - \D_\Psi W _ i \Psi \ ,
\ee
where $\D_\Psi$ is the scaling dimension of the field $\Psi$.\footnote{The Weyl transformation of a field has the form $\hat \Psi = e ^ {-\D_\Psi \s} \Psi$.}  We see that if it is possible to rewrite the Lagrangian of a theory in flat spacetime such that the time derivative appears only in the ``symmetric way'' 
$\bar \psi \overset{\leftrightarrow}{\p} _ t \psi$, then in curved space this leads to 
\be
\bar \psi \overset{\leftrightarrow}{D} _ t \psi = \bar \psi \overset{\leftrightarrow}{\nabla} _ t \psi \ ,
\ee
which is independent of $W _ t$. As a result, such a theory is going to be automatically Weyl invariant, for the time derivative is the only source of
$W _ t$.

It is interesting to note that, as in relativistic theories, the presence of Weyl symmetry guarantees that when the flat spacetime limit is considered, the resulting theory is conformal.  The opposite, however, is not true (see~\ref{ch:Weyl_vs_Conf}). In the context of  Galilei-invariant theories, the conformal transformations are defined analogously to the relativistic case as the diffeomorphisms preserving the vielbein up to a conformal factor $\Omega$,
\be
\label{nr-conf-viel}
n'_\m=\Omega^2\,n_\m \ ,~~~e^{'i}_\m = \Omega\l(\Lambda^i_je^i_\m +\Lambda^i n_\m \r) \ ,
\ee
where $\Lambda^i_j$ and $\Lambda^i$ are specific functions of the transformation parameters.

\subsection{Lifshitz algebra}
\label{sec:Lifshitz}

In the previous sections, we saw that the presence of boosts complicates the situation considerably, since a number of structures transform in a noncovariant way under them. Here, we investigate another type of nonrelativistic spacetime symmetry, the Lifshitz algebra,  which can be obtained from the Galilei one by discarding the boosts. By doing so, the presence of the $U (1)$ symmetry associated with the central extension becomes unnecessary, since it decouples from the spacetime generators and turns into an internal symmetry.

Now all the structures can be classified in terms of irreducible representations of the $\mathfrak{so}(n-1)$ algebra of spatial rotations. The corresponding transformation properties of the fields can be read from the tables in Sec.~\ref{sec:Galilei}, as well as from Eqs.~\eqref{scaling_gauge} and~\eqref{scaling_strength}. For the Lifshitz algebra,  $n_{\m\n}$, $\t^{ij}_{\m\n}$, and $W_{\m\n}$ are identical to the ones in~\eqref{mod_field_str}, whereas the spatial torsion reads
\be
\begin{aligned}
\label{field_str_2}
e ^ i _ {\m \n} & =  \p _ {\m}e ^ i _ {\n} - \p _ {\n}e ^ i _ {\m} + \t _ {\m} ^ {i j} e _ {\n  j} -\t _ {\n} ^ {i j} e _ {\m  j} 
+W _{\m} e ^ i _ {\n}-W _{\n} e ^ i _ {\m} \ .
\end{aligned}
\ee
Notice that all field strengths transform covariantly.

Imposing the following set of constraints,
\be
n _ {\m \n} E ^ \m _ i V ^ \n = 0\ ,~~e ^ i _ {\m \n}E ^ \m _ j E ^ \n _ k = 0 \ ,~~\l(e ^ {i } _ {\m \n} E ^ {j \m}-e ^ {j } _ {\m \n} E ^ {i \m}\r) V ^ \n = 0 \ ,~~e ^ {i} _ {\m \n} E ^ {\m} _ i V ^ \n = 0 \ ,
\ee
enables us to express in terms of the vielbein the connection that is once again given by~\eqref{spatial_theta_W}, and the Weyl gauge field whose spatial part is~\eqref{spatial_W}, whereas its temporal part reads
\be
W _ t=\frac{1}{2(n-1)}\l(\p_{\m}e_{\n}^{i }E^{j \m}+\p_{\m}e_{\n}^{j }E^{i \m}-\p_{\n}e_{\m}^{i }E^{j \m}-\p_{\n}e_{\m}^{j }E^{i \m}\r)V^\n \ .
\ee
The above results are completely analogous to the ones in the torsionful relativistic theory of Sec.~\ref{sec:altern}.

\section{Summary and Outlook}
\label{sec:conclusions3}

The aim of this chapter was to clarify certain issues related to the gauging of nonrelativistic symmetries. After presenting a systematic way of building a locally invariant Galilei theory from a globally invariant one, we found that within this approach the term corresponding to the interaction of a spin $s$ and the magnetic field is automatically included. In this case, no modification of the transformation properties of the $U(1)$ gauge field is needed in order to achieve invariance of the action under boosts. 

We demonstrated how the covariant constraints can be used in order to eliminate redundant (unnecessary) degrees of freedom. It should be emphasized once again that it is not consistent to set to zero the spatial torsion, unless the temporal torsion vanishes as well (provided no Goldstone bosons are introduced). 

We then turned to the question of how the addition of dilatations changes the situation. We showed that there are no Weyl invariant theories with vanishing temporal torsion, i.e. with global time. The condition of temporal torsionlessness is not covariant under local scale transformations.
On the contrary, when torsion is present, it is always possible to express the spatial part of the Weyl gauge field in terms of geometric data.
We showed, however, that for general backgrounds it is not possible to eliminate the temporal part of the Weyl vector. 
Nevertheless, as we saw, it may happen that the aforementioned field does not appear in the action. As a result, invariance under Weyl rescalings does not necessarily require the introduction of $W_t$.
  
Finally, we discussed Lifshitz-invariant theories. In this case, the field strengths transform covariantly, since we relaxed the requirement of having invariance under Galilei boosts. In these theories both the temporal and the spatial parts of the Weyl field can always be expressed in terms of the vielbein.

The fact that for the cases considered in the present chapter the Weyl vector can be (partly) eliminated in favor of other degrees of freedom, should not come as a surprise. This is nothing else than torsion playing the role of the Weyl gauge field. It would be interesting to carry out an analysis similar to the one in chapter~\ref{chapt:Poincaregrav} and investigate the behavior of the propagating modes when terms bilinear in the various field strengths are taken into account.

\chapter{Poincar\'e gravitational theory}
\label{chapt:Poincaregrav}

\section{Introduction}
\label{sec:Intro}

In the present chapter, based on the article~\cite{Karananas:2014pxa}, we abandon the ``minimalistic'' approach we have followed so far and we consider that vielbein and connection are independent degrees of freedom. Our purpose is to identify healthy subclasses of the Poincar\'e-invariant gravitational theory, with all possible parity-even as well as parity-odd terms that are at most quadratic in the field strengths $\omega$ and $T$.  
This clearly means that the action contains terms with \emph{two derivatives of the fields, at most}.  Let us explain why we restrict ourselves this way. From our point of view, the absence of terms with more than two derivatives is an essential requirement, since higher-derivative theories are usually plagued by ghosts. Since here vielbein and  connection are treated as independent fields, this theory should not be mistaken for a higher-derivative theory, but rather as ``gravity \`a la Yang-Mills''; this theory is dubbed Poincar\'e gauge theory of gravity (PGT) and it has been studied extensively in the literature~\cite{Neville:1978bk,Sezgin:1979zf,Hayashi:1979wj,Hayashi:1980av,Hayashi:1980ir,Hayashi:1980qp,Hayashi:1980bf,Hayashi:1981fx,Sezgin:1981xs,Neville:1981be,Nair:2008yh,Nikiforova:2009qr,Hernaski:2009wp}. An extensive review as well as historical details can be found for example in~\cite{Hehl:1976kj,Hehl:1994ue,Gronwald:1995em} and references therein. It is worth mentioning that PGT incorporates as simplest cases the Einstein-Cartan theory~\cite{Trautman:2006fp}, the teleparallel equivalent of GR~\cite{DeAndrade:2000sf,Obukhov:2002tm}, as well as GR in the absence of fermionic matter. Given the fact that GR has been extremely successful in the description of Nature at large scales, the fact that PGT is capable of reducing to GR in certain limiting cases is encouraging. 

The most straightforward way to accomplish our goal is to determine the particle spectrum of the theory around the flat spacetime. In the present work, we do not discuss how the dynamics is modified when arbitrary curved spacetimes are considered as backgrounds, which constitutes a complicated problem that deserves to be addressed separately. It is well known that once a theory is studied around backgrounds different from the Minkowski one, especially if it contains massive spin-2 modes, pathologies might appear; this is what happens for example in the Fierz-Pauli theory (Boulware-Deser effect). Notice though that this is not the case for certain subclasses of the PGT we consider here, which remain free from ghosts and tachyons  when studied around maximally symmetric backgrounds~\cite{Nair:2008yh}.

Investigating the behaviour of the physical propagator, we find constraints on the parameters of the action so that the propagating degrees of freedom are neither ghosts, nor tachyons. We believe that the reason we choose to proceed this way is clear: the poles of the propagator correspond to the masses of the particles the theory contains, whereas the sign of the residues evaluated at the poles determine whether or not the theory is ghost-free~\cite{Neville:1978bk,Sezgin:1979zf,Schwinger:1970xc}. 

Let us give some more details on the methodology we followed. First, we linearize the action around Minkowski spacetime and we retain only the bilinear in the fluctuations terms. We then employ the spin-projection operator formalism initially developed by Barnes~\cite{Barnes:1965} and Rivers~\cite{Rivers:1964}, see also Ref.~\cite{VanNieuwenhuizen:1973fi}. This framework is very powerful and ideally suited for such kind of problems, the main advantage being that the action for the excitations naturally breaks into independent spin sectors. Meanwhile, the coefficients of the expansion of the action in the projectors' basis can be conveniently arranged in matrices. This fact, together with the simple orthogonality relations the operators satisfy, makes the attainment of the propagator a straightforward exercise.

The inclusion of parity-odd terms, however, makes this exercise algebraically much more involved with respect to a number of interesting works that have appeared over a period of many years~\cite{Neville:1978bk,Sezgin:1979zf,Sezgin:1981xs,Neville:1981be,Hernaski:2009wp}. In these papers, the authors concentrated mainly on parity-even theories and studied in depth their particle dynamics. It is our purpose here to extend these works by including parity non-conserving invariants. We hope that by considering the effects of these terms in a systematic way could lead to new directions towards understanding questions that are of big significance in Cosmology, like the baryon asymmetry of the Universe~\cite{Sakharov:1967dj}.

There have been studies on PGT with parity-violating terms that are relevant to what we do here. 
The first one is work that has been carried out by Kuhfuss and Nitch in the 80's~\cite{Kuhfuss:1986rb}, in which the teleparallel equivalent of GR -- a certain sub-category of PGT with vanishing curvature -- was studied. They considered in addition to the three parity-even torsion terms, a parity-odd torsion term. Since vielbein is the only dynamical degree of freedom of this theory, projectors associated with the vielbein perturbations only were derived. The other one is a very interesting and relatively recent work by Hehl and collaborators~\cite{Baekler:2010fr}. In their paper, the authors allow for parity-odd pieces in a particular case of PGT that propagates only scalar degrees of freedom. This theory has interesting cosmological applications~\cite{Yo:2006qs,Shie:2008ms} and it has been argued that it remains consistent in the non-linear level as well~\cite{Yo:1999ex,Yo:2001sy}. 
The authors determine necessary and sufficient conditions on the parameters of their theory so that it is physically acceptable. Notice that they did not resort to linearization or the use of projection operators, but instead the initial Lagrangian was partially diagonalized  for the case where spin-2 torsion vanishes. Finally, we would like to mention that there has been some renewed interest in three-dimensional PGT and especially on the effect of the gravitational Chern-Simons term, see for example~\cite{HelayelNeto:2010jn} and references therein.

The present chapter is organized as follows. In Sec.~\ref{sec:Poincareg}, we introduce the 14-parameter theory under investigation and we present the linearized quadratic action for the perturbations. In Sec.~\ref{sec:spinproj}, we review the spin-projection formalism that is used to decompose the theory into independent spin sectors. Since we want to elucidate the role of parity-violating terms by treating them in the same footing as parity-preserving ones, we expand the original basis of projectors built in~\cite{Neville:1978bk,Sezgin:1979zf}, by introducing appropriate operators that allow us to work with terms that contain the totally antisymmetric tensor; most of them have never appeared before, as far as we know. In Sec.~\ref{sec:Particle Content}, we find the constraints on the parameters of the action so that it propagates only healthy degrees of freedom. This we achieve by requiring positive masses and residues of the propagators when evaluated at the poles. In~Sec.\ref{sec:concluspoinc} we present the concluding remarks. 

\section{The action}
\label{sec:Poincareg}

In four spacetime dimensions, the most general theory invariant under translations and local Lorentz transformations -- with terms that are at most quadratic in the field strengths -- contains all possible invariants built from the torsion and curvature tensor
\begin{align}
&T_{\m\n}^A= \p _ \m e _ \n ^ A - \p _ \n e _ \m ^ A - \omega _ {\m B} ^ {A}  e _ \n ^B + \omega _ {\n B} ^ {A}  e _ \m ^B \ , \\
&\omega_{\mu\nu}^{AB}=\partial_\mu \omega_\nu^{AB}-\partial_\nu \omega_\mu^{AB}-\omega_\mu^{AC}\omega_{\nu C}^{B}+\omega_\nu^{AC}\omega_{\mu C}^{B}  \ ,
\end{align}
For later convenience, we note that the above can be written in the tangent basis -- where indices are manipulated with the Minkowski metric -- with the help of the inverse vielbein
\begin{align}
&T_{ABC}=E^\mu_{A}E^\nu_{B}\eta_{CD} T_{\mu\nu}^{D} \ , \ \ \omega_{ABCD}= E^\mu_{A}E^\nu_{B}\eta_{CE}\eta_{DF}\omega_{\mu\nu}^{EF} \ . 
\end{align}
The Lagrangian of the theory reads~\cite{Neville:1978bk,Sezgin:1979zf,Obukhov:1987tz,Diakonov:2011fs,Baekler:2011jt}\footnote{The convention for the totally antisymmetric tensor $\epsilon^{ABCD}$ is $\epsilon^{0123}=-\epsilon_{0123}=1$. To keep the expressions as simple as possible, in this chapter we set $M_P=1$.}
\begin{align}
\label{acti}
{\mathscr L}&=\lambda \omega +\frac{1}{12}(4t_1+t_2+3\lambda)\,T_{ABC}\,T^{ABC} \nonumber\\
&-\frac{1}{3}(t_1-2t_3+3\lambda)\,T_{AB}^{\ \ \ B}\,T^{AC}_{\ \ \ C} \nonumber\\
&-\frac{1}{6}(2t_1-t_2+3\lambda)\,T_{ABC}\,T^{BCA}\nonumber\\
&-\frac{1}{12}(t_4+4t_5)\,\epsilon^{ABKL}\,T_{ABC}\,T_{KL}^{\ \ \ C}\nonumber\\
&-\frac{1}{3}(t_4-2t_5)\,\epsilon^{ABKL}\,T_{CAB}\,T^C_{\ \ KL}\nonumber\\
&+\frac{1}{6}(2r_1+r_2)\,\omega_{ABCD}\,\omega^{ABCD}\nonumber\\
&+\frac{2}{3}(r_1-r_2)\,\omega_{ABCD}\,\omega^{ACBD}\nonumber\\
&+\frac{1}{6}(2r_1+r_2-6r_3)\,\omega_{ABCD}\,\omega^{CDAB}\nonumber\\
&+(r_4+r_5)\,\omega_{AB}\, \omega^{AB} +(r_4-r_5)\,\omega_{AB}\,\omega^{BA}\nonumber\\
&-\frac{1}{6}(r_6-r_8)\,\epsilon^{ABKL}\omega\,\omega_{ABKL}\nonumber\\
&-\frac{1}{8}(r_7+r_8)\,\epsilon^{ABKL}\,\omega_{ABCD}\,\omega_{KL}^{\ \ \ CD}\nonumber\\
&+\frac{1}{4}(r_7-r_8)\,\epsilon^{ABKL}\,\omega_{ABCD}\,\omega^{CD}_{\ \ \ KL}  \ \ .
\end{align}
Here $\lambda,t_i,r_i$ are 14 arbitrary dimensionless constants and 
\begin{equation}
\omega_{AB}=\eta^{CD}\omega_{ACBD} \ , \ \ \omega=\eta^{AB}\omega_{AB} \ .
\end{equation}
We have allowed for parity-even $(\lambda,t_1, t_2, t_3, r_1, r_2, r_3, r_4, r_5)$ as well as parity-odd $(t_4, t_5, r_6, \allowbreak r_7, r_8)$ terms and we chose these peculiar combinations of coefficients because in this way the expressions that appear in the propagators simplify a lot. As it will turn out, these 5 new parity-violating parameters modify in a non-trivial way the conditions for the absence of ghost and tachyons. We will come back to this point in Section~\ref{sec:Particle Content}.

Some comments concerning our Lagrangian are in order at this point. First  of all, we have not written down a cosmological constant term; we want the field equations to admit Minkowski spacetime as solution. In addition to that, we have not included the following four terms 
\begin{equation}
\label{not-inc}
\bar\omega\ ,\ \ \epsilon^{ABCD}\, \omega_{ABCD} \  ,\ \ \omega^2 \ , \ \epsilon^{IJKL}\, \omega_{ABIJ}\, \omega^{AB}_{KL} \ ,
\end{equation}
where
\begin{align}
\label{barR}
&\bar\omega=E^\mu_{A}E^\nu_{B}\bar \omega _{\m\n}^{AB}\ ,\\
\label{barRiem}
&\bar \omega _{\m\n}^{AB}  = \p _ \m \bar\omega _ \n ^ {A B} - \p _ \n \bar\omega _ \m ^ {A B} 
-\bar \omega _ {\m C} ^ {A} \bar\omega _ \n ^{CB} +\bar \omega _ {\n C} ^ {A} \bar\omega _ \m ^{CB} \ ,
\end{align}
and $\bar \omega^{AB}_\m$ was defined in the previous section, see~\eqref{spin-con4}.\footnote{We have chosen to write the quantities with bar in the ``mixed'' basis for later convenience.}

The first two terms in~\eqref{not-inc} can be related to $\omega$ and/or torsion squared terms by virtue of
\begin{equation}
\begin{aligned}
\int d^4x\det e~\bar\omega&=\int d^4x\, \left[ \det e~\omega+\frac{1}{4}\,T_{ABC}\,T^{ABC}\r.\\
&\l.-\frac{1}{2}\,T_{ABC}\,T^{BCA}-T_{AB}^{\ \ \ B}\,T^{AC}_{\ \ \ C}  \right]  \ ,
\end{aligned}
\end{equation}
and up to a total derivative
\begin{equation}
\int d^4 x~\det e\, \epsilon^{ABKL}\,\omega_{ABKL}=-\frac{1}{2}\int d^4x~\det e\, \epsilon^{ABKL}\,T_{ABC}T_{KL}^{\ \ \ C} \ .
\end{equation}
As for the $\omega^2$ term, it is related to $\omega_{AB}\omega^{BA}$ and $\omega_{ABCD}\omega^{CDAB}$, by virtue of the Gauss-Bonnet theorem
\begin{equation}
\int d^4x~\det e\Bigg[\omega^2-4\, \omega_{AB}\,\omega^{BA}+\omega_{ABCD}\,\omega^{CDAB}\Bigg]=0 \ .
\end{equation}
Finally, the term $\epsilon^{IJKL}\, \omega_{ABIJ}\, \omega^{AB}_{KL}$ need not be included, since it is a total derivative.  

Before moving on, we would like to stress again that the PGT under consideration contains terms which are at most quadratic in the derivatives of the independent gauge fields $e_\mu^{A}$ and $\omega_\mu^{AB}$. Therefore, it should not be mistaken for a higher-derivative theory that usually suffer from unitarity issues. One notable exception is~\cite{Stelle:1977ry}
$$S=\int d^4x~\det e\Big[ \bar\omega+c\,  \bar\omega^2 \Big]  \ , $$ 
with $c$ a positive constant. This theory in addition to the graviton, contains one healthy scalar degree of freedom and provides a viable inflationary model able to describe the Universe evolution in its primordial stages~\cite{Starobinsky:1980te}. 

Let us now return to the theory under investigation and linearize the action~\eqref{acti} by considering the weak field approximation for the fields
\begin{equation}
\label{exp}
e_{\mu}^{A}\approx \d_{\mu}^{A}+h_\mu^{A} \ ,\ h_{\mu}^{A}\ll 1 \ \ \ \text{and} \ \ \ \omega_{\mu}^{AB}\ll  1\ .
\end{equation}
In this limit there is no need to keep the distinction between spacetime and Lorentz indices, so in what follows we will use only capital Latin letters for tensorial quantities. It is also convenient to split the vielbein excitations into symmetric and antisymmetric parts, i.e. 
\begin{equation}
h_{AB}=s_{AB}+a_{AB}\ ,
\end{equation}
with 
\begin{equation}
s_{AB}=\frac{1}{2}(h_{AB}+h_{BA}) \ \ \ \text{and}\ \ \ a_{AB}=\frac{1}{2}(h_{AB}-h_{BA}) \ .
\end{equation}
Using the decomposition \eqref{exp} in the action, expanding in powers of $h_\mu^{A}$ and $\omega_{\mu}^{AB}$ and retaining only the bilinear in perturbations parts,\footnote{The expression for the linearized action can be found in Appendix~\ref{app:lin_act}.}
the action can be recast into the following compact form\footnote{When convenient, we denote tensorial indices collectivelly by using Greek indices with acute accent $(\acute{\alpha},\acute{\beta},\ldots)$. This helps us to unclutter the notation and keep the expressions as short as possible.}
\begin{equation}
S_2=\frac{1}{2} \int d^4x \sum_{\acute{\alpha},\acute{\beta}} \phi_{\acute{\alpha}}~D_{\acute{\alpha}\acute{\beta}}~\phi_{\acute{\beta}} \ ,
\end{equation}
where the multiplet $\phi_{\acute{\alpha}}=(\omega_{CAB}, s_{AB}, a_{AB})$ contains the 40 components of the fields and the wave operator $D_{\acute{\alpha}\acute{\beta}}$ contains combinations of derivatives, the metric and the totally antisymmetric tensor. 

The quadratic action for the excitations~\eqref{lin-terms} has obviously inherited the linearised gauge symmetries of the original theory, i.e. it is invariant under 
\begin{equation}
\label{gaug-trans-1}
\delta h_{AB}=\partial_A \xi_B +\xi_{AB} \ ,\ \ \ \  \text{and}\ \ \ \  \delta \omega_{CAB}=-\partial_C \xi_{AB} \ ,
\end{equation}
where $\xi_A$ and $\xi_{AB}=-\xi_{BA}$ are the 10 gauge parameters of the Poincar\'e group. This fact has two important consequences. 

On one hand, since all fields appear with at most two derivatives in the action, it shows that 20 degrees of freedom are devoid of physical meaning and they can be set to zero by appropriately adjusting $\xi_A,\,\xi_{AB}$ and using the constraints. 
Therefore, out of the 40 independent fields we started with (16 in vielbein, 24 in connection), we are left with 20.\footnote{This is most easily seen in the canonical formalism, where the number of degrees of freedom is found by subtracting from the phase-space of the theory the number of constraints imposed by symmetries.} These are distributed among the different spin-sectors of the theory as follows: twelve are in the tensor part, which contains the massless graviton (two degrees of freedom) and two massive spin-2 fields (ten degrees of freedom). Six degrees of freedom are in the spin-1 part, which contains two massive vectors, whereas the remaining two comprise two massive scalar modes.

On the other hand, due to these symmetries, once we allow for the vielbein and connection to interact with appropriate external sources by introducing  
\begin{equation}
S_{sources}=\int d^4x~\Big[h^{AB}\,\tau_{AB}+\omega^{CAB}\,\sigma_{CAB} \Big] \ ,
\end{equation}
 we are immediately led to the following conservation laws 
\begin{equation}
\label{constr-1}
\partial^A \tau_{AB}= 0 \ , \ \ \ \ \text{and} \ \ \ \ \partial^C \sigma_{CAB}+\tau_{[AB]}= 0 \ .
\end{equation}
These 10 constraints on the sources will turn out to be very helpful in what follows.

\section{The spin-projection operator formalism}
\label{sec:spinproj}

In this section we lay the foundations in order to determine the spectrum of the theory in a systematic way. Our strategy is to study the behaviour of the (gauge-invariant)  \emph{saturated} propagator (i.e. the propagator sandwiched between conserved sources)
\begin{equation}
\label{prop1}
\Pi=-\sum_{\acute{\alpha},\acute{\beta}} j_{\acute{\alpha}}~D^{-1}_{\ \ \ \acute{\alpha}\acute{\beta}}~j_{\acute{\beta}} \ ,
\end{equation}
where the multiplet $j_{\acute{\alpha}}=(\sigma_{CAB}, \tau_{(AB)}, \tau_{[AB]})$ contains sources that couple only to the gauge-invariant components of the respective fields (physical sources).
We believe this is the most straightforward way to establish conditions on the parameters of the action, since the propagator contains all important information for the particle states predicted by the theory. First of all, the position of its poles correspond to the masses that have to be necessarily positive. Negative mass implies tachyonic behaviour. Also, the sign of the residues when evaluated at the poles determine whether or not the particles are ghosts. Negative residues correspond to negative contributions to the imaginary part of scattering amplitudes, which puts the unitarity of the theory under scrutiny.

In order to obtain the propagator, the wave operator has to be inverted and this is a rather non-trivial task. However, our goal is greatly facilitated when we take into account that vielbein and connection are reducible with respect to the three-dimensional rotations group. Therefore, they can be decomposed into subspaces of dimension $2J+1$ with definite spin $J$ and parity $P$.\footnote{Notice that this decomposition has nothing to do with the details of a theory. It simply follows from the construction of irreducible representations from tensorial quantities. Notice also that the classification of particle states according to their spin and parity has only meaning in the rest frame.} In the absence of parity-odd terms, the wave operator breaks into independent sectors that connect states with the same $J^P$ as follows:
\begin{center}
\begin{tabular}{|c|c|}
\hline
$J^P$ & sub-block dimension \\
\hline
$2^-$ & $1\times 1$ \\
\hline
$2^+$ & $2\times 2$ \\
\hline
$1^-$ & $4\times 4$ \\
\hline
$1^+$ & $3\times 3$ \\
\hline
$0^-$ & $1\times 1$ \\
\hline 
$0^+$ & $3\times 3$\\
\hline
\end{tabular}
\end{center}

To be able to proceed with this decomposition, it is very convenient to work in momentum space and employ the spin-projection operator formalism that was initially developed by Barnes~\cite{Barnes:1965} and Rivers~\cite{Rivers:1964}. The building blocks are the four-dimensional transverse and longitudinal projection operators; in momentum space these are respectively given by
\begin{equation}
\label{buildthetaom}
\Theta_{AB}=\eta_{AB}-\frac{k_A k_B}{k^2} \ \ \ \text{and} \ \ \ \Omega_{AB}=\frac{k_A k_B}{k^2} \ .
\end{equation}

In their seminal works, Neville~\cite{Neville:1978bk} and Sezgin-van Nieuwenhuizen~\cite{Sezgin:1979zf} studied the spectrum of the most general Poincar\'e-invariant theory with parity-even terms. To accomplish that, they used $\Theta$ and $\Omega$ to construct a covariant basis of projectors $P^{\phi\chi}_{ij}(J^P)_{\acute{\alpha}\acute{\beta}}$, which map between subspaces of fields $\phi,\chi$ with the same $J^P$. The lowercase Latin indices ($i,j,\dots$) denote the multiplicity of operators. This basis consists of 40 operators and is complete and orthogonal\footnote{Notice that the position of indices other than Lorentz ones is not important.}
\begin{eqnarray}
\label{comps}
&\displaystyle\sum_{\phi,i,J^P}P^{\phi\phi}_{ii}(J^P)_{\acute{\alpha}\acute{\beta}}=\mathbb I_{\acute{\alpha}\acute{\beta}} \ , &
\\
\vspace{.3cm}
\label{orths}
&P^{\phi \Sigma}_{i k}(I^P)_{\acute{\alpha}}^{\ \acute{\mu}}~P^{T \chi}_{l j}(J^Q)_{\acute{\nu}\acute{\beta}}=\delta_{\Sigma T}\delta_{IJ}\delta_{PQ}\delta_{kl}\delta^{\acute{\mu}}_{\acute{\nu}}P^{\phi\chi}_{ij}(J^P)_{\acute{\alpha}\acute{\beta}} \ .&
\end{eqnarray}

Let us move to the case of interest to us, i.e. the presence of parity-odd terms in the Lagrangian. The wave operator will now decompose into subspaces of same spin but not necessarily of same parity. A simple counting exercise reveals that the wave operator breaks into 3 independent spin sectors: one $3\times 3$ corresponding to spin-2 states, one $4\times 4$ corresponding to spin-0 states and a $7\times 7$ corresponding to spin-1 states.

It is obvious from the orthogonality conditions \eqref{orths} that the above-mentioned set of projectors is not able to handle the presence of terms that involve the totally antisymmetric tensor, since they cannot link states with same spin but different parity. It is therefore unavoidable to introduce new operators to take care of this; it turns out that in order to account for all possible mappings inside each spin sector, it is necessary to practically double in size the original basis built by Sezgin and van Nieuwenhuizen by adding 34 new operators. It is our understanding that this is the first time transition projectors that account for the parity-odd terms involving the connection is presented.\footnote{Kuhfuss and Nitsch~\cite{Kuhfuss:1986rb} introduced mixing projectors in order to study the interaction of states with different parity but only for the tertrad excitations.} 

 In our case, the completeness relation of eq. \eqref{comps} remains unchanged 
\begin{equation}
\label{compl}
\displaystyle\sum_{\phi,i,J}P^{\phi\phi}_{ii}(J)_{\acute{\alpha}\acute{\beta}}=\mathbb I_{\acute{\alpha}\acute{\beta}} \ , 
\end{equation}
whereas the orthogonality relation becomes 
\begin{equation}
\label{orth}
P^{\phi \Sigma}_{i k}(I)_{\acute{\alpha}}^{\ \acute{\mu}}~P^{T \chi}_{l j}(J)_{\acute{\nu}\acute{\beta}}=\delta_{\Sigma T}\delta_{IJ}\delta_{kl}\delta^{\acute{\mu}}_{\acute{\nu}}P^{\phi\chi}_{ij}(J)_{\acute{\alpha}\acute{\beta}} \ .
\end{equation}
Notice that we have suppressed the parity index. The full list of projectors as well as details on their derivation are given in the Appendix~\ref{app:spinproj} and~\ref{app:derivproj} respectively. 

In terms of the spin-projection operators, the action for the theory becomes 
\begin{equation}
S_2=\int d^4x \sum_{\phi,\chi,\acute{\alpha},\acute{\beta},i,j,J} c^{\phi\chi}_{ij}(J)~\phi_{\acute{\alpha}}~P^{\phi\chi}_{ij}(J)_{\acute{\alpha}\acute{\beta}}~\chi_{\acute{\beta}} \ ,
\end{equation}
where $c^{\phi\chi}_{ij}(J)$ are matrices that contain the coefficients of the expansion of the wave operator in the spin-projection operators basis. All ``physical information'' of the theory is contained in the $c^{\phi\chi}_{ij}(J)$ matrices: the zeros of their determinants correspond to the poles of the propagators, whereas their values at the poles correspond to the residues.

As we mentioned earlier, the action for the perturbations possesses certain gauge symmetries; namely it is invariant under the linearized form of general coordinate and local Lorentz transformations \eqref{gaug-trans-1}. These invariances manifest themselves in the spin-projectors language as well.  The way this happens is through degenerate coefficient matrices. Let us explain what this means. 

Assume that a matrix $M_{ij}(J)$ has dimension $(d\times d)$ and $\rank\left(M_{ij}(J)\right)=r$, so there exist $(d-r)$ right null eigenvectors $v_j^{R}(J)$ as well as $(d-r)$ left null eigenvectors $v_j^{L}(J)$. Consider the $n^{th}$ right null eigenvector $v_j^{(R,n)}(J)$ which satisfies
\begin{equation}
\sum_j M_{ij}(J)v_j^{(R,n)}(J)=0 \ .
\end{equation}
From the above we are led to the following gauge invariances
\begin{align}
\label{pro-gau-inv}
&\delta\phi_{\acute{\alpha}}=\sum_{J,i,\acute{\beta},n}v_i^{(R,n)}(J)P^{\phi\chi}_{ij}(J)_{\acute{\alpha}\acute{\beta}}f_{\acute{\beta}}(J) \ \ \ \ \text{for all $j$}  \ ,
\end{align}
with $f_{\acute{\alpha}}(J)$ an arbitrary element of the group.  On the other hand, for the $n^{th}$ left null eigenvector $v_j^{(L,n)}(J)$ we have
\begin{equation}
\sum_j v_j^{(L,n)}(J)M_{ji}(J)=0 \ ,
\end{equation}
and as result the sources are subject to the following constraints
\begin{align}
\label{pro-sour-con}
&\sum_{i,\acute{\beta}}v_i^{(L,n)}(J)P^{\phi\chi}_{ij}(J)_{\acute{\alpha}\acute{\beta}}S_{\acute{\beta}}=0 \ \ \ \ \text{for all $j$}  \ .
\end{align}

In the theory under consideration, the $7\times 7$ matrix that describes the sector associated to the vector perturbations of the theory is singular and of rank 4. In addition to that, the $4\times 4$ matrix for the spin-0 sector is also singular and of rank 3.
Using the explicit expressions for these matrices (given in Appendix~\ref{app:spinproj}), a direct calculation reveals that eqs. \eqref{pro-gau-inv} and \eqref{pro-sour-con} respectively yield 
\begin{equation}
\delta h_{AB}=\partial_A \xi_B +\xi_{AB} \ , \ \ \delta \omega_{CAB}=-\partial_C \xi_{AB} \ , 
\end{equation}
and
\begin{equation}
\label{pro-sour-con-2}
\partial^A \tau_{AB}= 0 \ ,\ \ \partial^C \sigma_{CAB}+\tau_{[AB]}= 0 \ .
\end{equation}
The above result is expected and should not come as a surprise. 

At this point we can  proceed with the inversion of the coefficient matrices and calculate the propagator. In order to do so and since some of the $c^{\phi\chi}_{ij}(J)$ are singular, we simply have to invert the largest non-singular sub-matrix $b^{\phi\chi}_{ij}(J)$ extracted from them~\cite{Sezgin:1979zf,VanNieuwenhuizen:1973fi,Berends:1979rv}. Deleting $(d-r)$ rows and columns, practically amounts to imposing $(d-r)$ gauge conditions. Notice however that the gauge invariance of the propagator is guaranteed due to the $(d-r)$ source constraints given in~\eqref{pro-sour-con}. By virtue of the completeness and orthogonality relations~\eqref{compl} and~\eqref{orth} that $P^{\phi\chi}_{ij}(J)$ obey, the saturated propagator~\eqref{prop1} is given by
\begin{equation}
\label{propgafull}
\Pi=-\sum_{J,\phi,\chi,\acute{\alpha},\acute{\beta},i,j} \left(b^{ \phi\chi}_{ij}(J)\right)^{-1}j_{\acute{\alpha}}^*~P^{\phi\chi}_{ij}(J)_{\acute{\alpha}\acute{\beta}}~j_{\acute{\beta}} \ .
\end{equation}

\section{Particle Content}
\label{sec:Particle Content}

In this section we apply the formalism presented previously and we determine the restrictions on the parameters of the action~\eqref{acti}. 

\subsection{Massless sector}

Let us start in an unorthodox way by analyzing first the massless sector of the theory. Since our result for the (massless) graviton must be proportional to the one that stems from Einstein's theory, this calculation provides a very useful check of our algebra. The projectors we use as a basis for expanding the wave operator are constructed with the use of $\Theta_{AB}$ and $\Omega_{AB}$ defined previously in~\eqref{buildthetaom}, as well as
\begin{equation}
\tilde k_A=\frac{k_A}{\sqrt{k^2}} \ .
\end{equation}
Subsequently, the limit $k^2=0$ has to be taken with some care. Apart from the genuine massless pole that corresponds to the graviton, we will also find $k^{-2n}~(n\ge 1)$ spurious singularities that originate from the operators and receive contributions from all spin sectors. Of course, the  propagator should be independent of the basis we use for the expansion. Therefore, all spurious singularities have to combine appropriately and cancel out in the final result, upon applying the source constraints. Since the expressions are rather involved and the calculations lengthy, we will omit them in what follows and we will only present the final results. The reader is referred to Appendix~\ref{app:spinproj} for the explicit form of the coefficient matrices and the projection operators.

After a considerable amount of calculations involving all 74 projectors, we find that the cancellations between all spin sectors indeed take place in an elegant way and the residue of the propagator at the $k^2=0$ pole is 
\begin{equation}
\begin{aligned} 
\text{Res}(\Pi; 0)&=-\frac{1}{\lambda}\left(\partial_C\sigma^{ABC} \ \tau^{AB}  \right)\left(\begin{array}{ccc}
4&2\\
2&1
\end{array}\right)\times\\
&\times\left(\eta_{AI}\eta_{BJ}+\eta_{AJ}\eta_{BI}-\eta_{AB}\eta_{IJ}\right)\left(\begin{array}{c}
\partial_K\sigma^{IJK}\\
\tau^{IJ}
\end{array}\right) \ ,
\end{aligned}
\end{equation}
as it should. The requirement for absence of ghosts in the massless sector of the theory is therefore 
\begin{equation}
\lambda>0 \ .
\end{equation}

\subsection{Massive sector}

For massive states, the propagator for each spin sector can be written as
\begin{equation}
\begin{aligned} 
\label{propga}
\Pi(J)&=-\frac{1}{(k^2-m_+(J)^2)(k^2-m_-(J)^2)}\times\\
&\times\sum_{\phi,\chi,\acute{\alpha},\acute{\beta},i,j} \left(b^{ \phi\chi}_{ij}(J)\right)^{-1}j_{\acute{\alpha}}~P^{\phi\chi}_{ij}(J)_{\acute{\alpha}\acute{\beta}}~j_{\acute{\beta}} \ ,
\end{aligned}
\end{equation}
by virtue of the completeness and orthogonality relations \eqref{compl} and \eqref{orth} that $P^{\phi\chi}_{ij}(J)$ obey. Here $b_{ij}^{\phi\chi}(J)$ is the residue matrix which is degenerate at the poles $k^2=m_\pm(J)^2$, with $m_\pm(J)$ the masses of the states. 
One might worry that the appearance of two poles in the propagator necessarily implies that one of the two states is ghost-like, since we can always write
\begin{equation}
\begin{aligned} 
\label{simp-fract}
\frac{1}{(k^2-m_+(J)^2)(k^2-m_-(J)^2)}&=\frac{1}{m_+(J)^2-m_-(J)^2}\times\\
&\times\left(\frac{1}{k^2-m_+(J)^2}-\frac{1}{k^2-m_-(J)^2}\right) \ .
\end{aligned}
\end{equation}
However, this is not always the case, for the coefficient matrices contribute rather non-trivially to the residues and their values at one of the poles can differ significantly from their values at the other. 

The requirement for absence of tachyons and ghosts corresponds to real masses and positive-definite residues at the poles, i.e. 
\begin{align} 
\label{reqs}
&m_\pm(J)^2>0\ , \\
\label{reqs1}
&\displaystyle \sum_{i}\left[\left(b_{ii}^{\phi\chi}(J)\right)^{-1} P_{ii}^{\phi\chi}(J)\right]_{k^2=m_\pm(J)^2}>0 \ ,
\end{align}
where we suppressed tensorial indices in the diagonal projection operators. Since at the pole $P_{ii}^{\phi\chi}(J)$ contribute only a sign depending on the number of longitudinal operators $n_\Theta$ they contain, the condition~\eqref{reqs1} can be written equivalently as
\begin{equation}
\label{reqs2}
\sum_i (-1)^{n_\Theta}\left(b_{ii}^{\phi\chi}(J)\right)^{-1}_{k^2=m_\pm(J)^2}>0 \ .
\end{equation}

After a tedious calculation involving the coefficient matrices of the various spin sectors given in Appendix~\ref{app:spinproj}, we apply~\eqref{reqs} and~\eqref{reqs2}, to find the following conditions on the parameters of the action for the absence of ghosts 
\begin{align*}
\label{conditions-full-s0}
\text{\hfill spin-0: \hfill}
&r_2<0\ ,~~~2r_2(r_1-r_3+2r_4)<-r_6^2\ ,\\
&r_1-r_3+2r_4>-\frac{r_6^2}{2r_2} \ , \numberthis
\\
\\
\label{conditions-full-s1}
\text{\hfill spin-1: \hfill}
&(r_1+r_4+r_5)<0\ ,~~~(r_1+r_4+r_5)(2r_3+r_5)<-r_7^2\ ,   \\
&\vphantom{\frac{r_7^2}{r_1+r_4+r_5}} 2r_3+r_5>-\frac{r_7^2}{r_1+r_4+r_5}\  , \numberthis 
\\ 
\\
\label{conditions-full-s2}
\text{\hfill spin-2: \hfill}
&r_1<0\ ,\ \ \ r_1(2r_1-2r_3+r_4)<-r_8^2\ , \\
&2r_1-2r_3+r_4>-\frac{r_8^2}{r1} \ , \numberthis
\end{align*}
and tachyons 
\begin{align*}
\label{conditions-full-s0}
\text{\hfill spin-0: \hfill}
&\vphantom{\frac{r_6^2}{2r_2} }  t_2(t_3-\lambda)+t_4^2>0\ ,~~~\left(t_2 t_3+t_4^2\right)\lambda(t_3-\lambda)>0 \numberthis
 \ ,\\
\\
\label{conditions-full-s1}
\text{\hfill spin-1: \hfill}
&t_2 t_3+t_4^2<0 \ ,~~~(t_1+t_2)(t_1+t_3)+(t_4-2t_5)^2>0\ ,\\
&\vphantom{\frac{r_7^2}{r_1+r_4+r_5}}t_1^2+4t_5^2>0\ ,~~~t_3(t_1^2+4t_5^2)> -t_1(t_2t_3+t_4^2) \ , \numberthis \\ 
\\
\label{conditions-full-s2}
\text{\hfill spin-2: \hfill}
&\vphantom{\frac{r_8^2}{r1}}   t_1\lambda(t_1+\lambda)<0 \ , \ \ \ t_1(t_1+\lambda)+4t_5^2 >0 \ .\numberthis
\end{align*}

Let us now comment on our results. First of all, when parity-mixing terms are absent, the expressions above reduce to the ones found by Sezgin-van Nieuwenhuizen~\cite{Sezgin:1979zf} and are presented below in~\eqref{conditions-sn-s0p}-\eqref{conditions-sn-s2m}. Meanwhile, it is apparent that the effect of the parameters corresponding to parity-odd invariants is indeed not-trivial: they are responsible for the fact that the inequalities we derived for the mass parameters can be simultaneously satisfied. Take as an example the tensor part of the theory (eq.~\eqref{conditions-full-s2}). We see that if $t_5=0$, there is a contradiction, since the two constraints 
\begin{equation}
t_1\lambda(t_1+\lambda)<0 \ \ \ \text{and} \ \ \ t_1\lambda(t_1+\lambda)>0 \ ,
\end{equation}
cannot be simultaneously satisfied. Therefore, if we want healthy behaviour in the spin-2 sector of the PGT, we have two options: either we consider the most general case by imposing $t_5\neq 0$, or if we insist on restricting the parameter space by considering $t_5= 0$, we also have to set $t_1+\lambda=0$, or $r_1=0$, or $2r_1-2r_3+r_4=0$. This would correspond to getting rid of the massive  $2^-$ or $2^+$ field respectively, even though in the parity-violating theory we investigate, this distinction is not entirely accurate.\footnote{Strictly speaking, the massive states predicted by the theory are not parity eigenstates, due to the presence of parity-odd terms in the Lagrangian. However, we used the label $J^P$ for convenience.}

However, the inequalities for the coefficients of the kinetic terms of the spin-1 and spin-2 sectors, boil down to\,\footnote{I am very grateful to James Nester for pointing out this contradiction.}
\begin{equation}
r_1>-\frac{r_7^2}{r_1+r_4+r_5}-\frac{r_8^2}{r_1}>0 \ ,~~~~~r_1<0 \ ,
\end{equation} 
which obviously cannot hold at the same time. As a result, even with the addition of the parity-odd invariants, vector or tensor ghost degrees of freedom are expected to be present in the most general quadratic in curvature and torsion gravitational theory based on the Poincar\'e group. The designation ``most general'' corresponds to the PGT whose action contains all possible parity-conserving and parity-violating  invariants, which are at most quadratic in the derivatives of the gauge fields $e$ and $\omega$. Notice, however, that there still exist boundaries of the extended parameter space where only healthy states may be present.

Having determined the restrictions the parameters of the theory should obey, it is useful at this point to see what happens if we consider a certain limiting case in the PGT we study.\footnote{Yet another limit that has been studied is the one of massless torsion, see~\cite{Karananas:2014pxa}.} Since this is the first time that an analysis on the full theory has been carried out, we believe that cross-checks on the results are crucial. Once we consider parity-preserving invariants only, we recover the results of Sezgin-van Nieuwenhuizen~\cite{Sezgin:1979zf} that read
\begin{align*}
\label{conditions-sn-s0p}
&\text{\hfill spin-0$^+$: \hfill}
r_1-r_3+2r_4>0 \ ,\ \ \ t_3\lambda(t_3-\lambda)>0 \numberthis
 \ ,\\
\\
\label{conditions-sn-s0m}
&\text{\hfill spin-0$^-$: \hfill}
r_2<0 \ ,\ \ \ t_2>0 \numberthis
 \ ,\\
\\
&\text{\hfill spin-1$^+$: \hfill}
2r_3+r_5>0\ ,\ \ \ t_1 t_2(t_1+t_2)<0 \ ,\numberthis \\
\\
&\text{\hfill spin-1$^-$: \hfill}
r_1+r_4+r_5<0\ ,\ \ \ t_1 t_3(t_1+t_3)>0 \ ,\numberthis \\
\\
\label{conditions-sn-s2p}
&\text{\hfill spin-2$^+$: \hfill}
2r_1-2r_3+r_4>0\ , \ \ \ t_1\lambda(t_1+\lambda)<0 \ , \numberthis \\
\\
\label{conditions-sn-s2m}
&\text{\hfill spin-2$^-$: \hfill}
r_1<0\ , \ \ \ t_1>0 \ . \numberthis
\end{align*}
Of course, all 12 healthy subclasses of the above theory found in~\cite{Sezgin:1979zf} and~\cite{Sezgin:1981xs} are also limiting cases of the theory we consider here. To name a couple, if we keep only the term linear in curvature (this amount to setting in the above $t_i=0\ ,i=1,\ldots,5$ and $r_j=0\ ,j=1,\ldots,8$) we recover General Relativity. If we assume that there are no torsion terms present ($t_1=-t_2=-t_3=-\lambda$, $t_4=2t_5=0$), we find that the only acceptable theory is given by $r_2<0$ and $r_i=0\ , i=1,\ldots,8$. Notice that the coefficients of the parity-odd curvature terms have to be chosen equal to zero in order to avoid higher order poles in the propagators. Another interesting case is the teleparallel limit~\cite{DeAndrade:2000sf,Obukhov:2002tm} of the PGT given in~\eqref{acti}, studied in detail in~\cite{Kuhfuss:1986rb}. To consider this particular subclass, one has to impose vanishing curvature with an appropriate Lagrange multiplier. As a result, the only dynamical degrees of freedom are contained in the vielbein field. Since the coefficient matrices in this case are very simple, after a straightforward calculation one can reproduce the results of Kuhfuss and Nitsch.

\section{Summary and Outlook}
\label{sec:concluspoinc}

In this chapter we presented a systematic study of the spectrum of the most general gravitational theory that emerges from the gauging of the  Poincar\'e group. We considered terms that are at most quadratic in the field strengths and allowed for the presence of all possible parity-even as well as parity-odd invariants.  Our purpose was to fill a gap in previous analyses of Poincar\'e-invariant theories and demonstrate the influence of parity-violating terms in the dynamics of the particle states. 

We derived necessary and sufficient conditions on the 14 parameters of the action so that all spin sectors of the theory are free from ghosts and tachyons and propagate simultaneously. This was made possible by examining the behaviour of the (gauge-invariant) propagator when sandwiched between conserved sources for the vielbein and connection. After linearizing the action around flat spacetime and moving to momentum space, we resorted to the spin-projection operator formalism that is used extensively for problems like the one addressed here. In order to account for terms that contain the totally antisymmetric tensor, we introduced in total 34 parity-violating projectors; most of them had never been constructed before. With the appropriate tools at hand we were able to decompose the action into 3 completely separate spin sectors and extract the corresponding coefficient matrices. Due to the presence of parity-odd terms, the computations concerning both massless and massive states was not as algebraically simple as in previous works. 

We considered first the massless sector of the theory that is a bit more involved in comparison to the massive one. Apart from the pole due to the graviton, the projection operators themselves introduce singularities at $k^2=0$. Since the choice of basis should not be of importance, we verified that these singularities are spurious and cancel in the final saturated propagator. We showed that the result for the graviton is identical to GR and at the same time we performed a non-trivial check of our algebra with this calculation.

We then turned our attention to the analysis of the massive degrees of freedom. Before inverting the coefficient matrices, we calculated the corresponding  determinants and specified what the physical masses of the particles are, i.e. where the poles of the propagators are located. Additionally, we found the residues of the propagators at the poles by inverting the coefficient matrices and evaluating them at the zeros of their determinants. 

Following that, we required: 
\vspace{-.3cm}
\begin{enumerate}[leftmargin=1.5cm]
\item Absence of negative masses, since they correspond to particles of tachyonic nature.

\item Positive-definite residues of the propagator at the poles; this guarantees that the particles' kinetic terms have the appropriate sign, therefore the theory is unitary.
\end{enumerate}

Imposing the above, we derived the constraints~\eqref{conditions-full-s0}-\eqref{conditions-full-s2} on the parameters of the theory, so that it contains only healthy states. As discussed in the main text, these inequalities cannot be satisfied simultaneously. Consequently, even though the massive spin-2, spin-1 and spin-0 fields do not exhibit tachyonic behaviour, it still contains ghosts. It should made clear though, that on the borders of the extended phase portrait, the resulting theory can be free from pathologies.

For example, among the many healthy subclasses of Poincar\'e gravitation (see~\cite{Sezgin:1979zf,Hayashi:1980qp,Sezgin:1981xs,Kuhfuss:1986rb} and the discussion in the main text) there are two that have been shown to be of great interest to the late Universe dynamics, since they can account for the present-day accelerated expansion. Let us shortly present them before concluding this section.

In the first, on top of the graviton, only a massive scalar and a pseudoscalar are present. This is achieved by completely eliminating the tensor and vector modes, i.e. by choosing the parameters that appear in the kinetic terms as $r_1=r_7=r_8=0,2r_3=r_4, r_5=-r_4$. A detailed analytical and numerical study of this case has been carried out in~\cite{Ho:2015ulu} and references therein.

The second interesting subclass contains -- in addition to the massless graviton -- one massive spin-2 field and a pseudoscalar. This particular model could be though of as the torsionful analog of  massive gravity and it is obtained by fixing $r_1=r_6=r_7=r_8=0, r_3=2r_4, t_2=t_3=-t_1$ and $t_4=t_5=0$. It should be noted that contrary to what happens in the Fierz-Pauli theory, the present  case apart from being ghost and tachyon free on the Minkowski background~\cite{Sezgin:1979zf,Hayashi:1980qp,Sezgin:1981xs}, it remains healthy also on Einstein manifolds~\cite{Nair:2008yh}.

\part{Phenomenology}
\label{part:pheno}

\chapter{Outline of Part II}

So far, we have  have solely discussed aspects of local (spacetime) symmetries and their significance for the physical systems that posses them. However, even if scale invariance is not gauged, its presence in a theory has far reaching ramifications for cosmological phenomenology as we will argue in the following.

It is now well accepted that the shortcomings of the hot big bang model can be solved in an elegant
way if we assume that the Universe underwent an inflationary period in
its early stages. The easiest way for this paradigm to be realized is
by a scalar field slowly rolling down towards the minimum of its
potential~\cite{Guth:1980zm,Linde:1981mu,Albrecht:1982wi,Linde:1983gd}. 

As discussed in~\cite{Bezrukov:2007ep}, inflation does not
necessarily  require the existence of a new degree of freedom. The
role of the inflaton can be played by the SM Higgs
field with its mass lying in the interval where the SM can be
considered a consistent effective field theory up to the inflationary
scale. More precisely, if the Higgs boson is non-minimally coupled to
gravity and the value of the corresponding coupling constant $\xi_h$
is sufficiently large, the model is able to provide a successful
inflationary period followed by a graceful exit to the standard hot
Big Bang theory~\cite{Bezrukov:2008ut,GarciaBellido:2008ab}. The
implications of this scenario have been  extensively studied in the
literature~\cite{Bezrukov:2008ej,
Barvinsky:2009fy,Barvinsky:2008ia,Bezrukov:2009db,Clark:2009dc,
Barvinsky:2009ii,Barvinsky:2009jd,Lerner:2010mq,Lerner:2009na,
Giudice:2010ka,Burgess:2009ea,Barbon:2009ya,Burgess:2010zq,Hertzberg:2010dc,  
Buck:2010sv,Lerner:2011it,Greenwood:2012aj}. Earlier studies of
non-minimally coupled scalar fields in the context of inflation can be
also found in~\cite{Spokoiny:1984bd,Salopek:1988qh,Fakir:1990eg}.

The Higgs inflation scenario can be easily incorporated into a larger framework, the Higgs-dilaton model~\cite{Shaposhnikov:2008xb,
GarciaBellido:2011de}. The key element of this extension is invariance under (global) scale transformations
\be
\label{scal-trans-fields-norm}
x^\m\rightarrow \a^{-1} x^\m\ ,~~~ g_{\m\n}(x)\rightarrow  g_{\m\n}(\a^{-1} x) \ ,~~~\text{and}~~~\Phi_i(x)\rightarrow \alpha^{d_i}\Phi_i(\alpha^{-1}x) \ ,
\ee
with $\a$ a constant, $\Phi_i$ the various (matter) fields and $d_i$ their scaling dimension. Thus, no dimensional parameters such as masses are
allowed to appear in the action, but instead, all the scales are induced
by the spontaneous breaking of this symmetry. As we have already mentioned, this can be achieved by the
introduction of the Goldstone boson related to the broken symmetry (the dilaton) which is
exactly massless. The coupling of the dilaton field to matter is weak
and takes place only through derivative couplings, not contradicting
therefore any 5th force experimental bounds~\cite{Kapner:2006si}.

Although the dilatation symmetry described above forbids the
introduction of a cosmological constant term, the ever-present
cosmological constant problem reappears associated to the fine-tuning
of the  dilaton self-interaction~\cite{Shaposhnikov:2008xb}. However,
if  the dilaton self-coupling $\beta$ is chosen to be zero (or
required to vanish due to some yet unknown reason), a slight
modification of GR, known as Unimodular Gravity
(UG), provides a dynamical dark energy (DE) stage responsible for the present day acceleration in good agreement
with observations. The scale-invariant  UG gives rise to a
symmetry-breaking ``run-away'' potential for the dilaton~\cite{Shaposhnikov:2008xb},
which plays the role of a quintessence field. The strength of such a
potential is determined by an integration constant $\Lambda_0$ that appears in the
Einstein equations of motion due to the unimodular constraint $\hat g\equiv -\det \l(\hat g_{\m\n}\r)
= 1$ on the metric determinant.  The common origin of the
inflationary and DE dominated stages in Higgs-dilaton inflation 
allowed to derive extra bounds on the initial inflationary
conditions,\footnote{The fine-tuning needed  to reproduce the present
dark energy abundance is transferred into the initial inflationary
conditions  for the fields at the beginning of inflation.} as well as a
potentially testable relation between the spectral tilt of scalar perturbations and the DE equation of state~\cite{GarciaBellido:2011de}.

When the model described above is rewritten in the
so-called Einstein frame, where the gravity part takes the usual
Einstein-Hilbert form, it becomes essentially non-polynomial and thus
non-renormalizable, even if the gravity part is dropped off.
Therefore, it should be understood as an effective field theory valid
only up to a certain ``cut-off'' scale. One should distinguish between
two different definitions of the ``cut-off''. Quite often  the cut-off
of the theory is understood as the energy at which  the tree level
unitarity in high-energy scattering processes is violated. A second
definition of the cut-off  is the energy associated to the onset of new
physics. As it was stressed in~\cite{Aydemir:2012nz},
the breaking of tree level unitarity does not imply the appearance of
new physics or extra degrees of freedom right above the corresponding
energy scale; it just signals that the perturbation theory in terms of
low-energy variables breaks down. For the case of Higgs and Higgs-dilaton inflation, the
tree-level scattering amplitudes {\emph{above the electroweak vacuum}}
appear to hit the perturbative unitarity bound at energies
$\Lambda\sim M_P/\xi_h$~\cite{Burgess:2009ea,Barbon:2009ya,Burgess:2010zq, Hertzberg:2010dc}.
Whether the theory requires an ultraviolet completion at these
energies or simply enters into the non-perturbative strong-coupling
regime with onset of new physics  at higher energies (which could be
as large as the Planck scale) is still an open question.
Nevertheless,  this scenario is {\emph self-consistent}, since the beginning of the strong coupling regime
(i.e. the cut-off scale according to the first definition which will be
used in this thesis) depends on the {\emph dynamical} expectation
values of the fields, which makes the theory weakly coupled for
all the relevant energy scales in the evolution of the Universe. 

It should be noted that even if the theory is unitary, this is not enough to guarantee that the tree-level results are robust against quantum effects. However, as we show in chapter~\ref{ch:HD_EFT}, if the symmetries of the theory are preserved at the quantum level as well, then the predictions of the Higgs-dilaton model are impervious to loop corrections. Thus, the connection between the early and late Universe observables
that the model predicts, remains unaltered.

In the Higgs-dilaton model, to achieve invariance under scale transformations, we were forced to introduce in an ad hoc manner an extra scalar field, the dilaton. This need not necessarily be the case, as we discuss in chapter~\ref{ch:SITDIFF}. It is well known that a self-consistent gravitational theory does not require invariance under the full group of diffeomorphisms~\cite{Buchmuller:1988wx,Alvarez:2006uu}. Rather, it is enough to consider the subgroup of the coordinate transformations with Jacobian equal to unity
\be
\label{tdiffs-def}
x'=F(x)\ ,~~~\text{such that}~~~J\equiv \left|\frac{\partial F}{\partial x}\right|=1\ ,
\ee
which constitute the \emph{transverse diffeomorphisms} (TDiffs), also called volume preserving diffeomorphisms. As one might expect, theories invariant under TDiffs contain -- in addition to the two polarizations of the massless graviton -- an extra propagating scalar mode associated with the determinant of the metric.\footnote{It is possible to eliminate this extra degree of freedom by forcing the determinant to take a constant value, like for example in the Higgs-dilaton model where it is fixed to be equal to one. In this case, we recover UG~\cite{vanderBij:1981ym,Unruh:1988in,Henneaux:1989zc}.} This minimalistic approach to gravitational dynamics, once combined with the requirement of exact scale invariance, results into an interesting class of theories (for which in what follows we will use the acronym SITDiff) in which the dilaton, being associated with the determinant of the metric, is already part of the gravitational sector~\cite{Blas:2011ac}.

When these theories are expressed in their diffeomorphism-invariant form, the action describing their dynamics includes an arbitrary integration constant that, in general, violates explicitly the scale symmetry. In the case of the Higgs-dilaton model, this is precisely what is behind the DE dominated stage. Notice that above we used ``in general'', because unlike a theory invariant under the full group of diffeomorphisms, the dimensionality of the metric plays a crucial role on whether the scale invariance of the system can be preserved. It turns out that when the metric carried dimension of area, then in the theory under consideration, dilatations are not broken.

Once the SM is coupled to this particular system, one can define a specific limit for the fields and their derivatives (associated with the ultraviolet domain) in which the only singular terms in the action correspond to the Higgs mass and the cosmological constant. It is very tempting to speculate that the self-consistency of the theory may require the regularity of the action, leading to the absence of these pathological terms. If this principle is to be taken at face value, one might attribute their presence at low energies to non-perturbative effects through some yet unknown mechanism.

\chapter{Higgs-dilaton cosmology}
\label{ch:HD_EFT}

\section{Introduction}

The first attempt to formulate a
viable scale-invariant theory non-minimally coupled to gravity  was done by Fujii
in~\cite{Fujii:1982ms}, although without establishing any 
connection to the SM Higgs. The role of dilatation symmetry in
cosmology  was  first considered by Wetterich in~\cite{Wetterich:1987fk,Wetterich:1987fm}. In these seminal papers, the
dynamical dark energy, associated with the  dilaton field, appears as
a consequence of the dilatation anomaly and is related to the 
breaking of SI by quantum effects. The present chapter, which has appeared in~\cite{Bezrukov:2012hx},
 has a number of
formal analogies and similarities regarding the cosmological
consequences for the late Universe with~\cite{Wetterich:1987fk,Wetterich:1987fm}. At the same time, our
approach to the source of dark energy is different from the one
adopted in~\cite{Wetterich:1987fk,Wetterich:1987fm}, as we
assume that SI is an exact (but spontaneously broken) symmetry at the
quantum level, leading therefore to a  massless dilaton. In~\cite{Wetterich:1987fk,Wetterich:1987fm}, both the cases of exact and
explicitly broken dilatation symmetry were considered. Our theory with
exact dilatation symmetry is different from that of~\cite{Wetterich:1987fk,Wetterich:1987fm} in two essential aspects.
First, in our work the Higgs field of the SM has non-minimal coupling
to gravity (it is absent in~\cite{Wetterich:1987fk,Wetterich:1987fm}), which is important for
the early Universe and leads to Higgs inflation. Second, the
unimodular character of gravity (as opposed to standard general
relativity used in~\cite{Wetterich:1987fk,Wetterich:1987fm}) leads to
an automatic and very particular type of dilatation symmetry  breaking, 
which results in dynamical dark energy due to the dilaton field
(absent in~\cite{Wetterich:1987fk,Wetterich:1987fm} for the case of
exact scale invariance).
 
Our purpose in this chapter is to study, following the approach of~\cite{Bezrukov:2010jz}, the self-consistency of the Higgs-Dilaton
model by adopting an effective field theory point of view. We will
estimate the field-dependent cut-offs associated to the different
interactions among scalars fields, gravity, vector bosons and
fermions. We will identify the lowest cut-off as a function of the
background fields and show that its value is higher than the typical
energy scales describing the Universe during its different epochs. The
issue concerning quantum corrections generated by the loop expansion
is also addressed.  Since the model is non-renormalizable, an infinite
number of counter-terms must be added in order to absorb the
divergences. It is important to stress at this point that, in the lack of a quantum theory for
gravity, the details of the regularization scheme to be used  cannot be univocally fixed. This means that the predictions
 of the model will be sensitive to the assumptions about the UV-completion of the theory 
(corresponding to different regularization prescriptions). We will adopt a ``minimal setup" that keeps intact the
exact and approximative symmetries of the classical action and does
not introduce any extra degrees of freedom.
 Within this approach, 
the relations connecting the inflationary and the dark energy
domination periods hold  even in the presence of quantum corrections.

The structure of the present chapter is as follows. In Sec.~\ref{sec:model}
we briefly review the Higgs-Dilaton model. In Sec~\ref{sec:cut-off} we calculate the cut-off of the theory in the Jordan
frame and compare it with the other relevant energy scales in the
evolution of the Universe. In Sec.~\ref{sec:divergences} we propose
a ``minimal setup'' which removes all the divergences and discuss the
sensitivity of the cosmological observables to radiative corrections.
Section~\ref{sec:conclusions5} contains the conclusions.

\section{Higgs-Dilaton cosmology}
\label{sec:model}
We start by reviewing the main results of~\cite{Shaposhnikov:2008xb,GarciaBellido:2011de}, where the
Higgs-Dilaton model was proposed and studied in detail. The two main
ingredients of the theory are outlined below. The first one is the 
invariance of the SM action under global scale transformations,
which leads to the absence of any dimensional parameters or scales. 

In order to achieve invariance under these transformations, we let the masses
and dimensional couplings in the theory to be dynamically induced by  a field. 
The simplest choice would be to use the SM Higgs, already present in the theory. Note however that this 
option is clearly incompatible with the experiment. As discussed in~\cite{Salopek:1988qh,CervantesCota:1995tz},
 the excitations of the Higgs field in this case
become massless and completely decoupled from the  SM
particles. 

The next simplest possibility  is to introduce a new scalar singlet under
the SM gauge group. We will refer to it as the dilaton $\chi$.  The coupling between the new 
field and the SM particles, with the exception of the Higgs boson, is
forbidden by quantum numbers. The corresponding Lagrangian is given by
\be
\label{general-theory}
\frac{\mathscr L}{\sqrt{g}}=
\frac{1}{2}(2\xi_h \varphi^\dagger \varphi+\xi_\chi \chi^2)R+
\mathscr L_{\text{SM}[\lambda\rightarrow0]}-
\frac{1}{2}g^{\mu\nu}\partial_\mu \chi\partial_\nu\chi-V(\varphi,\chi) \ ,
\ee
where $\varphi$ is the SM Higgs field doublet 
and $\xi_h\sim 10^3-10^5, \ \xi_\chi\sim 10^{-3}, $ are respectively the
non-minimal couplings of the Higgs and dilaton fields to gravity~\cite{GarciaBellido:2011de}. 
The term $\mathscr
L_{\text{SM}[\lambda\rightarrow0]}$ is the SM Lagrangian without the
Higgs potential, which in the present scale-invariant theory becomes
\be\label{general-potential}
V(\varphi,\chi)=\lambda\left(\varphi^\dagger \varphi-
\frac{\alpha}{2\lambda}\chi^2 \right)^2 +\beta \chi^4 \ ,
\ee
with $\lambda$ the self-coupling of the Higgs field. 

In order for this
theory to be phenomenologically viable, we demand the existence of
a symmetry-breaking ground state with non-vanishing background expectation
value for both\footnote{If $\bar\chi=0$
the Higgs field is massless, and if $\bar h=0$ there is no electroweak
symmetry breaking.}  the dilaton ($\bar\chi$) and the Higgs field in the unitary gauge
 ($\bar h$). This is given by
\be\label{ground-states}
\bar h^2=\frac{\alpha}{\lambda}\bar\chi^2 +\frac{\xi_h}{\lambda}R \ , 
\ \ \ \text{with} \ \ \  
R= \frac{4\beta\lambda}{\lambda\xi_\chi+\alpha\xi_h}\bar\chi^2 \,.
\ee
 All the physical scales are proportional to the non-zero background value of the
dilaton field. For instance, the SM Higgs mass is given by 
\be\label{Higgs-mass}
m_H^2=2\alpha M_P^2\frac{(1+6\xi_\chi)+
\frac{\alpha}{\lambda}(1+6\xi_h)}
{(1+6\xi_\chi)\xi_\chi+\frac{\alpha}{\lambda}(1+6\xi_h)\xi_h}
+\mathcal O(\beta) \ ,
\ee
with $M_P^2\equiv \xi_h \bar h^2+\xi_\chi \bar\chi^2\propto \bar\chi^2$
the effective Planck scale in the Jordan frame. The same happens with the effective cosmological 
constant
\be\label{cosm-const}
\Lambda=\frac{1}{4}M_P^2R=\frac{\beta M_P^4}
{(\xi_\chi+\frac{\alpha}{\lambda}\xi_h)^2+
4\frac{\beta}{\lambda}\xi_h^2} \ ,
\ee
which depending on the value of the dilaton self-coupling
$\beta$, gives rise to a
flat ($\beta=0$), deSitter ($\beta>0$) or anti-deSitter ($\beta <0$)
spacetime.
It is important to notice however that physical observables, corresponding to
dimensionless ratios between scales or masses, are independent of the
particular value of the background field $\bar\chi$. In order to reproduce the 
 ratio between the different energy scales, the parameters of the model 
must be properly fine-tuned. As shown in~\eqref{Higgs-mass}, the difference between the
electroweak and the Planck scale is encoded in the parameter\footnote{Note that the alternative choice 
$\xi_h \ggg 1$ is not compatible with CMB observations, see~\eqref{power} and Fig.~\ref{fig:nsxichiRG}.} $\alpha\sim 10^{-35}\lll 1$. Similarly, the hierarchy
 between the cosmological constant and the electroweak
scale~\eqref{cosm-const}, implies  $\beta\lll\alpha$. The smallness
 of these parameters, together with the tiny value of the non-minimal coupling
$\xi_\chi$, gives rise to an approximate shift symmetry for the dilaton field at the 
classical level, $\chi\to\chi+\text{const}$. As we will show in Sec.~\ref{sec:divergences}, this fact 
will will have important consequences for the analysis of the quantum effects.

The second ingredient of the Higgs-Dilaton cosmological model is the replacement of GR by Unimodular Gravity, which is just a 
particular case of the set of theories invariant under transverse diffeomorphisms. These
theories generically contain an extra scalar degree of freedom on top
of the massless graviton (for a general discussion see for instance~\cite{Blas:2011ac} and references therein). In UG
the number of dynamical components of the metric is effectively reduced to the standard value by 
requiring the metric determinant $\hat g$  to take some fixed constant value, conventionally
$\hat g= 1$. As shown in~\cite{Shaposhnikov:2008xb}, the equations of motion of a theory subject to
that constraint 
\be\label{unimod}
\mathscr L_{\text{UG}}= \mathscr L[\hat g_{\mu\nu},
\partial \hat g_{\mu\nu},\Phi,\partial\Phi] \ ,
\ee
coincide with those obtained from a diffeomorphism invariant theory 
 with modified action
\be\label{diff-inv}
\frac{\mathscr L}{\sqrt{g}}= 
\mathscr L[g_{\mu\nu},\partial g_{\mu\nu},
\Phi,\partial\Phi] +\Lambda_0 \ .
\ee
Note that, from the point of view of UG, the parameter $\Lambda_0$ is just a
conserved quantity associated to the unimodular constraint and it
should not be understood as a cosmological constant.

Since the two formulations are completely equivalent\footnote{As usual, there are some 
subtleties related to the quantum formulation of (unimodular) gravity. However, these will not play any role in the further 
developments. The interested reader is referred to the discussion in~\cite{Blas:2011ac} and references therein.}, we will stick to the diffeomorphism 
invariant language. Expressing the theory resulting from the
combination of the above ideas in the unitary gauge
$\varphi^T=(0,h/\sqrt{2})$ we get
\be\label{jord-theory}
\frac{\mathscr L}{\sqrt{g}}=\frac{1}{2}(\xi_h h^2+\xi_\chi \chi^2)R-
\frac{1}{2}(\partial \chi)^2-\frac{1}{2}(\partial h)^2-U(h,\chi) \ ,
\ee
where the potential includes now the UG integration constant $\Lambda_0$
\be\label{jord-pot}
U(h,\chi)\equiv V(h,\chi)+\Lambda_0=\frac{\lambda}{4}
\left(h^2-\frac{\alpha}{\lambda}\chi^2 \right)^2+\beta \chi^4+\Lambda_0 \ .
\ee
Notice that the Lagrangian given by~\eqref{jord-theory} and~\eqref{jord-pot} bears a clear
resemblance with the models  studied in~\cite{Wetterich:1987fk,Wetterich:1987fm}. In particular, it
coincides (up to the non-minimal coupling of the Higgs field to
gravity) with the Brans-Dicke theory with cosmological constant studied in~\cite{Wetterich:1987fk}. However, the interpretation  of the $\Lambda_0$ term is  different.  In our case this constant is not a 
fundamental parameter associated with the anomalous breaking of SI~\cite{Wetterich:1987fm}, but an automatic result of UG.

The phenomenological consequences of~\eqref{jord-theory} are more 
easily discussed in the Einstein frame. Let us then perform the following
redefinition of the metric $\tilde g_{\mu\nu}=\Omega^2g_{\mu\nu}$
with conformal factor  $\Omega^2=M_P^{-2}(\xi_h
h^2+\xi_\chi \chi^2)$. Using the standard relations
\be\label{confor2}
\sqrt{g}=\Omega^{-4}\sqrt{\tilde g}~~\text{and}~~R=\Omega^{2}\left(\tilde R+
6\tilde\square\log\Omega-6\tilde g^{\mu\nu}\partial_\mu
\log\Omega \ \partial_\nu \log\Omega\right) \ ,
\ee
we get
\be\label{einst-theory}
\frac{\mathscr L}{\sqrt{\tilde g}}=
\frac{M_P^2}{2}\tilde R-\frac{1}{2}\tilde K(h,\chi)-\tilde U(h,\chi) \ ,
\ee
where
\be\label{einst-pot}
\tilde U(h,\chi)\equiv\frac{U(h,\chi)}{\Omega^4}\equiv 
\frac{M_P^4}{(\xi_\chi \chi^2+\xi_h h^2)^2}
\left[ \frac{\lambda}{4}\left(h^2-\frac{\alpha}{\lambda}\chi^2 \right)^2+
\beta \chi^4	+\Lambda_0\right] \ ,
\ee
is the potential \eqref{jord-pot} in the new frame.
The non-canonical kinetic term in~\eqref{einst-theory} can be written as 
\be\label{kinetic}
\tilde K(h,\chi) = \kappa^E_{ij}
\tilde g^{\mu\nu}\partial_\mu\Phi^i\partial_\nu\Phi^j \ , 
\ee
where the quantity
\be\label{kmetric} 
\kappa^E_{ij}\equiv\frac{1}{\Omega^2}\left(\delta_{ij}+
\frac{3}{2}M_P^2\frac{\partial_i\Omega^2\partial_j\Omega^2}
{\Omega^2} \right) \ 
\ee
can be interpreted as the metric in the two-dimensional field space $(\Phi^1, \Phi^2)=(h,\chi)$ in the Einstein-frame. Note that, unlike the simplest Higgs inflationary
scenario~\cite{Bezrukov:2007ep}, expression~\eqref{kinetic} cannot be recast in canonical form by field redefinitions. In fact,  the Gaussian curvature
 associated to~\eqref{kmetric} does not identically vanish unless $\xi_h=\xi_\chi$, which, as shown
  in~\cite{GarciaBellido:2011de}, is not consistent with observations. Nevertheless, it is possible to write the 
  kinetic term in a quite simple diagonal form. As
shown in~\cite{GarciaBellido:2011de}, the whole inflationary period
takes place inside a field space domain in which the contribution
of the integration constant $\Lambda_0$ is completely negligible. We will refer to this domain as the
``scale invariant region'' and  assume that it is maintained even when the
radiative corrections are taken  into account (see Sec.~\ref{sec:divergences}). In this case, the dilatational 
Noether's current in the slow-roll approximation, $(1+6\xi_h)h^2+(1+6\xi_\chi)\chi^2$, 
 is approximately conserved, which suggests the definition of the set of variables
\be\label{var-rho-phi}
\rho= \frac{M_P}{2}\log\left[\frac{
(1+6\xi_h)h^2+(1+6\xi_\chi)\chi^2}{M_P^2} \right] \ ,\ \ \  
\tan\theta=\sqrt{\frac{1+6\xi_h}{1+6\xi_\chi}}\frac{h}{\chi} \ .
\ee
The physical interpretation of these variables is straightforward.
They are simply adequately rescaled polar variables in the $(h,\chi)$
plane. Expressed in terms of $\rho$ and $\theta$, the kinetic term \eqref{kinetic}
 turns out to be 
\be\label{polar-kin}
\tilde K= \left( \frac{1+6\xi_h}{\xi_h}\right)\frac{1}{\sin^2\theta+\varsigma\cos^2\theta}
(\partial \rho)^2\ +\frac{M_P^2 \ \varsigma}{\xi_\chi}
\frac{\tan^2\theta+\eta}{\cos^2\theta(\tan^2\theta+\varsigma)^2}(\partial \theta)^2 \ ,
\ee
with
\be\label{var-defs1}
\eta=\frac{\xi_\chi}{\xi_h} \ \
\ \text{and} \ \ \  \varsigma=\frac{(1+6\xi_h)\xi_\chi}{(1+6\xi_\chi)\xi_h} \ .
\ee 
The potential \eqref{einst-pot} is naturally divided into a scale-invariant part, depending only on the $\theta$ field, and
 a scale-breaking part, proportional to $\Lambda_0$ and depending on both $\theta$ and $\rho$. These are respectively given by 
\be\label{polar-pot}
\begin{aligned}
&\tilde U(\theta)= \frac{\lambda M_P^4}{4\xi_h^2}
\left(\frac{\sin^2\theta}
{\sin^2\theta+\varsigma\cos^2\theta}\right)^2 \ ,\\
&\tilde U_{\Lambda_0}(\rho,\theta)=\Lambda_0 \left( \frac{1+6\xi_h}{\xi_h}\right)^2
\frac{e^{-4\rho/M_P}}{(\sin^2\theta+\varsigma\cos^2\theta)^2} \ , 
\end{aligned}
\ee
where we have safely neglected the contribution of $\alpha$ and $\beta$ in~\eqref{einst-pot}.
Note that the non-minimal couplings of the fields to gravity with $\Lambda_0>0$ naturally generate a 
``run-away'' potential for the physical dilaton, similar  to those considered in the
 pioneering works on quintessence~\cite{Wetterich:1987fk,Wetterich:1987fm,Ratra:1987rm}.

The inflationary period of the expansion of the Universe 
takes place for field values $ \xi_h h^2 \gg \xi_\chi\chi^2$.   
From the definition of the angular variable $\theta$ in~\eqref{var-rho-phi}, this 
 corresponds to\footnote{Strictly speaking, the condition $\tan^2\theta\gg\eta$ holds beyond the inflationary
  region $ \xi_h h^2 \gg \xi_\chi\chi^2$ and includes also the reheating stage.}  $\tan^2\theta\gg\eta$. 
In that limit,
we can neglect the $\eta$ term in the kinetic term~\eqref{polar-kin} and perform an extra
field redefinition
\be\label{var-r-theta}
r= \gamma^{-1}\rho\ \ \ \ \text{and} \ \ \   
\vert\phi'\vert=\phi_0- \frac{M_P}{a}\tanh^{-1}\left[\sqrt{1-\varsigma}\cos\theta \ \right]  \ , 
\ee
where 
\be\label{var-defs2}
\gamma= \sqrt{\frac{\xi_\chi}{1+6\xi_\chi}} 
\ \ \ \text{and} \ \ \ a=\sqrt{\frac{\xi_\chi(1-\varsigma)}{\varsigma}} \ .
\ee
The variable $\phi'$ is periodic and defined in the
compact interval $\phi'
\in\left[-\phi_0,\phi_0\right]$, with $\phi_0=
M_P/a \ \tanh^{-1}\left[\sqrt{1-\varsigma}\ \right] $  the value of the 
field at the beginning of inflation. 
In terms of these variables the Lagrangian~\eqref{einst-theory} takes a
very simple form\footnote{Note that the definition of the angular variable 
$\phi$ used in this work is slightly different from that appearing in~\cite{GarciaBellido:2011de}. The new parametrization makes explicit the
 symmetry of the potential and shifts its minimum to make it coincide with that
  in Higgs-inflation.}
\be\label{angul-theory}
\frac{\mathscr L}{\sqrt{-\tilde g}}=\frac{M_P^2}{2}\tilde R -
\frac{\varsigma\cosh^2[a\phi/M_P]}{2}(\partial r)^2- 
\frac{1}{2}(\partial \phi)^2-\tilde U(\phi)-
\tilde U_{\Lambda_0}(r,\phi) \ ,
\ee
with $\phi= \phi_0-\vert\phi'\vert$. 
\begin{figure}
\centering
\includegraphics[scale=0.7]{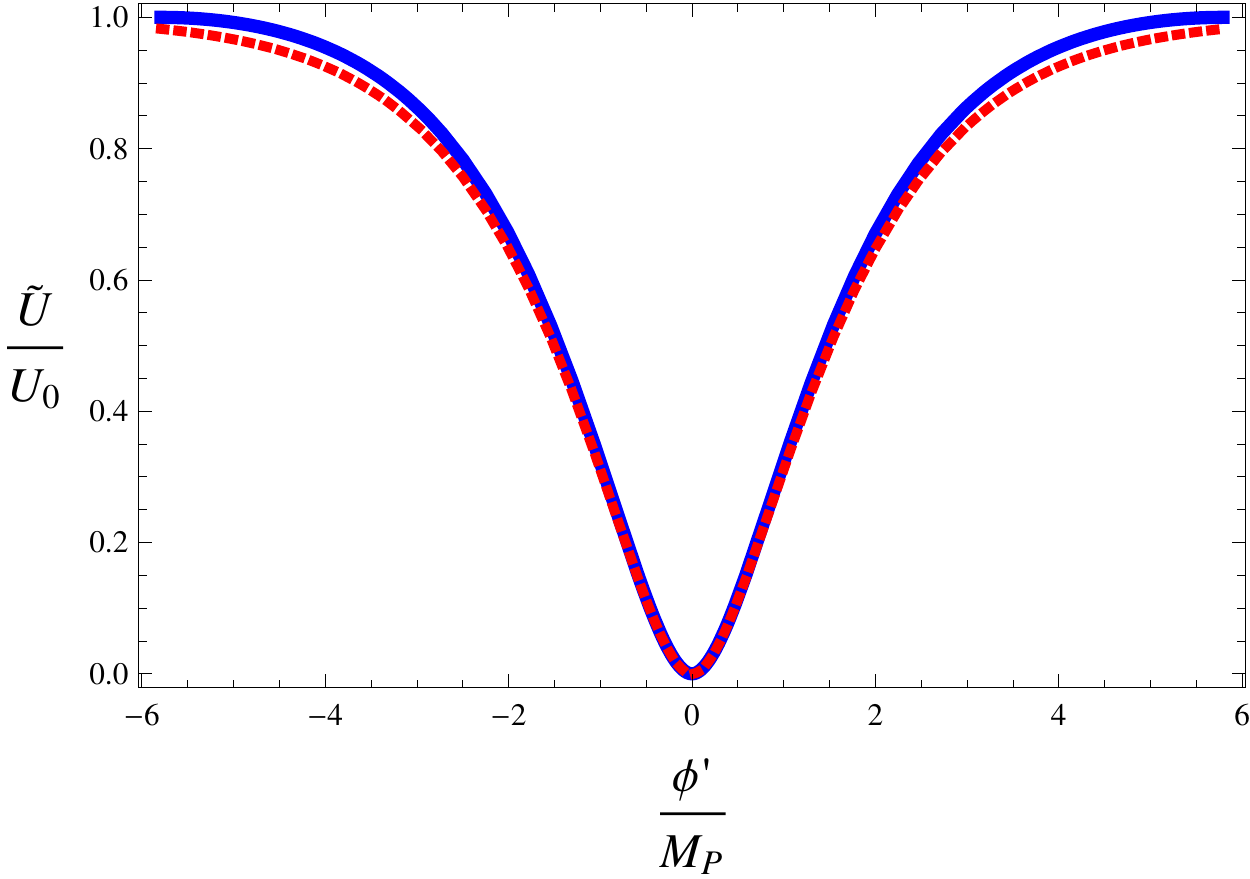}
\caption{Comparison between the Higgs-Dilaton inflationary potential
(blue continuous line) obtained from~\eqref{defin-pot} in the
scale-invariant region  and the corresponding one  for the Higgs
Inflation model (red dotted line). The amplitudes are normalized
to the asymptotic value $U_0=\frac{\lambda M_P^4}{4\xi_h^2}$.} 
\label{fig:pot-comparison}
\end{figure}
The potential~\eqref{polar-pot} becomes 
\be\label{defin-pot}
\begin{aligned}
&\tilde U(\phi)=\frac{\lambda M_P^4}{4\xi_h^2 (1-\varsigma)^2}
\left(1- \varsigma\cosh^2[a\phi/M_P]\right)^2 \ ,\\
&\tilde U_{\Lambda_0}(r,\phi)=\frac{\Lambda_0}{\gamma^4}\varsigma^2
\cosh^4[a\phi/M_P]e^{-4\gamma r/M_P} \ ,
\end{aligned}
\ee
whose scale-invariant part $\tilde U(\phi)$ resembles
 the potential of the simplest Higgs inflationary scenario~\cite{Bezrukov:2007ep}, see Fig.~\ref{fig:pot-comparison}.  
 The analytical expressions for the amplitude and the spectral tilt of scalar perturbations at 
order $\mathcal O(\xi_\chi,1/\xi_h,1/N^*)$
 can be easily calculated to obtain  \cite{GarciaBellido:2011de}
 \be\label{power}
P_\zeta(k_0)\simeq \frac{\lambda\sinh^2[4\xi_\chi N^*]}
{1152\pi^2\xi_\chi^2\xi_h^2} \ ,\hspace{10mm}
n_s(k_0)\simeq 1 -8\xi_\chi\coth(4\xi_\chi N^* ) \ ,
\ee
where $N^*$ denotes the number of e-folds between the moment at which 
the pivot scale $k_0/a_0=0.002 \ \text{Mpc}^{-1}$ exited the horizon 
and the end of inflation. Note that for  $1< 4\xi_\chi
N^*\ll 4N^*$, the expression for the tilt simplifies and becomes linear in $\xi_\chi$
\be\label{as-tilt}
n_s(k_0)\simeq 1-8\xi_\chi \ .
\ee

An interesting cosmological phenomenology arises with the peculiar
choice\footnote{Some arguments in favour of the $\beta=0$
case can be found in~\cite{Shaposhnikov:2008xi,
GarciaBellido:2011de,Blas:2011ac}.} $\beta=0$. In this case, the DE
dominated period in the late Universe depends only on  the dilaton
field $\rho$, which give rise to an intriguing relation between the inflationary and DE domination 
periods. Let us start by noticing that around the minimum
 of the potential the value of $\theta$ is very close to zero. In that limit, $\tan^2\theta\ll\eta$,  which prevents
 the use of the field redefinition~\eqref{var-r-theta}. The appropriate redefinitions needed to diagonalize the kinetic term~\eqref{polar-kin}
  in this case turn out to be
\be\label{low-en}
r=\gamma^{-1}\rho \ \ \ \text{and} \ \ \ \phi'\simeq  
\frac{M_P}{\sqrt{\xi_h \varsigma}}\theta \ .
\ee
Using~\eqref{polar-kin} and \eqref{polar-pot}, it is straightforward 
to show that the part of the theory associated to the Higgs field $\phi$
 simplifies to the SM one. The resulting scale-invariance breaking potential for
the dilaton is still of the ``run-away'' type
\be\label{darkener-pot}
\tilde U_{\Lambda_0}(r)=
\frac{\Lambda_0}{\gamma^4}e^{-4\gamma r/M_P} \ ,
\ee
 making it suitable for playing the role of quintessence. 
Let us assume that 
$\tilde U_{\Lambda_0}$ is negligible during the radiation
and matter dominated stages but responsible for the present
accelerated expansion of the Universe. In that case, it is possible to
write the following relation between the equation of state parameter
$\omega_r$ of the $r$ field and its  relative abundance $\Omega_r$
\cite{Scherrer:2007pu}
\be\label{eosp-abund}
1+\omega_r=\frac{16\gamma^2}{3}\left[ \frac{1}{\sqrt{\Omega_r}}-
\frac{1}{2}\left(\frac{1}{\Omega_r}-1\right)\log\frac{1+
\sqrt{\Omega_r}}{1-\sqrt{\Omega_r}}\right]^2 \ . 
\ee
For the present DE density $\Omega_\text{DE}=\Omega_r\simeq 0.74$, the above expression yields
\be\label{reduc}
1+\omega_\text{DE}=\frac{8}{3}\frac{\xi_\chi}{1+6\xi_\chi} \ .
\ee
Comparing~\eqref{as-tilt} and \eqref{reduc}, it follows that the
deviation of the scalar tilt $n_s$  from the scale-invariant one is
proportional to the deviation of the DE equation of state from a cosmological
constant\footnote{Outside this region of parameter space, the relation
connecting $n_s \ \text{to} \ \omega_\text{DE}$ is somehow more
complicated 
\be
\label{functional-rel-compl}\nonumber
n_s-1 \simeq -\frac{12(1+\omega_\text{DE})}
{4-9(1+\omega_\text{DE})}\coth\left[\frac{6N^*(1+
\omega_\text{DE})}{4-9(1+\omega_\text{DE})} \right] \ .
\ee}~\cite{GarciaBellido:2011de}
\be\label{functional-rel}
n_s-1 \simeq -3(1+\omega_\text{DE}), \ \ \ \text{for} \ \ \ 
  \frac{2}{3N^*} < 1+\omega_\text{DE}\ll 1 \ . 
\ee
The above condition is a non-trivial
prediction of Higgs-Dilaton cosmology, relating two a priori
completely independent periods in the history of the Universe. This
has interesting consequences from an observational point of
view\footnote{Similar consistency relations relating the rate of
change of the equation of state parameter $w(a) = w_0 + w_a (1 - a)$
with the logarithmic running of the scalar tilt can be also derived~\cite{GarciaBellido:2011de}. The practical relevance of those
consistence conditions is however much more limited than that of~\eqref{functional-rel}, given the small value of the running of the
scalar tilt in Higgs-driven scenarios.} and makes the Higgs-Dilaton
scenario rather unique. We will be back to this point
 in Sec.~\ref{sec:divergences}, where  we
will show that the consistency relation~\eqref{functional-rel} still holds even in  the
 presence of quantum corrections computed within the ``minimal setup''. 

\section{The dynamical cut-off scale}
\label{sec:cut-off}
Following~\cite{Bezrukov:2010jz}, we now turn to the determination of
the energy domain where the Higgs-Dilaton model can be considered as a
predictive effective field theory. This domain is bounded from above
by the field-dependent cut-off $\Lambda(\Phi)$, i.e. the energy where
perturbative tree-level unitarity is violated~\cite{Cornwall:1974km}.
At energies above that scale, the theory becomes strongly-coupled and
the standard perturbative methods fail. In order to determine this
 (background dependent) energy scale, two
related methods, listed below, can be used.
\begin{enumerate}
\item[(1)] Expand the generic fields of the theory around their background
values 
\be\label{split}
\Phi(\mathbf x,t)=\bar\Phi+\delta\Phi(\mathbf x,t) \ ,
\ee 
such that all kind of higher-dimensional non-renormalizable operators
\be\label{non-renorm-op}
c_n\frac{\mathcal O_n(\delta\Phi)}{[\Lambda(\bar\Phi)]^{n-4}} \ ,
\ee
with $c_n\sim\mathcal O(1)$ appear in the resulting action. These
operators are suppressed by appropriate powers of the field-dependent
coefficient $\Lambda(\bar\Phi)$, which can be identified as the
cut-off of the theory.  This procedure gives us only a lower estimate
of the cut-off, since it does not take into account the possible
cancelations that might occur between the different scattering diagrams.
\item[(2)] Calculate at which energy each of the N-particle scattering
amplitudes hit the unitarity bound. The cut-off will then be the
lowest of these scales. 
\end{enumerate}

In what follows we will apply these two methods to determine the
effective cut-off of the theory. We will start by applying the
method $(1)$ to compute the cut-off associated with the
gravitational and scalar interactions. The cut-off associated to the
gauge and fermionic sectors will be obtained via the method
$(2)$.

\subsection{Cut-off in the scalar-gravity sector}

We choose to work in the original Jordan frame where the Higgs and
dilaton fields are non-minimally coupled to gravity\footnote{A similar study in the Einstein frame can be found in Appendix~\ref{app:Einstein_cut}.}. Expanding these 
fields around a static background\footnote{Note that, in comparison with 
the analysis performed in~\cite{Lerner:2011it} for generalized Higgs inflationary models, both the dilaton and
 the Higgs field acquire a non-zero background expectation value, see Sec.~\ref{sec:model}. As we will see below, 
 this will give rise to a much richer cut-off structure.}
 \be\label{expansion}
g_{\mu\nu}=\bar g_{\mu\nu}+\delta g_{\mu\nu} \ , \hspace{.5cm}
\chi=\bar\chi+\delta\chi \ , \hspace{.5cm} h=\bar h+\delta h \ ,
\ee
we obtain the following kinetic term for the quadratic Lagrangian of
the gravity and scalar sectors
\be\label{quadratic}
\begin{aligned}
&\mathscr K_2^{\text{G+S}}=\frac{\xi_\chi \bar \chi^2 +\xi_h\bar
h^2}{8}\left(\delta g^{\mu\nu}\square \delta g_{\mu\nu}+2\partial_\nu
\delta g^{\mu\nu}\partial^\rho\delta g_{\mu\rho}-2\partial_\nu \delta
g^{\mu\nu}\partial_{\mu}\delta g \right.\\
&\left.-\delta g\square \delta g\right)-\frac{1}{2}(\partial \delta \chi)^2 -\frac{1}{2}(\partial \delta h)^2
+(\xi_\chi \bar\chi\delta\chi +\xi_h \bar h \delta
h)(\partial_\lambda\partial_\rho\delta g^{\lambda\rho}-\square\delta
g) \ .
\end{aligned}
\ee
The leading higher-order non-renormalizable operators obtained in this
way are given by
\be\label{interact}
 \xi_\chi(\delta \chi)^2\square \delta g \ , \ \ \  \xi_h(\delta
h)^2\square \delta g \ .
\ee
Note that these operators are written in terms of quantum excitations
with non-diagonal kinetic terms. In order to properly identify the
cut-off of the theory, we should determine the normal modes that
diagonalize the quadratic Lagrangian~\eqref{quadratic}. After doing that,
and using the equations of motion to eliminate artificial degrees of
freedom, we find that the metric perturbations depend on the scalar
fields perturbations, a fact that is implicit in the Lagrangian~\eqref{quadratic}. The gravitational part of  the above action 
can be recast into canonical form in terms of a new
metric perturbation $\delta\hat g_{\mu\nu}$ given by
\be\label{redefinition-metric}
\delta\hat g_{\mu\nu}=\frac{1}{\sqrt{\xi_\chi\bar \chi^2 +\xi_h \bar
h^2}}\left[(\xi_\chi\bar \chi^2 +\xi_h \bar h^2)\delta  g_{\mu\nu}
+2\bar g_{\mu\nu}(\xi_\chi\bar\chi \delta\chi+\xi_h\bar h \delta h)
\right] \ .
\ee
The cut-off scale associated to  purely gravitational interactions
becomes in this way the effective Planck scale in  the Jordan frame 
\be\label{gravcut-off}
\Lambda_{P}^2=\xi_\chi\bar \chi^2+ \xi_h \bar h^2 \ .
\ee
The remaining non-diagonal kinetic term for the scalar perturbations 
\\ $(\delta\Phi^1, \delta\Phi^2)=(\delta h,\delta \chi)$ is given
in compact matrix notation by
\be\label{matrixnot}
\mathscr K_2^{\text{S}}=-\frac{1}{2}\bar\kappa^J_{ij}\partial_\mu
\delta\Phi^i\partial^\mu\delta\Phi^j \ ,
\ee 
where $\bar \kappa^J_{ij}=\Omega^2 \bar \kappa^E_{ij}$ is the Jordan frame 
analogue of~\eqref{kmetric} and depends
only on the background values of the fields, i.e.
\be\label{kinetic-fields}
\bar\kappa^J_{ij}= \frac{1}{\xi_\chi\bar\chi^2+\xi_h\bar
h^2}\begin{pmatrix}\xi_\chi \bar\chi^2(1+6\xi_\chi) +\xi_h\bar h^2&
6\xi_\chi \bar\chi\xi_h\bar h\\ 6\xi_\chi \bar\chi\xi_h\bar h&\xi_\chi
\bar\chi^2 +\xi_h\bar h^2(1+6\xi_h)
\end{pmatrix} \ .
\ee
In order to diagonalize the above expression we make use of the
following set of variables
\be\label{redefinition-fields}
\begin{aligned}
&\delta\hat\chi=\sqrt{\frac{\xi_\chi\bar \chi^2(1+6\xi_\chi)
+\xi_h \bar h^2(1+6\xi_h)}{(\xi_\chi^2\bar \chi^2 +\xi_h^2 \bar
h^2)(\xi_\chi\bar \chi^2 +\xi_h \bar h^2)}}\left(\xi_\chi\bar\chi
\delta\chi+\xi_h\bar h \delta h\right) \
, \\ &\delta \hat h=\frac{1}{\sqrt{\xi_\chi^2\bar \chi^2 +\xi_h^2 \bar
h^2}}\left(-\xi_h\bar h \delta\chi+\xi_\chi\bar\chi\delta h\right) \ .
\end{aligned}
\ee
Note here that this is precisely the change of
variables (up to an appropriate rescaling with the conformal factor
$\Omega$) needed to diagonalize the kinetic terms for the scalar
perturbations in the Einstein frame. To see this, it is enough to
start from~\eqref{kinetic} and expand the fields around their
background values $\Phi^i\rightarrow \bar\Phi^i+\delta\Phi^i$. Keeping
the terms with the lowest power in the excitations, $\tilde K=
\bar\kappa^E_{ij}\partial_\mu\delta\Phi^i\partial^\mu\delta\Phi^j+\mathcal
O(\delta\Phi^3)$, it is straightforward to show that the previous
expression can be diagonalized in terms of 
\be\label{redefinition-fields-einstein}
\begin{aligned}
&\delta \hat\chi=\bar\Omega^{-1}\sqrt{\frac{\xi_\chi\bar \chi^2(1+6\xi_\chi) +\xi_h
\bar h^2(1+6\xi_h)}{(\xi_\chi^2\bar \chi^2 +\xi_h^2 \bar
h^2)(\xi_\chi\bar \chi^2 +\xi_h \bar h^2)}}\left(\xi_\chi\bar\chi
\delta\chi+\xi_h\bar h \delta h\right) \ , \\ 
&\delta\hat h=\bar\Omega^{-1}\frac{1}{\sqrt{\xi_\chi^2\bar \chi^2
+\xi_h^2 \bar h^2}}\left(-\xi_h\bar h
\delta\chi+\xi_\chi\bar\chi\delta h\right) \ .
\end{aligned}
\ee 
Written in terms of the canonically normalized
variables~\eqref{redefinition-metric} and~\eqref{redefinition-fields}
these operators read
\be\label{higherdimensional}
\frac{1}{\Lambda_1}(\delta\hat h)^2\square \delta \hat g \ , \ \ \ 
\frac{1}{\Lambda_2}(\delta\hat \chi)^2\square \delta \hat g \ , \ \ \ 
\frac{1}{\Lambda_3}(\delta\hat\chi)(\delta\hat h) \square \delta \hat
g \ ,
\ee
where the different cut-off scales are given by
\begin{align}
\centering
\label{cut-off1}
\Lambda_1&=\frac{\xi_\chi^2\bar \chi^2 +\xi_h^2 \bar
h^2}{\xi_\chi\xi_h\sqrt{\xi_\chi\bar \chi^2 +\xi_h \bar h^2}} \ ,\\
\label{cut-off2}
\Lambda_2&=\frac{(\xi_\chi^2\bar \chi^2 +\xi_h^2 \bar
h^2)(\xi_\chi\bar \chi^2(1+6\xi_\chi) +\xi_h \bar
h^2(1+6\xi_h))}{(\xi_\chi^3\bar \chi^2 +\xi_h^3 \bar
h^2)\sqrt{\xi_\chi\bar \chi^2 +\xi_h \bar h^2}} \ , \\
\label{cut-off3}
\Lambda_3&=\frac{(\xi_\chi^2\bar \chi^2 +\xi_h^2 \bar
h^2)(\xi_\chi\bar \chi^2(1+6\xi_\chi) +\xi_h \bar
h^2(1+6\xi_h))}{\xi_\chi\bar \chi\xi_h\bar h\left\vert \xi_h
-\xi_\chi\right\vert\sqrt{\xi_\chi\bar \chi^2 +\xi_h \bar h^2}} \ .
\end{align}
The effective cut-off of the scalar theory at a given value of the background
fields will be the lowest of the previous scales. We will be back to this point in 
Sec.~\ref{sec:comparison}.

\subsection{Cut-off in the gauge and fermionic sectors}\label{sec:gauge-cut-off}
Let us now move to the cut-off associated with the gauge sector. Since
we are working in the unitary gauge for the SM fields, it
is sufficient to look at the tree-level scattering of non-abelian
vector fields with longitudinal polarization. It is well known that in
the SM the ``good'' high energy behaviour of these processes is 
the result of cancellations that occur when we take into account the
interactions of the gauge bosons with the excitations $\delta h$ of
the Higgs field\footnote{In the absence of the Higgs field, the scattering
amplitudes grow as the square of the center-of-mass energy, due to
the momenta dependence of the longitudinal vectors $\sim q^\mu/m_W$.} ~\cite{Lee:1977yc,Lee:1977eg}. 

In our case, even though purely gauge interactions remain unchanged,
the graphs involving the Higgs field excitations are modified due to
the non-canonical kinetic term. This changes the pattern of  the
cancellations that occur in the standard Higgs mechanism, altering
therefore the asymptotic behaviour of these processes. As a result, the
energy scale where this part of the theory becomes strongly coupled becomes lower. 

To illustrate how this happens, let us consider the  $W_L
W_L\rightarrow W_L W_L$ scattering in the $s-$channel. The relevant
part of the Lagrangian is 
\be\label{lagr-gauge}
g\, m_{W}W_\mu^+ W^{-\mu}\delta h \ , 
\ee
where $m_W\sim g \bar h$. After diagonalizing the kinetic term for the
scalar fields with the change of variables~\eqref{redefinition-fields}, the above expression becomes
\be\label{lagr-gauge-diag}
g'  m_W W_\mu^+ W^{-\mu}\delta \hat h + g''  m_W W_\mu^+
W^{-\mu}\delta \hat \chi \ , 
\ee
where the effective coupling constants $g' \ \text{and} \ g''$ are
given by
\be\label{eff-coupl}
\begin{aligned}
&g' = g \frac{\xi_\chi\bar\chi}{\sqrt{\xi_\chi^2\bar\chi^2+\xi_h^2 \bar
h^2}} \ , \\ 
&g'' = g \frac{\xi_h\bar
h}{\sqrt{\xi_\chi^2\bar\chi^2+\xi_h^2 \bar
h^2}}\sqrt{\frac{\xi_\chi\bar\chi^2+\xi_h \bar
h^2}{\xi_\chi\bar\chi^2(1+6\xi_\chi)+\xi_h\bar h^2(1+6\xi_h)}} \ .
\end{aligned}
\ee

From the requirement of tree unitarity of the $S$-matrix, it is
straightforward to show that the scattering amplitude of this
interaction hits the perturbative unitarity bound at energies
\be\label{gauge-cut-off}
\Lambda_G\simeq\sqrt{
\frac{\xi_\chi\bar\chi^2(1+6\xi_\chi)+\xi_h\bar
h^2(1+6\xi_h)}{6\xi_h^2}} \ . 
\ee
It is interesting to compare the previous expression with the results for the gauge cut-off of the simplest
Higgs inflationary model~\cite{Bezrukov:2010jz}. In order to do that, let us consider two limiting cases: 
 the inflationary/high-energy period corresponding to field values $\xi_\chi \chi^2 \ll \xi_h h^2$ and the 
 low-energy regime at which  $\xi_\chi \chi^2 \gg \xi_h h^2$ . In these two cases, the above 
 expression simplifies to
\be\label{lim-gauge} 
\Lambda_G\simeq \Bigg\{ \begin{array}{cl} \bar h
&\mbox{for} \ \xi_\chi\bar\chi^2\ll \xi_h \bar h^2 \ , \\
\frac{\sqrt{\xi_\chi}\bar\chi}{\xi_h} &\mbox{for} \
\xi_\chi\bar\chi^2\gg \xi_h \bar h^2 \ , \\
\end{array} 
\ee
in agreement with the Higgs inflation model. 

To identify the cut-off of the fermionic part of the Higgs-Dilaton
model, we consider the chirality non-conserving process $\bar f
f\rightarrow W_L W_L$. This interaction receives contributions from
diagrams with $\gamma$ and $Z$ exchange  ($s-$channel) and from a
diagram with fermion exchange ($t-$channel). In the asymptotic
high-energy limit, the total amplitude of these graphs grows linearly
with the energy at the center of mass. Once again, the $s-$channel
diagram including the Higgs excitations unitarizes the associated amplitude~\cite{Chanowitz:1978uj,Chanowitz:1978mv,Appelquist:1987cf}.  Following
therefore the same steps as in the calculation of the gauge cut-off,
we find that this part of the theory enters into the strong-coupling
regime at energies
\be\label{ferm-cut-off}
\Lambda_F\simeq y^{-1}
\frac{\xi_\chi\bar\chi^2(1+6\xi_\chi)+\xi_h\bar
h^2(1+6\xi_h)}{6\xi_h^2 \bar h} \ , 
\ee
where $y$ is the Yukawa coupling constant. The above cut-off is higher
than that of the SM gauge interactions~\eqref{gauge-cut-off}
during the whole evolution of the Universe. 

\subsection{Comparison with the energy scales in the early and late Universe}\label{sec:comparison}

In this section we compare the cut-offs found above with the
characteristic energy scales in the different periods during the
evolution of the Universe. If the typical momenta involved in the
different processes are sufficiently small, the theory will remain in
the weak coupling limit, making the Higgs-Dilaton scenario
self-consistent.

\begin{figure}
\centering
\includegraphics[scale=.8]{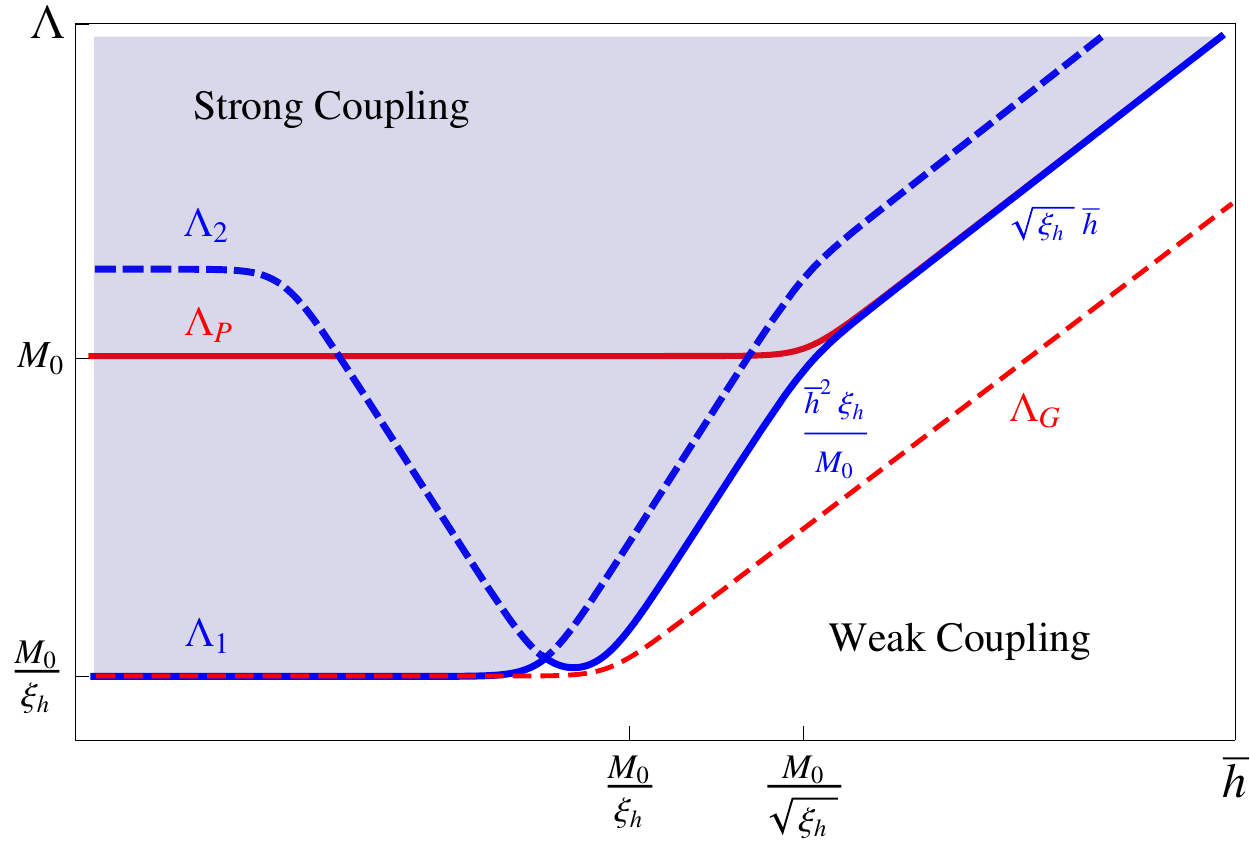}
\caption{Dependence of the different cut-off scales for a fixed value
of the dilaton field $\bar\chi$ as a function of the Higgs field $\bar
h$ in the Jordan frame. The cut-off~\eqref{cut-off3} is parametrically
above the other energy scales ($\Lambda_1, \ \Lambda_2,\
\Lambda_P,\  \Lambda_G \ \text{and} \ \Lambda_F$) during the
whole history and it is therefore not included in the figure. The
effective field theory description of scalar fields is applicable for
typical energies below the thick blue solid line, which correspond to
the minimum of the scalar cut-off scales at a given field value. This
is  given by  $\Lambda_2$ and $\Lambda_1$ in the scalar sector, for
large and small Higgs values respectively. The red solid line
correspond to the gravitational cut-off~\eqref{gravcut-off}, while the
red dashed one corresponds to the gauge cut-off~\eqref{gauge-cut-off}.
They coincide with the effective scalar cut-off for the limiting values of the Higgs field. 
 The scale $M_0$ is defined as
$M_0=\sqrt{\xi_\chi}\bar\chi$ and corresponds to the value of the effective
Planck mass at low energies.}
\label{fig:cut-off-comparison}
\end{figure}

Let us start by considering the inflationary period, characterized by 
$\xi_h\bar h^2 \gg \xi_\chi\bar\chi^2$. As shown in Fig.~\ref{fig:cut-off-comparison}, the 
lowest cut-off in this region is the
one associated with the gauge interactions $\Lambda_G$. The
typical momenta of the scalar perturbations produced during inflation
are of the order of the Hubble parameter at that time. This quantity
can be easily estimated in the Einstein frame, where it is basically
determined by the energy stored in the inflationary potential~\eqref{defin-pot}. We obtain $\tilde H\sim \sqrt{\lambda}M_P/\xi_h$.
When transformed to the Jordan frame ($H=\Omega \tilde H$) this
quantity becomes $H\sim \sqrt{\frac{\lambda}{\xi_h}}\bar h$, which is
significantly below the cut-off scale $\Lambda_G$ in that
region. The same conclusion is obtained for the total energy density,
which turns out to be much smaller than $\Lambda_G^4$.
Moreover, the cut-off $\Lambda_G$ exceeds the masses of all
particles in the Higgs background, allowing a
self-consistent estimate of radiative corrections (see Sec.~\ref{sec:divergences}). 

After the end of inflation, the field $\phi$ starts to oscillate
around the minimum of the potential with a decreasing amplitude, due to
the expansion of the Universe and particle production. This
amplitude varies between $M_0/\sqrt{\xi_h}$ and
$M_0/\xi_h$, where $M_0=\sqrt{\xi_\chi}\bar \chi$ is
the asymptotic Planck scale in the low energy regime. As shown in Fig.~\ref{fig:pot-comparison}, the 
curvature of the Higgs-Dilaton potential
around the minimum coincides (up to ${\cal{O}}(\xi_\chi)$ 
corrections) with that of the Higgs-inflation scenario.  All
the relevant physical scales, including the effective gauge and
fermion masses, agree, up to small corrections, with those in
Higgs-inflation~\cite{GarciaBellido:2012zu} . This allows us to directly apply the results of~\cite{Bezrukov:2008ut,GarciaBellido:2008ab,Bezrukov:2011sz} to the
Higgs-Dilaton scenario. According to these works, the typical momenta
of the gauge bosons produced at the minimum of the potential  in the
Einstein frame is of order $\tilde k\sim (\tilde m_A/M)^{2/3}M$, with
$\tilde m_A$ the mass of the gauge bosons in the Einstein frame and
$M=\sqrt{\lambda/3}M_P/\xi_h$ the curvature of the potential around
the minimum. After transforming to the Jordan frame we obtain $k\sim
\left(\frac{\lambda
g^4}{\xi_h}\right)^{1/6}\Lambda_G$, with $g$ the weak
coupling constant. The typical momentum of the created gauge bosons is
therefore parametrically below the gauge cut-off scale~\eqref{lim-gauge} in that region.

At the end of the reheating period, $\xi_\chi\bar\chi^2\gg \xi_h \bar h^2$, the system settles down to the
minimum of the potential $\tilde U(\phi)$, see~\eqref{defin-pot}. In that region the
effective Planck mass coincides with the value $M_0$. The cut-off
scale becomes  $\Lambda_1\simeq \sqrt{\xi_\chi}\bar\chi/\xi_h\simeq 
M_P/\xi_h$.
This value is much higher than the electroweak scale $m_H^2\sim
2\alpha/\xi_\chi M_P$ (see~\eqref{Higgs-mass}) where all the
physical processes take place. We conclude therefore that perturbative
unitarity is maintained for all the relevant processes 
during the whole evolution of the Universe.

\section{Quantum corrections}
\label{sec:divergences}

In this section we concentrate on the radiative corrections to the
inflationary potential  and on their influence on the predictions of
the model.

Our strategy is as follows. We regularize the quantum theory in such a
way that all multi-loop diagrams are finite, whereas the exact
symmetries of the chosen classical action (gauge, diffeomorphisms and
scale invariance) remain intact. Moreover, we will require the
regularization to respect the approximate  shift symmetry of the
dilaton field in the Jordan frame, see Sec.~\ref{sec:model}.  Then
we add to the classical action an infinite number of counter-terms
(including the finite parts as well) which remove all the divergences
from the theory and do not spoil the exact and approximate symmetries
of the classical action. Since the theory is not renormalizable, these
counter-terms will have a different structure  from that of the
classical action.  In particular, terms that are non-analytic with
respect to the Higgs and dilaton  fields will appear~\cite{Shaposhnikov:2009nk}. They can be considered as
higher-dimensional operators, suppressed by the field-dependent
cut-offs. For consistency with the analysis made earlier in this work,
we demand these cut-offs to exceed those found in Sec.~\ref{sec:cut-off}.

An example of the subtraction procedure which satisfies all the
requirements  formulated above has been constructed in~\cite{Shaposhnikov:2008xi} (see also earlier discussion in~\cite{Englert:1976ep}). It is based on dimensional regularization in
which the 't Hooft-Veltman normalization point $\mu$ is replaced by
some combination of the scalar fields with an appropriate dimension,
$\mu^2\to F(\chi,h)$ (we underline that we use the Jordan frame here
for all definitions). The infinite part of the counter-terms is
defined as in $\overline{MS}$ prescription, i.e. by subtracting the
pole terms in $\epsilon$, where the dimensionality of space-time is
$D=4-2\epsilon$. The finite part of the counter-terms has the same
operator structure as the infinite part, including the parametric
dependence on the coupling constants.  

Although the requirement of the structure of higher-dimensional
operators, formulated in the previous paragraphs puts important
constraints on the function $F(\chi,h)$, its precise form is not
completely determined~\cite{Shaposhnikov:2008xi,Shaposhnikov:2009nk,Codello:2012sn}, and the
physical results {\em do depend} on the choice of $F(\chi,h)$. This
somewhat mysterious fact from the point of view of uniquely defined
classical theory~\eqref{general-theory} becomes clear if we  recall
that we are dealing with a non-renormalizable theory. The quantization
of this kind of theories requires the choice of a particular classical
action together  with a set of subtraction rules. The ambiguity in the
choice of the field-dependent  normalization point $F(\chi,h)$ simply
reflects our ignorance about the proper set of rules. Different
subtractions prescriptions applied to  the same classical action do
produce unequal results. Sometimes this ambiguity is formulated as a
dependence of quantum theory on the choice of  conformally related
frames in scalar-tensor theories~\cite{Flanagan:2004bz}. The use of
the {\em same} quantization rules in different frames would lead to
quantum theories with different choices of  $F(\chi,h)$.

Among the many possibilities, the simplest  and most natural choice 
is to  identify the normalization point in the Jordan frame with the
gravitational cut-off~\eqref{gravcut-off},
\begin{equation}
\label{prescription1J}
\mu_I^2 \propto \xi_\chi
\chi^2 + \xi_h h^2,
\end{equation}
which corresponds to the scale-invariant
prescription of~\cite{Shaposhnikov:2008xi}. In the Einstein frame
the previous choice becomes standard (field-independent)
\be\label{prescription1E}
\tilde \mu_I^2 \propto M_P^2\,.
\ee
A second possibility is to choose the scale-invariant direction along
the dilaton field, i.e.
\begin{equation}
  \label{prescription2J}
  \mu_{II}^2\propto \xi_\chi \chi^2.
\end{equation}
When transformed to the Einstein frame it becomes  
\be\label{prescription2E}
\tilde \mu_{II}^2 \propto \frac{\xi_{\chi}\chi^2 M_P^2}{\xi_\chi \chi^2 + \xi_h h^2}\,,
\ee
and coincides with the prescription II of~\cite{Bezrukov:2009db} at the end of inflation.

In what follows we will use this ``minimal setup" for the analysis of the
radiative corrections. It will be  more convenient to work in the 
Einstein frame, where the coupling to gravity is minimal
and all non-linearities are moved to the matter sector.  The total action in the Einstein frame naturally divides into an Einstein- Hilbert (EH) part, a purely scalar piece involving only the Higgs and dilaton (HD) fields and a part corresponding to the chiral SM (CH) without the radial mode of the Higgs boson~\cite{Bezrukov:2009db,Dutta:2007st,Feruglio:1992wf}
  \begin{equation}
  S =S_\text{EH} +S_\text{HD}+ S_\text{CH}\,.
  \end{equation}
In the next section we estimate the contribution of the scalar sector to the effective inflationary potential, postponing the study of the chiral SM
to  Sec.~\ref{cheffpot}. All the computations will be performed in flat spacetime, since the inclusion of gravity does not modify the
 results \footnote{We recall that, in the Einstein frame, the coupling among SM particles and gravity is minimal.}.

\subsection{Scalar contribution to the effective inflationary potential}
\label{scalareffpot}

Let us start by reminding that the initial value of the dilaton field has to be sufficiently
large to keep its present contribution to DE at the appropriate observational
level~\cite{GarciaBellido:2011de}. The latter fact allows us to neglect
the exponentially suppressed contributions to the effective action stemming from 
$\tilde U_{\Lambda_0}$ in~\eqref{defin-pot}. As a result, the remaining corrections due 
to the dilaton field will emerge from its non-canonical kinetic term, whereas all the
 radiative corrections due to the Higgs field will emerge from the inflationary potential.

The construction of the effective action for the scalar sector of the theory is most
easily done in the following way: expand the action~\eqref{angul-theory} near the constant background of the dilaton and
the Higgs fields and drop the linear terms in perturbations. After
that, compute all the vacuum diagrams to account
 for the potential-type corrections and all the 
diagrams with external legs to account for the kinetic-type
corrections.

\subsubsection{Dilaton contribution}
Let us consider first the quantum corrections to the dilaton itself. 
Since our subtraction procedure respects the symmetries of the
classical action (in particular scale invariance, corresponding to the
shift symmetry of the dilaton field $r$ in the Einstein frame), no
potential terms for the dilaton can be generated. 
Thus, the loop expansion can only create two types 
of contributions, both stemming from its kinetic term. 
The first type are corrections to the
propagator of the field, and as we will show below they are
 effectively controlled by $(m_H/M_P)^{2k}$, with 
$m_H^2\equiv-\tilde U''(\phi)$ and $k$ the number 
of loops under consideration. The second type 
are operators with more derivatives of the field  
 suppressed by appropriate powers
of the scalar cut-off $M_P$. 
One should bear in mind that the appearance of these operators 
in the effective action is expected and
consistent. As discussed in the previous section, 
their presence does not affect the dynamics 
of the model, since the scalar cut-off is much larger than 
the characteristic momenta of the particles involved in
all physical processes throughout 
the whole history of the Universe. 

To demonstrate explicitly what
we described above, let us consider some of the associated diagrams.
Following the ideas of~\cite{Shaposhnikov:2008xi}, we perform the computations
in dimensional regularization in $D=4-2\epsilon$ dimensions. We avoid 
therefore the use of other regularizations schemes, such as cut-off 
regularization, where the scale invariance of the theory is badly 
broken at tree level\footnote{Similar arguments about the artifacts 
created by regularization methods that explicitly break scale 
invariance can be found for instance in~\cite{Bardeen:1995kv}.}. The magnitude 
of the corrections in dimensional regularization is of the order of the masses of the
 particles running in the loops, or in the case of the massless dilaton, its momentum. The
  structure of the corrections can be therefore guessed by simple power-counting and 
it becomes apparent already at the one-loop order. We get
\begin{figure}[H]
\centering
\includegraphics[scale=.75]{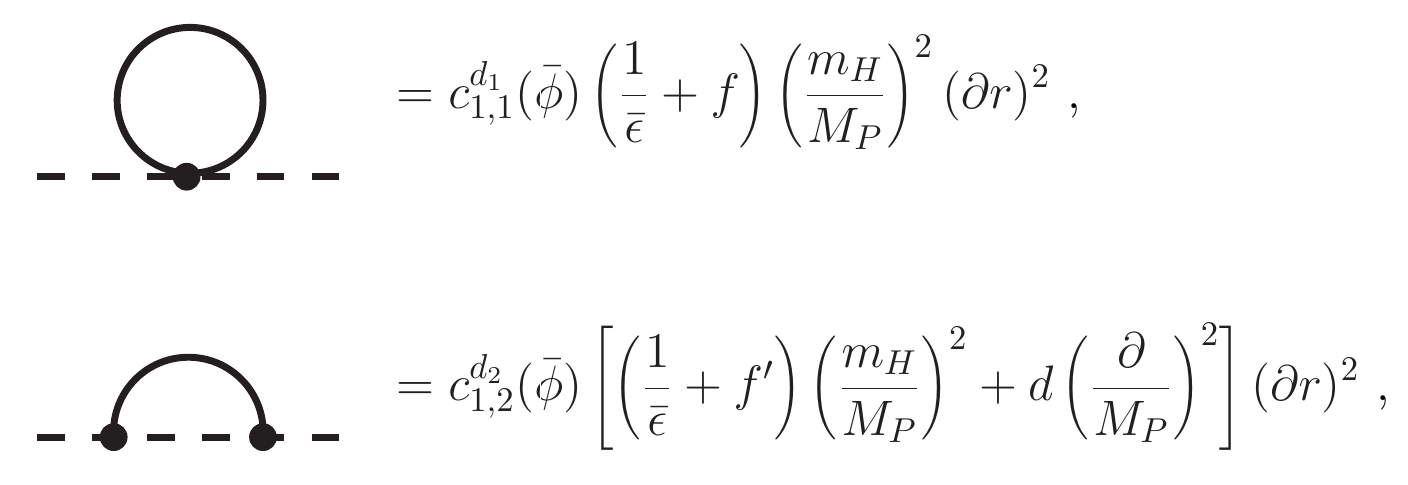}
\end{figure}
{\flushleft where} the Higgs and dilaton fields are represented by 
solid and dashed lines respectively. To keep the expressions as compact as possible
 we set $1/\bar \epsilon=1/\epsilon-\gamma+\log4\pi$ and denoted by $f \ \text{and} \ f'$ the finite parts 
of the diagrams, whose values depend on the normalization
point $\mu$. The higher-derivative operator in the second diagram is included for completion, but 
turns out to vanish accidentally  in this particular case.  Numerical factors are absorbed into
the background-dependent coefficients $c_{k,V}^{d_i}(\bar\phi)$,
which depend on the particular diagram $d_i$ under consideration, 
the number of loops $k$ and the number of vertices\footnote{We introduce the index  
$d_i$ to distinguish between the diagrams
with the same number of vertices but different combinations of hyperbolic 
functions that appear in higher loops.} $V$. Their values 
are always smaller than unity, and vary slightly with 
the background value $\bar\phi$. Their specific form of 
is presented in the Appendix~\ref{appendix2}.

\begin{figure}[H]
\centering
\includegraphics[scale=.75]{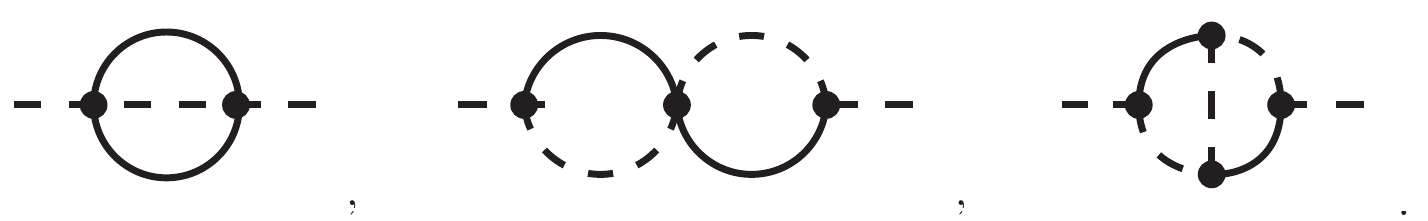}
\caption{Some of the two-loop diagrams
for the dilaton.}
\label{fig:dil-two-loop}
\end{figure}

In two-loops the situation is somehow similar.
The divergent (and finite) part of the corrections (consider
for example the diagrams presented in Fig.~\ref{fig:dil-two-loop}) 
is proportional to
\be\label{cor-2loop}
c_{2,V}^{d_i}(\bar\phi)
\left[\left(\dfrac{m_H}{M_P}\right)^4+
\left(\dfrac{m_H}{M_P}\right)^2
\left(\dfrac{\partial}{M_P}\right)^2+
\left(\dfrac{\partial}{M_P}\right)^4\right](\partial r)^2\ , \ \ \ V\le 4 \ . 
\ee
It is not difficult to convince oneself that this happens
in the higher order diagrams as well. The  
 structure of the corrections is therefore proportional to
\be\label{corr-dil1}
\begin{aligned}
c_{k,V}^{d_i}(\bar\phi)
&\left[\left(\dfrac{m_H}{M_P}\right)^{2k}+
\left(\dfrac{m_H}{M_P}\right)^{2k-2}
\left(\dfrac{\partial}{M_P}\right)^2+\ldots\right.\\
&\left.+\left(\dfrac{m_H}{M_P}\right)^2\left(\dfrac{\partial}{M_P}\right)^{2k-2}+
\left(\dfrac{\partial}{M_P}\right)^{2k}\right](\partial r)^2\ ,
\end{aligned}
\ee
 up to  $\mathcal{O}(1)$ numerical factors.  Notice that some operators involving higher
 derivatives were already present at lower orders, but they reappear with extra suppression factors  $(m_H/M_P)^2$ on top of the scalar cut-off $M_P$. 
The corrections from diagrams with
gauge bosons and fermions running inside the loops are
given also by~\eqref{corr-dil1}, by consistently replacing 
 $m_H$ by the mass of the particle considered. 

\subsubsection{Higgs contribution}
We now turn to the corrections to the Higgs field.
Once again we consider first the potential-type contributions.
The situation now is more complicated, since
the effective potential for the Higgs field $\phi$ will be modified
by terms stemming from the scale-invariant
part of the tree-level potential~\eqref{defin-pot} as well as from the non-canonical kinetic
term of the dilaton field $r$, with the latter starting from the second
order in perturbation theory. 

Let us start by considering the contributions due to the tree-level potential. To keep the 
notation as simple as possible, we express the scale-invariant
part of the potential~\eqref{defin-pot} in the following compact form 
\be\label{compact-infl-pot}
\tilde U(\phi)= \lambda U_0\left(u_0+\sum_{n=1}^{2}u_{n}\cosh [2n
a\phi/M_P]\right), \hspace{.5cm}
U_0=\frac{M_P^4}{4\xi_h^2(1-\varsigma)^2} \ ,
\ee
where, for completion, we have explicitly recovered the $\alpha$ and $\beta$ dependence and defined
\be\label{coeffs}
u_0=c^2-c\sigma+\frac{3\sigma^2}{8}+\frac{3\beta'}{2} \ ,\
u_1=\frac{\sigma^2}{2}-c\sigma-2\beta' \ ,\
u_2=\frac{\sigma^2}{8}+\frac{\beta'}{2} \ , 
\ee
with 
\be\label{coefs-pot}
c=1+\frac{\alpha}{\lambda}\frac{1+6\xi_h}{1+6\xi_\chi} \ , 
\ \ \ \sigma=\varsigma+\frac{\alpha}{\lambda}
\frac{1+6\xi_h}{1+6\xi_\chi} \ , \ \ \ \  \beta'\equiv\frac{\beta}{\lambda}\left(\frac{1+6\xi_h}{1+6\xi_\chi}\right)^2 \ .
\ee
Expanding the field around its background value $\bar \phi$, we get
 \be\label{correct-pot}
\begin{aligned}
\tilde U(\bar \phi+\delta\phi)&= \lambda U_0
\sum_{n=1}^{2}u_n\sum_{l=0}^{\infty} 
\frac{\cosh^{(l)}[2n a\bar\phi/M_P]}{l!}
\left(\frac{2na\delta\phi}{M_P}\right)^l \\
&=\lambda U_0\sum_{n=1}^2\sum_{l=0}^\infty u_n\left[c_{n,l} 
\cosh[2n a\bar\phi/M_P]
\left(\frac{a\delta \phi}{M_P}\right)^{2l}\right.\\
&\left.+d_{n,l} 
\sinh[2na\bar\phi/M_P]
\left(\frac{a\delta \phi}{M_P}\right)^{2l+1}\right] \ ,
\end{aligned}
\ee
where $c_{n,l} \ \text{and} \ d_{n,l}$ account for numerical coefficients
and combinatorial factors. Since the theory is non-renormalizable, the
 perturbative expansion creates terms which do not have the 
same background dependence of the original potential. 
Up to numerical factors, the contributions turn out to be of the form\footnote{To maintain the expressions as  compact as possible we decided not to express the result in terms
of $m_H/M_P$.} 
\be\label{div-form}
\frac{\lambda^{i+j}M_P^4}{[4\xi_h^2(1-\varsigma)^2]^{i+j}}
 \left[g\left(\frac{1}{\epsilon}\right)+f_{i,j}\right]\sum_{n,m}u_n^iu_m^
j\cosh^i[2na\bar\phi/M_P]\sinh^j[2ma\bar\phi/M_P] \ ,
\ee
where
$f_{i,j}$ denotes the (finite) integration constant, and $g(1/\epsilon)$ 
is a function of the divergent terms. Note that if we set $\beta=0$,
 we make sure that terms which contribute to the cosmological 
constant~\eqref{cosm-const} will not be generated 
by the loop expansion.  

By inspection of the structure of 
divergences, we can see that the leading corrections are 
those appearing with the lowest power in $\varsigma$. To gain 
insight on their contribution, we calculate the finite part of~\eqref{div-form}  for the maximal value of the hyperbolic functions.
 This corresponds to $\phi_\text{max}=\phi_0\equiv 
M_P/a\ \tanh^{-1}[\sqrt{1-\varsigma}]$. We get
\be\label{div-approx}
\begin{aligned}
&\frac{\lambda^{i+j}}{[4\xi_h^2(1-\varsigma)^2]^{i+j}}f_{i,j}\times\\
&\times\sum_{n,m}u_n^iu_m^j\cosh^i[2na\bar\phi/M_P]
\sinh^j[2ma\bar\phi/M_P]\Big\vert_{\bar\phi= 
\phi_\text{max}}\sim
\left(\frac{\lambda\varsigma}{4\xi_h^2}\right)^{i+j}f_{i,j} \ ,
\end{aligned}
\ee
which makes the corrections coming from the order $i+j+1$ negligible 
compared to the ones from $i+j$ order. In the last step we have simply
set $c=1,\ \sigma=\varsigma$, which, given the small value of 
the parameter $\alpha$ appearing in~\eqref{coefs-pot}, constitutes
a very good approximation.
\begin{figure}[t]
\centering
\begin{subfigure}{1\textwidth}
\centering
\includegraphics[scale=.75]{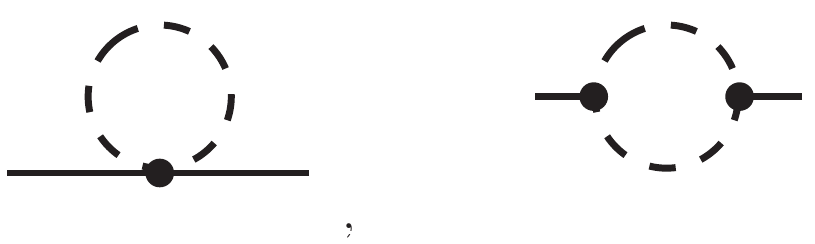}
\caption{}
\label{fig:oneloophig}
\end{subfigure}\\
\begin{subfigure}{1\textwidth}
\centering
\includegraphics[scale=.75]{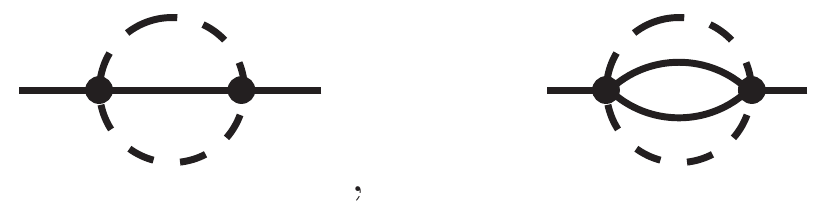}
\caption{}
\label{fig:twoloophig}
\end{subfigure}
\caption{Characteristic 
diagrams produced by the
non-canonical kinetic term of the dilaton field $r$.
Solid and dashed lines represent the Higgs and dilaton fields respectively.
The first one-loop diagram presented in (a)
vanishes in dimensional regularization due to the massless character 
of the dilaton field. On the other hand, the second diagram gives rise to 
higher derivative terms of the Higgs field. 
In (b) we consider two and three loop diagrams
which, apart from generating higher dimensional  operators,
 contribute to the effective potential once we amputate them.}
\end{figure}

As we mentioned earlier, potential-type corrections
to the Higgs field are also generated from diagrams
associated to the 
kinetic term of the dilaton $r$, starting from two loops.
This happens because the first order vacuum diagrams 
with dilaton running in the loop, vanish. If we consider higher loop 
diagrams, like those in Fig.~\ref{fig:twoloophig} but 
without momenta in the external legs, 
we see that even though the background
dependence of the corrections is complicated
due to the non-canonically normalized dilaton that 
runs inside the loops, their contributions
to the effective action are of the same order as 
 those in~\eqref{div-approx}. 

We now turn to the kinetic-type corrections to the Higgs
field. By that we mean corrections to the propagator, as well as terms with more derivatives of the field
suppressed by the scalar cut-off. 
The first type of contributions come only from the 
scale-invariant part of the potential given by~\eqref{compact-infl-pot}
, when 
the momenta associated to the external legs are considered. 
It is not difficult to show that these are precisely of the same
form as those in~\eqref{div-form}.
The second type of contributions, i.e. the higher dimensional
operators, are generated both from the Higgs potential at higher loops, as well as 
from the non-vanishing diagrams associated to the
 non-canonical kinetic term of the dilaton.
The terms we get are proportional to
\be\label{higherdim}
\frac{\partial^2}{M_P^2}(\partial\phi)^2 \ ,  \ \ \ 
\frac{\partial^4}{M_P^4}(\partial\phi)^2 \ \ldots \,,
\ee
and they can be safely neglected for the typical momenta involved in 
the different epochs of the evolution of the Universe.

Before moving on, we would like to comment on the 
appearance of mixing terms with derivatives of the fields.
These manifest themselves when we consider
diagrams with both fields in the external legs. 
They are higher dimensional operators, 
and it can be shown that they appear 
suppressed by the scalar cut-off of the theory, as before.

Since the kinetic-type operators do not modify
the dynamics, we will consider only potential-type corrections to estimate the change in the 
tree-level predictions of the model. At one-loop, the contribution of the scalar sector to the inflationary potential becomes~\cite{Jackiw:1974cv}
\be\label{one-loop-correction}
\begin{aligned}
 \Delta \tilde U_{HD}&\simeq \frac{U_0}{64\pi^2}
\frac{\lambda a^4}{\xi_h^2(1-\varsigma)^2 }
\left(\frac{1}{\bar\epsilon}+f_{2,0}\right)\left[\varsigma^2
\frac{1+\cosh[4a\bar\phi/M_P]}{2}+\mathcal{O}(\varsigma^3)\right]\ ,
\end{aligned}
\ee
where we just kept the leading contribution in $\varsigma$. The finite part $f_{2,0}$ in the previous expression is given by 
\be\label{lfinite-part}
\begin{aligned}
f_{2,0}&=\frac{3}{2}-\log\left[\frac{-\tilde U''(\bar\phi)}{\mu^2}\right]\\
&=\frac{3}{2}-\log\left[\frac{\lambda a^2 M_P^2}{\xi_h^2(1-\varsigma)^2\mu^2}\Bigg(
\varsigma \cosh[2a\bar\phi/M_P]+\mathcal{O}(\varsigma^2)\Bigg)\right] \ .
\end{aligned}
\ee
If we adopt the $\overline{MS}$ scheme, the remaining (logarithmic) corrections 
will be suppressed by an overall factor $\mathcal O(10^{-15})$ (apart from different powers of $\varsigma$) with respect to the tree-level 
potential~\eqref{compact-infl-pot}. The quantum contribution of the scalar sector to the effective inflationary 
potential is therefore completely negligible and rather insensitive to the particular choice of the renormalization point $\mu$. This allows us to approximate the value of $\phi$ at the end of inflation by its classical value $\phi_f\simeq M_P/a \ \tanh^{-1} \left[\sqrt{1-\varsigma}\cos(2\times 3^{1/4}\sqrt{\xi_\chi})\right]$, and compute analytically the spectral tilt $n_s$ of 
primordial scalar perturbations, which turns out to be
\be\label{spec-tilt}
n_s(k_0)-1 \simeq -8\xi_\chi +\frac{\lambda\xi_\chi^2}{96\pi^2\xi_h^2}f_{2,0} \ ,
 \ \ \ \text{for} \ \ \   1\lesssim 4\xi_\chi N^*\ll 4N^* \ .
\ee
We see therefore that the correction to the tree-level result is controlled by the effective self-coupling of the Higgs field in the Einstein frame $\lambda/\xi_h^2$. The small value of this parameter makes the scalar radiative contribution completely negligible and thus hardly modify the consistency relation~\eqref{functional-rel}. Note  however that there might be still a significant contribution to the inflationary potential coming from the SM particles, especially from those with a large coupling to the Higgs field. The study of this effect is the purpose of the next section. 

\subsection{Chiral SM contribution to the effective inflationary  potential.}\label{cheffpot}

The action for the SM fields during the inflationary stage is similar to that appearing in Higgs inflation~\cite{Bezrukov:2009db} and takes the form of a chiral SM with a nearly decoupled Higgs field. Its 
contribution to the effective potential can be analyzed by the methods presented in~\cite{Bezrukov:2009db}.  
The one-loop contribution during inflation reads\footnote{We neglect the contribution~\eqref{one-loop-correction}
 associated to the scalar sector, which, as shown in the previous section, turns out to be very small. }
\be\label{eq:Veff1}
  \Delta U_1 =
  \frac{6m_W^4}{64\pi^2}\left(\log\frac{m_W^2}{\mu^2}-\frac{5}{6}\right)
  + \frac{3m_Z^4}{64\pi^2}\left(\log\frac{m_Z^2}{\mu^2}-\frac{5}{6}\right)
  - \frac{3m_t^4 }{16\pi^2}\left(\log\frac{m_t^2}{\mu^2}-\frac{3}{2}\right)\ ,
\ee
where $m_W^2=g^2 h^2/2$, $m_Z^2=g^2h^2/2\cos^2\theta_W$ and
$m_t^2=y_t^2h^2/2$ stand for the effective $W, Z$ and top quark masses
in the Jordan frame.
The choice of the $\mu$ parameter here defines the renormalization prescription, as
described in the beginning of Sec.~\ref{sec:divergences}.  To retain
the possibility to use the RG equations to run between the
electroweak and inflationary scales we will write
$\mu^2=\frac{\hat{\mu}^2}{M_P^2}F(h,\chi)$.  Here the function $F(h,\chi)$
corresponds to the choice of the renormalization prescription and leads to
different physical results, while the parameter $\hat{\mu}$ plays the
role of the usual choice of momentum scale in the RG approach and
should disappear in the final result.
The conformal transformation to the Einstein frame $\Delta \tilde U_1 =\Delta U_1/\Omega^4$ acts only on the coefficients of the logarithmic terms in~\eqref{eq:Veff1}, leaving their arguments completely unchanged.  We obtain therefore
 \be\label{eq:Veff2}
\begin{aligned}
  \Delta \tilde U_1 &=
  \frac{6\tilde m_W^4}{64\pi^2}\left(\log\frac{m_W^2}{\hat{\mu}^2F(h,\chi)/M_P^2}-\frac{5}{6}\right)
  + \frac{3\tilde m_Z^4}{64\pi^2}\left(\log\frac{m_Z^2}{\hat{\mu}^2F(h,\chi)/M_P^2}-\frac{5}{6}\right)\\
  &- \frac{3\tilde m_t^4 }{16\pi^2}\left(\log\frac{m_t^2}{\hat{\mu}^2F(h,\chi)/M_P^2}-\frac{3}{2}\right)\ ,
\end{aligned}
\ee
where the Einstein-frame masses $\tilde m^2$ are proportional to the
effective vacuum expectation value of the Higgs field in the Einstein
frame\footnote{In particular we have $\tilde m_W^2(\phi)=\tilde
  m_Z^2(\phi) \cos^2\theta_w= g^2/2\cdot v^2(\phi)$ and $\tilde
  m_t^2(\phi)=y_t^2/2\cdot v^2(\phi)$.}, which is a slowly varying function during inflation,
\begin{equation}
  \label{eq:1}
  v^2(\phi) \equiv \frac{h^2}{\Omega^2}= \frac{M_P^2}{\xi_h(1-\varsigma)}
  \left(1-\varsigma\cosh^2\frac{a\phi}{M_p}\right).
\end{equation}
This fact
allows us to completely factor out the $\phi$ dependence in front of the logarithms in~\eqref{eq:Veff2} and perform the analysis  
below as if $v$ was a constant, $v\simeq M_P/\sqrt{\xi_h}$. 

Note that the explicit dependence on the  't Hooft-Veltman
normalization point $\hat{\mu}$ in~\eqref{eq:Veff1} is spurious
and is compensated by the running of the coupling constants
$\lambda(\hat{\mu})$, $\xi_h(\hat{\mu})$ in the
tree level part of the potential (see \cite{Bezrukov:2009db}).
 Once the RG running of the couplings is fixed,  it is convenient to
 choose the value of $\hat{\mu}$ in such a way that the logarithmic
 contribution   \ref{eq:Veff2}, for each given value $\phi$ of
 the Higgs field, is minimized, $\hat\mu^2\simeq \frac{y_t^2}{2}
 \frac{h^2}{F(h,\chi)/M_P^2}$.
 In that case, the RG enhanced (RGE) inflationary potential becomes
\begin{equation}
  \tilde U_\text{RGE}(\phi)=  \frac{\lambda(\hat{\mu}(\phi))}{4}
  \frac{M_P^4}{\xi^2_h(\hat{\mu}(\phi))(1-\varsigma)^2}
  \left(1-\varsigma\cosh^2\frac{a\phi}{M_p}\right)^2\,,
\end{equation}
which in fact suffices for practical purposes, with the corrections
form the 1-loop logarithms being rather small.

As discussed at the beginning of Sec.~\ref{sec:divergences}, the
different choices of $\mu$ correspond to different subtraction  rules
and  produce different results.  In what follows we will consider the
two most natural  choices. The first one is associated to  the scale
invariant prescription~\eqref{prescription1J}. The RG enhancement of 
the potential in this case dictates
\begin{equation}\label{PI}
  \hat\mu_\text{I}^2(\phi)
  = \frac{y_t^2}{2}\frac{M_p^2 h^2}{\xi_h h^2 + \xi_\chi\chi^2}
  = \frac{y_t^2}{2} v^2(\phi)\,,
\end{equation}
which is nothing else than the effective top mass in the Einstein frame. With this choice, the change in the shape of the 
potential is very small, given the insignificant variation of $v^2(\phi)$ during inflation. 
 The change in the inflationary observables $n_s$ and $r$ is therefore expected to be
 completely negligible. The second possibility that we will consider is associated to the prescription~\eqref{prescription2J}. In this 
 case the optimal choice of $\hat\mu$ is
\begin{equation}\label{PII}
  \hat\mu^2 _\text{II} (\phi)
  = \frac{y_t^2}{2}\frac{M_P^2 h^2}{\xi_\chi\chi^2}
  = \frac{y_t^2}{2} v^2(\phi)
    \frac{1-\varsigma}{\varsigma \sinh^2\left(a\phi/M_P\right)}\,,
\end{equation}
which, at the end of inflation, coincides with the effective top mass in the Jordan frame. This corresponds to the prescription II in~\cite{Bezrukov:2009db}. Note that contrary to the previous case, this choice strongly depends on the value of the $\phi$ field and noticeable contributions to the inflationary parameters are expected.  
\begin{figure}[!h]
  \begin{center}
\begin{subfigure}{1\textwidth}
\centering
\includegraphics[scale=0.5]{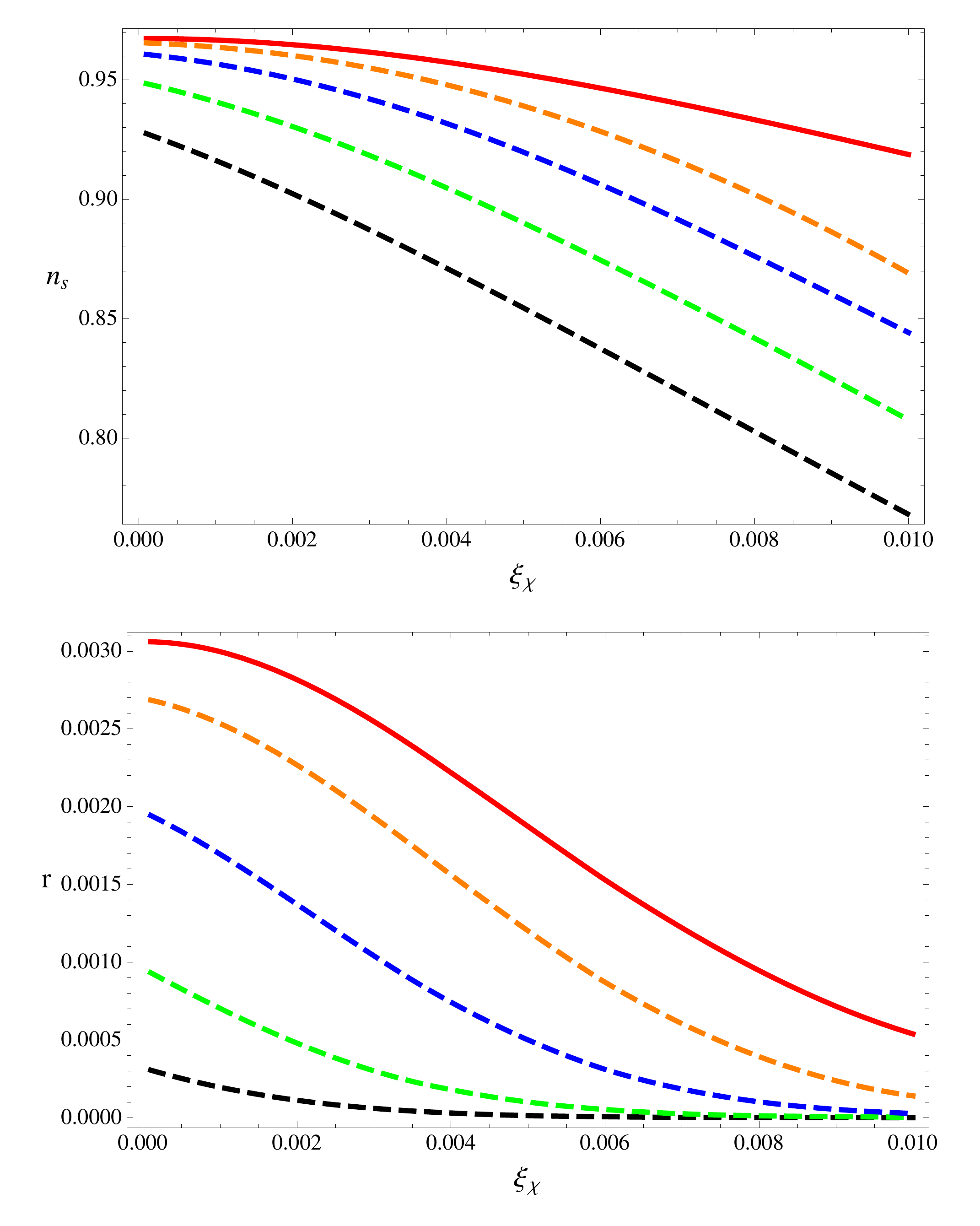}
\end{subfigure}\\
\fbox{\begin{subfigure}{1\textwidth}
\centering
\includegraphics[scale=0.6]{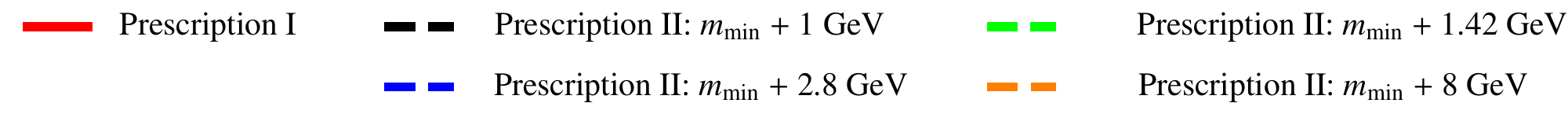}
\end{subfigure}}
  \end{center}
  \caption{The spectral index $n_s$ (top) and tensor to scalar ration $r$
    (bottom) as a function of the non-minimal coupling $\xi_\chi$.  The solid line corresponds to the
    quantization prescription I, which coincides with the tree level result. Dashed lines stand for the quantization choice II for different Higgs masses. The minimal Higgs boson
    mass $m_\text{min}$ can be obtained from~\cite{Bezrukov:2012sa}.}
  \label{fig:nsxichiRG}
\end{figure}

The calculation proceeds now along the same lines as those in~\cite{Bezrukov:2009db}, using the tree level RG enhanced potential and the one loop correction. The addition of the two loop effective potential does not significantly modify the result. The numerical outcome for the two prescriptions is shown in Fig.~\ref{fig:nsxichiRG}. As expected, the inflationary observables computed with the first prescription coincide with the tree level result. The only effect of the quantum corrections is setting a minimal value for the Higgs mass. This turns out to be  $m_H>m_\text{min}$, with $m_\text{min}\simeq 129.5\pm\text{5 GeV}$ (for details on the latest calculations of this value see~\cite{Bezrukov:2012sa,Degrassi:2012ry}). After the end of inflation and preheating, the system is outside
the scale-invariant region and the fields settle
down to the minimum of the potential.  From the expansion of
the potential~\eqref{darkener-pot}
 around the background, it is clear that all the
contributions to the effective action will be again suppressed by
powers of the exponent $e^{-\gamma r/M_P}$, in addition to powers
of $M_P$, not affecting therefore the predictions of the model
concerning the DE equation of state~\eqref{reduc}. Taking into account the above results, we conclude that the quantum corrections
computed with the prescription I do not modify the classical consistency relation~\eqref{functional-rel} characterizing Higgs-Dilaton cosmology. On the other hand, the inflationary observables computed using the prescription II clearly differ from the tree level result, especially for Higgs masses close to the critical value $m_{min}$ at large $\xi_\chi$. Note that in this prescription, the recent observation of a light Higgs-like state~\cite{Chatrchyan:2012xdj,Aad:2012tfa}, together with the present bounds on the spectral tilt $n_s$~\cite{Komatsu:2010fb}, further restrain the allowed $\xi_\chi$ interval. 

\section{Summary and Outlook}
\label{sec:conclusions5}

The purpose of this chapter was to study the self-consistency of the
Higgs-Dilaton cosmological model. We determined the field-dependent UV
cut-offs and studied their evolution in the different epochs
throughout the history of the Universe. We showed that the cut-off
value is higher than the relevant energy scales in the different
periods, making the model a viable effective field theory describing
inflation, reheating, and late-time acceleration of the Universe.
Since the theory is non-renormalizable, the loop expansion creates an
infinite number of divergences, something that may challenge the
classical predictions of the Higgs-Dilaton model. We argued that  this
is not the case if the UV-completion of the theory respects
scale-invariance  and the approximate shift symmetry for the dilaton
field. 

We computed within this framework the effective inflationary potential in the one-loop approximation and concluded that the dominant contribution comes from the chiral SM sector of the theory. We used two different regularizations prescriptions consistent with the symmetries of the model.  In the ``SI-prescription'' of ~\cite{Shaposhnikov:2008xi}, with a field-dependent normalization point proportional to the effective Planck scale in the Jordan frame, the effective potential turns out to coincide with the tree level one. This leaves practically intact the consistency relation~\eqref{functional-rel} which connects the inflationary spectral tilt to the deviation of the DE equation of state from a cosmological constant. This relation is however modified if the normalization point is chosen only along the dilaton's direction, especially for Higgs masses near the critical value $m_\text{min}\simeq 129.5\pm\text{5 GeV}$, which is amazingly close to the mass of the Higgs  particle observed at the LHC~\cite{Chatrchyan:2012xdj,Aad:2012tfa}. In the lack of a Planck scale UV completion, the proper choice of the normalization point $\mu$ can only be elucidated by improving the precision of the cosmological and particle physics observables.

\chapter{Scale-invariant alternatives to general relativity: dilaton properties}
\label{ch:SITDIFF}

\section{Introduction}
\label{sec:Intro}

In the SITDiff theories introduced and studied in~\cite{Blas:2011ac}, the scalar degree of freedom related to the metric determinant is identified with a massless dilaton that only couples derivatively and thus evades the fifth force constraints. Assuming that the metric is dimensionless and the Lagrangian contains up to two derivatives of the fields, the most general scalar-tensor theory that includes matter fields was presented. The form of the action can not be completely fixed;  rather, it involves arbitrary functions of the metric determinant (``theory defining functions''),  since this quantity behaves as a scalar under the restricted coordinate transformations. It was shown that the invariance of the system under dilatations, is explicitly broken at the level of the equations of motion by an arbitrary integration constant that appears because of TDiff rather than Diff invariance. This gives rise to a run-away potential for the dilaton. It was demonstrated that by appropriately choosing the theory defining functions, it is possible to get a theory which has interesting implications for particle physics and cosmology.  Its particle physics sector can be made identical to the Standard Model, whereas it is able to account for the inflationary period in the early Universe and provide a natural candidate for dynamical dark energy.

In this final chapter, whose findings were reported in~\cite{Karananas:2016grc}, we investigate what are the implications on the structure of these models when the metric tensor $g_{\m\n}$ has (arbitrary) mass dimension. Usually, it is somehow taken for granted that $g_{\m\n}$  is dimensionless, whereas the coordinates $x^\m$ carry dimensions of length. However, this is nothing more than a particular choice which follows ``naturally'' only when the Minkowski space-time is described in terms of cartesian coordinates. Notice that this choice is certainly not the most appropriate one when other coordinate systems are used, let alone when curved space-times are considered. 

Let us carry out some elementary dimensional analysis. Although what follows is in a sense trivial if the theory under consideration is diffeomorphism invariant, the situation changes considerably for SITDiff theories, since the metric determinant is a propagating degree of freedom that plays the role of the dilaton. By definition, $\left[g_{\m\n}dx^\m dx^\n\right]=\left[\text{GeV}\right]^{-2}$, so in principle, we have  the liberty to assign arbitrary dimensions -- also fractional -- both to $x^\m$ and $g_{\m\n}$, i.e. 
\be
\label{assign-dims}
\left[ x^\m\right]=\left[\text{GeV}\right]^{-p}\ ,~~~\left[g_{\m\n}\right]=\left[\text{GeV}\right]^{-2q} \ ,
\ee
as long as $p+q=1$. The dilatations now act on the coordinates and the metric as 
\be
\label{scal-trans-fields}
x^\m\rightarrow \a^{-p} x^\m~~~\text{and}~~~ g_{\m\n}(x)\rightarrow \a^{-2q} g_{\m\n}(\a^{-p} x) \ ,
\ee
since the scaling dimensions coincide with the mass dimensions.  Of a special interest  is the case in which $x^\m$ merely label events on the manifold and the metric carries dimensions of area
\be
\label{p0q1}
p=0~~\text{and}~~~q=1\ .
\ee

We will see that the class of  theories with $p\neq 0$ is equivalent to that already described in ~\cite{Blas:2011ac}. However, the case~\eqref{p0q1} is different. In particular,  a dilatation symmetry breaking potential for the dilaton will be shown to be  absent, an otherwise generic feature of the  theories with $p\neq 0$. Moreover, it is remarkable that by abandoning the prejudice of a dimensionless metric and requiring that there are no terms with more than two derivatives in the action, we can completely fix its form for pure gravity without matter fields.  It should be noted that, in principle, one can relax the requirement of having an action that contains terms which are at most quadratic in the derivatives. To ensure absence of ghosts, the starting point in this case should either be Horndeski theory~\cite{Horndeski:1974wa} or $f(R)$-gravity, see~\cite{Sotiriou:2008rp} and references therein.\footnote{The Horndeski theory is the most general scalar-tensor action with second order equations of motion. The scale- and Weyl-invariant subclasses of this theory have been identified in~\cite{Padilla:2013jza}. It would be interesting to understand what are the implications of having invariance under TDiff instead of the full group of diffeomorphisms, an investigation we leave for elsewhere.} For the latter, we will show that they can be used as the starting point for constructing biscalar SITDiff theories.

Next, we present how a scalar field can be incorporated in a consistent manner. If this field is identified with the Standard Model Higgs boson, we end up with a phenomenologically viable SITDiff theory. As we will demonstrate, the Higgs mass as well as the cosmological constant appear in the action in a peculiar way, different from the other terms.

Inspired by this, we formulate a set of rules that allows us to distinguish formally the Higgs mass and the cosmological constant from other contributions to the action based on their behaviour when the dilaton goes to zero. Since this field is related to the metric determinant that now carries dimension of length, this limit potentially corresponds to vanishing length and thus it is in a sense related to the UV regime.  More precisely, we notice that when the theory is expressed in terms of variables that are conjugate to the time and space derivatives of the fields (canonical four-momenta \cite{Schwinger:1951xk,Schwinger:1953tb}), then the only terms which  involve inverse powers of the dilaton --  and thus are presumably singular at the UV limit -- are the Higgs boson mass and the cosmological constant. Based on that, we speculate that their absence in the action may be a requirement of the self-consistency of the theory in the UV domain. The smallness of the observed  low energy  values of the Higgs mass and of the cosmological constant,  perhaps,  could be attributed to  some yet unknown nonperturbative mechanism. 

This chapter is structured as follows. In Sec.~\ref{sec:dil}, we construct the most general SITDiff theory that contains only the dilaton and  study its properties. In Sec.~\ref{sec:hig-dil}, we demonstrate how matter fields are introduced in this framework. We present a phenomenologically viable model that in addition to the dilaton contains an extra scalar field, that is identified with the Standard Model Higgs boson. In Sec.~\ref{sec:Regul}, we formulate the assumptions that make it possible to single out  the presence of certain terms in the action by requiring that the theory has a regular limit when determinant of the metric goes to zero. We present our conclusions in Sec.~\ref{sec:conclusSITDIFF}.

\section{Pure gravity}
\label{sec:dil}

As a warm-up exercise, we will write down the most general theory that contains at most two derivatives of the fields and is invariant under the restricted coordinate transformations and dilatations, which are given, respectively, by~\eqref{tdiffs-def} and~\eqref{scal-trans-fields}. The unique action that satisfies these requirements reads
\be
\label{TDSI-1}
S=\int d^4 x\sqrt{g}\left[ \frac{\z}{2} g^{\frac{1}{4(p-1)}} R - c_1\, g^{\frac{1}{4(p-1)}-2}g^{\m\n}\partial_\m g \partial_\n g -c_2 g^\frac{1}{2(p-1)} \right] \ ,
\ee
where  $\z, \ c_1,\ \text{and} \ c_2$ are dimensionless constants and the scalar curvature $R$ is defined in Appendix~\ref{sec:dim-an}. Observe that for  $p=1$, the above expression becomes singular. This is a manifestation of the fact that if we consider the standard mass (and scaling) dimension for the metric and coordinates, it is not possible to construct SITDiff theories with the metric determinant only. This was also realized in~\cite{Blas:2011ac}.

To get a better grasp on the dynamics of this theory, it is desirable to recast it in a form invariant under the full group of diffeomorphisms. Once we consider a coordinate transformation with $J\neq 1$, we obtain
\be
\label{TDSI-2}
S=\int d^4 x \sqrt{g}\left[\frac{\z}{2} \sigma^{\frac{1}{4(p-1)}} R - c_1\, \sigma^{\frac{1}{4(p-1)}-2}g^{\m\n}\partial_\m \sigma \partial_\n \sigma -c_2 \sigma^\frac{1}{2(p-1)}+c_3\,\sigma^{-1/2}\right] \ ,
\ee
where we defined the dilaton field $\sigma\equiv J^2g$, a scalar under diffeomorphisms. 
Some comments are in order at this point. First of all, when the theory is written this way, its particle spectrum can be read off immediately. It contains, in total, three degrees of freedom: the two graviton polarizations and an additional  scalar field which is associated with the determinant of the metric. Moreover, we notice the appearance of an extra term in the action proportional to the integration constant $c_3$, which emerged through the equations of motion, see for example~\cite{Alvarez:2006uu,Shaposhnikov:2008xb,Blas:2011ac} and references therein. It should be noted that for $p\neq 0$ (and equivalently $q\neq 1$), the resulting theories are all equivalent to the ones which were already considered in~\cite{Blas:2011ac}. In this case, the aforementioned constant necessarily carries  dimensions and consequently, its presence explicitly breaks the symmetry of the theory under dilatations and produces a run-away potential for the dilaton. This is a generic feature of these models. Hence, it seems that $p=0$ is a rather special point in the phase space of the theory, since $c_3$ is dimensionless and the theory under consideration is exactly scale invariant.\footnote{Actually, it  coincides with the induced gravity model introduced in~\cite{Zee:1978wi,Smolin:1979uz}. } 

Let us now introduce a field $\x$ with canonical dimensions
\be
\label{sigm-chi-1}
\x=\s^{\frac{1}{8(p-1)}} \ , 
\ee
and set 
\be
\label{c1}
c_1=\frac{1}{128(p-1)^2} \ ,
\ee
so that~\eqref{TDSI-2} is equivalently rewritten as 
\be
\label{TDSI-3}
S=\int d^4 x \sqrt{g}\left[\frac{\z}{2}  \x^2 R -\frac{1}{2} g^{\m\n}\partial_\m \x \partial_\n \x -c_2\,\x^4+c_3\,\x^{-4(p-1)}\right] \ . 
\ee

In order to eliminate the mixing between the field and the curvature, it is convenient to write the theory  such that the gravitational part takes the standard Einstein-Hilbert form and all nonlinearities are moved to the scalar sector. To this end, we perform the following change of variables,
\be
\label{conf-trans}
g_{\m\n}\rightarrow \omega^{-2}g_{\m\n} \ ,~~~\text{with}~~~\omega=\frac{\sqrt{\z}\x}{M_P} \ ,
\ee
where $M_P=2.4\times 10^{18} \ \text{GeV}$ is the Planck mass. A straightforward calculation gives us the action in the Einstein frame:
\be
\label{ein-fr-1}
S=\int d^4x\sqrt{g}\left[\frac{M_P^2}{2}R-\frac{M_P^2}{2\z}\left(1+6\z\right)\x^{-2} g^{\m\n}\partial_\m \x \partial_\n \x -\frac{c_2 M_P^4}{\z^2}+\frac{c_3 M_P^4}{\z^2}\x^{-4p} \right] \ .
\ee
To bring the kinetic term for the field into canonical form,  we define
\be
\label{chi-phi}
\x=e^{\frac{\gamma \phi}{M_P}} \ , \ \ \ \g=\sqrt{\frac{\z}{1+6\z}} \ ,
\ee
so that~\eqref{ein-fr-1} becomes
\be
\label{ein-fr-2}
S=\int d^4x\sqrt{g}\left[\frac{M_P^2}{2}R-\frac{1}{2}g^{\m\n}\partial_\m \phi\partial_\n \phi-\frac{c_2 M_P^4}{\z^2}+\frac{c_3 M_P^4}{\z^2}e^{\frac{-4p\gamma \phi}{M_P}}\right] \ .
\ee

We observe that for $p=0$, the theory in the Einstein frame boils down to that of a massless minimally coupled scalar field in curved spacetime,
\be
\label{ein-fr-3}
S=\int d^4x\sqrt{g}\left[\frac{M_P^2}{2}R-\frac{1}{2}g^{\m\n}\partial_\m \phi\partial_\n \phi-\frac{c M_P^4}{\z^2}\right] \ ,
\ee
where we denoted $c=c_2-c_3$. Notice that the (exact) scale invariance of the model in the Jordan frame has manifested itself as an (exact) shift symmetry,
\be
\label{shift-sym}
\phi\rightarrow \phi+\text{constant} \ ,
\ee
when the theory was written in the Einstein frame. Thus, instead of the typical symmetry-breaking exponential potential for the field, we got a contribution to the cosmological constant term. This is a novel feature of SITDiff theories with dimensionless coordinates.

At this point, it is worth taking a short detour and discussing the implications of allowing terms with more than two derivatives of the fields in the action, even though it lies outside the main scope of this chapter. In general, higher-derivative terms may put the self-consistency of a theory under scrutiny, since their presence often (but not always) leads to the appearance of ghostly degrees of freedom in the spectrum. One of the simplest examples of healthy theories that involve an arbitrary number of derivatives of the metric in the action is ``$f(R)$ gravity''~\cite{Sotiriou:2008rp}. It is based on the replacement of the Einstein-Hilbert term which is linear in the scalar curvature, by an arbitrary function of $R$, such that the action reads
\be
\label{f_R_1}
S=\frac{M_P^4}{2}\int d^4 x \sqrt{g}f\left(R\right) \ ,
\ee
where $f(R)$ need not be local and for dimensional reasons can only depend on $R/M_P^2$. This modification to general relativity is motivated both from theory and phenomenology. Since gravity is an effective field theory, curvature corrections are expected to be present and play significant role when quantum effects are taken into account. Also, with an appropriate choice of the  function, it is possible to get interesting cosmological consequences for the early and late Universe.\footnote{The succesful Starobinsky model of inflation~\cite{Starobinsky:1980te} is a higher-derivative theory with 
$$
f\left(R\right)=\frac{R}{M_P^2}+ \frac{2\a R^2}{M_P^4} \ ,
$$
and $\alpha>0$ is a dimensionless constant.}

As is customary when dealing with these theories, it is convenient to express the above in a way that the dynamics of the extra degree(s) of freedom is separated from the gravitational sector. Performing a Legendre transformation, we can cast~\eqref{f_R_1} into the following equivalent form,
\be
\label{f_R_2}
S=\frac{M_P^4}{2}\int d^4 x \sqrt{g}\Big[f'(\chi) R-V(\chi)\Big] \ ,
\ee
where prime denotes derivative with respect to $\chi$ and we define 
\be
\label{f_R_pot}
V(\chi)=\chi\, f'(\chi)-f(\chi) \ . 
\ee
Note that the absence of ghosts forces us to impose $f'(\chi)>0$, and we have to require $f''(\chi)\neq 0$ such that $\chi=R$. 

To make the kinetic term for $f'(\chi)$ appear explicitly in the action, we Weyl-rescale the metric as 
\be
\label{star_infl_5}
g_{\m\n}\rightarrow \frac{1}{M_P^2f'(\chi)}g_{\m\n} \ ,
\ee
to obtain
\be
\label{star_infl_6}
S=\int d^4 x \sqrt{g}\left[\frac{M_P^2}{2}R-\frac{3M_P^2}{4f^{'2}(\chi)}\partial_\m f'(\chi)\partial^\m f'(\chi)-\frac{V(\chi)}{2f^{'2}(\chi)} \right] \ . 
\ee
Finally, we introduce
\be
\label{star_infl_7}
\varphi=\sqrt{\frac{3}{2}}M_P \log\left[M_P^2f'(\chi)\right] \ ,
\ee
in terms of which the action takes its ``standard'' form,
\be
\label{star_infl_8}
S=\int d^4 x \sqrt{g}\left[\frac{M_P^2}{2}R-\frac{1}{2}(\partial_\m \varphi)^2-U(\varphi) \right] \ ,
\ee
with
\be
U(\varphi)=\frac{V[\chi(\varphi)]}{2f^{'2}[\chi(\varphi)]} \ .
\ee

The above procedure can be straightforwardly generalized to the class of theories that we are considering here, something that will lead to biscalar theories. For the purposes of illustration, it suffices to stick to the ``special'' case $p=0$. Requiring invariance under dilatations and TDiff fixes the action as
\be
\label{f_R_Tdiff}
S=\int d^4 x\Big[ f\left(R\right)- c_1\, g^{-7/4}g^{\m\n}\partial_\m g \partial_\n g -c_2 \Big]\ ,
\ee
where, for the function $f$ to be dimensionless, the scalar curvature must only appear multiplied by $g^{1/4}$. Repeating the steps outlined previously and restoring the invariance under general coordinate transformations, we can write the above as
\be
S=\int d^4 x\sqrt{g}\Big[\s^{-1/4} f'\left(\chi\right)R-\s^{-1/4}V(\chi)- c_1\, \s^{-9/4}g^{\m\n}\partial_\m \s \partial_\n \s -c\, \s^{-1/2}\Big]\ ,
\ee
where, as before, $\s=J^2g$, $c=c_2-c_3$, and the ``potential'' $V(\chi)$ was presented in~\eqref{f_R_pot}. As expected, we ended up with a scalar-tensor theory that contains -- on top of the graviton -- two propagating fields. Choosing the function in~\eqref{f_R_Tdiff} appropriately, it is possible to construct a vast number of models with interesting cosmological phenomenology.

\section{Including matter fields}
\label{sec:hig-dil}

In the present section we wish to generalize the SITDiff theory we constructed previously by showing how matter fields can be incorporated into this setup. Let us start by introducing another scalar $h$ with canonical mass dimensions.

We saw that the theory presented previously was completely determined by requiring invariance under TDiff and scale transformations; see~\eqref{tdiffs-def} and~\eqref{scal-trans-fields}, respectively. When we bring into the game an extra scalar field, the situation changes. The dimensionless quantity,
\be
\label{dim-quant}
h^2g^{-\frac{1}{4(p-1)}}\ ,
\ee
is invariant under both TDiff and dilatations. Therefore, arbitrary functions of the above can, in principle, appear in the action. As in the previous section, we restrict ourselves to terms that are, at most, quadratic in the derivatives of the various fields. Dimensional analysis dictates that the gravitational and scalar sectors of the action that possess the desired properties read
\be
\begin{aligned}
\label{TD-SI2-full-1}
S&=\int d^4x \sqrt{g} \left[\frac{\z}{2}g^{\frac{1}{4(p-1)}}F_1\left(h^2g^{-\frac{1}{4(p-1)}}\right)R\right.\\
&\left.-c_1 g^{\frac{1}{4(p-1)}-2}F_2\left(h^2g^{-\frac{1}{4(p-1)}}\right)g^{\m\n}\partial_\m g \partial_\n g\right.\\
&\left.-\frac{1}{2} F_3\left(h^2g^{-\frac{1}{4(p-1)}}\right)g^{\m\n}\partial_\m h \partial_\n h\right.\\
&\left.+\d g^{-1}h F_4\left(h^2g^{-\frac{1}{4(p-1)}}\right)g^{\m\n}\partial_\m g \partial_\n h\right.\\
&\left.-c_2 g^{\frac{1}{2(p-1)}}V\left(h^2g^{-\frac{1}{4(p-1)}}\right)\right] \ .
\end{aligned}
\ee
Here $F_i$ and $V$ are arbitrary functions that can only depend on the dimensionless combination~\eqref{dim-quant}.  For later convenience, we have also included the constants $\z,c_1,c_2,$ and $\d$. We now consider a transformation with $J\neq 1$ and introduce $\sigma=J^2 g$ to recast the action into its diffeomorphism-invariant form
\be
\begin{aligned}
\label{TD-SI2-full-2}
S&=\int d^4x \sqrt{g}\left[\frac{\z}{2}\s^{\frac{1}{4(p-1)}}F_1\left(h^2\s^{-\frac{1}{4(p-1)}}\right)R\right.\\
&\left.-c_1 \s^{\frac{1}{4(p-1)}-2}F_2\left(h^2\s^{-\frac{1}{4(p-1)}}\right)g^{\m\n}\partial_\m \s \partial_\n \s\right.\\
&\left.-\frac{1}{2} F_3\left(h^2\s^{-\frac{1}{4(p-1)}}\right)g^{\m\n}\partial_\m h \partial_\n h\right.\\
&\left.+\d\s^{-1}h F_4\left(h^2\s^{-\frac{1}{4(p-1)}}\right)g^{\m\n}\partial_\m \s \partial_\n h\right.\\
&\left.-c_2 \s^{\frac{1}{2(p-1)}}V\left(h^2\s^{-\frac{1}{4(p-1)}}\right)+c_3\,\sigma^{-1/2}\right] \ .
\end{aligned}
\ee
We should stress, once again, that unless $p=0$, the above theory is completely analogous to the one presented in~\cite{Blas:2011ac}, in which the term proportional to $c_3$ explicitly violates the invariance of the theory under scale transformations. Also, like in the purely gravitational theory, the limit $p=1$ is peculiar. In the two-field case, however, the presence of the extra scalar makes it possible to construct SITDiff theories even if the dimensionality of the metric is zero.

Before moving on, we would like to mention that the inclusion of gauge fields and fermions in the present framework goes along the same lines as in~\cite{Blas:2011ac}. Since here we are interested solely on the gravitational and scalar sectors of the SITDiff theories, the interested reader is referred to this work for an extensive discussion on the subject.

\subsection{Higgs-dilaton cosmology from TDiff} 

The presence of gravity in the theory under consideration makes it nonrenormalizable. Hence, it should be thought  of as an effective field theory which is valid up to some energy scale. Let us assume that for energies well below this cutoff, $h\ll \s^{\frac{1}{8(p-1)}}$. In this case, if the the various functions are analytic in their argument, we can Taylor expand them as
\be
\begin{aligned}
F_i(h^2\s^{-\frac{1}{4(p-1)}})&\approx 1+f_i\, h^2\s^{-\frac{1}{4(p-1)}}+\ldots \ ,\\
V(h^2\s^{-\frac{1}{4(p-1)}})&\approx 1+\tilde\a h^2\s^{-\frac{1}{4(p-1)}}+\tilde\b h^4\s^{-\frac{1}{2(p-1)}}+\ldots \ ,
\end{aligned}
\ee
where the ellipses denote higher order terms, and $f_i,\tilde\a,\tilde\b$ are constants that depend on the structure of the particular function. Plugging the above into~\eqref{TD-SI2-full-2} and keeping the leading terms, we see that for $p=0$, the action becomes
\be
\begin{aligned}
\label{TD-SI2-3}
S&=\int d^4x \sqrt{g}\left[\frac{\z \s^{-\frac{1}{4}}+\xi_h h^2}{2}R-\frac{1}{128}\, \s^{-\frac{9}{4}}g^{\m\n}\partial_\m \s\partial_\n \s-\frac{1}{2}g^{\m\n}\partial_\m h \partial_\n h\right. \\
&\left.+\frac{\d}{8}\, \s^{-1}h g^{\m\n}\partial_\m \s \partial_\n h-\frac{\lambda}{4} h^4+\frac{\alpha}{2}\,\s^{-\frac{1}{4}}h^2-c\, \s^{-\frac{1}{2}}\right] \ ,
\end{aligned}
\ee
with
\be
\label{const-defs}
\xi_h=\frac{\z f_1}{2} \ ,~~~\a=-2c_2\tilde\a\ ,~~~\lambda=4c_2\tilde \b \ ,~~~c=c_2-c_3 \ .
\ee
Making use of~\eqref{sigm-chi-1}, we can express the above in a more familiar form:
\be
\begin{aligned}
\label{TD-SI2-4}
S&=\int d^4x \sqrt{g}\left[\frac{\z\x^2+\xi_h h^2}{2}R-\frac{1}{2}g^{\m\n}\partial_\m \x\partial_\n \x-\frac{1}{2}g^{\m\n}\partial_\m h \partial_\n h\right. \\
&\left.-\d\, \x^{-1}h g^{\m\n}\partial_\m \x \partial_\n h-\frac{\lambda}{4} h^4+\frac{\a}{2}\,\x^2h^2-c\,\x^4\right] \ .
\end{aligned}
\ee

Notice that once we identify the scalar field $h$ with the Higgs boson (in the unitary gauge), then for $\d=0$ (and renaming $\z=\xi_\chi$), the above bears resemblance to the phenomenologically viable Higgs-dilaton cosmological model that was presented and studied in detail in the previous chapter, see also~\cite{Shaposhnikov:2008xb,Shaposhnikov:2008xi,GarciaBellido:2011de,GarciaBellido:2012zu,Bezrukov:2012hx,Rubio:2014wta}. There are, however, certain differences which should be pointed out. First of all, in the present context, we need not introduce the field $\chi$ ad hoc, since this degree of freedom is already present in the gravitational sector. Moreover, as we mentioned before, a symmetry-breaking potential is absent. This means that contrary to what happens in theories for which $p\neq 0$, the scale symmetry of the system remained intact when it was cast into a form invariant under the full group of diffeomorphisms. Finally, it is interesting to note that the way  this theory is derived here is much simpler as compared to the conventional SITDiff, where complicated theory-defining functions have to be chosen~\cite{Blas:2011ac}.

Once we have identified $h$ with the Higgs field, we have to make sure that the theory has satisfactory particle physics as well as cosmological phenomenology, which puts constraints on the various parameters that appear in the action~\eqref{TD-SI2-4}. To start with, we observe that we have to set $\lambda\sim \mathcal O(1)$, in order for the model to be compatible with the SM predictions. Also, if $h$ is responsible for the inflationary expansion in the early Universe, then the nonminimal coupling has to satisfy $\xi_h\approx 47000\sqrt{\lambda}$, such that the amplitude of the primordial fluctuations agree with the observations~\cite{Bezrukov:2007ep}. 

Moreover, since $\a$ accounts for the difference between the Higgs boson mass and the Planck mass, it should be fixed at order $\mathcal O(10^{-30})$. In addition, we have to impose $c\sim\mathcal O(10^{-120})$ to reproduce the hierarchy between the value of the cosmological constant and the Planck scale. In the next section, we will present a conjecture about why these two parameters might be zero at the classical level.

\section{Regularity?}
\label{sec:Regul}

The fact that the Higgs boson mass and the cosmological constant terms are much smaller with respect to the Planck scale, might be an indication that at the  level of fundamental action both of them are zero. It is reasonable to wonder whether it exists some underlying principle or mechanism that forbids the presence of these terms in the action. 

Inspection of~\eqref{TD-SI2-3} reveals that due to the peculiar way the dilaton appears, all terms in the action that involve this field seem to be ill defined when $\sigma\rightarrow 0$, arguably related to the high energy limit.  As we will demonstrate in this section,  this is not the case if the theory is expressed in terms of  variables conjugate to space and time derivatives of the fields. These momentum densities were first introduced by Schwinger~\cite{Schwinger:1951xk,Schwinger:1953tb} (see also~\cite{Heinzl:2000ht}) and should be thought of as the covariant counterparts of canonical momenta.  
For a theory described by a Lagrangian $\mathscr L[\phi_i, \partial_\m \phi_i]$ which depends on a set of fields $\phi_i$ and their derivatives $\partial_\m \phi_i$, these quantities are defined as 
\be
\label{momenta-1}
\pi^{\ \m}_i\equiv\frac{\d\mathscr L}{\d \partial_\m\phi_i} \ .
\ee
Let us focus now on~\eqref{TD-SI2-3} and set $\d=0$, such that there is no kinetic mixing between the Higgs and the dilaton. This is purely for convenience, since the results will not be qualitatively different from the case where the mixing term is present, whereas the manipulations simplify considerably.  For our purposes, it is necessary to cast the action in such a way that it only contains first derivatives of the metric. A straightforward calculation, along the lines of the one in~\cite{landau_fields} for the Einstein-Hilbert action, gives us
\be
\label{TD-SI2-5}
S=\int d^4x \sqrt{g}\,\mathscr L \ ,
\ee
where the Lagrangian $\mathscr L$ is
\be
\begin{aligned}
\label{lagr-1}
\mathscr L&=\frac{\z \s^{-\frac{1}{4}}+\j h^2}{2}T^{\a\b\g\k\lambda\m}\G_{\a\b\g}\G_{\k\lambda\m}\\
&+\left(\j h \partial_\n h-\frac{\z \s^{-\frac{5}{4}}}{8}\partial_\n \s\right) S^{\k\lambda\m\n}\G_{\k\lambda\m}\\
&-\frac{1}{128}\, \s^{-\frac{9}{4}}g^{\m\n}\partial_\m \s\partial_\n \s-\frac{1}{2}g^{\m\n}\partial_\m h \partial_\n h\\
&-\frac{\lambda}{4} h^4+\frac{\a}{2}\,\s^{-\frac{1}{4}}h^2-c\, \s^{-\frac{1}{2}} \ .
\end{aligned}
\ee
Here 
\be
\label{chr-down}
\G_{\lambda\m\n}=\frac{1}{2}\left(\partial_\n g_{\m\lambda}+\partial_\m g_{\lambda\n}-\partial_\lambda g_{\m\n}\right) \ ,
\ee
and we introduce the tensors
\be
\label{S-T-tensor}
S^{\k\lambda\m\n}= g^{\k\lambda}g^{\m\n}-g^{\n\k}g^{\lambda\m}~~~\text{and}~~~T^{\a\b\g\k\lambda\m}= g^{\a\lambda}g^{\b\k}g^{\g\m}-g^{\a\b}g^{\g\k}g^{\lambda\m} \ .
\ee
Using~\eqref{momenta-1}, we find that  Schwinger's  ``momenta,''
\be
\label{momenta-2}
\pi_h^{ \ \n}= \frac{\d\mathscr L}{\d \partial_\n h} \ ,~~~\pi_\sigma^{ \ \n}= \frac{\d\mathscr L}{\d \partial_\n \sigma} \ ,~~~\text{and}~~~\rho^{\lambda\m\n}=\frac{\d\mathscr L}{\d \G_{\lambda\m\n}} \ ,
\ee
are given by 
\be
\label{momenta-3}
\begin{aligned}
&\pi_h^{\ \n}=\j h S^{\k\lambda\m\n}\G_{\k\lambda\m}-\partial^\n h \ , \\
&\pi_\s^{\ \n}=-\frac{1}{8}\left(\z \s^{-\frac{5}{4}} S^{\k\lambda\m\n}\G_{\k\lambda\m}+\frac{1}{8} \s^{-\frac{9}{4}}\partial^\n \s\right) \ ,
\end{aligned}
\ee
and
\be
\begin{aligned}
\label{momenta-4}
\rho^{\lambda\m\n}&=\frac{\z \s^{-\frac{1}{4}}+\j h^2}{2}\left(T^{\a\b\g\lambda(\m\n)}+T^{\lambda(\m\n)\a\b\g}\right)\G_{\a\b\g}\\
&+S^{\lambda(\m\n)\a}\left(\j h \partial_\a h-\frac{\z}{8}\s^{-\frac{5}{4}}\partial_\a \s\right) \ ,
\end{aligned}
\ee
where the parentheses $(\ldots)$ denote symmetrization of the corresponding indices. Using the relations~\eqref{momenta-3}, the above can be rewritten as
\be
\begin{aligned}
\label{momenta-4}
\rho^{\lambda\m\n}&=\frac{\z \s^{-\frac{1}{4}}+\j h^2}{2}\left(T^{\a\b\g\lambda(\m\n)}+T^{\lambda(\m\n)\a\b\g}\right)\G_{\a\b\g}\\
&+S^{\lambda\m\n\d}S^{\a\b\g}_{\ \ \ \ \d}\G_{\a\b\g}\left(\z^2\s^{-\frac{1}{4}}+\j^2h^2\right)-S^{\lambda\m\n}_{\ \ \ \ \k}\left(\j h \pi_h^{\ \k}+8\z \s \pi_\s^{\ \k} \right)\ .
\end{aligned}
\ee
In terms of $\rho,\ \pi_h, \ \text{and} \ \pi_\s$, we find that~\eqref{lagr-1} becomes
\be
\begin{aligned}
\label{momenta-rho-explicit}
\mathscr L&= \left(\frac{1}{\z \s^{-1/4}+\j h^2}\right)\left(\frac{(1+4\z)\z \s^{-1/4}+(1+4\j)\j h^2}{(1+6\z)\z \s^{-1/4}+(1+6\j)\j h^2}\right)\times\\
&\times \left(\frac{1}{4}\rho_{\k\lambda}^{\ \ \lambda}\rho^{\k\m}_{\ \ \m}+\frac{1}{3((1+4\z)\z \s^{-1/4}+(1+4\j)\j h^2)}\rho_\k^{\ \k\lambda}\rho^{\m}_{\ \m\lambda}-\rho_\k^{\ \k\lambda}\rho_{\lambda\m}^{\ \ \m}\right)\\
&+\left(\frac{1}{\z \s^{-1/4}+\j h^2}\right)\left(\rho_{\lambda\m\n}\rho^{\m\n\lambda}-\frac{1}{2}\rho_{\lambda\m\n}\rho^{\lambda\m\n}\right)\\
&+\frac{2\j h}{((1+6\z)\z \s^{-1/4}+(1+6\j)\j h^2)}\left(\rho_\k^{\ \k\lambda}-\frac{1}{2}\rho^{\lambda\k}_{\ \ \k}\right)\pi_{h\,\lambda}\\
&-\frac{16\z \s}{((1+6\z)\z \s^{-1/4}+(1+6\j)\j h^2)}\left(\rho_\k^{\ \k\lambda}-\frac{1}{2}\rho^{\lambda\k}_{\ \ \k}\right)\pi_{\s\,\lambda}\\
&-\frac{48\z\j \s h}{((1+6\z)\z \s^{-1/4}+(1+6\j)\j h^2)}\pi_h^{\, \m}\pi_{\s\,\m}\\
&-\frac{1}{2}\left(\frac{(1+6\z)\z \s^{-1/4}+\j h^2}{(1+6\z)\z \s^{-1/4}+(1+6\j)\j h^2}\right)\pi_{h\,\m}\pi_h^{\,\m} \\
&-32\s^{9/4}\left(\frac{\z \s^{-1/4}+(1+6\j)\j h^2}{(1+6\z)\z \s^{-1/4}+(1+6\j)\j h^2}\right)\pi_{\s\,\m}\pi_\s^{\,\m}\\ 
&-\frac{\lambda}{4} h^4+\frac{\a}{2}\,\s^{-\frac{1}{4}}h^2-c\, \s^{-\frac{1}{2}} \ .
\end{aligned}
\ee 
It is convenient to introduce at this point
\be
\begin{aligned}
\label{momenta-5}
&P_{\lambda\m\n}\equiv\rho_{\lambda\m\n}-\frac{1}{2}\left(\frac{(1+4\z)\z  \s^{-\frac{1}{4}}+(1+4\j)\j h^2}{(1+6\z)\z  \s^{-\frac{1}{4}}+(1+6\j)\j h^2}\right)g_{\m\n}\rho_{\lambda\k}^{\ \ \k}\\
&-\frac{1}{3}\left(g_{\lambda\m}\rho^\k_{\ \k\n}+g_{\lambda\n}\rho^\k_{\ \k\m}-\frac{\z  \s^{-\frac{1}{4}}+\j h^2}{(1+6\z)\z  \s^{-\frac{1}{4}}+(1+6\j)\j h^2}g_{\m\nu}\rho^\k_{\ \k\lambda}\right)\\
&+\frac{(\z  \s^{-\frac{1}{4}}+\j h^2)}{(1+6\z)\z  \s^{-\frac{1}{4}}+(1+6\j)\j h^2}g_{\m\n}\left(\j h\pi_{h\,\lambda}-8\z \s \pi_{\s\,\lambda}\right) \ ,
\end{aligned}
\ee
such that~\eqref{momenta-rho-explicit} simplifies considerably and reads
\be
\begin{aligned}
\label{theory-5}
\mathscr L&=\frac{1}{(\j h^2+\z \s^{-\frac{1}{4}})}\left(P_{\lambda\m\n}P^{\m\n\lambda}-\frac{1}{2}P_{\lambda\m\n}P^{\lambda\m\n}\right)\\
&-\frac{(1+4\j)\j h^2+(1+4\z)\z \s^{-\frac{1}{4}}}{(\j h^2+\z \s^{-\frac{1}{4}})^2}\left(P_\k^{\ \k\lambda}P_{\lambda\m}^{\ \ \m}-\frac{1}{2}P_{\k\lambda}^{\ \ \lambda}P^{\k\m}_{\ \ \m}\right)\\
&+\frac{2(\j^2 h^2+\z^2 \s^{-\frac{1}{4}})}{(\j h^2+\z \s^{-\frac{1}{4}})^2} P_\k^{\ \k\lambda}P^\m_{\ \m\lambda}-\frac{1}{2}\pi_{h\, \m}\pi_h^{\ \m} -32 \s^{\frac{9}{4}} \pi_{\s\, \m}\pi_\s^{\ \m}\\
&-\frac{\lambda}{4} h^4+\frac{\a}{2}\,\s^{-\frac{1}{4}}h^2-c\, \s^{-\frac{1}{2}} \ .
\end{aligned}
\ee
Observe that in the limit where the four-momenta $P$ (or equivalently $\rho$) are kept fixed while $\s$ tends to zero, i.e., for
\be
\begin{aligned}
\s &\rightarrow 0 \ , \\
\pi_h \,,\pi_\s&\,,\,P~\text{or}~\rho\rightarrow\text{fixed} \ ,
\end{aligned}
\ee
the only terms that blow up are the Higgs mass and the cosmological constant. Therefore, it is tempting to speculate that both these terms should not be included in the action in the first place if we want the theory to remain regular at the UV limit. It is interesting to note that the pathological behavior of the cosmological constant persists for arbitrary metric dimensions, the reason being that it is always proportional to $\s^{-\frac{1}{2}}$. On the other hand, if $p$ is not chosen to be equal to zero, the Higgs mass term, as well as the term proportional to $\s^{\frac{1}{2(p-1)}}$ (which does not feed into the cosmological constant unless $p=0$), are singular only if $p<1$.

Even though we do not have an answer to what is the origin of this selection rule, it could be a manifestation of some yet unknown mechanism at very high energies. Notice that if a scale-invariant regularization scheme is used (see for example~\cite{Shaposhnikov:2008xi}), then these terms cannot be generated at any order in perturbation theory. It may well be the case that they emerge from nonperturbative physics, something that can explain their smallness.

\section{Summary and Outlook}
\label{sec:conclusSITDIFF}

The purpose of this chapter was to investigate a previously unexplored region of the parameter space of theories with dilatational symmetry whose gravitational sector is constructed by requiring invariance under the group of transverse diffeomorphisms. Due to the invariance under this restricted group of coordinate transformations, the determinant of the metric becomes a dynamical degree of freedom which can be thought of as a dilaton. 

We argued that the most appropriate and natural option for the description of arbitrary coordinate systems is for the metric to have dimensionality of area. We demonstrated that the particular setup is distinct from the ordinary theories in a number of aspects. The form of the pure gravitational action is completely fixed and, moreover, once diffeomorphism invariance is restored via the St\"uckelberg mechanism, the scale symmetry remains intact. As a result, there is no runaway potential for the dilaton. 

Next, we investigated the form of the action of a model that on top of the dilaton contains an extra scalar field which we identified with the Standard Model Higgs boson. Based on the way the dilaton appears and interacts with the Higgs field, we observed that the Higgs mass and cosmological constant are the only singular terms in the specific limit (fixing the proper variables which we define) involving a metric determinant going to zero.  An appealing hypothesis is  that these terms should not be included in the fundamental theory, but rather their low-energy presence should result from nonperturbative effects through some yet unknown mechanism.

It would be interesting to understand how these considerations can be applied to theories without Lorentz invariance, such as, for example in Ho\v{r}ava-Lifshitz gravity (see for example~\cite{Horava:2009uw,Blas:2009qj}), a version of which has recently been proven to be renormalizable~\cite{Barvinsky:2015kil}.

\chapter{Concluding remarks}
\label{ch:Conclusions}

Theories that are invariant under scale and conformal transformations are of utmost interest. In this thesis, we dealt with various aspects -- purely of theoretical but also of phenomenological nature -- related to them. 

In chapter~\ref{sec:coset_constr}, we presented the necessary modifications that have to be made such that coset construction can be used to gauge spacetime symmetries. We argued that -- even when a (spacetime) symmetry is linearly realized -- this technique  provides us with the appropriate machinery for studying these systems.  To understand the logic behind this framework, we first considered the gauging of the Poincar\'e group and we showed how, by imposing covariant conditions, redundant degrees of freedom can be consistently eliminated. These should be considered equivalent to the inverse Higgs constraints that are a standard tool when a symmetry is spontaneously broken. They are used to eliminate the Goldstone modes which are unnecessary and thus, account for the fact that their number is smaller than the number of the broken generators of the group under consideration. 

In chapter~\ref{ch:Weyl_Ricci}, we employed the coset construction in order to gauge the Poincar\'e group plus dilatations. We showed that in the absence of torsion, an analog of the inverse Higgs constraint allows to trade a certain configuration of Weyl gauge field for  the Schouten tensor. Thus, Ricci gauging appears naturally in the framework of the coset construction. We determined that even higher-derivative theories can be coupled to a curved spacetime in a Weyl invariant way, without the introduction of extra degrees of freedom. This means that the range of applicability of this method is much larger than what was previously thought.  We illustrated that the quartic in derivatives conformal theory of a scalar field in an arbitrary number of spacetime dimensions $n>2$ can be made Weyl invariant using this procedure. As we showed, the presence of more than one derivatives of the field brings some complications, nevertheless, Ricci gauging can be carried out consistently. Meanwhile, once the requirement of having a torsionless theory is dropped, then inverse Higgs constraint dictates that the role of the gauge field associated with the dilatations can be played by one of the irreducible pieces of the torsion tensor.

In chapter~\ref{ch:Weyl_vs_Conf}, we started by demonstrating in a pedagogic way the difference between conformal and Weyl symmetries. We then took a closer look at the higher-derivative theory constructed previously and we investigated what happens for $n=2$. Even though the starting point was a conformal theory, it turns out that it was not possible to be made Weyl-invariant in two spacetime dimensions. But this was just the ``tip of the iceberg'', since this seems to be the case for a whole class of higher-derivative theories invariant under the conformal group (both in curved and flat manifolds), which do not allow for Weyl invariant generalizations.

In chapter~\ref{ch:NR_Weyl}, we turned our attention to nonrelativistic spacetime symmetries and we discussed how they can be gauged in the context of the coset construction. We showed that for a nonrelativistic field theory to be made Weyl invariant, torsion must not vanish. Considering first the centrally extended Galilei algebra (which is a contraction of the Poincar\'e one), we demonstrated that for a certain subclass of these models (the \emph{twistless torsionful theories}), it is always possible to express the spatial components of the Weyl vector in terms of torsion. We then focused on the Lifshitz algebra and we found that any scale-invariant theory in flat spacetime can be coupled to a curved background in a Weyl-invariant way, with torsion acting as the Weyl gauge field.

Even though it is tangent to the philosophy of the present thesis, in chapter~\ref{chapt:Poincaregrav}, we allowed for connection and vielbein to be independent degrees of freedom and we investigated the particle dynamics of the Poincar\'e gravitational theory with terms that are at most quadratic in the field strengths. In order to carry out the analysis, we employed the spin projection operator formalism and extended it in order to determine the effect of terms that do not preserve parity. Most of the operators that we constructed had not been presented previously. We derived constraints that the various parameters of the theory must satisfy, so that it contains only healthy modes. We showed that the parity-odd invariants might prevent the presence of tachyons, but unfortunately ghosts are still present. Nevertheless, there exist torsionful theories in which the extra degrees of freedom are neither ghosts nor tachyons and have vast cosmological applications. As we have argued, parity-odd terms are non-trivial modifications to the dynamics of the theory. Detailed analysis has to be made in order to see what the effects beyond the linear order are, or what happens when the theory is considered on backgrounds different from flat.

The last two chapters of the thesis were devoted to more phenomenological aspects related to global scale invariance. Namely, in chapter~\ref{ch:HD_EFT}, we studied the self-consistency of the Higgs-dilaton model -- a particular case of scale-invariant systems invariant under transverse diffeomorphisms (SITDiff)-- from an effective field theory point of view. Taking into account the influence of the dynamical background fields, we determine the effective cut-off of the theory, which turned out to be parametrically larger than all the relevant energy scales from inflation to the present epoch. We formulated a set of assumptions needed to estimate the amplitude of the quantum corrections in a systematic way and showed that the connection between the tilt of scalar perturbations and the DE equation of state remains unaltered if these assumptions are satisfied. 

In chapter~\ref{ch:SITDIFF}, we considered SITDiff theories and we showed that if the metric carries mass dimension $[$GeV$]^{-2}$, the scale invariance of the system is preserved, unlike the situation in theories in which the metric has mass dimension different from $-2$. We speculated that for the action to have a well defined high-energy limit, the one should not include the bare Higgs mass and cosmological constant in the action. It is reasonable to wonder if a non-perturbative mechanism could be responsible for their smallness.

\part{Appendices}
\label{part:append}

\appendix

\chapter{Christoffel symbols and covariant derivatives \label{Christoffel}}

The coset construction allows us to write a covariant derivatives for internal symmetries (having introduced the fields $y ^ A$, we have made spacetime translations effectively internal), meaning that it acts only on Lorentz indices  $A,B,\ldots$. However, the procedure does not produce the covariant derivative for fields with spacetime indices or, in particular, for the vielbein. Nevertheless, one can introduce the analog of Christoffel symbols,\footnote{However, one should be careful, since the new symbols depend explicitly on the scaling dimension of fields they act on.} so that the covariant derivative is consistent with interchanging the Lorentz and spacetime indices. Namely, using the vector with scaling dimension $\D _ V$,
\be
V ^ A = e ^ A _ \m V ^ \m \ ,
\ee
one defines
\be
D _ \m V ^ \n = \p _ \m V ^ \n + G ^ \n _ {\m \lambda} V ^ \lambda = E _ A^\n e ^ B _ \m D _ B V ^ A \ .
\ee
Using the expression for $\omega$ from~\eqref{spin-con3}, it is not difficult to show that in this case
\be
G ^ \s _ {\m \n} = \G ^ \s _ {\m \n} + \d G ^ \s _ {\m \n} \ ,
\ee
with $\G$ being the standard Christoffel symbols
\be
\G ^ \s _ {\m \n} = - E^ \s _ A \l (  \bar \omega ^ {A} _ {\m B} e ^ B _ \n + \p _ {\m} e ^ A _ \n \r ) \ ,
\ee
which are compatible with the metric and thus satisfy
\be
\nabla _ \m V ^ \n \equiv \p _ \m V ^ \n + \G ^ \n _ {\m \lambda} V ^ \lambda = E _ A ^ \n  (\p _ \m V ^ A - \bar \omega ^ A _ {\m B} V ^ B ) \ ,
\ee
 with $\nabla$ the standard covariant derivative. Meanwhile
\be
\d G ^ \s _ {\m \n} = - \D _ V W _ \m \d ^ \s _ \n + W _ \n \d ^ \s _ \m - W ^ \s g _ {\m \n} \ .
\ee
Using the fact that the covariant derivative for a field $V ^ \m$ with scaling dimension $\D _ V + 1$ can be written as
\be
D _ \m V ^ \s = \nabla _ \m V ^ \s +  \underbrace{( W _ \m \d ^ \s _ \n + W _ \n \d ^ \s _ \m - W ^ \s g _ {\m \n})} _ {\d \G ^ \s _ {\m \n}}  V ^ \n - ( \D _ V + 1 ) W _ \m V ^ \s \ ,
\ee 
it is straightforward to show that the covariant derivative $\nabla _ \m$ can be made Weyl covariant, provided all partial derivatives 
are substituted by
\be
\p _ \m \to \p _ \m - \D W _ \m \ ,
\ee
where $\D$ is the scaling dimension of the field the partial derivative $\p _ \m$ acts on. For instance
\be
\p _ \m g _ {\lambda \s} \to \p _ \m g _ {\lambda \s} + 2 W _ \m g _ {\lambda \s}  \ .
\ee

\chapter{Conformal algebra \label{conf_group}}

The conformal group in $n \neq 2$ dimensions is an extension of the Poincar\'e group. On top of the momenta $P _ A$ (translations) and the Lorentz generators $J_{AB}$, it contains dilatations $D$ and special conformal transformations (SCT) $K_A$, also called conformal boosts. Overall, there are $n(n+1)/2$ generators with the following nonzero commutation relations~\cite{Rychkov:2016iqz,DiFrancesco:1997nk}
\be
\begin{aligned}
\label{conf_cr}
\l [ D, P _ A \r ] & =  - i P _ A \ ,  \\
\l [ J _ {AB}, P _ C \r ] & =  i \l ( \eta _ {BC} P _ A - \eta _ {AC} P _ B \r ) \ ,  \\
\l [ K _ A, P _ B \r ] & =  - 2 i \l ( \eta _ {AB} D + J _ {AB} \r ) \ ,  \\
\l [ D, K _ A \r ] & =  i K _ A\ , \\
\l [ J _ {AB}, J _ {CD} \r ] & =  i \l ( J _ {AD} \eta _ {BC} + J _ {BC} \eta _ {AD} -
J _ {BD} \eta _ {AC} - J _ {AC} \eta _ {BD} \r ) \ ,  \\
\l [ J _ {AB}, K _ C \r ] & =  i \l ( \eta _ {BC} K _ A - \eta _ {AC} K _ B \r ) \ . 
\end{aligned}
\ee

For completeness, let us briefly describe what would happen if the full conformal group was gauged instead of just Poincar\'e and dilatations. It is straightforward to repeat the steps of the coset construction using the commutation relations for the conformal group. This leads to the following transformation rules for the gauge fields

\begin{table}[H]
\centering
\scalebox{0.85}{
\bt{c | cccc}
$ $ &$e ^ {'A} _ {\m}$ & $\omega ^ {' AB} _ {\m} $ & $W' _ {\m}$ & $B ^ {'A} _ {\m}$ \\
\hline
$J$ & $ e _ {\m} ^ B \Lambda _ B ^ {~A}$ & $\omega ^ {CD} _ {\m} \Lambda _ {C} ^ {~A} \Lambda _ {D} ^ {~B}+\l(\Lambda\p_\m \Lambda^{-1}\r)^{AB}$ & $W _ {\m}$ & $ B _ {\m} ^ B \Lambda _ B ^ {~A}$ \\
$D$ & $ e ^ {-\a} e _ {\m} ^ A $ & $\omega ^ {AB} _ {\m}$ & $W _ {\m} + \p _ \m \a $ & $ e ^ {\a} B _ {\m} ^ A $ \\
$K$ & $ e _ {\m} ^ A $ & $\omega _ {\m} ^ {AB} +  e _ {\m} ^ {A } \a ^ {B } $ & $W _ {\m} - 2 e _ {\m} ^ C \a _ C$ & $ B _ {\m} ^ A 
+ \a ^ {A} _ B e _ {\m} ^ B $\\
& & $-e _ {\m} ^ {B } \a ^ {A }$& & $- \omega _ {\m} ^ {AB} \a _ B - \a ^ A W _ {\m} + \p _ \m \a ^ A$
\et
}

\end{table}

\noindent
Notice that we introduced the new gauge fields $B ^ A_ \m$, associated with  SCT. The corresponding field strengths are found to be
\begin{align}
\centering
\label{tensor-defs1}
e _{ \m \n } ^ A & =  \p _ \m e _ \n ^ A - \p _ \n e _ \m ^ A - \omega _ {\m B} ^ {A}  e _ \n ^B + \omega _ {\n B} ^ {A}  e_ \m ^B +
W _ \m e _ \n ^ A - W _ \n e _ \m ^ A\ ,  \\
\label{tensor-defs2}
\omega ^ {AB} _ {\m \n} & =  \p _ \m \omega _ \n ^ {AB} - \p _ \n \omega _ \m ^ {AB} 
- \omega _ {\m C} ^ {A} \omega _ \n ^{CB} + \omega _ {\n C} ^ {A} \omega _ \m ^{CB}\nn \\
&+ 
2 \l ( B ^ A _ \m e ^ B _ \n - B ^ A _ \n e ^ B _ \m - B ^ B _ \m e ^ A _ \n + B ^ B _ \n e ^ A _ \m \r )\ ,  \\
\label{tensor-defs3}
W _ {\m \n} & =  \p _ \m W_ \n - \p _ \n W _ \m + 2 \l ( B ^ A _ \m e _ {\n A} - B ^ A _ \n e _ {\m A} \r )\ ,  \\
\label{tensor-defs4}
B _{ \m \n } ^ A & =  \p _ \m B _ \n ^ A - \p _ \n B _ \m ^ A - \omega _ {\m B} ^ {A}  B _ \n ^B + \omega _ {\n B} ^ {A}  B _ \m ^B -
W _ \m B _ \n ^ A + W _ \n B _ \m ^ A \ .
\end{align}
\noindent
Their transformations have the following form

\begin{table}[H]
\centering
\bt{c | cccc}
$ $ &$e ^ {'A} _ {\m \n}$ & $\omega ^ {' AB} _ {\m \n} $ & $W' _ {\m \n}$ & $B ^ {'A} _ {\m \n}$ \\
\hline
$J$ & $ e _ {\m \n} ^ B \Lambda _ B ^ {~A}$ & $\omega ^ {CD} _ {\m \n} \Lambda _ {C} ^ {~A} \Lambda _ {D} ^ {~B}$ & $W _ {\m \n}$ & $ B _ {\m \n} ^ B \Lambda _ B ^ {~A}$ \\
$D$ & $ e ^ {-\a} e _ {\m \n} ^ A $ & $\omega ^ {AB} _ {\m \n}$ & $W _ {\m \n}$ & $ e ^ {\a} B _ {\m \n} ^ A $ \\
$K$ & $ e _ {\m \n} ^ A $ & $\omega _ {\m \n} ^ {AB} +  e _ {\m \n} ^ {A } \a ^ { B } $ & $W _ {\m \n} - 2 e _ {\m \n} ^ C \a _ C$ & $ B _ {\m \n} ^ A + \a ^ {A} _ B e _ {\m \n} ^ B $\\
& & $-e _ {\m \n} ^ {B } \a ^ { A } $ & & $~~ - \omega _ {\m \n} ^ {AB} \a _ B - \a ^ A W _ {\m \n} $
\et

\end{table}

\noindent
We notice that under SCT, the gauge fields mix with the vielbein $e _ \m ^ A$. The origin of this unordinary behavior is the specific form of the commutation relations. According to the rules of the coset construction, the momenta and all the nonlinearly realized generators should form a representation of the group formed by the rest of the generators. Clearly, this condition is broken by the commutation relation between the momenta and conformal boosts~\eqref{conf_cr}.

The transformation properties of the gauge fields would create an obstacle on the way to introducing the covariant derivative for matter fields. However, looking at the transformations of the field strengths,  we see that the expressions simplify considerably once $e _ {\m \n} ^ A = 0$ is imposed. Therefore, as long as pure gravity is concerned, the coset construction produces a sensible result.

The constraint $e _ {\m \n} ^ A$ has the same solution as in the main text; see~\eqref{spin-con3}-\eqref{spin-con5}. The changes appear when one uses also the constraint $E^\n_B\omega ^ {A B}_ {\m \n} = 0$, which can now be solved algebraically in favor of $B_\m^A$. This leads to
\be
B _ \m ^ A e _ {\n A} = (n-2) \l ( \mc R _ {\m \n} - \f {1} {2 (n-1)}g _ {\m \n} \mc R \r ) \ , 
\ee
where $\mc R _ {\m \n} =  \mc R _ {\m \s} ^ {A B} E _ {B} ^ {\s} e _ {A \n}$ and $\mc R=g^{\m\n}\mc R_{\m\n}$ are contractions of the curvature tensor 
\be
\mc R _ {\m \n} ^ {AB}\equiv \bar \omega ^ {AB} _ {\m \n} + \d \omega ^ {AB} _ {\m \n} \ ,
\ee
with $\bar \omega ^ {AB} _ {\m \n} $ and $ \d \omega ^ {AB} _ {\m \n}$ given by~\eqref{barR} and~\eqref{del_omega}.

To obtain the condition for Ricci gauging~\eqref{Ricci_Weyl}, we have to force $B _ \m ^ A$ to vanish. However, it is clear that this constraint is not consistent with SCT. Therefore, in one way or another, we have to dispense of SCT and consider only the gauging of the Poincar\'e group plus dilatations.

\newpage

\chapter{Irreducible decomposition of torsion \label{irred_decomp}}

We defined the torsion tensor as
\be
\label{tors-append}
T_{\m\n}^A\equiv \p _ \m e _ \n ^ A - \p _ \n e _ \m ^ A - \omega _ {\m B} ^ {A}  e _ \n ^B + \omega _ {\n B} ^ {A}  e _ \m ^B \ , 
\ee
and since it is antisymmetric in $\m$ and $\n$, it has $\frac{n^2(n-1)}{2}$ independent components in an $n$-dimensional spacetime. Under the action of the Lorentz group $SO(1,n-1)$, it can be decomposed into three irreducible quantities:\footnote{In fact, every tensor with the same symmetries as $T_{\m\n}^A$ admits this decomposition.}
\begin{itemize}
\item The vector $\upsilon_\m$
\be
\upsilon_\m=E_{A}^\n T_{\m\n}^A=E^\n_A\l (\p _ \m e _ \n ^ A - \p _ \n e _ \m ^ A + \omega _ {\n B} ^ {A}  e _ \m ^B \r) \ ,
\ee
with $n$ independent components. 

\item The totally antisymmetric ``dual'' tensor 
\be
\a^{\s_1\s_2\cdots \s_{n-3}}=\frac{1}{n \det e}\ep^{\s_1\s_2\cdots \s_{n-3}\m\n\lambda}e_{\lambda A}T_{\m\n}^A \ ,
\ee
with $\frac{n(n-1)(n-2)}{6}$ independent components.
\item The traceless $\frac{n(n^2-4)}{3}$ - component reduced torsion tensor $\tau_{\m\n}^A$
\be
\begin{aligned}
\tau_{\m\n}^A & =T_{\m\n}^A-\frac{3}{2(n-1)}\l(\upsilon_\m e_\n^A-\upsilon_\n e^A_\m \r)\\
&-\frac{1}{2}E^{\lambda A}\l(T^B_{\lambda\m} e_{\n B}-T^B_{\lambda\n} e_{\m B}\r) \ ,
\end{aligned}
\ee
which is subject to the following $n+\frac{n(n-1)(n-2)}{6}$ constraints 
\be
\label{constraints-tors}
E^\n_A\tau_{\m\n}^A=0 \ \ \  \text{and} \ \ \ \ep^{\s_1 \s_2 \cdots \s_{n-3}\m\n\lambda} e_{\lambda A}\tau_{\m\n}^A =0 \ .
\ee
\end{itemize}
It is a straightforward exercise to show that~\eqref{tors-append} can be written in terms of the irreducible pieces we presented above as
\be
\begin{aligned}
\label{tors-append2}
T_{\m\n}^A&=\frac{n}{6(n-3)!}\det e~ E^{\lambda A}\ep_{\s_1\s_2\ldots \s_{n-3}\m\n\lambda}\a^{\s_1\s_2\cdots \s_{n-3}}\\
&+\frac{1}{n-1}\l(\upsilon_\m e_\n^A-\upsilon_\n e^A_\m \r)+\frac{2}{3}\tau_{\m\n}^A \ .
\end{aligned}
\ee
Notice that these expressions for  $n=4$ boil down to the ones in~\cite{Diakonov:2011fs}.

\chapter{Paneitz-Riegert operator}
\label{riegert}

The  Weyl covariant generalization of $\Box ^ 2$ is the Paneitz operator whose form in $n$ dimensions $(n\neq 2)$ was given in~\eqref{pan-rieg}. Using the definition of the Schouten tensor~\eqref{tensor_R}, this operator can be written in a more familiar form as
\be
\begin{aligned}
\mc Q (g) & = \nabla ^ 2 + \nabla ^ \m  \l[\l(\frac{4} {n-2} R _ {\m \n} - \frac{n^2-4n+8}{2(n-1)(n-2)} g _ {\m \n}R \r ) \nabla ^ \n\r]  \\
&- \frac{n-4} {4 (n-1)} \nabla ^ 2 R- \frac{n-4}{(n-2)^2} R _ {\m \n} R ^ {\m \n} \\
&+ \frac{(n-4)(n^3-4n^2+16n-16)}{16 (n-1)^2(n-2)^2} R ^ 2 \ .
\end{aligned}
\ee
It is interesting to note that for $n=4$, the above expression simplifies considerably
\be
\mc Q (g) \to \nabla ^ 2 +2 \nabla ^ \m  \l[\l(R _ {\m \n} - \frac{1}{3} g _ {\m \n}R \r ) \nabla ^ \n\r] \ ,
\ee
and is also known as the Paneitz-Riegert operator~\cite{Fradkin:1981jc,Fradkin:1981iu,Fradkin:1982xc,Riegert:1984kt}.

\chapter{The linearized action}
\label{app:lin_act}
The linearized action for the PGT in chapter~\ref{chapt:Poincaregrav} can be expressed as the sum of several terms that contain pure connection and vielbein excitations, as well as their mixings
\begin{equation}
\label{lin-terms}
S_2=S_2(\omega,\omega)+S_2(s,s)+S_2(a,a)+S_2(\omega,s)+S_2(\omega,a)+S_2(s,a) \ .
\end{equation}

A lengthy calculation reveals that each of the above terms reads
\small
\begin{eqnarray}
S_2(\omega,\omega)&=&\displaystyle\frac{1}{12}\int d^4x~\Big\{4(2r_1-2r_2+3r_4+3r_5)\partial^{B}\omega_{CAB}\partial_D\omega^{CAD}\nonumber \\
&-&\vphantom{\frac{a}{b}}12(r_7-r_8)\epsilon^{ABIK}\partial_I\omega_{CAB}\partial_J\omega_K^{\ CJ}-3(r_7+r_8)\epsilon^{ABIJ}\partial_D\omega_{CAB}\partial^D\omega^{CIJ} \nonumber \\
&+&\vphantom{\frac{a}{b}}24(r_4+r_5)\partial^C\omega_B^{\ BA}\partial^D\omega_{CAD}+3(r_7+r_8)\epsilon^{ABIJ}\partial_C\omega_{CAB}\partial^K\omega_{KIJ} \nonumber  \\
&-&\vphantom{\frac{a}{b}}8(r_6-r_8)\epsilon^{ABCD}\partial_D\omega_{ABC}\partial_I\omega_K^{\ KI} +16(r_1-r_2)\partial^C\omega_{CAB}\partial_D\omega^{ABD} \nonumber \\
&-&\vphantom{\frac{a}{b}}4(2r_1+r_2)\partial^C\omega_{CAB}\partial_D\omega^{DAB}+4(2r_1+r_2)\partial_D \omega_{CAB}\partial^D\omega^{CAB}\nonumber \\
&+&\vphantom{\frac{a}{b}}8(r_1-r_2)\partial_D \omega_{CAB}\partial^D\omega^{ACB}+12(r_4+r_5)\partial_D\omega_B^{\ BA}\partial^D\omega^C_{\ CA}\nonumber \\
&+&\vphantom{\frac{a}{b}}4(4r_1+2r_2-4r_3+3r_4-3r_5)\partial^{B}\omega_{CAB}\partial_D\omega^{ACD}\nonumber  \\
&+&\vphantom{\frac{a}{b}}12(r_4-r_5)\partial_A\omega_C^{\ CA}\partial^B\omega_D^{\ DB}-24t_5\epsilon^{ACIK}\omega_{CAB}\omega_{KI}^{\ \ B}\nonumber  \\
&-&\vphantom{\frac{a}{b}}4(t_1-2t_3)\omega_B^{\ BA}\omega^C_{\ CA}+4(t_1+t_2)\omega_{CAB}\omega^{CAB}\nonumber \\
&-&\vphantom{\frac{a}{b}}8(t_4-2t_5)\epsilon^{ABIK}\omega_{CAB}(2\omega_{KI}^{\ \ B}+\omega^B_{\ KI}) \nonumber \\
&-&\vphantom{\frac{a}{b}}4(t_1-2t_2)\omega_{CAB}\omega^{ACB}\Big\} \ ,\\
\nonumber \\
S_2(s,s)&=&\displaystyle\frac{1}{3}\int d^4x~\Big\{3(t_1+\lambda)\partial_C s_{AB}\partial^C s^{AB} -(t_1-2t_3+3\lambda)\times\nonumber\\
&\times&\vphantom{\frac{a}{b}}(\partial_A s\partial^A s-2\partial_A s\partial_B s^{AB} )-2(2t_1-t_3+3\lambda)\partial_B s_A^{\ B} \partial_C s^{AC}\Big\}
\ , \\
\nonumber \\
S_2(a,a)&=&\displaystyle\frac{1}{3}\int d^4x~\Big\{(t_1+t_2)\partial_C a_{AB}\partial^C a^{AB}-2(t_2-t_3)\partial_B a_A^{\ B}\partial_C a^{AC}\nonumber \\
&+&\vphantom{\frac{a}{b}}(t_4-2t_5)\epsilon^{ABKL}\left(\partial_C a_{AB}\partial^C a_{KL}-2\partial_C a_A^{\ C} \partial_L a_{BK}\right)\Big\}
\ , \\
\nonumber \\
S_2(\omega,s)&=&\displaystyle\frac{2}{3}\int d^4x \Big\{t_1\,\omega^{CAB}\partial_B s_{CA}+(t_1-2t_3)\omega_{C}^{\ CA}\left(\partial_B s_A^{\ B}-\partial_A s\right)\,\nonumber \\
&+&\vphantom{\frac{a}{b}}2(t_4+t_5)\epsilon^{AKLM}\omega_{KLB}\partial_M s_A^{\ B} \nonumber \\
&+&\vphantom{\frac{a}{b}}(t_4-2t_5)\epsilon^{AKLM}\omega_{BKL}\partial_M s_A^{\ B} \Big\} \ ,\\
\nonumber \\
S_2(\omega,a)&=&\displaystyle\frac{2}{3}\int d^4x\Big\{(t_1-2t_3)\omega_C^{\ CA}\partial_B a_A^{\ B}-(t_1-2t_2)\omega^{CAB}\partial_B a_{CA} \nonumber \\
&+&\vphantom{\frac{a}{b}}(t_1+t_2)\omega^{ABC}\partial_A a_{BC}+6t_5\epsilon^{AKLM}\omega_{KLB}\partial_M a_A^{\ B} 
\nonumber \\
&+&\vphantom{\frac{a}{b}}(t_4-2t_5)\epsilon^{ABKL}\left(\omega_{CKL}+\omega_{KLC}\right)\partial^C a_{AB}\nonumber \\
&-&\vphantom{\frac{a}{b}}(t_4-2t_5)\epsilon^{AKLM}\omega_{BKL}\partial_M a_A^{\ B}\Big\} \ ,  \\
\nonumber \\
S_2(s,a)&=&\displaystyle\frac{2}{3}\int d^4x\Big\{2(t_1+t_3)\partial_B s_A^{\ B}\partial_C a^{AC}\\
&+&\vphantom{\frac{a}{b}}(t_4-2t_5)\epsilon^{ABKL}\partial_C s_A^{\ C}\partial_B a_{KL}\Big\}\ .
\end{eqnarray}
\normalsize

\chapter{Spin-projection operators I}
\label{app:spinproj}

In this Appendix, we first give the full set of spin-projection operators $P^{\phi\chi}_{ij}(J)_{\acute{\alpha}\acute{\beta}}$ that we used as a basis to break the theory into spin sub-blocks. We then present the coefficient matrices, as well as their inverses. We have arranged matters in such a way that the upper left sub-matrices always correspond to the negative parity states. When parity-violating terms are not present in the action, the matrices acquire block-diagonal form, so they can be inverted separately. This enables us to check our algebra easily by comparing with the results of Sezgin and van Nieuwenhuizen~\cite{Sezgin:1979zf}.
Finally, by looking at the zeros of the determinants, we write down the masses of the particles related to each spin sector.

In what follows, we denote with $\Theta_{AB}$ the transverse and with $\Omega_{AB}$ the longitudinal projection operators. In momentum space they are respectively given by
\begin{equation}
\Theta_{AB}=\eta_{AB}-\frac{k_A k_B}{k^2} \ , \ \ \ \text{and} \ \ \ \Omega_{AB}=\frac{k_A k_B}{k^2} \ .
\end{equation}
We also denote $\tilde k_A=k_A/\sqrt{k^2}$. It is understood that the projectors have to be symmetrized or antisymmetrized in their $(A,B)$ and $(I,J)$ indices, depending on the symmetries of the fields they act on. For example, $P^{\omega\omega}_{ij}(J)_{CABKIJ}$ have to be antisymmetrized in both $(A,B)$ and  $(I,J)$, whereas $P^{\omega s}_{ij}(J)_{CABIJ}$ have to be antisymmetrized in $(A,B)$ and symmetrized in $(I,J)$. 

The tensorial manipulations that are involved are quite tedious and prone to algebraic mistakes. For that reason, we have cross-checked extensively our calculations with \emph{Mathtensor}~\cite{Mathtensor:1994}.

\section{Spin-0}

The 16 operators corresponding to the scalar part of the theory are
\begin{equation}
\label{s0op}
\begin{aligned}
&\displaystyle P^{\omega\omega}_{11}(0)_{CABKIJ}=\frac{1}{3} \Theta_{CK}\Theta_{AI}\Theta_{BJ}+\frac{2}{3} \Theta_{AK}\Theta_{BI}\Theta_{CJ}\ , \\
&\displaystyle P^{\omega\omega}_{12}(0)_{CABKIJ}=\frac{1}{3}\epsilon_{ABCD}\Omega^D_I\Theta_{JK}\ ,\\
&\displaystyle P^{\omega s}_{13}(0)_{CABIJ}=\frac{1}{3\sqrt{2}}\epsilon_{ABCD}\tilde k^D \Theta_{IJ}\ , \\
&\displaystyle P^{\omega s}_{14}(0)_{CABIJ}=\frac{1}{\sqrt{6}}\epsilon_{ABCD}\tilde k^D \Omega_{IJ} \ ,\\
&\displaystyle P^{\omega\omega}_{21}(0)_{CABKIJ}=-\frac{1}{3}\epsilon_{IJKL}\Omega^L_A\Theta_{BC}\ ,\\
&\displaystyle P^{\omega\omega}_{22}(0)_{CABKIJ}=\frac{2}{3}\Theta_{BC}\Omega_{AI}\Theta_{JK}\ , \\
&\displaystyle P^{\omega s}_{23}(0)_{CABIJ}=\frac{\sqrt{2}}{3}\tilde k_B\Theta_{CA} \Theta_{IJ} \ ,\\
&\displaystyle P^{\omega s}_{24}(0)_{CABIJ}=\sqrt{\frac{2}{3}}\tilde k_B \Theta_{CA} \Omega_{IJ}\ , \\
&\displaystyle P^{s\omega }_{31}(0)_{ABKIJ}=-\frac{1}{3\sqrt{2}}\epsilon_{IJKL}\tilde k^L \Theta_{AB}\ ,\\
&\displaystyle P^{s\omega }_{32}(0)_{ABKIJ}=\frac{\sqrt{2}}{3}\tilde k_J\Theta_{KI} \Theta_{AB}\ ,\\
&\displaystyle P^{ss}_{33}(0)_{ABIJ}=\frac{1}{3}\Theta_{AB}\Theta_{IJ}\ ,\\
&\displaystyle P^{ss}_{34}(0)_{ABIJ}=\sqrt{\frac{1}{3}}\Theta_{AB}\Omega_{IJ}\ , \\
&\displaystyle P^{s\omega }_{41}(0)_{ABKIJ}=-\frac{1}{\sqrt{6}}\epsilon_{IJKL}\tilde k^L \Omega_{AB}\ ,\\
&\displaystyle P^{s\omega }_{42}(0)_{ABKIJ}=\sqrt{\frac{2}{3}}\tilde k_J\Theta_{K I} \Omega_{AB}\ ,\\
&\displaystyle P^{ss}_{43}(0)_{ABIJ}=\sqrt{\frac{1}{3}}\Theta_{IJ}\Omega_{AB}\ ,\\
&\displaystyle P^{ss}_{44}(0)_{ABIJ}=\Omega_{AB}\Omega_{IJ}\ .
\end{aligned}
\end{equation}

\newpage

Using the above projectors we derived the $4\times 4$ coefficient matrix for the spin-0 sector that reads 
\begin{equation}
\label{s0dm}
c_{ij}^{\phi\chi}(0)=\begin{blockarray}{ccccc}
\scp\omega^-&\scp\omega^+&\scp s^+&\scp s^+\\
\begin{block}{(cccc) c}
c_{11} & c_{12}  & c_{13} &c_{14} &\scp\omega^- \\
c_{21} & c_{22}  & c_{23} &c_{24} &\scp\omega^+ \\
c_{31} & c_{32}  & c_{33} &c_{34}&\scp s^+ \\
c_{41} & c_{42}  & c_{43} &c_{44}&\scp s^+ \\
\end{block}
\end{blockarray}\ , 
\end{equation}
\begin{equation*}
\begin{aligned}
&c_{11}=k^2r_2+t_2\ ,\\ 
&c_{12}=k^2r_6-t_4\ ,\\ 
&c_{13}=-i\sqrt{2k^2}  t_4\ ,\\ 
&c_{14}=0\ , \\
&c_{21}=-k^2r_6+t_4\ , \\
&c_{22}=2k^2(r_1-r_3+2r_4)+t_3\ , \\ 
&c_{23}=i\sqrt{2k^2}t_3\ , \\
&c_{24}=0\ ,\\
&c_{31}=-i\sqrt{2k^2} t_4\ ,\\
&c_{32}=-i\sqrt{2k^2}t_3\ , \\
&c_{33}=2k^2(t_3-\lambda)\ , \\
&c_{34}=0\ ,\\
&c_{41}=0\ , \\
&c_{42}=0\ , \\
&c_{43}=0\ , \\
&c_{44}=0\ .
\end{aligned}
\end{equation*}

Several comments concerning the above coefficient matrix are in order. First of all, the matrix is not Hermitian, something that can create confusion at first sight. This fact is simply a consequence of the normalization of the corresponding parity-mixing projection operators. As discussed in detail in Appendix~\ref{app:derivproj}, operators which connect the same states but contain the totally antisymmetric tensor, are required to have opposite signs. This is because we want them to obey the simple orthogonality relations given in eq.~\eqref{orth}, so that the inversion of the wave operator becomes straightforward. Obviously, the action is still Hermitian. 

In addition to that, the matrix~\eqref{s0dm} is clearly degenerate and of rank 3. This is expected due to the gauge invariances of the theory. To proceed with the attainment of the propagator we delete the last row and column of~\eqref{s0dm}. Denoting with $b_{ij}^{\phi\chi}(0)$ the resulting matrix, we perform the inversion to find
\begin{equation}
\label{invs0dm}
\left(b_{ij}^{\phi\chi}(0)\right)^{-1}=\frac{k^2}{\det\left(b_{ij}^{\phi\chi}(0)\right)} \left(
\begin{array}{ccc}
B_{11}&B_{12}&B_{13}\\
B_{21}&B_{22}&B_{23}\\
B_{31}&B_{32}&B_{33}
\end{array} \right)\ ,
\end{equation} 
\begin{equation*}
\begin{aligned}
&\displaystyle B_{11}=4k^2(r_1-r_3+2r_4)(t_3-\lambda)-2 t_3\lambda\ ,\\
&\displaystyle B_{12}=-2k^2 r_6(t_3-\lambda)-2\lambda t_4\ , \\
&\displaystyle B_{13}=i\sqrt{2k^2}\left(r_6t_3+2(r_1-r_3-2r_4)t_4 \right)\ ,\\
&\displaystyle B_{21}=2k^2 r_6(t_3-\lambda)+2\lambda t_4\ , \\
&\displaystyle B_{22}=2k^2 r_2 (t_3-\lambda)+2\left(t_2(t_3-\lambda)+t_4^2\right)\ ,  \\
&\displaystyle B_{23}=-i\sqrt{\frac{2}{k^2}}\left(k^2(r_2 t_3-r_6t_4)+t_2t_3+t_4^2\right)\ ,\\
&\displaystyle B_{31}=i\sqrt{2k^2}\left(r_6t_3+2(r_1-r_3-2r_4)t_4 \right)\ , \\
&\displaystyle B_{32}=i\sqrt{\frac{2}{k^2}}\left(k^2(r_2 t_3-r_6t_4)+t_2t_3+t_4^2\right)\ ,\\
&\displaystyle B_{33}=k^2\left(2r_2(r_1-r_3+2r_4)+r_6^2\right)+2(r_1-r_3+2r_4)t_2\\
&\hspace{1cm}\displaystyle+r_2t_3-2r_6t_4+\frac{1}{k^2}\left(t_2t_3-t_4^2\right)\ .
\end{aligned}
\end{equation*}

The determinant of the matrix can be written conveniently as
\begin{equation}
\begin{aligned}
\det\left(b_{ij}^{\phi\chi}(0)\right)&=2\left(2r_2(r_1-r_3+2r_4)+r_6^2\right)\times\\
&\hspace{1cm}\times(t_3-\lambda)k^2(k^2-m_+(0)^2)(k^2-m_-(0)^2) \ ,
\end{aligned}
\end{equation}
where the masses of the spin-0 states $m_{\pm}(0)^2$, are given by
\begin{equation}
\begin{aligned}
m_{\pm}(0)^2&=\frac{1}{2\left(2r_2(r_1-r_3+2r_4)+r_6^2\right)(t_3-\lambda)}\times\\
&\times\Bigg\{\left(2(r_1-r_3+2r_4)t_2+r_2 t_3-2r_6t_4\right)\lambda\\
&\pm \Bigg[4\left(2r_2(r_1-r_3+2r_4)+r_6^2\right)(t_2 t_3+t_4^2)(t_3-\lambda)\lambda \\
&+\left[\left(2(r_1-r_3+2r_4)t_2+r_2 t_3-2r_6t_4\right)\lambda\right.\\
&\left.-2(r_1-r_3+2r_4)\left(t_2 t_3+t_4^2\right)\right]^2 \Bigg]^{\frac{1}{2}} \Bigg\} \ .
\end{aligned}
\end{equation}
The notation we chose for the zeros of the determinant leaves no room for confusion; they correspond to the poles of the propagator, i.e. the physical masses of the spin-0 particle states of the theory. Therefore, they have to obey
\begin{equation}
m_+(0)^2 > 0 \ \ \ \text{and} \ \ \ m_-(0)^2 > 0 \ .
\end{equation}

In order to simplify as much as possible the calculations for the residue of the massless graviton, we found it helpful to isolate the $k^2=0$ pole in the spin-0 (and spin-2) sector of the theory. To do so, we rewrite the inverse of the coefficient matrix given above as

\small
\begin{equation}
\begin{aligned}
\left(b_{ij}^{\phi\chi}(0)\right)^{-1}=&-\frac{1}{2\lambda k^2}\left(\begin{array}{ccc}
0&0&0\\
0&2k^2&-i \sqrt{2k^2}\\
0&i \sqrt{2k^2}&1
\end{array}\right)   \\
\vphantom{\frac{a}{b}} \\
&\hspace{-2cm}+\frac{1}{t_2\, t_3+t_4^2}\left(\begin{array}{ccc}
t_3&t_4&0\\
-t_4&t_2&0\\
0&0&-(2\lambda)^{-1}\Big(2(r_1-r_3+2r_4)t_2+r_2 t_3-2r_6t_4\Big)\end{array}\right) \\
\vphantom{\frac{a}{b}} \\
&\hspace{-2cm}+\frac{k^2}{2\big(2r_2(r_1-r_3+2r_4)+ r_6^2\big)(t_3-\lambda)\big(m_+(0)^2-m_-(0)^2\big)}\times \\
\vphantom{\frac{a}{b}} \\
&\hspace{-2cm}\times \left(\frac{1}{m_+(0)^2(k^2-m_+(0)^2)}-\frac{1}{m_-(0)^2(k^2-m_-(0)^2)}\right)\times\\
\vphantom{\frac{a}{b}} \\
&\hspace{-2cm}\times\left(\begin{array}{ccc}
B_{11}&B_{12}&B_{13}\\
B_{21}&B_{22}&B_{23}\\
B_{31}&B_{32}&B_{33}
\end{array} \right) \ ,
\end{aligned}
\end{equation}
\normalsize
where the matrix elements $B_{ij}$ were given in~\eqref{invs0dm}. 

\newpage

\section{Spin-1}

The 49 operators corresponding to the vector part of the theory are

\begin{align*}
\label{s1op}
&P^{\omega\omega}_{11}(1)_{CABKIJ}=\Theta_{CB}\Theta_{AI}\Theta_{JK}\ , \\
&P^{\omega\omega}_{12}(1)_{CABKIJ}=\sqrt{2}~\Theta_{CB}\Theta_{AI}\Omega_{JK} \ ,\\
&P^{\omega s}_{13}(1)_{CABIJ}=\sqrt{2}~\tilde k_J \Theta_{CB}\Theta_{AI} \ ,\\
&P^{\omega a}_{14}(1)_{CABIJ}=\sqrt{2}~\tilde k_J \Theta_{CB}\Theta_{AI} \ ,\\
&P^{\omega\omega}_{15}(1)_{CABKIJ}=\epsilon_{AJKL}\Omega_I^L\Theta_{BC} \ ,\\
&P^{\omega\omega}_{16}(1)_{CABKIJ}=-\frac{1}{\sqrt{2}}~\epsilon_{AIJL}\Omega_K^L\Theta_{BC}\ ,\\
&P^{\omega a}_{17}(1)_{CABIJ}=\frac{1}{\sqrt{2}}\epsilon_{AIJL}\tilde k^L\Theta_{BC} \ ,\\
&P^{\omega\omega}_{21}(1)_{CABKIJ}=\sqrt{2}~\Omega_{CB}\Theta_{AI}\Theta_{JK}\ , \\
&P^{\omega\omega}_{22}(1)_{CABKIJ}=2~\Omega_{CB}\Theta_{AI}\Omega_{JK}  \ ,\\
&P^{\omega s}_{23}(1)_{CABIJ}= 2~\tilde k_B \Theta_{AI}\Omega_{CJ} \ , \\
&P^{\omega a}_{24}(1)_{CABIJ}=2~\tilde k_B \Theta_{AI}\Omega_{CJ} \ , \\
&P^{\omega\omega}_{25}(1)_{CABKIJ}=\sqrt{2}~\epsilon_{AJKL}\Omega_I^L\Omega_{BC} \ ,\\
&P^{\omega\omega}_{26}(1)_{CABKIJ}=-\epsilon_{AIJL}\Omega_K^L\Omega_{BC}\ , \\
&P^{\omega a}_{27}(1)_{CABIJ}=\epsilon_{AIJL}\tilde k^L\Omega_{BC}\ ,  \\
&P^{s\omega }_{31}(1)_{ABKIJ}=\sqrt{2}~\tilde k_B \Theta_{KJ}\Theta_{AI}\ , \\
&P^{s\omega }_{32}(1)_{ABKIJ}=2~\tilde k_J \Theta_{AI}\Omega_{KB} \ ,\\
&P^{ss}_{33}(1)_{ABIJ}=2~\Theta_{AI}\Omega_{BJ} \ , \\
&P^{sa}_{34}(1)_{ABIJ}=2~\Theta_{AI}\Omega_{BJ} \ , \\
&P^{s\omega }_{35}(1)_{ABKIJ}=\sqrt{2}\epsilon_{IKAD}\tilde k^D\Omega_{BJ}\ , \\
&P^{s\omega }_{36}(1)_{ABKIJ}=\epsilon_{IJAD}\tilde k^D\Omega_{BK}\ , \\
&P^{s a}_{37}(1)_{ABIJ}=\epsilon_{AIJL}\Omega_{B}^L\ , \\
&P^{a\omega }_{41}(1)_{ABKIJ}=\sqrt{2}~\tilde k_B \Theta_{KJ}\Theta_{AI}\ ,\\
&P^{a\omega }_{42}(1)_{ABKIJ}=2~\tilde k_J \Theta_{AI}\Omega_{KB}\ , \\
&P^{as}_{43}(1)_{ABIJ}=2~\Theta_{AI}\Omega_{BJ}\ ,  \\
&P^{aa}_{44}(1)_{ABIJ}=2~\Theta_{AI}\Omega_{BJ}\ ,  
\end{align*}

\newpage

\begin{align*}
&P^{a\omega }_{45}(1)_{ABKIJ}=\sqrt{2}\epsilon_{IKAD}\tilde k^D\Omega_{BJ}\ ,\\
&P^{a\omega }_{46}(1)_{ABKIJ}=\epsilon_{IJAD}\tilde k^D\Omega_{BK}\ , \\
&P^{a a}_{47}(1)_{ABIJ}=\epsilon_{AIJL}\Omega_{B}^L\ , \\
&P^{\omega\omega}_{51}(1)_{CABKIJ}=-\epsilon_{IBCD}\Omega_A^D\Theta_{JK} \ ,\\
&P^{\omega\omega}_{52}(1)_{CABKIJ}=-\sqrt{2}~\epsilon_{IBCD}\Omega_A^D\Omega_{JK}\ , \\
&P^{\omega s}_{53}(1)_{CABIJ}=-\sqrt{2}\epsilon_{ACIL}\tilde k^L\Omega_{BJ} \ , \\
&P^{\omega a}_{54}(1)_{CABIJ}=-\sqrt{2}\epsilon_{ACIL}\tilde k^L\Omega_{BJ}\ , \\
&P^{\omega\omega}_{55}(1)_{CABKIJ}=\Theta_{CK}\Theta_{AI}\Omega_{BJ}+\Theta_{AK}\Omega_{BI}\Theta_{CJ}\ ,\\
&P^{\omega\omega}_{56}(1)_{CABKIJ}=-\sqrt{2}~\Omega_{AK}\Theta_{BI}\Theta_{CJ}\ ,\\
&P^{\omega a}_{57}(1)_{CABKIJ}=\sqrt{2}~\tilde k_B \Theta_{AI}\Theta_{CJ} \ , \\
&P^{\omega\omega}_{61}(1)_{CABKIJ}=\frac{1}{\sqrt{2}}~\epsilon_{IABD}\Omega_C^D\Theta_{JK} \ ,\\
&P^{\omega\omega}_{62}(1)_{CABKIJ}=\epsilon_{IABD}\Omega_C^D\Theta_{JK}  \ ,\\
&P^{\omega s}_{63}(1)_{CABIJ}=-\epsilon_{ABIL}\tilde k^L\Omega_{CJ} \ , \\
&P^{\omega a}_{64}(1)_{CABIJ}=-\epsilon_{ABIL}\tilde k^L\Omega_{CJ} \ , \\
&P^{\omega\omega}_{65}(1)_{CABKIJ}=-\sqrt{2}~\Omega_{CI}\Theta_{AJ}\Theta_{BK}\ ,\\
&P^{\omega\omega}_{66}(1)_{CABKIJ}= \Omega_{CK}\Theta_{AI}\Theta_{BJ}\ ,\\
&P^{\omega a}_{67}(1)_{CABKIJ}= \tilde k_C \Theta_{AI}\Theta_{BJ}\ , \\
&P^{a\omega }_{71}(1)_{ABKIJ}=-\frac{1}{\sqrt{2}}\epsilon_{IABD}\tilde k^D\Theta_{JK}\ ,\\
&P^{a\omega }_{72}(1)_{ABKIJ}=-\epsilon_{IABD}\tilde k^D\Omega_{JK} \ , \\
&P^{a s }_{73}(1)_{ABIJ}=-\epsilon_{IABD}\Omega_{J}^D\ , \\
&P^{a a }_{74}(1)_{ABIJ}=-\epsilon_{IABD}\Omega_{J}^D\ ,\\
&P^{a\omega}_{75}(1)_{CABKIJ}= \sqrt{2}~\tilde k_J \Theta_{AI}\Theta_{BK}\ , \\
&P^{a\omega}_{76}(1)_{CABKIJ}=\tilde k_K \Theta_{AI}\Theta_{BJ} \ , \\
&P^{aa}_{77}(1)_{ABIJ}=\Theta_{AI}\Theta_{BJ} \ .
\end{align*}

The $7\times 7$ coefficient matrix corresponding to spin-1 sector is found to be
\begin{equation}
\label{s1dm}
c_{ij}^{\phi\chi}(1)=\begin{blockarray}{cccccccc}
\scp\omega^-&\scp\omega^-&\scp s^-&\scp a^-&\scp\omega^+&\scp\omega^+&\scp a^+ \\
\begin{block}{(ccccccc) c}
c_{11}&c_{12}&c_{13}&c_{14}&c_{15}&c_{16}&c_{17}&\scp\omega^- \\
c_{21}&c_{22}&c_{23}&c_{24}&c_{25}&c_{26}&c_{27}&\scp\omega^- \\
c_{31}&c_{32}&c_{33}&c_{34}&c_{35}&c_{36}&c_{37}&\scp s^- \\
c_{41}&c_{42}&c_{43}&c_{44}&c_{45}&c_{46}&c_{47}&\scp a^- \\
c_{51}&c_{52}&c_{53}&c_{54}&c_{55}&c_{56}&c_{57}&\scp \omega^+ \\
c_{61}&c_{62}&c_{63}&c_{64}&c_{65}&c_{66}&c_{67}&\scp \omega^+ \\
c_{71}&c_{72}&c_{73}&c_{74}&c_{75}&c_{76}&c_{77}&\scp a^+ \\
\end{block} 
\end{blockarray}  \ , 
\end{equation}

\begin{align*}
&c_{11}=k^2(r_1+r_4+r_5)+\frac{1}{6}(t_1+4t_3) \ ,\\
&c_{12}=\frac{-1}{3\sqrt{2}}(t_1-2t_3)\ ,\\
&c_{13}=\frac{i}{3}\sqrt{\frac{k^2}{2}}(t_1-2t_3)\ , \\
&c_{14}=\frac{i}{3}\sqrt{\frac{k^2}{2}}(t_1-2t_3) \ , \\
&c_{15}=-k^2r_7+\frac{1}{3}(2t_4-t_5)\ , \\
&c_{16}=-\frac{\sqrt{2}}{3}(t_4+t_5) \ , \\
&c_{17}=\frac{i}{3}\sqrt{2k^2}(t_4+t_5) \ ,\\
&c_{21}=\frac{-1}{3\sqrt{2}}(t_1-2t_3) \ , \\
&c_{22}=\frac{1}{3}(t_1+t_3)\ ,\\
&c_{23}=-\frac{i}{3}\sqrt{k^2}(t_1+t_3) \\
&c_{24}=-\frac{i}{3}\sqrt{k^2}(t_1+t_3)\ ,\\
&c_{25}=\frac{\sqrt{2}}{3}(t_4-2t_5)\ ,\\
&c_{26}=-\frac{1}{3}(t_4-2t_5) \ ,\\
&c_{27}=\frac{i}{3}\sqrt{k^2}(t_4-2t_5)\ ,\\
&c_{31}=-\frac{i}{3}\sqrt{\frac{k^2}{2}}(t_1-2t_3) \ ,\\
&c_{32}=\frac{i}{3}\sqrt{k^2}(t_1+t_3)\ , \\
&c_{33}=\frac{1}{3}k^2(t_1+t_3) \ , \\
&c_{34}=\frac{1}{3}k^2(t_1+t_3) \ ,\\
&c_{35}=\frac{i}{3}\sqrt{2k^2}(t_4+t_5)\ ,\\
&c_{36}=-\frac{i}{3}\sqrt{k^2}(t_4-2t_5) \ , \\
&c_{37}=-\frac{1}{3}k^2(t_4-2t_5) \ ,\\
&c_{41}=-\frac{i}{3}\sqrt{\frac{k^2}{2}}(t_1-2t_3)\ ,\\
&c_{42}=\frac{i}{3}\sqrt{k^2}(t_1+t_3)\ ,\\
&c_{43}= \frac{1}{3}k^2(t_1+t_3)\ ,\\
&c_{44}=\frac{1}{3}k^2(t_1+t_3)\ , \\
&c_{45}=\frac{i}{3}\sqrt{2k^2}(t_4+t_5)\ ,\\
&c_{46}= \frac{i}{3}\sqrt{k^2}(t_4-2t_5) \ ,\\
&c_{47}=-\frac{1}{3}k^2(t_4-2t_5) \ ,\\
&c_{51}=k^2r_7-\frac{1}{3}(2t_4-t_5)\ ,\\
&c_{52}=-\frac{\sqrt{2}}{3}(t_4-2t_5)\ ,\\
&c_{53}=\frac{i}{3}\sqrt{2k^2}(t_4+t_5) \ ,\\
&c_{54}=\frac{i}{3}\sqrt{2k^2}(t_4+t_5)\ ,\\
&c_{55}=k^2(2r_3+r_5)+\frac{1}{6}(t_1+4t_2)\ ,\\
&c_{56}=\frac{1}{3\sqrt{2}}(t_1-2t_2) \ ,\\
&c_{57}=-\frac{i}{3}\sqrt{\frac{k^2}{2}}(t_1-2t_2) \ ,\\
&c_{61}=\frac{\sqrt{2}}{3}(t_4+t_5)\ ,\\
&c_{62}=\frac{1}{3}(t_4-2t_5)\ ,\\
&c_{63}=-\frac{i}{3}\sqrt{k^2}(t_4-2t_5)\ , \\
&c_{64}=-\frac{i}{3}\sqrt{k^2}(t_4-2t_5)\ ,\\
&c_{65}=\frac{1}{3\sqrt{2}}(t_1-2t_2)\ ,\\
&c_{66}=\frac{1}{3}(t_1+t_2) \ ,\\
&c_{67}=-\frac{i}{3}\sqrt{k^2}(t_1+t_2) \ ,\\
&c_{71}=\frac{i}{3}\sqrt{2k^2}(t_4+t_5)\ , \\
&c_{72}=\frac{i}{3}\sqrt{k^2}(t_4-2t_5)\ ,\\
&c_{73}=\frac{1}{3}k^2(t_4-2t_5) \ , \\
&c_{74}=\frac{1}{3}k^2(t_4-2t_5)\ ,\\
&c_{75}=\frac{i}{3}\sqrt{\frac{k^2}{2}}(t_1-2t_2)\ ,\\
&c_{76}=\frac{i}{3}\sqrt{k^2}(t_1+t_2)  \ ,\\
&c_{77}=\frac{1}{3}k^2(t_1+t_2) \ .
\end{align*}

As was the case in the spin-0 sector, the above matrix is not Hermitian because of the normalization of the projectors that connect states with different parity. Also, due to the gauge invariances of the theory we expect this matrix to be singular. It turns out that the rank of the largest non-degenerate sub-matrix extracted from~\eqref{s1dm} is actually 4. We consider only the coefficients associated to connection excitations by dropping rows (and columns) 3, 6 and 7. We work with this particular sub-matrix purely for convenience. Clearly, this is not a unique choice. However, the propagator does not depend on what (regular) sub-matrix of rank 4 we study; its gauge invariance is guaranteed from the source constraints that we obtain.

To avoid confusion with the spin-0 sector, we denote the resulting matrix with $\widetilde b_{ij}^{\phi\chi}(1)$. It reads
\begin{equation}
\label{s1dm}
\widetilde b_{ij}^{\phi\chi}(1)= \left(
\begin{array}{cccc}
\widetilde b_{11}&\widetilde b_{12}&\widetilde b_{13}&\widetilde b_{14}\\
\widetilde b_{21}&\widetilde b_{22}&\widetilde b_{23}&\widetilde b_{24}\\
\widetilde b_{31}&\widetilde b_{32}&\widetilde b_{33}&\widetilde b_{34}\\
\widetilde b_{41}&\widetilde b_{42}&\widetilde b_{43}&\widetilde b_{44}\\
\end{array} \right)\ ,  
\end{equation}
\begin{align*}
&\widetilde b_{11}=k^2(r_1+r_4+r_5)+\frac{1}{6}(t_1+4t_3) \ ,\\
&\widetilde b_{12}=\frac{-1}{3\sqrt{2}}(t_1-2t_3)\ ,\\
&\widetilde b_{13}=-k^2r_7+\frac{1}{3}(2t_4-t_5) \\
&\widetilde b_{14}=-\frac{\sqrt{2}}{3}(t_4+t_5) \ ,\\
&\widetilde b_{21}=\frac{-1}{3\sqrt{2}}(t_1-2t_3)\ , \\
&\widetilde b_{22}=\frac{1}{3}(t_1+t_3)\ , \\
&\widetilde b_{23}=\frac{\sqrt{2}}{3}(t_4-2t_5) \\
&\widetilde b_{24}=\frac{1}{3}(t_4-2t_5) \ ,\\
&\widetilde b_{31}=k^2r_7-\frac{1}{3}(2t_4-t_5) \ ,\\
&\widetilde b_{32}=-\frac{\sqrt{2}}{3}(t_4-2t_5) \ , \\
&\widetilde b_{33}=k^2(2r_3+r_5)+\frac{1}{6}(t_1+4t_2)\ ,\\
&\widetilde b_{34}=\frac{1}{3\sqrt{2}}(t_1-2t_2) \ ,\\
&\widetilde b_{41}=\frac{\sqrt{2}}{3}(t_4+t_5) \ ,\\
&\widetilde b_{42}=\frac{1}{3}(t_4-2t_5)\ ,\\
&\widetilde b_{43}=\frac{1}{3\sqrt{2}}(t_1-2t_2) \ , \\
&\widetilde b_{44}=\frac{1}{3}(t_1+t_2) \ .
\end{align*}
The inverse of the above matrix can be written as
\begin{equation}
\label{invs1dm}
\left(\widetilde b_{ij}^{\phi\chi}(1)\right)^{-1}=\frac{1}{\det\left(\widetilde b_{ij}^{\phi\chi}(1)\right)} \text{adj}\left(\widetilde b_{ij}^{\phi\chi}(1)\right) \ ,
\end{equation}
where $ \text{adj}\left(\widetilde b_{ij}^{\phi\chi}(1)\right)$ is the adjoint of matrix~\eqref{s1dm}, whose elements are found to be
\begin{equation}
\text{adj}\left(\widetilde b_{ij}^{\phi\chi}(1)\right)=\left(\begin{array}{cccc}
\widetilde B_{11}&\widetilde B_{12}&\widetilde B_{13}&\widetilde B_{14}\\
\widetilde B_{21}&\widetilde B_{22}&\widetilde B_{23}&\widetilde B_{24}\\
\widetilde B_{31}&\widetilde B_{32}&\widetilde B_{33}&\widetilde B_{34}\\
\widetilde B_{41}&\widetilde B_{42}&\widetilde B_{43}&\widetilde B_{44}\\
\end{array} \right)\ ,  
\end{equation}
\small
\begin{equation*} 
\renewcommand\arraystretch{2}
\arraycolsep=.5cm
\begin{array}{l}
\begin{aligned}
\displaystyle
\widetilde B_{11}&=\frac{1}{18}\left\{\vphantom{\frac{a}{b}}2k^2(2r_3+r_5)\left((t_1+t_2)(t_1+t_3)+(t_4-2t_5)^2\right)\right.\\
&\left.+3\left(t_1^2 t_2+t_1(t_2 t_3+t_4^2)+4t_2t_5^2\right) \vphantom{\frac{a}{b}}\right\} \ ,\\
\widetilde B_{12}&=\frac{1}{18\sqrt{2}}\left\{\vphantom{\frac{a}{b}}2k^2\left[(2r_3+r_5)\left((t_1+t_2)(t_1-2t_3)-2(t_4-2t_5)(t_4+t_5)\right)\right.\right.\\
&\left.\left.+3r_7(t_1t_4+2t_2t_5)\right]+3\left(t_1^2t_2-2t_1(t_2t_3+t_4^2)+4t_2t_5^2\right) \vphantom{\frac{a}{b}}\right\} \ ,
\end{aligned}\\
\displaystyle
\widetilde B_{13}=\frac{1}{18}\left\{\vphantom{\frac{a}{b}}2k^2 r_7\left((t_1+t_2)(t_1+t_3)+(t_4-2t_5)^2\right)-3\left[t_1^2t_4-2\left(t_2\right)\right] \right\}  \ ,\\
\begin{aligned}
\widetilde B_{14}&=\frac{1}{18\sqrt{2}}\left\{ \vphantom{\frac{a}{b}}
2k^2\left[ 
3(2r_3+r_5) \left(t_1t_4+2t_3t_5 \right)-r_7\left((t_1-2t_2)(t_1+t_3)\right.\right.\right.\\
&\left.\left.\left.-2(t_4-2t_5)(t_4+t_5)\right)\right]
+3t_1^2t_4+12\left(t_2t_3+(t_4+t_5)t_4\right)t_5
\vphantom{\frac{a}{b}}\right\} \ ,
\end{aligned}\\
\begin{aligned}
\widetilde B_{21}&=
\frac{1}{18\sqrt{2}}\left\{\vphantom{\frac{a}{b}}2k^2\left[(2r_3+r_5)\left((t_1+t_2)(t_1-2t_3)-2(t_4-2t_5)(t_4+t_5)\right)\right.\right.\\
&\left.\left.+3r_7(t_1t_4+2t_2t_5)\right]+3\left(t_1^2t_2-2t_1(t_2t_3+t_4^2)+4t_2t_5^2\right)\vphantom{\frac{a}{b}} \right\} \ ,
\end{aligned}\\
\begin{aligned}
\widetilde B_{22}&=\frac{1}{36}\Big\{2k^4\left[(2r_3+r_5)(r_1+r_4+r_5)+r_7^2)(t_1+t_2)\right]\\
&+2k^2[r_5\left((t_1+10t_2)t_1+4(t_1+t_2)t_3+4(t_4+t_5)^2\right)\\
&+2r_3\left((t_1+t_2)(t_1+4t_3)+4(t_4+t_5)^2\right)\\ 
&+3\left(3(r_1+r_4)t_1t_2-4r_7(t_4t_1-t_5t_2)\right)]\\
&+3\left[ t_1^2t_2+4(t_2t_3+t_4^2)+4t_2t_5\right]\Big\} \ ,
\end{aligned}\\
\begin{aligned}
\widetilde B_{23}&=-\frac{1}{18\sqrt{2}}\left\{\vphantom{\frac{a}{b}}2k^2\left[3(r_1+r_4+r_5)(t_1t_4+2t_2t_5)-r_7\left((t_1+t_2)(t_1-2t_3)\right.\right. \right.\\
&\left.\left.\left.-2(t_4-2t_5)(t_4+t_5)\right)\right]+3\left(t_1^2t_4+4(t_2t_3+(t_4+t_5)t_4)\right)t_5\vphantom{\frac{a}{b}}\right\} \ ,
\end{aligned}\\
\end{array}
\end{equation*}
\begin{equation*} 
\renewcommand\arraystretch{2}
\arraycolsep=.5cm
\begin{array}{l}
\begin{aligned}
\hspace{-.9cm}\widetilde B_{24}&=-\frac{1}{36}\left\{-\vphantom{\frac{a}{b}}12k^4\left[\left((2r_3+r_5)(r_1+r_4+r_5)+r_7^2\right)(t_4-2t_5)\right]\right.\\
&\left.+2k^2\left[3\left[-(r_1+2r_3+r_4+2r_5)t_1t_4+4(r_1+r_3+r_5)t_2t_5+4(2r_3+r_5)t_3t_5\right]\right.\right.\\
&\left.+r_7\left(t_1^2-2t_1(t_2+t_3)+4(t_2t_3+t_4^2-7t_4t_5+t_5^2+3(t_4+4t_5))\right]\vphantom{\frac{a}{b}}\right\} \ ,
\end{aligned}\\
\hspace{-.5cm}\displaystyle
\widetilde B_{31}=-\frac{1}{18}\left\{\vphantom{\frac{a}{b}}2k^2 r_7\left((t_1+t_2)(t_1+t_3)+(t_4-2t_5)^2\right)-3\left[t_1^2t_4-2t_2\right] \right\} \ , \\
\begin{aligned}
\hspace{-.9cm}\widetilde B_{32}&=\frac{1}{18\sqrt{2}}\left\{\vphantom{\frac{a}{b}}2k^2\left[3(r_1+r_4+r_5)(t_1t_4+2t_2t_5)-r_7\left((t_1+t_2)(t_1-2t_3)\right.\right. \right.\\
&\left.\left.\left.-2(t_4-2t_5)(t_4+t_5)\right)\right]+3\left(t_1^2t_4+4(t_2t_3+(t_4+t_5)t_4)\right)t_5\vphantom{\frac{a}{b}}\right\} \ ,
\end{aligned}\\
\begin{aligned}
\hspace{-.9cm}\widetilde B_{33}&=\frac{1}{18}\left\{\vphantom{\frac{a}{b}}2k^2\left[(r_1+r_4+r_5)\left((t_1+t_2)(t_1+t_3)+(t_4-2t_5)^2\right)\right]\right.\\
&\left.+3\left(t_1^2t_3+t_1\left(t_2t_3+t_4^2\right)+4t_3t_5^2\right)\vphantom{\frac{a}{b}}\right\} \ ,
\end{aligned}\\
\begin{aligned}
\hspace{-.9cm}\widetilde B_{34}&=\frac{1}{18\sqrt{2}}\left\{\vphantom{\frac{a}{b}}2k^2\left[(2r_1+2r_4-r_5)\left((t_1-2t_2)(t_1+t_3)-2(t_4-2t_5)(t_4+t_5)\right)\right.\right.\\
&\left.\left. -6r_7\left(t_1t_4+2t_3t_5\right)\right]+3\left(t_1^2t_3-2t_1(t_2t_3+t_4^2)+4t_3t_5^2\right)\vphantom{\frac{a}{b}}\right\} \ ,
\end{aligned}\\
\begin{aligned}
\hspace{-.9cm}\widetilde B_{41}&=-\frac{1}{18\sqrt{2}}\left\{ \vphantom{\frac{a}{b}}
2k^2\left[ 
3(2r_3+r_5) \left(t_1t_4+2t_3t_5 \right)-r_7\left((t_1-2t_2)(t_1+t_3)\right.\right.\right.\\
&\left.\left.\left.-2(t_4-2t_5)(t_4+t_5)\right)\right]
+3t_1^2t_4+12\left(t_2t_3+(t_4+t_5)t_4\right)t_5
\vphantom{\frac{a}{b}}\right\} \ ,
\end{aligned}\\
\begin{aligned}
\hspace{-.9cm}\widetilde B_{42}&=\frac{1}{36}\left\{-\vphantom{\frac{a}{b}}12k^4\left[\left((2r_3+r_5)(r_1+r_4+r_5)+r_7^2\right)(t_4-2t_5)\right]\right.\\
&\left.+2k^2\left[3\left[-(r_1+2r_3+r_4+2r_5)t_1t_4+4(r_1+r_3+r_5)t_2t_5+4(2r_3+r_5)t_3t_5\right]\right.\right.\\
&\left.\vphantom{\frac{a}{b}}+r_7\left(t_1^2-2t_1(t_2+t_3)+4(t_2t_3+t_4^2-7t_4t_5+t_5^2+3(t_4+4t_5))\right]\right\} \ ,
\end{aligned}\\ 
\begin{aligned}
\hspace{-.9cm}\widetilde B_{43}&=\frac{1}{18\sqrt{2}}\left\{\vphantom{\frac{a}{b}}2k^2\left[(2r_1+2r_4-r_5)\left((t_1-2t_2)(t_1+t_3)-2(t_4-2t_5)(t_4+t_5)\right)\right.\right.\\
&\left.\left. -6r_7\left(t_1t_4+2t_3t_5\right)\right]+3\left(t_1^2t_3-2t_1(t_2t_3+t_4^2)+4t_3t_5^2\right)\vphantom{\frac{a}{b}}\right\} \ ,
\end{aligned}\\
\begin{aligned}
\hspace{-.9cm}\widetilde B_{44}&=\frac{1}{36}\left\{\vphantom{\frac{a}{b}}12k^4\left[((2r_3+r_5)(r_1+r_4+r_5))(t_1+t_3)\right]\right.\\
&\left.+2k^2\left[ \vphantom{\frac{a}{b}}2(r_1+r_4+r_5)\left((t_1+4t_2)(t_1+t_3)+4(t_4+t_5)^2\right)\right. \vphantom{\frac{a}{b}}\right.\\
&\left.\left.\vphantom{\frac{a}{b}}+9(4r_3+r_5)t_1t_3-24r_7\left(t_1t_4-t_3t_5\right)\right]+3\left(t_1^2t_3+4t_1(t_2t_3+t_4^2+4t_3t_5^2)\right)\right\} \ .
\end{aligned}
\end{array}
\end{equation*}
\normalsize

The determinant in eq.~\eqref{invs1dm} can be written as
\small
\begin{equation}
\begin{aligned}
&\det\left(\widetilde b_{ij}^{\phi\chi}(1)\right)=\frac{1}{9}\left((2r_3+r_5)(r_1+r_4+r_5)+r_7^2\right)\times\\
&\hspace{.5cm}\times\left((t_1+t_2)(t_1+t_3)+(t_4-2t_5)^2\right)(k^2-m_+(1)^2)(k^2-m_-(1)^2)  \ ,
\end{aligned}
\end{equation}
\normalsize
where $m_\pm(1)^2$ are given by the following
\small
\begin{align*}
&\hspace{-.9cm}m_{\pm}(1)^2=-\frac{3}{4\left((2r_3+r_5)(r_1+r_4+r_5)+r_7^2\right)\left((t_1+t_2)(t_1+t_3)+(t_4-2t_5)^2\right)}\times\\
&\hspace{-.9cm}\times\Bigg\{(r_1+r_4+r_5)\vphantom{\frac{a}{b}}\left(t_1^2t_2+t_1(t_2t_3+t_4^2))+4t_2t_5^2\right)+(2r_3+r_5)\left(t_1^2t_3+t_1(t_2t_3+t_4^2))\right. \\
&\vphantom{\frac{A}{B}}\left.+4t_3t_5^2\right)-2r_7\left(t_1^2t_4-2(t_2t_3+(t_4-2t_5)t_4)t_5\right)\pm \Bigg[-4 \left((2r_3+r_5)\times\right.\\
&\left.\vphantom{\frac{a}{b}}\times(r_1+r_4+r_5)+r_7^2\right)\left((t_1+t_2)(t_1+t_3)+(t_4-2t_5)^2\right)(t_2t_3+t_4^2)(t_1^2+4t_5^2) \\
&\vphantom{\frac{a}{b}}+\left[(r_1+r_4+r_5)\left(t_1^2t_2+t_1(t_2t_3+t_4^2))+4t_2t_5^2\right)+(2r_3+r_5)\times \right.\\
&\left.\vphantom{\frac{A}{B}}\times\left(t_1^2t_3+t_1(t_2t_3+t_4^2))+4t_3t_5^2\right)-2r_7\left(t_1^2t_4\right.\right.\\
&\left.\left.-2(t_2t_3+(t_4-2t_5)t_4)t_5\right) \right]^2\vphantom{\frac{a}{b}}\Bigg]^\frac{1}{2} \Bigg\} \ . \numberthis
\end{align*}
\normalsize

\newpage

\section{Spin-2}

The 9 operators corresponding to the tensor part of the theory are
\begin{equation}
\begin{aligned}
\label{s2op}
&P^{\omega\omega}_{11}(2)_{CABKIJ}=\frac{4}{3}\, \Theta_{K(C}\Theta_{A)I}\Theta_{BJ}-\Theta_{CB}\Theta_{AI}\Theta_{JK}\ , \\
&P^{\omega\omega}_{12}(2)_{CABKIJ}=\frac{2}{3}\left(\epsilon_{ABD(J}\Theta_{K)C}-\epsilon_{BCD(J}\Theta_{K)A}\right)\Omega^D_I \ , \\
&P^{\omega s}_{13}(2)_{CABIJ}=\frac{2\sqrt{2}}{3}\epsilon_{ADJ(B}\Theta_{C)I}\tilde k^D\ ,   \\
&P^{\omega\omega}_{21}(2)_{CABKIJ}=\frac{2}{3}\left(\epsilon_{IJL(B}\Theta_{C)K}-\epsilon_{JKL(B}\Theta_{C)I}\right)\Omega^L_A \ ,\\
&P^{\omega\omega}_{22}(2)_{CABKIJ}=2\,\Theta_{K(C}\Theta_{A)I}\Omega_{BJ}-\frac{2}{3}\, \Theta_{CB}\Omega_{AI}\Theta_{JK} \ , \   \\ 
&P^{\omega s}_{23}(2)_{CABIJ}=\sqrt{2}~\tilde k_B \left(\Theta_{CI}\Theta_{AJ}-\frac{1}{3}\Theta_{CA}\Theta_{IJ}\right) \ , \\
&P^{s\omega }_{31}(2)_{ABKIJ}=-\frac{2\sqrt{2}}{3}\epsilon_{ILB(J}\Theta_{K)A}\tilde k^L\ ,\\
&P^{s\omega }_{32}(2)_{ABKIJ}=\sqrt{2}~\tilde k_J \left(\Theta_{KA}\Theta_{IB}-\frac{1}{3}\Theta_{KI}\Theta_{AB}\right) \ , \\
&P^{ss}_{33}(2)_{ABIJ}=\Theta_{AI}\Theta_{BJ}-\frac{1}{3}\Theta_{AB}\Theta_{IJ} \ .
\end{aligned}
\end{equation}
The coefficient matrix for the spin-2 sector is found to be
\begin{equation}
\label{s2dm}
c_{ij}^{\phi\chi}(2)=\begin{blockarray}{cccc}
\scp\omega^-&\scp\omega^+&\scp s^+\\
\begin{block}{(ccc) c}
c_{11}&c_{12}&c_{13}&\scp\omega^- \\
c_{21}&c_{22}&c_{23}&\scp\omega^+ \\
c_{31}&c_{32}&c_{33}&\scp s^+ \\
\end{block} \numberthis
\end{blockarray}\ , 
\end{equation}
\vspace{-1.5cm}
\begin{align*}
&c_{11}= k^2 r_1+\frac{1}{2}t_1\ , \\
&c_{12}=k^2 r_8+t_5\ ,  \\
& c_{13}=i \sqrt{2k^2}t_5\ ,\\
&c_{21}=-k^2 r_8-t_5\ ,\\
&c_{22}=k^2(2r_1-2r_3+r_4)+\frac{1}{2}t_1\ , \\
&c_{23}=i \sqrt{\frac{k^2}{2}}t_1\ , \\
&c_{31}=i \sqrt{2k^2}t_5  \ , \\
&c_{32}=-i \sqrt{\frac{k^2}{2}}t_1\ , \\
&c_{33}=k^2(t_1-\lambda) \ .
\end{align*}

Since the above is not a singular matrix, we can immediately calculate its inverse 
\begin{equation}
\label{invs2dm}
\left(c_{ij}^{\phi\chi}(2)\right)^{-1}=\frac{k^2}{\det\left(c_{ij}^{\phi\chi}(2)\right)} \left(
\begin{array}{ccc}
C_{11}&C_{12}&C_{13}\\
C_{21}&C_{22}&C_{23}\\
C_{31}&C_{32}&C_{33}
\end{array} \right)\ ,
\end{equation}
\vspace{-.5cm}
\begin{align*} 
&C_{11}=k^2(2r_1-2r_3+r_4)(t_1+\lambda)-\frac{1}{2}t_1\lambda\ , \\
&C_{12}=-k^2 r_8(t_1+\lambda)-\lambda t_5 \ ,\\
&C_{13}=i\sqrt{\frac{k^2}{2}}\left(r_8t_1-2(2r_1-2r_3+r_4)t_5\right)\ , \\
&C_{21}=k^2 r_8(t_1+\lambda)+\lambda t_5 \ ,\\
&C_{22}=\frac{1}{2}\left((k^2 r_1+t_1 )(t_1+\lambda)+2t_5^2\right)\ , \\
&C_{23}=-\frac{i}{2}\sqrt{\frac{1}{2k^2}}\left(2k^2(r_1 t_1+2r_8t_5)+t_1^2+4t_5^2\right)\ , \\
&C_{31}=i\sqrt{\frac{k^2}{2}}\left(r_8t_1-2(2r_1-2r_3+r_4)t_5\right) \ , \\
&C_{32}=\frac{i}{2}\sqrt{\frac{1}{2k^2}}\left(2k^2(r_1 t_1+2r_8t_5)+t_1^2+4t_5^2\right) \ ,\\
&C_{33}=k^2\left(r_1(2r_1-2r_3+r_4)+r_8^2\right)+\frac{1}{2}(3r_1-2r_3+r_4)t_1\\
&\hspace{1cm}+2r_8t_5+\frac{1}{k^2}\left(\frac{1}{4}t_1^2+t_5^2\right)  \ .
\end{align*}
The determinant of the matrix $c_{ij}^{\phi\chi}(2)$ reads
\begin{equation}
\begin{aligned}
\det\left(c_{ij}^{\phi\chi}(2)\right)&=\left(r_1(2r_1-2r_3+r_4)+r_8^2\right)\times\\
&\hspace{1cm}\times(t_1+\lambda)k^2(k^2-m_+(2)^2)(k^2-m_-(2)^2) \ ,
\end{aligned}
\end{equation}
with $m_{\pm}(2)^2$ given by
\small
\begin{align*}
m_{\pm}(2)^2&=\frac{1}{4\left(r_1(2r_1-2r_3+r_4)+r_8^2\right)(t_1+\lambda)}\times\\
&\Bigg\{-(2r_1-2r_3+r_4)\left(t_1^2+4 t_5^2\right)\\
&-(3r_1-2r_3+r_4)t_1\lambda+4r_8 t_5\lambda\\
&\pm \Bigg[-4\left(r_1(2r_1-2r_3+r_4)+r_8^2\right)(t_1^2+t_5^2)(t_1+\lambda)\lambda \\
&+\left[(2r_1-2r_3+r_4)\left(t_1^2+t_5^2\right)\right.\\
&\left.+\left( (3r_1-2r_3+r_4)t_1+4r_8 t_5\right)\lambda\right]^2 \Bigg]^{\frac{1}{2}} \Bigg\}  \ , \numberthis
\end{align*}
\normalsize
where once again, we require the masses to be positive. 

Like in the scalar sector of the theory, it is very convenient to write the inverse coefficient matrix~\eqref{invs2dm} as
\small
\begin{equation}
\begin{aligned}
&\left(c_{ij}^{\phi\chi}(2)\right)^{-1}=-\frac{1}{\lambda k^2}\left(\begin{array}{ccc}
0&0&0\\
0&-2k^2&i \sqrt{2k^2}\\
0&-i \sqrt{2k^2}&-1
\end{array}\right)   \\
\vphantom{\frac{a}{b}} \\
&+\frac{2}{t_1^2+4t_5^2}
\left(\begin{array}{ccc}
t_1&-2t_5&0\\
2t_5&t_1&0\\
0&0&\lambda^{-1}\left((3r_1-2r_3+r_4)t_1+8r_8 t_5\right)
\end{array}\right) \\
\vphantom{\frac{a}{b}} \\
&+\frac{1}{4\big((t_1+\lambda)(r_1(2r_1-2r_3+r_4)+ r_8^2)\big)\big(k^2-m_{2+}^{\ \ 2}\big)\big(k^2-m_{2-}^{\ \ 2}\big)}\times\\
\vphantom{\frac{a}{b}} \\
&\times\left(\begin{array}{ccc}
C_{11}&C_{12}&C_{13}\\
C_{21}&C_{22}&C_{23}\\
C_{31}&C_{32}&C_{33}
\end{array} \right)  \ .
\end{aligned}
\end{equation}
\normalsize
The matrix elements $C_{ij}$ can be found above in eq.~\eqref{invs2dm}.

\newpage

\chapter{Spin-projection operators II}
\label{app:derivproj}

In an attempt to make this thesis as self-contained as possible, we would like to give some details on the way the projectors used to decompose the theory into spin sectors are obtained. The operators are classified into two categories. The first contains the ``diagonal'' projectors $P^{\phi\phi}_{ii}(J)$, which correspond to the decomposition of the fields into irreducible representations of the three-dimensional rotations group. Their derivation amounts to addition of angular momenta, since with respect to $SO(3)$
\begin{equation*} 
\begin{aligned}
&\omega_{CAB}\rightarrow \  2^-\oplus 2^+\oplus 1^-\oplus 1^-\oplus 1^+\oplus 1^+\oplus 0^-\oplus 0^+ \ , \\
&h_{AB}\rightarrow \ 2^+\oplus 1^-\oplus 1^-\oplus 1^+\oplus 0^+\oplus 0^+ \ .  
\end{aligned}
\end{equation*}
In terms of $\Theta$ and $\Omega$, this decomposition of the fields can be written in covariant form as
\small
\begin{align*}
\hspace{-1cm}\omega_{CAB}&=\left[\underbrace{\frac{4}{3}\, \Theta_{K(C}\Theta_{A)I}\Theta_{BJ}-\Theta_{CB}\Theta_{AI}\Theta_{JK}}_{P^{\omega\omega}_{11}(2)}+\underbrace{2\,\Theta_{K(C}\Theta_{A)I}\Omega_{BJ}-\frac{2}{3}\, \Theta_{CB}\Omega_{AI}\Theta_{JK}}_{P^{\omega\omega}_{22}(2)}\right.\\
&\left.\hspace{.5cm}+\underbrace{\vphantom{\frac{A}{B}}\Theta_{CB}\Theta_{AI}\Theta_{JK}}_{P^{\omega\omega}_{11}(1)}+\underbrace{\vphantom{\frac{a}{b}}2~\Omega_{CB}\Theta_{AI}\Omega_{JK}}_{P^{\omega\omega}_{22}(1)}+\underbrace{\vphantom{\frac{a}{b}}\Theta_{CK}\Theta_{AI}\Omega_{BJ}+\Theta_{AK}\Omega_{BI}\Theta_{CJ}}_{P^{\omega\omega}_{55}(1)}\right. \\
&\left.\hspace{.5cm}+\underbrace{\vphantom{\frac{a}{b}}\Omega_{CK}\Theta_{AI}\Theta_{BJ}}_{P^{\omega\omega}_{66}(1)}+\underbrace{\frac{1}{3} \Theta_{CK}\Theta_{AI}\Theta_{BJ}+\frac{2}{3} \Theta_{AK}\Theta_{BI}\Theta_{CJ}}_{P^{\omega\omega}_{11}(0)}\right.\\
&\left.\hspace{.5cm}+\underbrace{\frac{2}{3}\Theta_{BC}\Omega_{AI}\Theta_{JK}}_{P^{\omega\omega}_{22}(0)}\right]\omega^{KIJ}\numberthis \ ,
\end{align*}
\normalsize
and 
\small
\begin{align*}
h_{AB}=&\left[\underbrace{\Theta_{AI}\Theta_{BJ}-\frac{1}{3}\Theta_{AB}\Theta_{IJ}}_{P^{ss}_{33}(2)}+\underbrace{\vphantom{\frac{a}{b}}2~\Theta_{AI}\Omega_{BJ} }_{P^{ss}_{33}(1)}+\underbrace{\vphantom{\frac{a}{b}}2~\Theta_{AI}\Omega_{BJ}}_{P^{aa}_{44}(1)}\right.\\
&\hspace{3cm}\left.+\underbrace{\vphantom{\frac{a}{b}}\Theta_{AI}\Theta_{BJ}}_{P^{aa}_{77}(1)}+\underbrace{\frac{1}{3}\Theta_{AB}\Theta_{IJ}
}_{P^{ss}_{33}(0)}+\underbrace{\vphantom{\frac{a}{b}}\Omega_{AB}\Omega_{IJ} }_{P^{ss}_{44}(0)}\right]h^{IJ}\numberthis \ .
\end{align*}
\normalsize

The second category contains the ``off-diagonal'' operators $P^{\phi\chi}_{ij}(J), \ \text{with} \ i\neq j$; they implement mappings between the same spin subspaces of the fields. They connect states with the same spin and same parity, as well as states with the same spin but different parity if the totally antisymmetric tensor is present. 

Consider the following mixing term between the symmetric part of the vielbein and the connection that contributes only to the scalar part of the theory
 \begin{equation}
k^B\eta^{CA}\eta^{DE}\omega_{CAB}\, s_{DE} \ .
\end{equation}
We wish to find the off-diagonal projectors that link the $J^P=0^+$ component of connection (projected out by $P^{\omega\omega}_{22}(0)$) to one of the $J^P=0^+$ components of the vielbein, for example $P^{ss}_{33}(0)$. Plugging the expressions for the operators from eq. \eqref{s0op} into the above, we find after some algebra that the mixing operators are proportional to
\begin{equation}
P^{\omega s}_{23}(0)_{CABIJ}=c(k)~k^B \Theta^{CA}\Theta^{IJ}\ , P^{s \omega }_{32}(0)_{ABKIJ}=c(k)~k^J \Theta^{KI}\Theta^{AB} \ .
\end{equation}
Here $c(k)$ is a coefficient that depends on momentum and is determined from the orthogonality relations~\eqref{orths}. In particular, for these operators we have
\begin{align} 
&P^{\omega s}_{23}(0)_{CABDE}\, P^{s \omega}_{32}(0)^{DE}_{\ \ \ KIJ}=P^{\omega\omega}_{22}(0)_{CABKIJ} \ ,\\
&P^{s \omega}_{32}(0)_{ABDEF}\,P^{\omega s}_{23}(0)^{DEF}_{\ \ \ \ IJ}=P^{ss}_{33}(0)_{ABIJ} \ ,
\end{align}
so we immediately find
\begin{equation}
c(k)=\frac{1}{3}\sqrt{\frac{2}{k^2}} \ .
\end{equation}

The construction of operators that are capable of handling terms that contain the totally antisymmetric symbol follows pretty much the same reasoning as in the previous example. A term like $\epsilon^{ABCD}a_{AB}a_{CD}$, mixes the $J^P=1^- \ (P^{aa}_{44}(1))$ with the $J^P=1^+\ (P^{aa}_{77}(1))$ states of the vielbein excitation. A straightforward computation reveals that the corresponding projectors read
\begin{equation}
P^{aa}_{47}(1)_{ABIJ}= c\, \epsilon_{AIJL}\Omega_B^L \ \ \ \text{and} \ \ \ \ P^{aa}_{74}(1)_{ABIJ}=c'\, \epsilon_{IABD}\Omega_J^D \ ,
\end{equation}
where in this case it is necessary to introduce two normalization coefficients $c \ \text{and} \ c
'$, that do not depend on momentum. The orthogonality relations read
\begin{align}
&P^{aa}_{47}(1)_{ABCD}\,P^{aa}_{74}(1)^{CD}_{\ \ \ IJ}= P^{aa}_{44}(1)_{ABIJ} \ , \\
& P^{aa}_{74}(1)_{ABCD}\,P^{aa}_{47}(1)^{CD}_{\ \ \ IJ}= P^{aa}_{77}(1)_{ABIJ} \ , 
\end{align}
and in order for them to hold, we are required to set $c=-c'=1$. The fact that the projectors involving the totally antisymmetric tensor differ in sign is something that holds for all operators that connect states with opposite parities.   

Let us close with a technical remark. Terms that contain the totally antisymmetric tensor are responsible for the appearance of mixing between states with (same spin but) different parity. Obviously, they must not affect the mixing of states with same parity. It is indeed easy to show explicitly that their contribution vanishes by using the Schouten identity 
\begin{equation} 
\epsilon^{ABCD}k^E+\epsilon^{BCDE}k^A+\epsilon^{CDEA}k^B+\epsilon^{DEAB}k^C+\epsilon^{EABC}k^D=0 \ .
\end{equation}

\chapter{Einstein frame cut-offs}
\label{app:Einstein_cut}

Now, we will briefly discuss the computation of the effective cut-off  in the Einstein frame. As before, the cut-off is understood as the energy at which perturbative unitarity is violated and not necessarily as the onset of new physics. As shown in Eq.~\eqref{einst-theory}, the gravitational part of the action in the transformed frame takes the usual Einstein-Hilbert form, which allows us to directly identify the gravitational cut-off with the reduced Planck mass $M_P$.  The cut-off associated to the gauge sector can be also easily determined by looking at the scattering of gauge bosons with longitudinal polarization. Since the kinetic terms for the gauge fields are invariant under the conformal rescaling, the only modification comes through their coupling to the Higgs field $h$. The interaction under consideration can be schematically written as
\be\label{coup-app}
g^2 h^2 W_\mu^+ W^{-\mu}\rightarrow g^2 \frac{h^2}{\Omega^2}\tilde W_\mu^+ \tilde W^{-\mu} \ .
\ee
where we have rescaled the gauge boson fields in the Einstein frame with the corresponding conformal weight, $\tilde W^{\pm}=W^{\pm}/\Omega$.
Expanding \eqref{coup-app} around the background value of the Higgs field, $h\rightarrow \bar h +\delta h$, we find the following interaction term
\be\label{}
g\, \frac{m_W}{\bar\Omega^2} \tilde W_\mu^+ \tilde W^{-\mu}\delta h \ , 
\ee
where $m_W\sim g \bar h$ is the mass of the $W$ bosons in the Jordan frame and the conformal factor $\bar\Omega$ depends now on the background values of the Higgs and dilaton fields.
Taking into account the canonically normalized perturbations of the Higgs field  \eqref{redefinition-fields-einstein}, together with the unitarity of the S-matrix, we find that the cut-off scale associated to the gauge sector is given by 
\be\label{einst-gauge-app}
\tilde \Lambda_G\simeq \bar\Omega^{-1}\sqrt{
\frac{\xi_\chi\bar\chi^2(1+6\xi_\chi)+\xi_h\bar
h^2(1+6\xi_h)}{6\xi_h^2}} \ .
\ee
For the two limiting cases discussed in~\ref{sec:gauge-cut-off}, the previous expression becomes
\be\label{lim-gauge} 
\tilde \Lambda_G\simeq \Bigg\{ 
\begin{array}{cl} \frac{M_P}{\sqrt{\xi_h}}
&\mbox{for} \ \xi_\chi\bar\chi^2\ll \xi_h \bar h^2 \ ,\vspace{.1cm}
 \\
\frac{M_P}{\xi_h} &\mbox{for} \
\xi_\chi\bar\chi^2\gg \xi_h \bar h^2 \ .\\
\end{array} 
\ee 
where we have identified $\sqrt{\xi_\chi}\chi=M_P$. As expected, the gauge cut-off in the Einstein frame  is nothing else that the conformal rescaling of the Jordan frame cut-off, $\tilde\Lambda_G=\Lambda_G/\Omega$.

The computation of the scalar cut-off in the Einstein frame is more complicated than in the single field case~\cite{Bezrukov:2010jz}. Although all the non-linearities of the initial frame are moved to the matter sector of the theory,  the existence of non-minimal couplings to gravity give rise to a non-trivial kinetic mixing for the scalar fields in the Einstein frame (cf. Eq.~\eqref{kinetic}). This fact substantially complicates the treatment of the problem in terms of the original $(h,\chi)$ variables, especially in the high energy region. Therefore, in order to compute the scalar cut-off at large energies, we choose to recast the kinetic terms~\eqref{kinetic} in a diagonal form by means of the angular variables defined in~\eqref{var-r-theta}. Expanding the resulting inflationary potential\,\footnote{Equivalently we could consider higher order terms arising from the non-canonical kinetic term of the dilaton.} in  Eq.~\eqref{defin-pot} around the background value of the Higgs field $\bar\phi$ we obtain a  series of  terms of the form (cf. Eq.~\eqref {correct-pot})
\be\label{exp-pot-app}
c_{n,l} 
\cosh[2n a\bar\phi/M_P]
\left(\frac{a\delta \phi}{M_P}\right)^{2l}+d_{n,l} 
\sinh[2na\bar\phi/M_P]
\left(\frac{a\delta \phi}{M_P}\right)^{2l+1}\ . 
\ee
The scalar cut-off  during inflation and reheating can be directly read from the previous expression. Note however that a direct comparison of the previous result with those obtained in the Jordan frame is only possible in some limiting cases. The angular perturbation $\delta\phi$ depends on both of the original field perturbations and only coincides with the Higgs perturbation $\delta h$ in the very high energy regime. Indeed, at the beginning of inflation\,\footnote{The background value of the field $\phi$ is very close to zero. Remember that $\phi$ is defined as $\phi=\phi_0-\vert\phi'\vert$.} the angular dependence on the background field in Eq.~\eqref{exp-pot-app} becomes negligible. We are left therefore with a series of higher order operators suppressed by the reduced Planck mass  $M_P$, which coincides with the conformally transformed Jordan frame cut-off in the corresponding regime, $\tilde \Lambda\simeq \Lambda/\Omega \simeq \sqrt{\xi_h}h/\Omega$.

The determination of the scalar cut-off in the low-energy regime, $\xi_h h^2\ll  \xi_\chi\chi^2$, is also non-trivial, since the field redefinition~\eqref{var-r-theta} is no longer applicable. Fortunately, the kinetic mixing between the Higgs and dilaton fields can be neglected at low energies and Eq.~\eqref{kinetic} simplifies to
\be\label{kinetic-app-le}
\tilde K(\chi,h) \simeq (\partial \chi)^2+\left(1+\frac{\xi_h^2h^2}{M_P^2}\right)(\partial h)^2 \ ,\ee
where again we identified  $\sqrt{\xi_\chi}\chi=M_P$.
The kinetic term for the Higgs field can be recast into canonical form in terms of 
\be\label{chvar-app-le}
\hat h= h\left(1+\frac{\xi_h^2 h^2}{M_P^2}+\ldots\right)=h\left(1+\sum_{n=1}c_n\left(\frac{\xi_h^2h^2}{M_P^2}\right)^n\right) \ ,
\ee
where $c_n$ are numerical factors. Inverting the above relation and plugging it to the potential in this limit
\be\label{pot-le-app}
\tilde U(h)\simeq \frac{\lambda}{4}h^4 \ ,
\ee
we see that the cut-off is proportional to $M_P/\xi_h$, in agreement with the Jordan frame result.\footnote{Notice that in the low energy regime the conformal factor is approximately equal to one.}
\clearpage

\chapter{Feynman rules for the dilaton}\label{appendix2}
In this Appendix, we gather the Feynman rules as well as the expressions for the coefficients appearing in the one-loop diagrams in~\ref{scalareffpot}. We denote with a dashed (solid) line the dilaton (Higgs) and perform the calculations in dimensional regularization in $D=4-2\epsilon$ dimensions.  After expanding the fields around their background values and normalizing the kinetic term for the dilaton, we find the following Feynman rules stemming from its kinetic term   
\begin{figure}[!h]
\centering
\includegraphics[scale=.75]{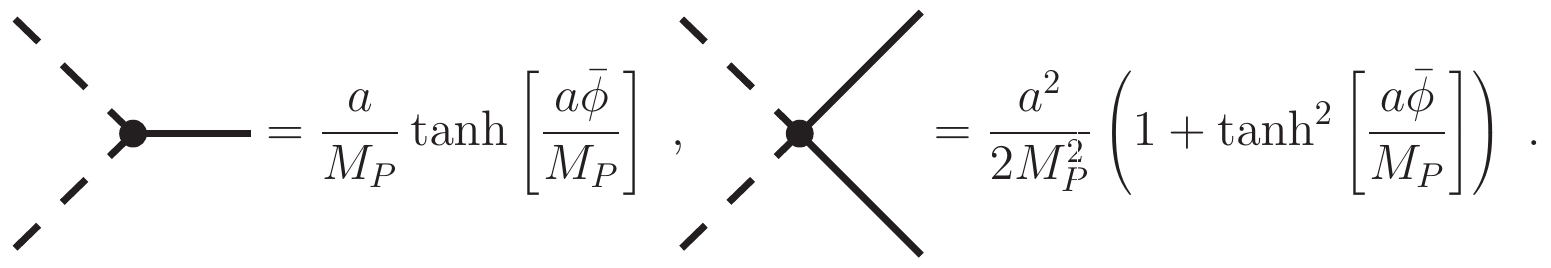}  
\end{figure}

Using the above expression, we can calculate the coefficients appearing in the different diagrams. Let us start by considering the simplest diagram $d_1$ . We obtain

\begin{figure}[!h]
\centering 
\includegraphics[scale=.75]{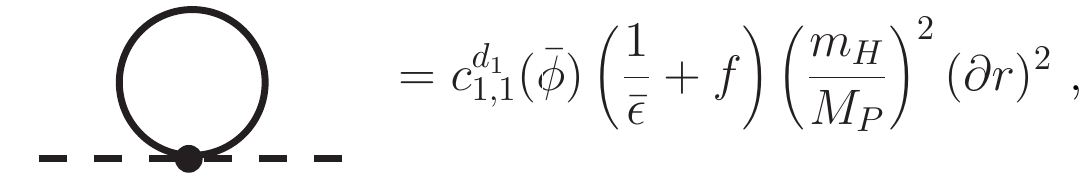}
\end{figure}

{\flushleft with} $1/\bar\epsilon=1/\epsilon-\gamma+\log4\pi$, and  
\be\label{appeq1} 
c^{d1}_{1,1}(\bar\phi)=\frac{a^2}{64\pi^2}\left(1+\tanh^2\left [\frac{a\bar\phi}{M_P}\right] \right) \ ,  \ \ \
f=-\log\left[\frac{m_H^2}{\mu^2}\right] \ .
\ee
Let us move to the more complicated diagram $d_2$. We find

\begin{figure}[!h]
\centering
\includegraphics[scale=.75]{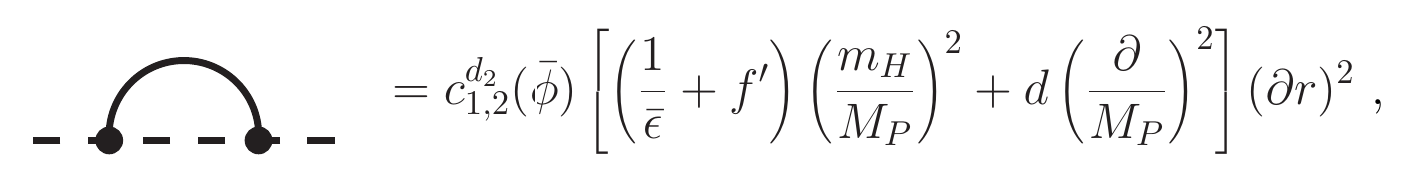}
\end{figure}

{\flushleft where}
\be\label{appeq2} 
c^{d2}_{1,2}(\bar\phi)=\frac{a^2}{16\pi^2}\tanh^2\left [\frac{a\bar\phi}{M_P}\right] \ ,  \ \ \ 
f'=\frac{1}{2}-\log\left[\frac{m_H^2}{\mu^2}\right]  \ \ \ \text{and} \ \ \ d=0 \ .
\ee
Note that in this particular diagram, the coefficient $d$ is coincidentally zero. As we argued in~\ref{scalareffpot},  this kind 
of terms are expected to appear by simple power-counting arguments in higher-loop diagrams.  We see that in both diagrams, 
for the maximal value of the hyperbolic tangent,  the corrections are suppressed by loop factors as well as powers of $M_P$.

\chapter{Dimensional Analysis}
\label{sec:dim-an}

When the metric $g_{\m\n}$ is dimensionful, the operation of lowering and raising indices has to be done with some care, since covariant and contravariant tensors carry different dimensions. For example, the inverse metric $g^{\m\n}$ has dimensions of $\left[\text{GeV}\right]^{2q}$. Moreover, for the metric determinant $g\equiv-\det(g_{\m\n})>0$, we obtain
\be
\label{metdet-dim}
\left[g\right]=\left[\text{GeV}\right]^{-8q} \ ,
\ee
whereas from~\eqref{assign-dims}, it follows that 
\be
\label{der-dim}
\left[\partial_\m\right]=\left[\text{GeV}\right]^{p} \ .
\ee
We are now in a position to determine the dimensionality of various geometrical quantities. First of all, for the Christoffel symbols which are defined as
\be
\label{chris-def}
\G^\lambda_{\m\n}=\frac{1}{2}g^{\k\lambda}\left(\partial_\n g_{\m\k}+\partial_\m g_{\k\n}-\partial_\k g_{\m\n}\right) \ ,
\ee
we obtain
\be
\label{chris-dim}
\left[\G^\lambda_{\m\n}\right]=\left[\text{GeV}\right]^p \ ,
\ee
in accordance with~\eqref{der-dim}.
Consequently, for the curvatures
\be
\label{riem-def}
R^\k_{\lambda\m\n}=\partial_\m\G^\k_{\lambda\n}-\partial_\n\G^\k_{\lambda\m}+\G^\rho_{\lambda\n}\G^\k_{\rho\m}-\G^\rho_{\lambda\m}\G^\k_{\rho\n} \ ,\ R_{\m\n}= R^\kappa_{\m\kappa\n} \ ,\ R= g^{\m\n}R_{\m\n} \ ,
\ee
we see that
\be
\label{riem-dim}
\left[R^\kappa_{\ \lambda\m\n}\right]=\left[\text{GeV}\right]^{2p} \ ,~~\left[R_{\m\n}\right]=\left[\text{GeV}\right]^{2p} \ ,~~\left[R\right]=\left[\text{GeV}\right]^{2\left(p+q\right)} =\left[\text{GeV}\right]^2\ .
\ee

\backmatter
\cleardoublepage
\phantomsection
\addcontentsline{toc}{chapter}{Bibliography}
\bibliographystyle{utphys}
\bibliography{bibliography}

\providecommand{\href}[2]{#2}\begingroup\raggedright\begin{thebibliography}{100}

\bibitem{DiFrancesco:1997nk}
P.~Di~Francesco, P.~Mathieu, and D.~Senechal,
  \href{http://dx.doi.org/10.1007/978-1-4612-2256-9}{{\em {Conformal Field
  Theory}}}.
\newblock Graduate Texts in Contemporary Physics. Springer-Verlag, New York,
1997.
\newblock

\bibitem{Rychkov:2016iqz}
S.~Rychkov, ``{EPFL Lectures on Conformal Field Theory in $D \ge 3$
  Dimensions},''
\href{http://arxiv.org/abs/1601.05000}{{\ttfamily arXiv:1601.05000 [hep-th]}}.

\bibitem{Komargodski:2011vj}
Z.~Komargodski and A.~Schwimmer, ``{On Renormalization Group Flows in Four
  Dimensions},'' \href{http://dx.doi.org/10.1007/JHEP12(2011)099}{{\em JHEP}
  {\bfseries 12} (2011) 099},
\href{http://arxiv.org/abs/1107.3987}{{\ttfamily arXiv:1107.3987 [hep-th]}}.

\bibitem{Komargodski:2011xv}
Z.~Komargodski, ``{The Constraints of Conformal Symmetry on RG Flows},''
  \href{http://dx.doi.org/10.1007/JHEP07(2012)069}{{\em JHEP} {\bfseries 07}
  (2012) 069},
\href{http://arxiv.org/abs/1112.4538}{{\ttfamily arXiv:1112.4538 [hep-th]}}.

\bibitem{Luty:2012ww}
M.~A. Luty, J.~Polchinski, and R.~Rattazzi, ``{The $a$-theorem and the
  Asymptotics of 4D Quantum Field Theory},''
  \href{http://dx.doi.org/10.1007/JHEP01(2013)152}{{\em JHEP} {\bfseries 01}
  (2013) 152},
\href{http://arxiv.org/abs/1204.5221}{{\ttfamily arXiv:1204.5221 [hep-th]}}.

\bibitem{Dymarsky:2013pqa}
A.~Dymarsky, Z.~Komargodski, A.~Schwimmer, and S.~Theisen, ``{On Scale and
  Conformal Invariance in Four Dimensions},''
\href{http://arxiv.org/abs/1309.2921}{{\ttfamily arXiv:1309.2921 [hep-th]}}.

\bibitem{ElShowk:2012ht}
S.~El-Showk, M.~F. Paulos, D.~Poland, S.~Rychkov, D.~Simmons-Duffin, and
  A.~Vichi, ``{Solving the 3D Ising Model with the Conformal Bootstrap},''
  \href{http://dx.doi.org/10.1103/PhysRevD.86.025022}{{\em Phys. Rev.}
  {\bfseries D86} (2012) 025022},
\href{http://arxiv.org/abs/1203.6064}{{\ttfamily arXiv:1203.6064 [hep-th]}}.

\bibitem{Englert:1976ep}
F.~Englert, C.~Truffin, and R.~Gastmans, ``{Conformal Invariance in Quantum
  Gravity},''
\href{http://dx.doi.org/10.1016/0550-3213(76)90406-5}{{\em Nucl. Phys.}
  {\bfseries B117} (1976) 407--432}.

\bibitem{Shaposhnikov:2008xi}
M.~Shaposhnikov and D.~Zenhausern, ``{Quantum scale invariance, cosmological
  constant and hierarchy problem},''
  \href{http://dx.doi.org/10.1016/j.physletb.2008.11.041}{{\em Phys. Lett.}
  {\bfseries B671} (2009) 162--166},
\href{http://arxiv.org/abs/0809.3406}{{\ttfamily arXiv:0809.3406 [hep-th]}}.

\bibitem{Armillis:2013wya}
R.~Armillis, A.~Monin, and M.~Shaposhnikov, ``{Spontaneously Broken Conformal
  Symmetry: Dealing with the Trace Anomaly},''
  \href{http://dx.doi.org/10.1007/JHEP10(2013)030}{{\em JHEP} {\bfseries 10}
  (2013) 030},
\href{http://arxiv.org/abs/1302.5619}{{\ttfamily arXiv:1302.5619 [hep-th]}}.

\bibitem{Gretsch:2013ooa}
F.~Gretsch and A.~Monin, ``{Perturbative conformal symmetry and dilaton},''
  \href{http://dx.doi.org/10.1103/PhysRevD.92.045036}{{\em Phys. Rev.}
  {\bfseries D92} no.~4, (2015) 045036},
\href{http://arxiv.org/abs/1308.3863}{{\ttfamily arXiv:1308.3863 [hep-th]}}.

\bibitem{Yang:1954ek}
C.-N. Yang and R.~L. Mills, ``{Conservation of Isotopic Spin and Isotopic Gauge
  Invariance},''
\href{http://dx.doi.org/10.1103/PhysRev.96.191}{{\em Phys. Rev.} {\bfseries 96}
  (1954) 191--195}.

\bibitem{Utiyama:1956sy}
R.~Utiyama, ``{Invariant theoretical interpretation of interaction},''
\href{http://dx.doi.org/10.1103/PhysRev.101.1597}{{\em Phys. Rev.} {\bfseries
  101} (1956) 1597--1607}.

\bibitem{Brodsky:1961}
A.~Brodsky, D.~Ivanenko, and G.~Sokolik {\em JETPH} {\bfseries 41} (1961) 1307.

\bibitem{Sciama:1962}
D.~W. Sciama, {\em Recent Developments in General Relativity}.
\newblock 1962.

\bibitem{Kibble:1961ba}
T.~W.~B. Kibble, ``{Lorentz invariance and the gravitational field},''
\href{http://dx.doi.org/10.1063/1.1703702}{{\em J. Math. Phys.} {\bfseries 2}
  (1961) 212--221}.

\bibitem{Delacretaz:2014oxa}
L.~V. Delacr\'etaz, S.~Endlich, A.~Monin, R.~Penco, and F.~Riva,
  ``{(Re-)Inventing the Relativistic Wheel: Gravity, Cosets, and Spinning
  Objects},'' \href{http://dx.doi.org/10.1007/JHEP11(2014)008}{{\em JHEP}
  {\bfseries 11} (2014) 008},
\href{http://arxiv.org/abs/1405.7384}{{\ttfamily arXiv:1405.7384 [hep-th]}}.

\bibitem{Baume:2014rla}
F.~Baume, B.~Keren-Zur, R.~Rattazzi, and L.~Vitale, ``{The local
  Callan-Symanzik equation: structure and applications},''
  \href{http://dx.doi.org/10.1007/JHEP08(2014)152}{{\em JHEP} {\bfseries 08}
  (2014) 152},
\href{http://arxiv.org/abs/1401.5983}{{\ttfamily arXiv:1401.5983 [hep-th]}}.

\bibitem{Iorio:1996ad}
A.~Iorio, L.~O'Raifeartaigh, I.~Sachs, and C.~Wiesendanger, ``{Weyl gauging and
  conformal invariance},''
  \href{http://dx.doi.org/10.1016/S0550-3213(97)00190-9}{{\em Nucl. Phys.}
  {\bfseries B495} (1997) 433--450},
\href{http://arxiv.org/abs/hep-th/9607110}{{\ttfamily arXiv:hep-th/9607110
  [hep-th]}}.

\bibitem{O'Raifeartaigh:1996hf}
L.~O'Raifeartaigh, I.~Sachs, and C.~Wiesendanger, ``{Weyl gauging and curved
  space approach to scale and conformal invariance},'' in {\em {Meeting on 70
  Years of Quantum Mechanics Calcutta, India, January 29-February 2, 1996}}.
\newblock 1996.
\newblock
\url{http://lss.fnal.gov/archive/other/dias-stp-96-06.pdf}.
\newblock

\bibitem{Obukhov:1982zn}
{\relax Yu}.~N. Obukhov, ``{Conformal Invariance and Space-time Torsion},''
\href{http://dx.doi.org/10.1016/0375-9601(82)90037-8}{{\em Phys. Lett.}
  {\bfseries A90} (1982) 13--16}.

\bibitem{Son:2005rv}
D.~T. Son and M.~Wingate, ``{General coordinate invariance and conformal
  invariance in nonrelativistic physics: Unitary Fermi gas},''
  \href{http://dx.doi.org/10.1016/j.aop.2005.11.001}{{\em Annals Phys.}
  {\bfseries 321} (2006) 197--224},
\href{http://arxiv.org/abs/cond-mat/0509786}{{\ttfamily arXiv:cond-mat/0509786
  [cond-mat]}}.

\bibitem{Son:2008ye}
D.~T. Son, ``{Toward an AdS/cold atoms correspondence: A Geometric realization
  of the Schrodinger symmetry},''
  \href{http://dx.doi.org/10.1103/PhysRevD.78.046003}{{\em Phys. Rev.}
  {\bfseries D78} (2008) 046003},
\href{http://arxiv.org/abs/0804.3972}{{\ttfamily arXiv:0804.3972 [hep-th]}}.

\bibitem{Ho:2011qn}
F.-H. Ho and J.~M. Nester, ``{Poincar\'e gauge theory with even and odd parity
  dynamic connection modes: isotropic Bianchi cosmological models},''
  \href{http://dx.doi.org/10.1088/1742-6596/330/1/012005}{{\em
  J.Phys.Conf.Ser.} {\bfseries 330} (2011) 012005},
\href{http://arxiv.org/abs/1105.5001}{{\ttfamily arXiv:1105.5001 [gr-qc]}}.

\bibitem{Ho:2011xf}
F.-H. Ho and J.~M. Nester, ``{Poincar\'e Gauge Theory With Coupled Even And Odd
  Parity Dynamic Spin-0 Modes: Dynamic Equations For Isotropic Bianchi
  Cosmologies},'' \href{http://dx.doi.org/10.1002/andp.201100101,
  10.1142/S0218271811020391}{{\em Annalen Phys.} {\bfseries 524} (2012)
  97--106},
\href{http://arxiv.org/abs/1106.0711}{{\ttfamily arXiv:1106.0711 [gr-qc]}}.

\bibitem{Ho:2015ulu}
F.-H. Ho, H.~Chen, J.~M. Nester, and H.-J. Yo, ``{General Poincaré Gauge
  Theory Cosmology},'' \href{http://dx.doi.org/10.6122/CJP.20151014}{{\em Chin.
  J. Phys.} {\bfseries 53} (2015) 110109},
\href{http://arxiv.org/abs/1512.01202}{{\ttfamily arXiv:1512.01202 [gr-qc]}}.

\bibitem{Karananas:2016hrm}
G.~K. Karananas and A.~Monin, ``{Gauging nonrelativistic field theories using
  the coset construction},''
  \href{http://dx.doi.org/10.1103/PhysRevD.93.064069}{{\em Phys. Rev.}
  {\bfseries D93} (2016) 064069},
\href{http://arxiv.org/abs/1601.03046}{{\ttfamily arXiv:1601.03046 [hep-th]}}.

\bibitem{Karananas:2015eha}
G.~K. Karananas and A.~Monin, ``{Weyl and Ricci gauging from the coset
  construction},'' \href{http://dx.doi.org/10.1103/PhysRevD.93.064013}{{\em
  Phys. Rev.} {\bfseries D93} no.~6, (2016) 064013},
\href{http://arxiv.org/abs/1510.07589}{{\ttfamily arXiv:1510.07589 [hep-th]}}.

\bibitem{Coleman:1969sm}
S.~R. Coleman, J.~Wess, and B.~Zumino, ``{Structure of phenomenological
  Lagrangians. 1.},''
\href{http://dx.doi.org/10.1103/PhysRev.177.2239}{{\em Phys. Rev.} {\bfseries
  177} (1969) 2239--2247}.

\bibitem{Callan:1969sn}
C.~G. Callan, Jr., S.~R. Coleman, J.~Wess, and B.~Zumino, ``{Structure of
  phenomenological Lagrangians. 2.},''
\href{http://dx.doi.org/10.1103/PhysRev.177.2247}{{\em Phys. Rev.} {\bfseries
  177} (1969) 2247--2250}.

\bibitem{Ivanov:1975zq}
E.~A. Ivanov and V.~I. Ogievetsky, ``{The Inverse Higgs Phenomenon in Nonlinear
  Realizations},''
\href{http://dx.doi.org/10.1007/BF01028947}{{\em Teor. Mat. Fiz.} {\bfseries
  25} (1975) 164--177}.

\bibitem{Low:2001bw}
I.~Low and A.~V. Manohar, ``{Spontaneously broken space-time symmetries and
  Goldstone's theorem},''
  \href{http://dx.doi.org/10.1103/PhysRevLett.88.101602}{{\em Phys. Rev. Lett.}
  {\bfseries 88} (2002) 101602},
\href{http://arxiv.org/abs/hep-th/0110285}{{\ttfamily arXiv:hep-th/0110285
  [hep-th]}}.

\bibitem{Endlich:2013vfa}
S.~Endlich, A.~Nicolis, and R.~Penco, ``{Ultraviolet completion without
  symmetry restoration},''
  \href{http://dx.doi.org/10.1103/PhysRevD.89.065006}{{\em Phys. Rev.}
  {\bfseries D89} no.~6, (2014) 065006},
\href{http://arxiv.org/abs/1311.6491}{{\ttfamily arXiv:1311.6491 [hep-th]}}.

\bibitem{Brauner:2014aha}
T.~Brauner and H.~Watanabe, ``{Spontaneous breaking of spacetime symmetries and
  the inverse Higgs effect},''
  \href{http://dx.doi.org/10.1103/PhysRevD.89.085004}{{\em Phys. Rev.}
  {\bfseries D89} no.~8, (2014) 085004},
\href{http://arxiv.org/abs/1401.5596}{{\ttfamily arXiv:1401.5596 [hep-ph]}}.

\bibitem{Wigner:1939cj}
E.~P. Wigner, ``{On Unitary Representations of the Inhomogeneous Lorentz
  Group},'' \href{http://dx.doi.org/10.2307/1968551}{{\em Annals Math.}
  {\bfseries 40} (1939) 149--204}.
[Reprint: Nucl. Phys. Proc. Suppl.6,9(1989)].

\bibitem{Weinberg:1995mt}
S.~Weinberg, {\em {The Quantum theory of fields. Vol. 1: Foundations}}.
\newblock Cambridge University Press,
2005.
\newblock

\bibitem{Ivanov:1981wn}
E.~A. Ivanov and J.~Niederle, ``{Gauge Formulation of Gravitation Theories. 1.
  The Poincare, De Sitter and Conformal Cases},''
\href{http://dx.doi.org/10.1103/PhysRevD.25.976}{{\em Phys. Rev.} {\bfseries
  D25} (1982) 976}.

\bibitem{Ivanov:1981wm}
E.~A. Ivanov and J.~Niederle, ``{Gauge Formulation of Gravitation Theories. 2.
  The Special Conformal Case},''
\href{http://dx.doi.org/10.1103/PhysRevD.25.988}{{\em Phys. Rev.} {\bfseries
  D25} (1982) 988}.

\bibitem{Buchmuller:1988wx}
W.~Buchmuller and N.~Dragon, ``{Einstein Gravity From Restricted Coordinate
  Invariance},''
\href{http://dx.doi.org/10.1016/0370-2693(88)90577-1}{{\em Phys. Lett.}
  {\bfseries B207} (1988) 292}.

\bibitem{Alvarez:2006uu}
E.~Alvarez, D.~Blas, J.~Garriga, and E.~Verdaguer, ``{Transverse Fierz-Pauli
  symmetry},'' \href{http://dx.doi.org/10.1016/j.nuclphysb.2006.08.003}{{\em
  Nucl. Phys.} {\bfseries B756} (2006) 148--170},
\href{http://arxiv.org/abs/hep-th/0606019}{{\ttfamily arXiv:hep-th/0606019
  [hep-th]}}.

\bibitem{Blas:2011ac}
D.~Blas, M.~Shaposhnikov, and D.~Zenhausern, ``{Scale-invariant alternatives to
  general relativity},''
  \href{http://dx.doi.org/10.1103/PhysRevD.84.044001}{{\em Phys. Rev.}
  {\bfseries D84} (2011) 044001},
\href{http://arxiv.org/abs/1104.1392}{{\ttfamily arXiv:1104.1392 [hep-th]}}.

\bibitem{Paneitz:1983_2008}
S.~M. Paneitz, ``{A Quartic Conformally Covariant Differential Operator for
  Arbitrary Pseudo-Riemannian Manifolds (Summary)},''
  \href{http://dx.doi.org/10.3842/SIGMA.2008.036}{{\em SIGMA, Symmetry,
  Integrability and Geometry: Methods and Applications} {\bfseries 4} (2008)
  3}, \href{http://arxiv.org/abs/0803.4331}{{\ttfamily arXiv:0803.4331
  [math.DG]}}.

\bibitem{Fradkin:1981jc}
E.~S. Fradkin and A.~A. Tseytlin, ``{One Loop Beta Function in Conformal
  Supergravities},''
\href{http://dx.doi.org/10.1016/0550-3213(82)90481-3}{{\em Nucl. Phys.}
  {\bfseries B203} (1982) 157}.

\bibitem{Fradkin:1981iu}
E.~S. Fradkin and A.~A. Tseytlin, ``{Renormalizable asymptotically free quantum
  theory of gravity},''
\href{http://dx.doi.org/10.1016/0550-3213(82)90444-8}{{\em Nucl. Phys.}
  {\bfseries B201} (1982) 469--491}.

\bibitem{Fradkin:1982xc}
E.~S. Fradkin and A.~A. Tseytlin, ``{Asymptotic freedom in extended conformal
  supergravities},''
\href{http://dx.doi.org/10.1016/0370-2693(82)91018-8}{{\em Phys. Lett.}
  {\bfseries B110} (1982) 117--122}.

\bibitem{Riegert:1984kt}
R.~J. Riegert, ``{A Nonlocal Action for the Trace Anomaly},''
\href{http://dx.doi.org/10.1016/0370-2693(84)90983-3}{{\em Phys. Lett.}
  {\bfseries B134} (1984) 56--60}.

\bibitem{Karananas:2015ioa}
G.~K. Karananas and A.~Monin, ``{Weyl vs. Conformal},''
  \href{http://dx.doi.org/10.1016/j.physletb.2016.04.001}{{\em Phys. Lett.}
  {\bfseries B757} (2016) 257--260},
\href{http://arxiv.org/abs/1510.08042}{{\ttfamily arXiv:1510.08042 [hep-th]}}.

\bibitem{Wess:conf}
J.~Wess, ``{Conformal Invariance in Quantum Field Theory},'' {\em Nuovo Cim}
  {\bfseries 18} (1960) 1086.

\bibitem{Polchinski:1987dy}
J.~Polchinski, ``{Scale and Conformal Invariance in Quantum Field Theory},''
\href{http://dx.doi.org/10.1016/0550-3213(88)90179-4}{{\em Nucl. Phys.}
  {\bfseries B303} (1988) 226}.

\bibitem{Nicolis:2008in}
A.~Nicolis, R.~Rattazzi, and E.~Trincherini, ``{The Galileon as a local
  modification of gravity},''
  \href{http://dx.doi.org/10.1103/PhysRevD.79.064036}{{\em Phys. Rev.}
  {\bfseries D79} (2009) 064036},
\href{http://arxiv.org/abs/0811.2197}{{\ttfamily arXiv:0811.2197 [hep-th]}}.

\bibitem{Goon:2012dy}
G.~Goon, K.~Hinterbichler, A.~Joyce, and M.~Trodden, ``{Galileons as
  Wess-Zumino Terms},'' \href{http://dx.doi.org/10.1007/JHEP06(2012)004}{{\em
  JHEP} {\bfseries 06} (2012) 004},
\href{http://arxiv.org/abs/1203.3191}{{\ttfamily arXiv:1203.3191 [hep-th]}}.

\bibitem{GJMS}
C.~R. Graham, R.~Jenne, L.~J. Mason, and G.~A.~J. Sparling, ``{Conformally
  Invariant Powers of the Laplacian I: Existence},''
  \href{http://dx.doi.org/10.1112/jlms/s2-46.3.557}{{\em J. London Math. Soc.}
  {\bfseries s2-46 (3)} (1992) 557--565}.

\bibitem{GJMS_2}
C.~R. Graham, ``{Conformally Invariant Powers of the Laplacian II:
  Nonexistence},'' \href{http://dx.doi.org/10.1112/jlms/s2-46.3.566}{{\em J.
  London Math. Soc.} {\bfseries s2-46 (3)} (1992) 566--576}.

\bibitem{GJMS_3}
A.~R. Gover and K.~Hirachi, ``{Conformally Invariant Powers of the Laplacian --
  A Complete Nonexistence Theorem},'' {\em Journa of the American Mathematical
  Society} {\bfseries 17} no.~2, (2004) 389--405.

\bibitem{Nakayama:2013is}
Y.~Nakayama, ``{Scale invariance vs conformal invariance},''
  \href{http://dx.doi.org/10.1016/j.physrep.2014.12.003}{{\em Phys. Rept.}
  {\bfseries 569} (2015) 1--93},
\href{http://arxiv.org/abs/1302.0884}{{\ttfamily arXiv:1302.0884 [hep-th]}}.

\bibitem{Branson:1985}
T.~Branson, ``{Differential Operators Canonically Associated to a Conformal
  Structure},'' {\em Math.Scand.} {\bfseries 57} (1985) 293.

\bibitem{Osborn:2015rna}
H.~Osborn and A.~Stergiou, ``{Structures on the Conformal Manifold in Six
  Dimensional Theories},''
  \href{http://dx.doi.org/10.1007/JHEP04(2015)157}{{\em JHEP} {\bfseries 04}
  (2015) 157},
\href{http://arxiv.org/abs/1501.01308}{{\ttfamily arXiv:1501.01308 [hep-th]}}.

\bibitem{Horava:2009uw}
P.~Horava, ``{Quantum Gravity at a Lifshitz Point},''
  \href{http://dx.doi.org/10.1103/PhysRevD.79.084008}{{\em Phys. Rev.}
  {\bfseries D79} (2009) 084008},
\href{http://arxiv.org/abs/0901.3775}{{\ttfamily arXiv:0901.3775 [hep-th]}}.

\bibitem{Balasubramanian:2008dm}
K.~Balasubramanian and J.~McGreevy, ``{Gravity duals for non-relativistic
  CFTs},'' \href{http://dx.doi.org/10.1103/PhysRevLett.101.061601}{{\em Phys.
  Rev. Lett.} {\bfseries 101} (2008) 061601},
\href{http://arxiv.org/abs/0804.4053}{{\ttfamily arXiv:0804.4053 [hep-th]}}.

\bibitem{Son:2013rqa}
D.~T. Son, ``{Newton-Cartan Geometry and the Quantum Hall Effect},''
\href{http://arxiv.org/abs/1306.0638}{{\ttfamily arXiv:1306.0638
  [cond-mat.mes-hall]}}.

\bibitem{Geracie:2014nka}
M.~Geracie, D.~T. Son, C.~Wu, and S.-F. Wu, ``{Spacetime Symmetries of the
  Quantum Hall Effect},''
  \href{http://dx.doi.org/10.1103/PhysRevD.91.045030}{{\em Phys. Rev.}
  {\bfseries D91} (2015) 045030},
\href{http://arxiv.org/abs/1407.1252}{{\ttfamily arXiv:1407.1252
  [cond-mat.mes-hall]}}.

\bibitem{Iorio:2014pwa}
A.~Iorio, ``{Curved Spacetimes and Curved Graphene: a status report of the
  Weyl-symmetry approach},''
  \href{http://dx.doi.org/10.1142/S021827181530013X}{{\em Int. J. Mod. Phys.}
  {\bfseries D24} no.~05, (2015) 1530013},
\href{http://arxiv.org/abs/1412.4554}{{\ttfamily arXiv:1412.4554 [hep-th]}}.

\bibitem{Andringa:2010it}
R.~Andringa, E.~Bergshoeff, S.~Panda, and M.~de~Roo, ``{Newtonian Gravity and
  the Bargmann Algebra},''
  \href{http://dx.doi.org/10.1088/0264-9381/28/10/105011}{{\em Class. Quant.
  Grav.} {\bfseries 28} (2011) 105011},
\href{http://arxiv.org/abs/1011.1145}{{\ttfamily arXiv:1011.1145 [hep-th]}}.

\bibitem{Christensen:2013lma}
M.~H. Christensen, J.~Hartong, N.~A. Obers, and B.~Rollier, ``{Torsional
  Newton-Cartan Geometry and Lifshitz Holography},''
  \href{http://dx.doi.org/10.1103/PhysRevD.89.061901}{{\em Phys. Rev.}
  {\bfseries D89} (2014) 061901},
\href{http://arxiv.org/abs/1311.4794}{{\ttfamily arXiv:1311.4794 [hep-th]}}.

\bibitem{Bergshoeff:2014uea}
E.~A. Bergshoeff, J.~Hartong, and J.~Rosseel, ``{Torsional Newton-Cartan
  geometry and the Schroedinger algebra},''
  \href{http://dx.doi.org/10.1088/0264-9381/32/13/135017}{{\em Class. Quant.
  Grav.} {\bfseries 32} no.~13, (2015) 135017},
\href{http://arxiv.org/abs/1409.5555}{{\ttfamily arXiv:1409.5555 [hep-th]}}.

\bibitem{Hartong:2015zia}
J.~Hartong and N.~A. Obers, ``{Horava-Lifshitz gravity from dynamical
  Newton-Cartan geometry},''
  \href{http://dx.doi.org/10.1007/JHEP07(2015)155}{{\em JHEP} {\bfseries 07}
  (2015) 155},
\href{http://arxiv.org/abs/1504.07461}{{\ttfamily arXiv:1504.07461 [hep-th]}}.

\bibitem{Hartong:2015wxa}
J.~Hartong, E.~Kiritsis, and N.~A. Obers, ``{Field Theory on Newton-Cartan
  Backgrounds and Symmetries of the Lifshitz Vacuum},''
  \href{http://dx.doi.org/10.1007/JHEP08(2015)006}{{\em JHEP} {\bfseries 08}
  (2015) 006},
\href{http://arxiv.org/abs/1502.00228}{{\ttfamily arXiv:1502.00228 [hep-th]}}.

\bibitem{Hartong:2014pma}
J.~Hartong, E.~Kiritsis, and N.~A. Obers, ``{Schrodinger Invariance from
  Lifshitz Isometries in Holography and Field Theory},''
  \href{http://dx.doi.org/10.1103/PhysRevD.92.066003}{{\em Phys. Rev.}
  {\bfseries D92} (2015) 066003},
\href{http://arxiv.org/abs/1409.1522}{{\ttfamily arXiv:1409.1522 [hep-th]}}.

\bibitem{Hartong:2014oma}
J.~Hartong, E.~Kiritsis, and N.~A. Obers, ``{Lifshitz space-times for
  Schrodinger holography},''
  \href{http://dx.doi.org/10.1016/j.physletb.2015.05.010}{{\em Phys. Lett.}
  {\bfseries B746} (2015) 318--324},
\href{http://arxiv.org/abs/1409.1519}{{\ttfamily arXiv:1409.1519 [hep-th]}}.

\bibitem{Brauner:2014jaa}
T.~Brauner, S.~Endlich, A.~Monin, and R.~Penco, ``{General coordinate
  invariance in quantum many-body systems},''
  \href{http://dx.doi.org/10.1103/PhysRevD.90.105016}{{\em Phys. Rev.}
  {\bfseries D90} no.~10, (2014) 105016},
\href{http://arxiv.org/abs/1407.7730}{{\ttfamily arXiv:1407.7730 [hep-th]}}.

\bibitem{Jensen:2014aia}
K.~Jensen, ``{On the coupling of Galilean-invariant field theories to curved
  spacetime},''
\href{http://arxiv.org/abs/1408.6855}{{\ttfamily arXiv:1408.6855 [hep-th]}}.

\bibitem{Banerjee:2014pya}
R.~Banerjee, A.~Mitra, and P.~Mukherjee, ``{A new formulation of
  non-relativistic diffeomorphism invariance},''
  \href{http://dx.doi.org/10.1016/j.physletb.2014.09.004}{{\em Phys. Lett.}
  {\bfseries B737} (2014) 369--373},
\href{http://arxiv.org/abs/1404.4491}{{\ttfamily arXiv:1404.4491 [gr-qc]}}.

\bibitem{Banerjee:2014nja}
R.~Banerjee, A.~Mitra, and P.~Mukherjee, ``{Localization of the Galilean
  symmetry and dynamical realization of Newton-Cartan geometry},''
  \href{http://dx.doi.org/10.1088/0264-9381/32/4/045010}{{\em Class. Quant.
  Grav.} {\bfseries 32} no.~4, (2015) 045010},
\href{http://arxiv.org/abs/1407.3617}{{\ttfamily arXiv:1407.3617 [hep-th]}}.

\bibitem{Geracie:2015xfa}
M.~Geracie, K.~Prabhu, and M.~M. Roberts, ``{Fields and fluids on curved
  non-relativistic spacetimes},''
  \href{http://dx.doi.org/10.1007/JHEP08(2015)042}{{\em JHEP} {\bfseries 08}
  (2015) 042},
\href{http://arxiv.org/abs/1503.02680}{{\ttfamily arXiv:1503.02680 [hep-th]}}.

\bibitem{Geracie:2015dea}
M.~Geracie, K.~Prabhu, and M.~M. Roberts, ``{Curved non-relativistic
  spacetimes, Newtonian gravitation and massive matter},''
  \href{http://dx.doi.org/10.1063/1.4932967}{{\em J. Math. Phys.} {\bfseries
  56} no.~10, (2015) 103505},
\href{http://arxiv.org/abs/1503.02682}{{\ttfamily arXiv:1503.02682 [hep-th]}}.

\bibitem{Inonu:1953sp}
E.~Inonu and E.~P. Wigner, ``{On the Contraction of groups and their
  represenations},''
\href{http://dx.doi.org/10.1073/pnas.39.6.510}{{\em Proc. Nat. Acad. Sci.}
  {\bfseries 39} (1953) 510--524}.

\bibitem{Andreev:2013qsa}
O.~Andreev, M.~Haack, and S.~Hofmann, ``{On Nonrelativistic Diffeomorphism
  Invariance},'' \href{http://dx.doi.org/10.1103/PhysRevD.89.064012}{{\em Phys.
  Rev.} {\bfseries D89} (2014) 064012},
\href{http://arxiv.org/abs/1309.7231}{{\ttfamily arXiv:1309.7231 [hep-th]}}.

\bibitem{Karananas:2014pxa}
G.~K. Karananas, ``{The particle spectrum of parity-violating Poincar\'e
  gravitational theory},''
  \href{http://dx.doi.org/10.1088/0264-9381/32/5/055012}{{\em Class. Quant.
  Grav.} {\bfseries 32} no.~5, (2015) 055012},
\href{http://arxiv.org/abs/1411.5613}{{\ttfamily arXiv:1411.5613 [gr-qc]}}.

\bibitem{Neville:1978bk}
D.~E. Neville, ``{A Gravity Lagrangian With Ghost Free Curvature**2 Terms},''
\href{http://dx.doi.org/10.1103/PhysRevD.18.3535}{{\em Phys.Rev.} {\bfseries
  D18} (1978) 3535}.

\bibitem{Sezgin:1979zf}
E.~Sezgin and P.~van Nieuwenhuizen, ``{New Ghost Free Gravity Lagrangians with
  Propagating Torsion},''
\href{http://dx.doi.org/10.1103/PhysRevD.21.3269}{{\em Phys. Rev.} {\bfseries
  D21} (1980) 3269}.

\bibitem{Hayashi:1979wj}
K.~Hayashi and T.~Shirafuji, ``{Gravity from Poincare Gauge Theory of the
  Fundamental Particles. 1. Linear and Quadratic Lagrangians},''
\href{http://dx.doi.org/10.1143/PTP.64.866}{{\em Prog.Theor.Phys.} {\bfseries
  64} (1980) 866}.

\bibitem{Hayashi:1980av}
K.~Hayashi and T.~Shirafuji, ``{Gravity From Poincare Gauge Theory of the
  Fundamental Particles. 2. Equations of Motion for Test Bodies and Various
  Limits},''
\href{http://dx.doi.org/10.1143/PTP.64.883}{{\em Prog.Theor.Phys.} {\bfseries
  64} (1980) 883}.

\bibitem{Hayashi:1980ir}
K.~Hayashi and T.~Shirafuji, ``{Gravity From Poincare Gauge Theory of the
  Fundamental Particles. 3. Weak Field Approximation},''
\href{http://dx.doi.org/10.1143/PTP.64.1435}{{\em Prog.Theor.Phys.} {\bfseries
  64} (1980) 1435}.

\bibitem{Hayashi:1980qp}
K.~Hayashi and T.~Shirafuji, ``{Gravity From Poincare Gauge Theory of the
  Fundamental Particles. 4. Mass and Energy of Particle Spectrum},''
\href{http://dx.doi.org/10.1143/PTP.64.2222}{{\em Prog.Theor.Phys.} {\bfseries
  64} (1980) 2222}.

\bibitem{Hayashi:1980bf}
K.~Hayashi and T.~Shirafuji, ``{Gravity From Poincare Gauge Theory of the
  Fundamental Particles. 5. The Extended Bach-lanczos Identity},''
\href{http://dx.doi.org/10.1143/PTP.65.525}{{\em Prog.Theor.Phys.} {\bfseries
  65} (1981) 525}.

\bibitem{Hayashi:1981fx}
K.~Hayashi and T.~Shirafuji, ``{Gravity From Poincare Gauge Theory of the
  Fundamental Particles. 6. Scattering Amplitudes},''
\href{http://dx.doi.org/10.1143/PTP.66.318}{{\em Prog.Theor.Phys.} {\bfseries
  66} (1981) 318}.

\bibitem{Sezgin:1981xs}
E.~Sezgin, ``{Class of Ghost Free Gravity Lagrangians With Massive or Massless
  Propagating Torsion},''
\href{http://dx.doi.org/10.1103/PhysRevD.24.1677}{{\em Phys. Rev.} {\bfseries
  D24} (1981) 1677--1680}.

\bibitem{Neville:1981be}
D.~Neville, ``{Spin-2 Propagating Torsion},''
\href{http://dx.doi.org/10.1103/PhysRevD.23.1244}{{\em Phys.Rev.} {\bfseries
  D23} (1981) 1244--1249}.

\bibitem{Nair:2008yh}
V.~P. Nair, S.~Randjbar-Daemi, and V.~Rubakov, ``{Massive Spin-2 fields of
  Geometric Origin in Curved Spacetimes},''
  \href{http://dx.doi.org/10.1103/PhysRevD.80.104031}{{\em Phys. Rev.}
  {\bfseries D80} (2009) 104031},
\href{http://arxiv.org/abs/0811.3781}{{\ttfamily arXiv:0811.3781 [hep-th]}}.

\bibitem{Nikiforova:2009qr}
V.~Nikiforova, S.~Randjbar-Daemi, and V.~Rubakov, ``{Infrared Modified Gravity
  with Dynamical Torsion},''
  \href{http://dx.doi.org/10.1103/PhysRevD.80.124050}{{\em Phys. Rev.}
  {\bfseries D80} (2009) 124050},
\href{http://arxiv.org/abs/0905.3732}{{\ttfamily arXiv:0905.3732 [hep-th]}}.

\bibitem{Hernaski:2009wp}
C.~A. Hernaski, A.~A. Vargas-Paredes, and J.~A. Helayel-Neto, ``{A Discussion
  on Massive Gravitons and Propagating Torsion in Arbitrary Dimensions},''
  \href{http://dx.doi.org/10.1103/PhysRevD.80.124012}{{\em Phys. Rev.}
  {\bfseries D80} (2009) 124012},
\href{http://arxiv.org/abs/0905.1068}{{\ttfamily arXiv:0905.1068 [hep-th]}}.

\bibitem{Hehl:1976kj}
F.~W. Hehl, P.~Von Der~Heyde, G.~D. Kerlick, and J.~M. Nester, ``{General
  Relativity with Spin and Torsion: Foundations and Prospects},''
\href{http://dx.doi.org/10.1103/RevModPhys.48.393}{{\em Rev. Mod. Phys.}
  {\bfseries 48} (1976) 393--416}.

\bibitem{Hehl:1994ue}
F.~W. Hehl, J.~D. McCrea, E.~W. Mielke, and Y.~Ne'eman, ``{Metric affine gauge
  theory of gravity: Field equations, Noether identities, world spinors, and
  breaking of dilation invariance},''
  \href{http://dx.doi.org/10.1016/0370-1573(94)00111-F}{{\em Phys. Rept.}
  {\bfseries 258} (1995) 1--171},
\href{http://arxiv.org/abs/gr-qc/9402012}{{\ttfamily arXiv:gr-qc/9402012
  [gr-qc]}}.

\bibitem{Gronwald:1995em}
F.~Gronwald and F.~W. Hehl, ``{On the gauge aspects of gravity},'' in {\em
  {Quantum gravity. Proceedings, International School of Cosmology and
  Gravitation, 14th Course, Erice, Italy, May 11-19, 1995}}.
\newblock 1995.
\newblock \href{http://arxiv.org/abs/gr-qc/9602013}{{\ttfamily
  arXiv:gr-qc/9602013 [gr-qc]}}.
\newblock
\url{http://alice.cern.ch/format/showfull?sysnb=0217544}.
\newblock

\bibitem{Trautman:2006fp}
A.~Trautman, ``{Einstein-Cartan theory},''
\href{http://arxiv.org/abs/gr-qc/0606062}{{\ttfamily arXiv:gr-qc/0606062
  [gr-qc]}}.

\bibitem{DeAndrade:2000sf}
V.~C. De~Andrade, L.~C.~T. Guillen, and J.~G. Pereira, ``{Teleparallel gravity:
  An Overview},'' in {\em {Recent developments in theoretical and experimental
  general relativity, gravitation and relativistic field theories. Proceedings,
  9th Marcel Grossmann Meeting, MG'9, Rome, Italy, July 2-8, 2000. Pts. A-C}}.
\newblock 2000.
\newblock \href{http://arxiv.org/abs/gr-qc/0011087}{{\ttfamily
  arXiv:gr-qc/0011087 [gr-qc]}}.
\newblock
\url{http://alice.cern.ch/format/showfull?sysnb=2235348}.
\newblock

\bibitem{Obukhov:2002tm}
{\relax Yu}.~N. Obukhov and J.~G. Pereira, ``{Metric affine approach to
  teleparallel gravity},''
  \href{http://dx.doi.org/10.1103/PhysRevD.67.044016}{{\em Phys. Rev.}
  {\bfseries D67} (2003) 044016},
\href{http://arxiv.org/abs/gr-qc/0212080}{{\ttfamily arXiv:gr-qc/0212080
  [gr-qc]}}.

\bibitem{Schwinger:1970xc}
J.~Schwinger,
``{Particles, sources, and fields. Volume 1},''.

\bibitem{Barnes:1965}
K.~J. Barnes, ``{Lagrangian Theory for the Second-Rank Tensor Field},'' {\em J.
  Math. Phys.} {\bfseries 6} (1965) 788.

\bibitem{Rivers:1964}
R.~J. Rivers, ``{Lagrangian Theory for Neutral Massive Spin-2 Fields},'' {\em
  Il Nuovo Cimento} {\bfseries 34} (1964) 387.

\bibitem{VanNieuwenhuizen:1973fi}
P.~Van~Nieuwenhuizen, ``{On ghost-free tensor lagrangians and linearized
  gravitation},''
\href{http://dx.doi.org/10.1016/0550-3213(73)90194-6}{{\em Nucl. Phys.}
  {\bfseries B60} (1973) 478--492}.

\bibitem{Sakharov:1967dj}
A.~D. Sakharov, ``{Violation of CP Invariance, c Asymmetry, and Baryon
  Asymmetry of the Universe},''
  \href{http://dx.doi.org/10.1070/PU1991v034n05ABEH002497}{{\em Pisma Zh. Eksp.
  Teor. Fiz.} {\bfseries 5} (1967) 32--35}.
[Usp. Fiz. Nauk161,61(1991)].

\bibitem{Kuhfuss:1986rb}
R.~Kuhfuss and J.~Nitsch, ``{Propagating Modes in Gauge Field Theories of
  Gravity},''
\href{http://dx.doi.org/10.1007/BF00763447}{{\em Gen. Rel. Grav.} {\bfseries
  18} (1986) 1207}.

\bibitem{Baekler:2010fr}
P.~Baekler, F.~W. Hehl, and J.~M. Nester, ``{Poincare gauge theory of gravity:
  Friedman cosmology with even and odd parity modes. Analytic part},''
  \href{http://dx.doi.org/10.1103/PhysRevD.83.024001}{{\em Phys.Rev.}
  {\bfseries D83} (2011) 024001},
\href{http://arxiv.org/abs/1009.5112}{{\ttfamily arXiv:1009.5112 [gr-qc]}}.

\bibitem{Yo:2006qs}
H.-J. Yo and J.~M. Nester, ``{Dynamic Scalar Torsion and an Oscillating
  Universe},'' \href{http://dx.doi.org/10.1142/S0217732307025303}{{\em Mod.
  Phys. Lett.} {\bfseries A22} (2007) 2057--2069},
\href{http://arxiv.org/abs/astro-ph/0612738}{{\ttfamily arXiv:astro-ph/0612738
  [astro-ph]}}.

\bibitem{Shie:2008ms}
K.-F. Shie, J.~M. Nester, and H.-J. Yo, ``{Torsion Cosmology and the
  Accelerating Universe},''
  \href{http://dx.doi.org/10.1103/PhysRevD.78.023522}{{\em Phys. Rev.}
  {\bfseries D78} (2008) 023522},
\href{http://arxiv.org/abs/0805.3834}{{\ttfamily arXiv:0805.3834 [gr-qc]}}.

\bibitem{Yo:1999ex}
H.-j. Yo and J.~M. Nester, ``{Hamiltonian analysis of Poincare gauge theory
  scalar modes},'' \href{http://dx.doi.org/10.1142/S021827189900033X}{{\em Int.
  J. Mod. Phys.} {\bfseries D8} (1999) 459--479},
\href{http://arxiv.org/abs/gr-qc/9902032}{{\ttfamily arXiv:gr-qc/9902032
  [gr-qc]}}.

\bibitem{Yo:2001sy}
H.-J. Yo and J.~M. Nester, ``{Hamiltonian analysis of Poincare gauge theory:
  Higher spin modes},'' \href{http://dx.doi.org/10.1142/S0218271802001998}{{\em
  Int. J. Mod. Phys.} {\bfseries D11} (2002) 747--780},
\href{http://arxiv.org/abs/gr-qc/0112030}{{\ttfamily arXiv:gr-qc/0112030
  [gr-qc]}}.

\bibitem{HelayelNeto:2010jn}
J.~A. Helayel-Neto, C.~A. Hernaski, B.~Pereira-Dias, A.~A. Vargas-Paredes, and
  V.~J. Vasquez-Otoya, ``{Chern-Simons Gravity with (Curvature)$^{2}$- and
  (Torsion)$^{2}$-Terms and A Basis of Degree-of-Freedom Projection
  Operators},'' \href{http://dx.doi.org/10.1103/PhysRevD.82.064014}{{\em Phys.
  Rev.} {\bfseries D82} (2010) 064014},
\href{http://arxiv.org/abs/1005.3831}{{\ttfamily arXiv:1005.3831 [hep-th]}}.

\bibitem{Obukhov:1987tz}
Y.~Obukhov, V.~Ponomarev, and V.~Zhytnikov, ``{Quadratic Poincare Gauge Theory
  of Gravity: A Comparison With the General Relativity Theory},''
\href{http://dx.doi.org/10.1007/BF00763457}{{\em Gen.Rel.Grav.} {\bfseries 21}
  (1989) 1107--1142}.

\bibitem{Diakonov:2011fs}
D.~Diakonov, A.~G. Tumanov, and A.~A. Vladimirov, ``{Low-energy General
  Relativity with torsion: A Systematic derivative expansion},''
  \href{http://dx.doi.org/10.1103/PhysRevD.84.124042}{{\em Phys. Rev.}
  {\bfseries D84} (2011) 124042},
\href{http://arxiv.org/abs/1104.2432}{{\ttfamily arXiv:1104.2432 [hep-th]}}.

\bibitem{Baekler:2011jt}
P.~Baekler and F.~W. Hehl, ``{Beyond Einstein-Cartan gravity: Quadratic torsion
  and curvature invariants with even and odd parity including all boundary
  terms},'' \href{http://dx.doi.org/10.1088/0264-9381/28/21/215017}{{\em
  Class.Quant.Grav.} {\bfseries 28} (2011) 215017},
\href{http://arxiv.org/abs/1105.3504}{{\ttfamily arXiv:1105.3504 [gr-qc]}}.

\bibitem{Stelle:1977ry}
K.~S. Stelle, ``{Classical Gravity with Higher Derivatives},''
\href{http://dx.doi.org/10.1007/BF00760427}{{\em Gen. Rel. Grav.} {\bfseries 9}
  (1978) 353--371}.

\bibitem{Starobinsky:1980te}
A.~A. Starobinsky, ``{A New Type of Isotropic Cosmological Models Without
  Singularity},''
\href{http://dx.doi.org/10.1016/0370-2693(80)90670-X}{{\em Phys. Lett.}
  {\bfseries B91} (1980) 99--102}.

\bibitem{Berends:1979rv}
F.~A. Berends, J.~W. van Holten, P.~van Nieuwenhuizen, and B.~de~Wit, ``{On
  Field Theory for Massive and Massless Spin 5/2 Particles},''
\href{http://dx.doi.org/10.1016/0550-3213(79)90514-5}{{\em Nucl. Phys.}
  {\bfseries B154} (1979) 261--282}.

\bibitem{Guth:1980zm}
A.~H. Guth, ``{The Inflationary Universe: A Possible Solution to the Horizon
  and Flatness Problems},''
\href{http://dx.doi.org/10.1103/PhysRevD.23.347}{{\em Phys. Rev.} {\bfseries
  D23} (1981) 347--356}.

\bibitem{Linde:1981mu}
A.~D. Linde, ``{A New Inflationary Universe Scenario: A Possible Solution of
  the Horizon, Flatness, Homogeneity, Isotropy and Primordial Monopole
  Problems},''
\href{http://dx.doi.org/10.1016/0370-2693(82)91219-9}{{\em Phys. Lett.}
  {\bfseries B108} (1982) 389--393}.

\bibitem{Albrecht:1982wi}
A.~Albrecht and P.~J. Steinhardt, ``{Cosmology for Grand Unified Theories with
  Radiatively Induced Symmetry Breaking},''
\href{http://dx.doi.org/10.1103/PhysRevLett.48.1220}{{\em Phys. Rev. Lett.}
  {\bfseries 48} (1982) 1220--1223}.

\bibitem{Linde:1983gd}
A.~D. Linde, ``{Chaotic Inflation},''
\href{http://dx.doi.org/10.1016/0370-2693(83)90837-7}{{\em Phys. Lett.}
  {\bfseries B129} (1983) 177--181}.

\bibitem{Bezrukov:2007ep}
F.~L. Bezrukov and M.~Shaposhnikov, ``{The Standard Model Higgs boson as the
  inflaton},'' \href{http://dx.doi.org/10.1016/j.physletb.2007.11.072}{{\em
  Phys. Lett.} {\bfseries B659} (2008) 703--706},
\href{http://arxiv.org/abs/0710.3755}{{\ttfamily arXiv:0710.3755 [hep-th]}}.

\bibitem{Bezrukov:2008ut}
F.~Bezrukov, D.~Gorbunov, and M.~Shaposhnikov, ``{On initial conditions for the
  Hot Big Bang},'' \href{http://dx.doi.org/10.1088/1475-7516/2009/06/029}{{\em
  JCAP} {\bfseries 0906} (2009) 029},
\href{http://arxiv.org/abs/0812.3622}{{\ttfamily arXiv:0812.3622 [hep-ph]}}.

\bibitem{GarciaBellido:2008ab}
J.~Garcia-Bellido, D.~G. Figueroa, and J.~Rubio, ``{Preheating in the Standard
  Model with the Higgs-Inflaton coupled to gravity},''
  \href{http://dx.doi.org/10.1103/PhysRevD.79.063531}{{\em Phys. Rev.}
  {\bfseries D79} (2009) 063531},
\href{http://arxiv.org/abs/0812.4624}{{\ttfamily arXiv:0812.4624 [hep-ph]}}.

\bibitem{Bezrukov:2008ej}
F.~L. Bezrukov, A.~Magnin, and M.~Shaposhnikov, ``{Standard Model Higgs boson
  mass from inflation},''
  \href{http://dx.doi.org/10.1016/j.physletb.2009.03.035}{{\em Phys. Lett.}
  {\bfseries B675} (2009) 88--92},
\href{http://arxiv.org/abs/0812.4950}{{\ttfamily arXiv:0812.4950 [hep-ph]}}.

\bibitem{Barvinsky:2009fy}
A.~O. Barvinsky, A.~{\relax Yu}. Kamenshchik, C.~Kiefer, A.~A. Starobinsky, and
  C.~Steinwachs, ``{Asymptotic freedom in inflationary cosmology with a
  non-minimally coupled Higgs field},''
  \href{http://dx.doi.org/10.1088/1475-7516/2009/12/003}{{\em JCAP} {\bfseries
  0912} (2009) 003},
\href{http://arxiv.org/abs/0904.1698}{{\ttfamily arXiv:0904.1698 [hep-ph]}}.

\bibitem{Barvinsky:2008ia}
A.~O. Barvinsky, A.~{\relax Yu}. Kamenshchik, and A.~A. Starobinsky,
  ``{Inflation scenario via the Standard Model Higgs boson and LHC},''
  \href{http://dx.doi.org/10.1088/1475-7516/2008/11/021}{{\em JCAP} {\bfseries
  0811} (2008) 021},
\href{http://arxiv.org/abs/0809.2104}{{\ttfamily arXiv:0809.2104 [hep-ph]}}.

\bibitem{Bezrukov:2009db}
F.~Bezrukov and M.~Shaposhnikov, ``{Standard Model Higgs boson mass from
  inflation: Two loop analysis},''
  \href{http://dx.doi.org/10.1088/1126-6708/2009/07/089}{{\em JHEP} {\bfseries
  07} (2009) 089},
\href{http://arxiv.org/abs/0904.1537}{{\ttfamily arXiv:0904.1537 [hep-ph]}}.

\bibitem{Clark:2009dc}
T.~E. Clark, B.~Liu, S.~T. Love, and T.~ter Veldhuis, ``{The Standard Model
  Higgs Boson-Inflaton and Dark Matter},''
  \href{http://dx.doi.org/10.1103/PhysRevD.80.075019}{{\em Phys. Rev.}
  {\bfseries D80} (2009) 075019},
\href{http://arxiv.org/abs/0906.5595}{{\ttfamily arXiv:0906.5595 [hep-ph]}}.

\bibitem{Barvinsky:2009ii}
A.~O. Barvinsky, A.~{\relax Yu}. Kamenshchik, C.~Kiefer, A.~A. Starobinsky, and
  C.~F. Steinwachs, ``{Higgs boson, renormalization group, and naturalness in
  cosmology},'' \href{http://dx.doi.org/10.1140/epjc/s10052-012-2219-3}{{\em
  Eur. Phys. J.} {\bfseries C72} (2012) 2219},
\href{http://arxiv.org/abs/0910.1041}{{\ttfamily arXiv:0910.1041 [hep-ph]}}.

\bibitem{Barvinsky:2009jd}
A.~O. Barvinsky, A.~{\relax Yu}. Kamenshchik, C.~Kiefer, and C.~F. Steinwachs,
  ``{Tunneling cosmological state revisited: Origin of inflation with a
  non-minimally coupled Standard Model Higgs inflaton},''
  \href{http://dx.doi.org/10.1103/PhysRevD.81.043530}{{\em Phys. Rev.}
  {\bfseries D81} (2010) 043530},
\href{http://arxiv.org/abs/0911.1408}{{\ttfamily arXiv:0911.1408 [hep-th]}}.

\bibitem{Lerner:2010mq}
R.~N. Lerner and J.~McDonald, ``{A Unitarity-Conserving Higgs Inflation
  Model},'' \href{http://dx.doi.org/10.1103/PhysRevD.82.103525}{{\em Phys.
  Rev.} {\bfseries D82} (2010) 103525},
\href{http://arxiv.org/abs/1005.2978}{{\ttfamily arXiv:1005.2978 [hep-ph]}}.

\bibitem{Lerner:2009na}
R.~N. Lerner and J.~McDonald, ``{Higgs Inflation and Naturalness},''
  \href{http://dx.doi.org/10.1088/1475-7516/2010/04/015}{{\em JCAP} {\bfseries
  1004} (2010) 015},
\href{http://arxiv.org/abs/0912.5463}{{\ttfamily arXiv:0912.5463 [hep-ph]}}.

\bibitem{Giudice:2010ka}
G.~F. Giudice and H.~M. Lee, ``{Unitarizing Higgs Inflation},''
  \href{http://dx.doi.org/10.1016/j.physletb.2010.10.035}{{\em Phys. Lett.}
  {\bfseries B694} (2011) 294--300},
\href{http://arxiv.org/abs/1010.1417}{{\ttfamily arXiv:1010.1417 [hep-ph]}}.

\bibitem{Burgess:2009ea}
C.~P. Burgess, H.~M. Lee, and M.~Trott, ``{Power-counting and the Validity of
  the Classical Approximation During Inflation},''
  \href{http://dx.doi.org/10.1088/1126-6708/2009/09/103}{{\em JHEP} {\bfseries
  09} (2009) 103},
\href{http://arxiv.org/abs/0902.4465}{{\ttfamily arXiv:0902.4465 [hep-ph]}}.

\bibitem{Barbon:2009ya}
J.~L.~F. Barbon and J.~R. Espinosa, ``{On the Naturalness of Higgs
  Inflation},'' \href{http://dx.doi.org/10.1103/PhysRevD.79.081302}{{\em Phys.
  Rev.} {\bfseries D79} (2009) 081302},
\href{http://arxiv.org/abs/0903.0355}{{\ttfamily arXiv:0903.0355 [hep-ph]}}.

\bibitem{Burgess:2010zq}
C.~P. Burgess, H.~M. Lee, and M.~Trott, ``{Comment on Higgs Inflation and
  Naturalness},'' \href{http://dx.doi.org/10.1007/JHEP07(2010)007}{{\em JHEP}
  {\bfseries 07} (2010) 007},
\href{http://arxiv.org/abs/1002.2730}{{\ttfamily arXiv:1002.2730 [hep-ph]}}.

\bibitem{Hertzberg:2010dc}
M.~P. Hertzberg, ``{On Inflation with Non-minimal Coupling},''
  \href{http://dx.doi.org/10.1007/JHEP11(2010)023}{{\em JHEP} {\bfseries 11}
  (2010) 023},
\href{http://arxiv.org/abs/1002.2995}{{\ttfamily arXiv:1002.2995 [hep-ph]}}.

\bibitem{Buck:2010sv}
M.~Buck, M.~Fairbairn, and M.~Sakellariadou, ``{Inflation in models with
  Conformally Coupled Scalar fields: An application to the Noncommutative
  Spectral Action},'' \href{http://dx.doi.org/10.1103/PhysRevD.82.043509}{{\em
  Phys. Rev.} {\bfseries D82} (2010) 043509},
\href{http://arxiv.org/abs/1005.1188}{{\ttfamily arXiv:1005.1188 [hep-th]}}.

\bibitem{Lerner:2011it}
R.~N. Lerner and J.~McDonald, ``{Unitarity-Violation in Generalized Higgs
  Inflation Models},''
  \href{http://dx.doi.org/10.1088/1475-7516/2012/11/019}{{\em JCAP} {\bfseries
  1211} (2012) 019},
\href{http://arxiv.org/abs/1112.0954}{{\ttfamily arXiv:1112.0954 [hep-ph]}}.

\bibitem{Greenwood:2012aj}
R.~N. Greenwood, D.~I. Kaiser, and E.~I. Sfakianakis, ``{Multifield Dynamics of
  Higgs Inflation},'' \href{http://dx.doi.org/10.1103/PhysRevD.87.064021}{{\em
  Phys. Rev.} {\bfseries D87} (2013) 064021},
\href{http://arxiv.org/abs/1210.8190}{{\ttfamily arXiv:1210.8190 [hep-ph]}}.

\bibitem{Spokoiny:1984bd}
B.~L. Spokoiny, ``{Inflation and Generation of Perturbations in Broken
  Symmetric Theory of Gravity},''
\href{http://dx.doi.org/10.1016/0370-2693(84)90587-2}{{\em Phys. Lett.}
  {\bfseries B147} (1984) 39--43}.

\bibitem{Salopek:1988qh}
D.~S. Salopek, J.~R. Bond, and J.~M. Bardeen, ``{Designing Density Fluctuation
  Spectra in Inflation},''
\href{http://dx.doi.org/10.1103/PhysRevD.40.1753}{{\em Phys. Rev.} {\bfseries
  D40} (1989) 1753}.

\bibitem{Fakir:1990eg}
R.~Fakir and W.~G. Unruh, ``{Improvement on cosmological chaotic inflation
  through nonminimal coupling},''
\href{http://dx.doi.org/10.1103/PhysRevD.41.1783}{{\em Phys. Rev.} {\bfseries
  D41} (1990) 1783--1791}.

\bibitem{Shaposhnikov:2008xb}
M.~Shaposhnikov and D.~Zenhausern, ``{Scale invariance, unimodular gravity and
  dark energy},'' \href{http://dx.doi.org/10.1016/j.physletb.2008.11.054}{{\em
  Phys. Lett.} {\bfseries B671} (2009) 187--192},
\href{http://arxiv.org/abs/0809.3395}{{\ttfamily arXiv:0809.3395 [hep-th]}}.

\bibitem{GarciaBellido:2011de}
J.~Garcia-Bellido, J.~Rubio, M.~Shaposhnikov, and D.~Zenhausern,
  ``{Higgs-Dilaton Cosmology: From the Early to the Late Universe},''
  \href{http://dx.doi.org/10.1103/PhysRevD.84.123504}{{\em Phys. Rev.}
  {\bfseries D84} (2011) 123504},
\href{http://arxiv.org/abs/1107.2163}{{\ttfamily arXiv:1107.2163 [hep-ph]}}.

\bibitem{Kapner:2006si}
D.~J. Kapner, T.~S. Cook, E.~G. Adelberger, J.~H. Gundlach, B.~R. Heckel, C.~D.
  Hoyle, and H.~E. Swanson, ``{Tests of the gravitational inverse-square law
  below the dark-energy length scale},''
  \href{http://dx.doi.org/10.1103/PhysRevLett.98.021101}{{\em Phys. Rev. Lett.}
  {\bfseries 98} (2007) 021101},
\href{http://arxiv.org/abs/hep-ph/0611184}{{\ttfamily arXiv:hep-ph/0611184
  [hep-ph]}}.

\bibitem{Aydemir:2012nz}
U.~Aydemir, M.~M. Anber, and J.~F. Donoghue, ``{Self-healing of unitarity in
  effective field theories and the onset of new physics},''
  \href{http://dx.doi.org/10.1103/PhysRevD.86.014025}{{\em Phys. Rev.}
  {\bfseries D86} (2012) 014025},
\href{http://arxiv.org/abs/1203.5153}{{\ttfamily arXiv:1203.5153 [hep-ph]}}.

\bibitem{vanderBij:1981ym}
J.~J. van~der Bij, H.~van Dam, and Y.~J. Ng, ``{The Exchange of Massless Spin
  Two Particles},''
\href{http://dx.doi.org/10.1016/0378-4371(82)90247-3}{{\em Physica} {\bfseries
  116A} (1982) 307--320}.

\bibitem{Unruh:1988in}
W.~G. Unruh, ``{A Unimodular Theory of Canonical Quantum Gravity},''
\href{http://dx.doi.org/10.1103/PhysRevD.40.1048}{{\em Phys. Rev.} {\bfseries
  D40} (1989) 1048}.

\bibitem{Henneaux:1989zc}
M.~Henneaux and C.~Teitelboim, ``{The Cosmological Constant and General
  Covariance},''
\href{http://dx.doi.org/10.1016/0370-2693(89)91251-3}{{\em Phys. Lett.}
  {\bfseries B222} (1989) 195--199}.

\bibitem{Fujii:1982ms}
Y.~Fujii, ``{Origin of the Gravitational Constant and Particle Masses in Scale
  Invariant Scalar - Tensor Theory},''
\href{http://dx.doi.org/10.1103/PhysRevD.26.2580}{{\em Phys. Rev.} {\bfseries
  D26} (1982) 2580}.

\bibitem{Wetterich:1987fk}
C.~Wetterich, ``{Cosmologies With Variable Newton's 'Constant'},''
\href{http://dx.doi.org/10.1016/0550-3213(88)90192-7}{{\em Nucl. Phys.}
  {\bfseries B302} (1988) 645--667}.

\bibitem{Wetterich:1987fm}
C.~Wetterich, ``{Cosmology and the Fate of Dilatation Symmetry},''
\href{http://dx.doi.org/10.1016/0550-3213(88)90193-9}{{\em Nucl. Phys.}
  {\bfseries B302} (1988) 668}.

\bibitem{Bezrukov:2012hx}
F.~Bezrukov, G.~K. Karananas, J.~Rubio, and M.~Shaposhnikov, ``{Higgs-Dilaton
  Cosmology: an effective field theory approach},''
  \href{http://dx.doi.org/10.1103/PhysRevD.87.096001}{{\em Phys. Rev.}
  {\bfseries D87} no.~9, (2013) 096001},
\href{http://arxiv.org/abs/1212.4148}{{\ttfamily arXiv:1212.4148 [hep-ph]}}.

\bibitem{Bezrukov:2010jz}
F.~Bezrukov, A.~Magnin, M.~Shaposhnikov, and S.~Sibiryakov, ``{Higgs inflation:
  consistency and generalisations},''
  \href{http://dx.doi.org/10.1007/JHEP01(2011)016}{{\em JHEP} {\bfseries 01}
  (2011) 016},
\href{http://arxiv.org/abs/1008.5157}{{\ttfamily arXiv:1008.5157 [hep-ph]}}.

\bibitem{CervantesCota:1995tz}
J.~L. Cervantes-Cota and H.~Dehnen, ``{Induced gravity inflation in the
  standard model of particle physics},''
  \href{http://dx.doi.org/10.1016/0550-3213(95)00128-X}{{\em Nucl. Phys.}
  {\bfseries B442} (1995) 391--412},
\href{http://arxiv.org/abs/astro-ph/9505069}{{\ttfamily arXiv:astro-ph/9505069
  [astro-ph]}}.

\bibitem{Ratra:1987rm}
B.~Ratra and P.~J.~E. Peebles, ``{Cosmological Consequences of a Rolling
  Homogeneous Scalar Field},''
\href{http://dx.doi.org/10.1103/PhysRevD.37.3406}{{\em Phys. Rev.} {\bfseries
  D37} (1988) 3406}.

\bibitem{Scherrer:2007pu}
R.~J. Scherrer and A.~A. Sen, ``{Thawing quintessence with a nearly flat
  potential},'' \href{http://dx.doi.org/10.1103/PhysRevD.77.083515}{{\em Phys.
  Rev.} {\bfseries D77} (2008) 083515},
\href{http://arxiv.org/abs/0712.3450}{{\ttfamily arXiv:0712.3450 [astro-ph]}}.

\bibitem{Cornwall:1974km}
J.~M. Cornwall, D.~N. Levin, and G.~Tiktopoulos, ``{Derivation of Gauge
  Invariance from High-Energy Unitarity Bounds on the s Matrix},''
  \href{http://dx.doi.org/10.1103/PhysRevD.10.1145,
  10.1103/PhysRevD.11.972}{{\em Phys. Rev.} {\bfseries D10} (1974) 1145}.
[Erratum: Phys. Rev.D11,972(1975)].

\bibitem{Lee:1977yc}
B.~W. Lee, C.~Quigg, and H.~B. Thacker, ``{The Strength of Weak Interactions at
  Very High-Energies and the Higgs Boson Mass},''
\href{http://dx.doi.org/10.1103/PhysRevLett.38.883}{{\em Phys. Rev. Lett.}
  {\bfseries 38} (1977) 883--885}.

\bibitem{Lee:1977eg}
B.~W. Lee, C.~Quigg, and H.~B. Thacker, ``{Weak Interactions at Very
  High-Energies: The Role of the Higgs Boson Mass},''
\href{http://dx.doi.org/10.1103/PhysRevD.16.1519}{{\em Phys. Rev.} {\bfseries
  D16} (1977) 1519}.

\bibitem{Chanowitz:1978uj}
M.~S. Chanowitz, M.~A. Furman, and I.~Hinchliffe, ``{Weak Interactions of
  Ultraheavy Fermions},''
\href{http://dx.doi.org/10.1016/0370-2693(78)90024-2}{{\em Phys. Lett.}
  {\bfseries B78} (1978) 285}.

\bibitem{Chanowitz:1978mv}
M.~S. Chanowitz, M.~A. Furman, and I.~Hinchliffe, ``{Weak Interactions of
  Ultraheavy Fermions. 2.},''
\href{http://dx.doi.org/10.1016/0550-3213(79)90606-0}{{\em Nucl. Phys.}
  {\bfseries B153} (1979) 402--430}.

\bibitem{Appelquist:1987cf}
T.~Appelquist and M.~S. Chanowitz, ``{Unitarity Bound on the Scale of Fermion
  Mass Generation},'' \href{http://dx.doi.org/10.1103/PhysRevLett.59.2405}{{\em
  Phys. Rev. Lett.} {\bfseries 59} (1987) 2405}.
[Erratum: Phys. Rev. Lett.60,1589(1988)].

\bibitem{GarciaBellido:2012zu}
J.~Garcia-Bellido, J.~Rubio, and M.~Shaposhnikov, ``{Higgs-Dilaton cosmology:
  Are there extra relativistic species?},''
  \href{http://dx.doi.org/10.1016/j.physletb.2012.10.075}{{\em Phys. Lett.}
  {\bfseries B718} (2012) 507--511},
\href{http://arxiv.org/abs/1209.2119}{{\ttfamily arXiv:1209.2119 [hep-ph]}}.

\bibitem{Bezrukov:2011sz}
F.~Bezrukov, D.~Gorbunov, and M.~Shaposhnikov, ``{Late and early time
  phenomenology of Higgs-dependent cutoff},''
  \href{http://dx.doi.org/10.1088/1475-7516/2011/10/001}{{\em JCAP} {\bfseries
  1110} (2011) 001},
\href{http://arxiv.org/abs/1106.5019}{{\ttfamily arXiv:1106.5019 [hep-ph]}}.

\bibitem{Shaposhnikov:2009nk}
M.~E. Shaposhnikov and F.~V. Tkachov, ``{Quantum scale-invariant models as
  effective field theories},''
\href{http://arxiv.org/abs/0905.4857}{{\ttfamily arXiv:0905.4857 [hep-th]}}.

\bibitem{Codello:2012sn}
A.~Codello, G.~D'Odorico, C.~Pagani, and R.~Percacci, ``{The Renormalization
  Group and Weyl-invariance},''
  \href{http://dx.doi.org/10.1088/0264-9381/30/11/115015}{{\em Class. Quant.
  Grav.} {\bfseries 30} (2013) 115015},
\href{http://arxiv.org/abs/1210.3284}{{\ttfamily arXiv:1210.3284 [hep-th]}}.

\bibitem{Flanagan:2004bz}
E.~E. Flanagan, ``{The Conformal frame freedom in theories of gravitation},''
  \href{http://dx.doi.org/10.1088/0264-9381/21/15/N02}{{\em Class. Quant.
  Grav.} {\bfseries 21} (2004) 3817},
\href{http://arxiv.org/abs/gr-qc/0403063}{{\ttfamily arXiv:gr-qc/0403063
  [gr-qc]}}.

\bibitem{Dutta:2007st}
S.~Dutta, K.~Hagiwara, Q.-S. Yan, and K.~Yoshida, ``{Constraints on the
  electroweak chiral Lagrangian from the precision data},''
  \href{http://dx.doi.org/10.1016/j.nuclphysb.2007.08.017}{{\em Nucl. Phys.}
  {\bfseries B790} (2008) 111--137},
\href{http://arxiv.org/abs/0705.2277}{{\ttfamily arXiv:0705.2277 [hep-ph]}}.

\bibitem{Feruglio:1992wf}
F.~Feruglio, ``{The Chiral approach to the electroweak interactions},''
  \href{http://dx.doi.org/10.1142/S0217751X93001946}{{\em Int. J. Mod. Phys.}
  {\bfseries A8} (1993) 4937--4972},
\href{http://arxiv.org/abs/hep-ph/9301281}{{\ttfamily arXiv:hep-ph/9301281
  [hep-ph]}}.

\bibitem{Bardeen:1995kv}
W.~A. Bardeen, ``{On naturalness in the standard model},'' in {\em {Ontake
  Summer Institute on Particle Physics Ontake Mountain, Japan, August
  27-September 2, 1995}}.
\newblock 1995.
\newblock
\url{http://lss.fnal.gov/cgi-bin/find_paper.pl?conf-95-391}.
\newblock

\bibitem{Jackiw:1974cv}
R.~Jackiw, ``{Functional evaluation of the effective potential},''
\href{http://dx.doi.org/10.1103/PhysRevD.9.1686}{{\em Phys. Rev.} {\bfseries
  D9} (1974) 1686}.

\bibitem{Bezrukov:2012sa}
F.~Bezrukov, M.~{\relax Yu}. Kalmykov, B.~A. Kniehl, and M.~Shaposhnikov,
  ``{Higgs Boson Mass and New Physics},''
  \href{http://dx.doi.org/10.1007/JHEP10(2012)140}{{\em JHEP} {\bfseries 10}
  (2012) 140},
\href{http://arxiv.org/abs/1205.2893}{{\ttfamily arXiv:1205.2893 [hep-ph]}}.

\bibitem{Degrassi:2012ry}
G.~Degrassi, S.~Di~Vita, J.~Elias-Miro, J.~R. Espinosa, G.~F. Giudice,
  G.~Isidori, and A.~Strumia, ``{Higgs mass and vacuum stability in the
  Standard Model at NNLO},''
  \href{http://dx.doi.org/10.1007/JHEP08(2012)098}{{\em JHEP} {\bfseries 08}
  (2012) 098},
\href{http://arxiv.org/abs/1205.6497}{{\ttfamily arXiv:1205.6497 [hep-ph]}}.

\bibitem{Chatrchyan:2012xdj}
{\bfseries CMS} Collaboration, S.~Chatrchyan {\em et~al.}, ``{Observation of a
  new boson at a mass of 125 GeV with the CMS experiment at the LHC},''
  \href{http://dx.doi.org/10.1016/j.physletb.2012.08.021}{{\em Phys. Lett.}
  {\bfseries B716} (2012) 30--61},
\href{http://arxiv.org/abs/1207.7235}{{\ttfamily arXiv:1207.7235 [hep-ex]}}.

\bibitem{Aad:2012tfa}
{\bfseries ATLAS} Collaboration, G.~Aad {\em et~al.}, ``{Observation of a new
  particle in the search for the Standard Model Higgs boson with the ATLAS
  detector at the LHC},''
  \href{http://dx.doi.org/10.1016/j.physletb.2012.08.020}{{\em Phys. Lett.}
  {\bfseries B716} (2012) 1--29},
\href{http://arxiv.org/abs/1207.7214}{{\ttfamily arXiv:1207.7214 [hep-ex]}}.

\bibitem{Komatsu:2010fb}
{\bfseries WMAP} Collaboration, E.~Komatsu {\em et~al.}, ``{Seven-Year
  Wilkinson Microwave Anisotropy Probe (WMAP) Observations: Cosmological
  Interpretation},'' \href{http://dx.doi.org/10.1088/0067-0049/192/2/18}{{\em
  Astrophys. J. Suppl.} {\bfseries 192} (2011) 18},
\href{http://arxiv.org/abs/1001.4538}{{\ttfamily arXiv:1001.4538
  [astro-ph.CO]}}.

\bibitem{Karananas:2016grc}
G.~K. Karananas and M.~Shaposhnikov, ``{Scale invariant alternatives to general
  relativity. II. Dilaton properties},''
  \href{http://dx.doi.org/10.1103/PhysRevD.93.084052}{{\em Phys. Rev.}
  {\bfseries D93} no.~8, (2016) 084052},
\href{http://arxiv.org/abs/1603.01274}{{\ttfamily arXiv:1603.01274 [hep-th]}}.

\bibitem{Horndeski:1974wa}
G.~W. Horndeski, ``{Second-order scalar-tensor field equations in a
  four-dimensional space},''
\href{http://dx.doi.org/10.1007/BF01807638}{{\em Int. J. Theor. Phys.}
  {\bfseries 10} (1974) 363--384}.

\bibitem{Sotiriou:2008rp}
T.~P. Sotiriou and V.~Faraoni, ``{f(R) Theories Of Gravity},''
  \href{http://dx.doi.org/10.1103/RevModPhys.82.451}{{\em Rev. Mod. Phys.}
  {\bfseries 82} (2010) 451--497},
\href{http://arxiv.org/abs/0805.1726}{{\ttfamily arXiv:0805.1726 [gr-qc]}}.

\bibitem{Padilla:2013jza}
A.~Padilla, D.~Stefanyszyn, and M.~Tsoukalas, ``{Generalised Scale Invariant
  Theories},'' \href{http://dx.doi.org/10.1103/PhysRevD.89.065009}{{\em Phys.
  Rev.} {\bfseries D89} no.~6, (2014) 065009},
\href{http://arxiv.org/abs/1312.0975}{{\ttfamily arXiv:1312.0975 [hep-th]}}.

\bibitem{Schwinger:1951xk}
J.~S. Schwinger, ``{The Theory of quantized fields. 1.},''
\href{http://dx.doi.org/10.1103/PhysRev.82.914}{{\em Phys. Rev.} {\bfseries 82}
  (1951) 914--927}.

\bibitem{Schwinger:1953tb}
J.~S. Schwinger, ``{The Theory of quantized fields. 2.},''
\href{http://dx.doi.org/10.1103/PhysRev.91.713}{{\em Phys. Rev.} {\bfseries 91}
  (1953) 713--728}.

\bibitem{Zee:1978wi}
A.~Zee, ``{A Broken Symmetric Theory of Gravity},''
\href{http://dx.doi.org/10.1103/PhysRevLett.42.417}{{\em Phys. Rev. Lett.}
  {\bfseries 42} (1979) 417}.

\bibitem{Smolin:1979uz}
L.~Smolin, ``{Towards a Theory of Space-Time Structure at Very Short
  Distances},''
\href{http://dx.doi.org/10.1016/0550-3213(79)90059-2}{{\em Nucl. Phys.}
  {\bfseries B160} (1979) 253}.

\bibitem{Rubio:2014wta}
J.~Rubio and M.~Shaposhnikov, ``{Higgs-Dilaton cosmology: Universality versus
  criticality},'' \href{http://dx.doi.org/10.1103/PhysRevD.90.027307}{{\em
  Phys. Rev.} {\bfseries D90} (2014) 027307},
\href{http://arxiv.org/abs/1406.5182}{{\ttfamily arXiv:1406.5182 [hep-ph]}}.

\bibitem{Heinzl:2000ht}
T.~Heinzl, ``{Light cone quantization: Foundations and applications},''
  \href{http://dx.doi.org/10.1007/3-540-45114-5_2}{{\em Lect. Notes Phys.}
  {\bfseries 572} (2001) 55--142},
\href{http://arxiv.org/abs/hep-th/0008096}{{\ttfamily arXiv:hep-th/0008096
  [hep-th]}}.

\bibitem{landau_fields}
L.~Landau and E.~Lifshitz, {\em The Classical Theory of Fields}, vol.~2.
\newblock Pergamon, Amsterdam, fourth~ed., 1975.

\bibitem{Blas:2009qj}
D.~Blas, O.~Pujolas, and S.~Sibiryakov, ``{Consistent Extension of Horava
  Gravity},'' \href{http://dx.doi.org/10.1103/PhysRevLett.104.181302}{{\em
  Phys. Rev. Lett.} {\bfseries 104} (2010) 181302},
\href{http://arxiv.org/abs/0909.3525}{{\ttfamily arXiv:0909.3525 [hep-th]}}.

\bibitem{Barvinsky:2015kil}
A.~O. Barvinsky, D.~Blas, M.~Herrero-Valea, S.~M. Sibiryakov, and C.~F.
  Steinwachs, ``{Renormalization of Hořava gravity},''
  \href{http://dx.doi.org/10.1103/PhysRevD.93.064022}{{\em Phys. Rev.}
  {\bfseries D93} no.~6, (2016) 064022},
\href{http://arxiv.org/abs/1512.02250}{{\ttfamily arXiv:1512.02250 [hep-th]}}.

\bibitem{Mathtensor:1994}
L.~Parker and S.~M. Christensen, {\em {MathTensor: A system for doing tensor
  analysis by computer}}.
\newblock Addison-Wesley, Redwood City, CA, 1994.

\end{thebibliography}\endgroup

\end{document}